\documentclass[11pt]{thesis91e}
\usepackage{textcomp}
\usepackage{float}
\usepackage{amsmath}
\usepackage{amsfonts}
\usepackage[numbers]{natbib}
\usepackage{tensor}
\usepackage{calc}
\usepackage{mathrsfs}
\usepackage{amssymb}
\usepackage{amsmath}
\usepackage{calligra}
\usepackage{tensor}
\usepackage{color}
\usepackage{bm}

\usepackage[draft=false,ps2pdf,colorlinks, anchorcolor=black,linkcolor=black,citecolor=black]{hyperref}
\usepackage[final,dvips]{graphicx}

\title{Motion of small bodies in general relativity: foundations and implementations of the self-force}
\author{Adam Pound}
\advisor{Professor Eric Poisson}
\degree{Doctor of Philosophy}

\newcommand{\orbit}{z^{\alpha}}
\newcommand{\geo}{z_G^{\alpha}}
\newcommand{\diff}[2]{\frac{d #1}{d #2}}
\newcommand{\pdiff}[2]{\frac{\partial #1}{\partial #2}}
\newcommand{\ddiff}[2]{\frac{d^2 #1}{\ d #2^2}}
\newcommand{\pddiff}[2]{\frac{\partial^2 #1}{\ \partial #2^2}}
\newcommand{\Chr}[3]{\Gamma^{#1}_{\ #2#3}}

\newcommand{\e}{\varepsilon}
\newcommand{\A}[2]{\mathscr{A}_{#1}\coeff{#2}}
\newcommand{\B}[2]{\mathscr{B}_{#1}\coeff{#2}}
\newcommand{\C}[2]{\mathscr{C}_{#1}\coeff{#2}}
\newcommand{\D}[2]{\mathscr{D}_{#1}\coeff{#2}}
\newcommand{\E}[2]{\mathscr{E}_{#1}\coeff{#2}}
\newcommand{\F}[2]{\mathscr{F}_{#1}\coeff{#2}}
\newcommand{\G}[2]{\mathscr{G}_{#1}\coeff{#2}}
\renewcommand{\H}[2]{\mathscr{H}_{#1}\coeff{#2}}
\newcommand{\I}[2]{\mathscr{I}_{#1}\coeff{#2}}
\newcommand{\K}[2]{\mathscr{K}_{#1}\coeff{#2}}
\newcommand{\nhat}{\hat{n}}
\DeclareMathAlphabet{\mathcalligra}{T1}{calligra}{m}{n}
\DeclareFontShape{T1}{calligra}{m}{n}{<->s*[2.75]callig15}{}
\renewcommand{\r}{\ensuremath{\mathcalligra{r}}\hspace{0.45 mm}}
\newcommand{\size}{\ell}
\newcommand{\Size}{\mathscr{L}}
\newcommand{\rad}{\mathscr{R}}
\newcommand{\retr}{r_{\text{ret}}}
\newcommand{\advr}{r_{\rm adv}}
\newcommand{\h}{\mathfrak{h}}
\renewcommand{\sb}{\bar\sigma}%\sigma(x,\bar x)
\newcommand{\spb}{\bar\sigma'}%\sigma(x',\bar x)
\newcommand{\spp}{\sigma''}%\sigma(x,x'')
\newcommand{\av}[1]{\left\langle #1 \right\rangle}
\newcommand{\tail}{h^{{}^{{\!\scriptstyle\text{tail}}}}}
\newcommand{\order}[1]{O\!\left(#1\right)}
\newcommand{\etide}{\mathcal{E}}
\newcommand{\ein}{\widetilde{\mathcal{E}}}
\newcommand{\btide}{\mathcal{B}}
\newcommand{\bin}{\widetilde{\mathcal{B}}}
\newcommand{\U}{\mathfrak{U}}
\newcommand{\past}{I^{\;{}^\text{-}}}
\newcommand{\hmn}[2]{h^{^{\!\text{(#2)}}}_{#1}}
\newcommand{\hbarmn}[2]{\bar h^{^{\!\text{(#2)}}}_{#1}}
\newcommand{\del}[1]{\nabla_{\!\!#1}}
\newcommand{\Lie}[1]{\pounds_{\!#1}}

\DeclareMathOperator{\STF}{STF}
\newcommand{\trans}{\psi}
\newcommand{\map}{\varphi}
\newcommand{\man}{\mathcal{M}}
\newcommand{\exact}[1]{\mathsf{#1}}
\newcommand{\expand}{\Phi}

\newcommand{\gauge}[3]{L\coeff{#2}_{#1}\!\left[#3\right]}
\newcommand{\ddR}[3]{\delta^2 R\coeff{#2}_{#1}\!\left[#3\right]}
\newcommand{\coeff}[1]{^{\scriptscriptstyle{\text{(#1)}}}}
\newcommand{\dcoeff}[1]{_{\scriptscriptstyle{\text{(#1)}}}}
\newcommand{\an}[1]{a^{\scriptscriptstyle{(#1)}}}
\newcommand{\zn}[1]{z_{\scriptscriptstyle{(#1)}}}
\newcommand{\un}[1]{\dot{z}_{\scriptscriptstyle{(#1)}}}
\renewcommand{\t}{\tilde t}
\newcommand{\R}{\tilde R}

\begin{document} 
      \prepages
      \maketitle
      \begin{abstract}
Extreme mass-ratio inspirals, in which solar-mass compact bodies spiral into supermassive black holes, are an important potential source for gravitational wave detectors. Because of the extreme mass-ratio, one can treat this problem using perturbation theory. However, in order to relate the motion of the small body to the emitted waveform, one requires a model that is accurate on extremely long timescales. Additionally, in order to avoid intractable divergences, one requires a model that treats the small body as asymptotically small rather than exactly pointlike. Both of these difficulties can be resolved by using techniques of singular perturbation theory. I begin this dissertation with an analysis of singular perturbation theory on manifolds, including the common techniques of matched asymptotic expansions and two-timescale expansions. I then formulate a systematic asymptotic expansion in which the metric perturbation due to the body is expanded while a representative worldline is held fixed, and I contrast it with a regular expansion in which both the metric and the worldline must be expanded. This results in an approximation that is potentially uniformly accurate on long timescales. I illustrate the utility of the expansion for an exact point particle; I then generalize it to an asymptotically small body. The equation of motion for the body's fixed worldline is determined by performing a local-in-space expansion in the neighbourhood of the body. Using this local expansion as boundary data, I then construct a global solution to the perturbative Einstein equations. As a means of concretely characterizing particular orbits, I next devise a relativistic generalization of the Newtonian method of osculating orbits. However, the equations of motion for the body and its metric perturbation are purely formal; a concrete calculation of a particular orbit and waveform brings further difficulties. In order to sidestep these difficulties, earlier authors have suggested making use of adiabatic approximations that capture the dissipative dynamics of the system while ignoring the conservative dynamics. I test the utility of some such approximations in two test cases, making use of the method of osculating orbits and two-timescale expansions. 
\end{abstract}

      \begin{acknowledgements}
I wish to thank Eric Poisson for the many helpful discussions we had over the course of my graduate studies, as well as for supporting me in the final difficult months of my degree. I also wish to thank Roland Haas and Aron Pasieka for bearing my incessant complaints about worldtubes and matched asymptotic expansions in diffeomorphism-invariant theories. Lastly, I wish to thank the members of my advisory and examining committees: Achim Kempf, Luis Lehner, Bernie Nickel, and Bob Wald.
\end{acknowledgements}

\begin{history}
Much of the material in this dissertation was previously presented in Refs.~\cite{perturbation_techniques, my_paper, osculating_paper, other_paper}. Specifically, the material of Chapters~\ref{approximations} and \ref{matching} was presented in slightly different form in Ref.~\cite{perturbation_techniques}; that of Chapters~\ref{point_particle}, \ref{extended_body}, \ref{buffer_region}, and \ref{perturbation calculation}, in Ref.~\cite{my_paper}; and that of Chapters~\ref{osculating} and \ref{adiabatic}, in Ref.~\cite{osculating_paper}. Parts of Chapter \ref{adiabatic} appeared in a significantly different form in Ref.~\cite{other_paper}.
\end{history}

\begin{notation}
I work in geometrical units in which $G=c=1$, use the sign conventions of Ref.~\cite{MTW}, and frequently omit indices for simplicity.
\begin{table}[h] 
\begin{tabular*}{\textwidth}{cl}
\hline\hline
$\alpha,\beta,\gamma,...$ & coordinate indices running from 0 to 3 \\
$I,J,K,...$ & orthonormal indices running from 0 to 3 \\
$a,b,c,...$ & either coordinate or orthonormal spatial indices running from 1 to 3 \\
$A,B,C,...$ & angular-coordinate indices running from 1 to 2 \\
$(t,x^i)$ and $(t,r,\theta^A)$ & in Chs.~\ref{approximations}--\ref{perturbation calculation}, coordinates centered on worldline of small body\\
& in Chs.~\ref{osculating} and \ref{adiabatic}, coordinates centered on large body\\
$t_\alpha,x^a_\alpha$ & the one-forms $t_\alpha\equiv \partial_\alpha t$ and $x^a_\alpha\equiv\partial_\alpha x^a$\\
$n_i$ & the unit vector $n_i=\partial_i r$\\
$n_L$ & the product $n_L\equiv n_{i_1...i_\ell}\equiv n_{i_1}...n_{i_\ell}$\\
$\nhat_L$ & the symmetric trace-free part of $n_L$, $\nhat_L\equiv n_{\langle L\rangle}$\\
$\exact{g},\exact{R},\exact{T},...$ & sans-serif symbols denote exact quantities to be expanded\\
${}^{\exact{g}\!}\del{\nu}$ & the covariant derivative compatible with $\exact{g}$\\
$\nabla$ or ; & the covariant derivative compatible with a background metric\\
$\mathcal{R}$ & the typical lengthscale of an external spacetime \\
$\e$ & a small quantity, typically $m/\mathcal{R}$ \\
$E_{\mu\nu}$ & the relativistic wave operator $E_{\mu\nu}[h] = \left(g^\rho_\mu g^\sigma_\nu\nabla^\gamma\del{\gamma} +2R\indices{_\mu^\rho_\nu^\sigma}\right)\!h_{\rho\sigma}$\\
$G_{\mu\nu\mu'\nu'}(x,x')$ & the Green's function for $E_{\mu\nu}$\\
$L_\mu$ & the Lorenz-gauge operator $L_\mu[h] =\left(g^\rho_\mu g^{\sigma\gamma}-\tfrac{1}{2}g^\gamma_\mu g^{\rho\sigma}\right)\!\del{\gamma}h_{\rho\sigma}$\\
$\gamma$ & the worldline of the small body \\
$a^\mu$ & the acceleration of $\gamma$ with respect to $g$\\
$\rad$ & the radius of a worldtube $\Gamma$ around the small body \\
$\etide_{ab}$ & the gravitoelectric tidal quadrupole field of $g$ on $\gamma$ \\
$\btide_{ab}$ & the gravitomagnetic tidal quadrupole field of $g$ on $\gamma$\\
$g_{E\alpha\beta}$ & the outer expansion of $\exact{g}$\\
$g_{\alpha\beta}$ & the background metric in outer expansion;\\
 & sometimes the background metric in a generic expansion\\
$\hmn{E\mu\nu}{\emph{n}}$ & $n$th-order term in outer, fixed-worldline expansion of metric \\
$\hmn{\mu\nu}{\emph{n}}$ & the approximation to $\hmn{E\mu\nu}{\emph{n}}$ obtained by setting $a^\mu=0$; \\
 & sometimes the $n$th-order term in generic expansion \\
$\A{L}{\emph{n}},\B{L}{\emph{n}},...$ & Cartesian STF tensors in STF decomposition of $\hmn{E}{\emph{n}}$\\
$h^R_{\alpha\beta}$ & the Detweiler-Whiting regular field \\
$g_{I\alpha\beta}$ & the inner expansion of $\exact{g}$\\
$g_{B\alpha\beta}$ & the background metric in the inner expansion\\
$H_{\mu\nu}\coeff{\emph{n}}$ & the $n$th-order term in the inner expansion\\
$\delta^n G_{\alpha\beta}[h]$ & the terms containing $n$ powers of $h$ in the expansion of $\exact{G}_{\alpha\beta}$\\
$\bar f_{\alpha\beta}$ & the trace-reversal of $f_{\alpha\beta}$, $\bar f_{\alpha\beta}\equiv f_{\alpha\beta}-\tfrac{1}{2}g_{\alpha\beta}g^{\mu\nu}f_{\mu\nu}$
\end{tabular*}
\end{table}
  
\end{notation}

      \tableofcontents
			\listoffigures
			\listoftables
%------------------------------------------------
			\mainbody 
			\chapter{Introduction}
\section{Relativistic motion and gravitational wave detectors}
The problem of motion is of tremendous historical importance in General Relativity (GR), both theoretically and experimentally. In conceiving of the theory, Einstein was fundamentally concerned with explaining the motion of bodies solely in terms of the geometric relationships between them. And much of the observational evidence for GR---e.g., the deflection of light by massive objects, the post-Newtonian effects in the motion of planets, and the slow decay of binary pulsar orbits due to the emission of gravitational waves---is tied to analyses of motion.

Despite its historical importance, theoretical research in this area has largely been confined to two limiting regimes: first, the Newtonian limit of weak gravity and slow motion, in which Newton's laws of motion and relativistic corrections to them can be derived for widely separated bodies; and second, the point particle limit, in which the geodesic equation and corrections to it can be derived for bodies of asymptotically small mass and size. Study of the Newtonian limit was pioneered by Einstein, Infeld, and Hoffmann \cite{Einstein, Einstein2} and is now fully developed in post-Newtonian theory (see the reviews \cite{300_years,Futamase_review,Blanchet_review} and references therein). Study of the point particle limit is less well developed, and it has typically focused on proving that at leading order, a small body behaves as a test particle, moving on a geodesic of some background spacetime (see, e.g., Refs.~\cite{Infeld, Kates_motion, DEath_paper, DEath, Geroch_particle1, Geroch_particle2}).\footnote{In the case of an electrically \emph{charged} body, the test particle moves on a worldline determined by the Lorentz force law, rather than a geodesic.}

In recent years, the scope of this research has been rapidly broadened by the advent of gravitational wave detectors such as the ground-based Laser Interferometer Gravitational Wave Observatory (LIGO)~\cite{LIGO} and the European detector VIRGO~\cite{VIRGO}, which are currently operating, and the space-based Laser Interferometer Space Antenna (LISA), which is planned to be launched within the next decade~\cite{LISA}. Since gravitational waveforms encode information about their sources, and carry that information undisturbed over great distances, these detectors have the potential to accurately measure the dynamics of bodies in previously unprobed regions of very strong gravity. This will provide the first tests of GR in its strong-field regime \cite{gravity_tests}; it will also make gravitational-wave astronomy possible, allowing us to study the behavior of objects with strong mutual and internal gravity \cite{parameter_extraction,parameter_estimation2}. Because of this potential, there is now a pressing need for accurate predictions of the waveforms produced by these highly relativistic systems. Such predictions require theoretical treatments of the problem of motion that go beyond either the post-Newtonian or test-particle approximation.

Consider, for example, the canonical problem of motion in celestial mechanics: the two-body problem. In a relativistic binary, the two bodies will emit gravitational radiation and thereby lose energy and angular momentum, slowly circularizing and spiraling into one another. See Ref.~\cite{binary_review} for an overview of such systems. One can capture the early stages of the inspiral, while the two bodies are widely separated, via a post-Newtonian expansion; however, once the two bodies are very near one another, this approximation breaks down. 

If the two bodies are of comparable mass, then their motion during the final stages of inspiral must be determined via a numerical integration of the full Einstein field equations. Much progress has been made in this problem in recent years, and numerical evolutions of comparable-mass neutron-star--neutron-star, neutron-star--black-hole, and black-hole--black-hole binaries are now producing astrophysically relevant predictions. (See Refs.~\cite{numerical_review, Pretorius_review} for reviews of numerical techniques.) A complete description of the inspiral is then constructed from the combination of post-Newtonian theory for the early stages and full numerical relativity for the late stages. Such comparable-mass systems are expected to be the principal source of wave-signals for LIGO.

However, if one of the bodies is much less massive than the other, then the entire inspiral can be treated analytically, rather than numerically, by utilizing the point particle limit. One such system is called an extreme mass-ratio inspiral (EMRI), in which a compact body (such as a neutron star or black hole) of mass $m\sim M_\odot$ spirals into a supermassive Kerr black hole of mass $M\sim (10^4\text{--}10^9)M_\odot$ lying at the center of a galaxy. See Refs.~\cite{Drasco_review, EMRI_review} for an overview of these systems. Detectable EMRIs are likely to occur with a frequency of several-to-one-hundred per year \cite{event_rate, event_rate2}, and they are a potentially important source of wave-signals for LISA.

For an EMRI, an expansion in the point particle limit roughly corresponds to an expansion in powers of the mass ratio $m/M\sim\e$; since both the small and large bodies are compact, their respective linear dimensions $\size$  and $\Size$ are of the same order as their masses, meaning that the expansion could instead be thought of in terms of the size-ratio $\size/\Size$. At leading order in this expansion, the small body moves on a geodesic of the spacetime of the large body, while simultaneously emitting gravitational waves. Obviously this approximation breaks down after a brief time, since it implies that the body will travel forever on a geodesic even as it emits waves that carry off energy and angular momentum. Thus, one must proceed beyond the leading-order, geodesic approximation. At sub-leading order, the metric perturbations produced by the small body force it onto an accelerated worldline that slowly spirals into the large body. The acceleration of this worldline, caused by the body's interaction with its own gravitational field, is called the gravitational self-force. Along with all other corrections to the test-particle approximation, it will be the principal subject of this dissertation. Although inspiraling orbits in Kerr are the primary worldlines of interest for EMRIs, most of my treatment applies to orbits in arbitrary spacetimes.

\section{The Einstein equation for an asymptotically small body}
\subsection{Non-systematic expansion of the field equations}
I now briefly overview the expansion of the field equations in the limit $\e\to0$. Before proceeding to the expansion, I will first take note of the properties of the equations that are being expanded. In GR, the dynamical variables are the metric $\exact{g}_{\mu\nu}$, which describes the geometry of spacetime, and a set of matter fields $\exact{\Psi}$, which describe, e.g., mass and charge distributions and scalar and electromagnetic fields. The evolution of these fields is governed by the Einstein Field Equation (EFE)
\begin{equation}
\exact{G}_{\mu\nu}[\exact{g}]=8\pi\exact{T}_{\mu\nu}[\exact{g},\exact{\Psi}],
\end{equation}
which is a set of highly nonlinear, coupled partial differential equations for the components of the metric and matter fields. Here $\exact{G}$ is the Einstein curvature tensor of the spacetime, and $\exact{T}$ is the stress-energy tensor of the matter fields in that spacetime.

Unlike in other field theories such as electrodynamics, in GR the evolution of the source fields $\exact{\Psi}$ is not independent of the field equation---in fact, the evolution equation for the source is an integrability condition for the Einstein field equation, following from the restriction imposed by the Bianchi identity \cite{Einstein,Detweiler_review}. Explicitly, the Bianchi identity ${}^{\exact{g}\!}\del{\nu}\exact{G}^{\mu\nu}=0$, when combined with the EFE, immediately implies the conservation equation ${}^{\exact{g}\!}\del{\nu}\exact{T}^{\mu\nu}=0$, which provides a field equation for the matter fields $\exact{\Psi}$; this might be, for example, the Klein-Gordon equation for a scalar field or the Maxwell equations for an electromagnetic field, along with an equation of motion for the charge distribution that generates the field. Hence, in GR the matter evolution equations are partially determined by the EFE; in some cases, such as those just mentioned, the matter field equations are \emph{entirely} determined by the EFE. This fact has important consequences in the point particle limit, where the details of the body's composition become irrelevant, and so the Einstein equation entirely determines the body's mean motion. Only at some relatively high order in the expansion in powers of small size does one require a separate matter field equation modeling the interior of the body.

Turning now to the expansion, assume that the metric and the matter fields depend on $\e$, such that in the limit of $\e\to 0$ we can expand them as $\exact{g}=g+h+\order{\e^2}$ and $\exact{\Psi}=\Psi+\psi+\order{\e^2}$. This determines a \emph{background metric} $g$, which must satisfy $G_{\mu\nu}[g]=8\pi T_{\mu\nu}[g,\Psi]$; throughout this dissertation, I assume that in the spacetime region of interest, the small body provides the only source of matter (i.e., $\Psi\equiv0$), such that the background stress-energy tensor $T_{\mu\nu}[g,\Psi]$ vanishes and the background metric satisfies the vacuum Einstein equation $G_{\mu\nu}=0$. Once the background metric is determined, it defines the geometry of the background spacetime, and it is used to raise and lower indices. The metric perturbation $h$ is then a field on this background geometry. At linear order in $\e$, the perturbations obey the \emph{linearized} Einstein equation
\begin{equation}\label{linear_EFE}
\delta G_{\mu\nu}[h] = 8\pi T_{\mu\nu}[g,\psi],
\end{equation}
where the linearized Einstein tensor $\delta G[h]$ is a linear functional of $h$ and its derivatives, given explicitly in Appendix \ref{general_expansions}, and $T[g,\psi]$ is linear in $\psi$ (and its derivatives). As in the non-perturbative problem, the linearized Bianchi identity $\nabla^\nu \delta G_{\mu\nu}=0$ partially determines the evolution of the matter via the linearized conservation equation
\begin{equation}
\nabla^\nu T_{\mu\nu}[g,\psi]=0.
\end{equation}

Note that perturbation theory in General Relativity has an important gauge freedom. If we make a near-identity coordinate transformation of the form $x^\mu\to x^\mu-\e\xi^\mu(x)+...$, we can opt to transfer its $\order{\e}$ effect on the background metric into the perturbation, leaving the background and its coordinates unchanged; the effect on the leading-order perturbation is then given by
\begin{equation}
h_{\mu\nu}\to h_{\mu\nu}+\xi_{\mu;\nu}+\xi_{\nu;\mu}.
\end{equation}
In other words, a small coordinate transformation is equivalent to a redefinition of the metric perturbation in the original coordinates. But since this redefinition is merely a coordinate transformation, no physical observables are affected by it; hence, it is a gauge freedom. A convenient choice of gauge is the \emph{Lorenz gauge}, defined by the condition $\nabla^\nu\bar h_{\mu\nu}=0$, and named in analogy to electromagnetism. Here an overbar indicates trace reversal: $\bar h\equiv h_{\mu\nu}-\tfrac{1}{2}g_{\mu\nu}g^{\alpha\beta}h_{\alpha\beta}$. In a vacuum background, imposition of the Lorenz gauge splits the linearized EFE into a wave equation, which I write as
\begin{equation}\label{linear_wave_eqn}
E_{\mu\nu}[\bar h] = -16\pi T_{\mu\nu}[g,\Psi],
\end{equation}
and the gauge condition, which I write as
\begin{equation}
L_{\mu}[h] = 0,
\end{equation}
where $E_{\mu\nu}$ and $L_\mu$ are linear operators defined by
\begin{align}
E_{\mu\nu}[f] & = \left(g^\rho_\mu g^\sigma_\nu\nabla^\gamma\del{\gamma} +2R\indices{_\mu^\rho_\nu^\sigma}\right)\!f_{\rho\sigma},\label{wave op def}\\
L_\mu[f] & =\left(g^\rho_\mu g^{\sigma\gamma}-\tfrac{1}{2}g^\gamma_\mu g^{\rho\sigma}\right)\!\del{\gamma}f_{\rho\sigma}\label{gauge op def},
\end{align}
for arbitrary $f_{\alpha\beta}$. Equation~\eqref{linear_wave_eqn} is a hyperbolic equation with well-defined retarded and advanced solutions. Taking the divergence of it, we find that the gauge condition is satisfied if and only if the conservation equation $\nabla^\nu T_{\mu\nu}=0$ holds; hence, one can determine the equation of motion of the source from either the gauge condition or the conservation equation. Note that a solution to the wave equation is also a solution to the linearized Einstein equation if and only if it also satisfies the gauge condition. In other words, the Einstein equation constrains the motion of the source, as expected.

In an asymptotically flat spacetime, at sufficiently large distance from all sources, the background metric approaches that of flat spacetime, and its Riemann tensor vanishes. One can also make a refinement of the Lorenz gauge by specifying that $h$ is both transverse, such that $h_{\mu t}=0$, and trace-free, such that $h_{\mu\nu}=\bar h_{\mu\nu}$. The linearized Einstein equation then asymptotically approaches the ordinary flat-spacetime wave equation $\eta^{\mu\nu}\partial_\mu\partial_\nu h_{\alpha\beta}=0$, where $\eta_{\mu\nu}=\text{diag}(-1,1,1,1)$ is the Minkowski metric. The solutions to this wave equation are the gravitational waves that LISA and LIGO are designed to detect. In gravitational wave physics, the eventual goal is to determine the precise relationship between these waves and the distant system that generated them. Specifically, for EMRIs, one hopes to generate reliable, accurate waveform templates that allow one to extract from an observed signal both the motion of the small body and the details of the background spacetime in which it moves. In order to extract those parameters, one requires a model with a fractional error smaller than $\sim\e\sim 10^{-5}$, such that the predicted waveforms remain accurate to within one wave-cycle over the course of an inspiral that produces $\sim 1/\e\sim 10^5$ cycles. 

\subsection{The point particle source and the self-force}\label{intro_to_self_force}
Next, we assume that at linear order in the point particle limit, the small body can be described as a point particle in the background spacetime; this will be justified in Chapter~\ref{perturbation calculation}. If the body is uncharged, then this assumption means that $\psi$ is a delta-function mass-distribution, and the stress-energy tensor of the body is given by
\begin{equation}\label{stress_energy_point_particle}
T^{\mu\nu}(x) = \int_\gamma m u^\mu u^\nu \delta(x,z(t))dt,
\end{equation}
where $\delta(x,z(t))=\delta^4(x^\alpha-z^\alpha(t))/\sqrt{|g|}$ is a covariant delta function with support on the worldline $\gamma=\lbrace z^\alpha(t):-\infty<t<\infty\rbrace$, $t$ and $u^\mu\equiv \diff{z^\mu}{t}$ are the proper time and four-velocity on $\gamma$, and $|g|$ denotes the absolute value of the determinant of $g_{\alpha\beta}$. This stress-energy tensor describes a mass concentrated at a point that traces out a curve in the background spacetime. If the small body is also charged, then the stress-energy tensor will include a (scalar or electric) point-charge term and a (scalar or electromagnetic) field term.

Now, in each case, the particle generates a metric perturbation $h$ and possibly a scalar or electromagnetic field. When the body moves in a curved spacetime (or accelerates in a flat spacetime), these fields that it creates become asymmetrically distributed around it. Because of the asymmetry, the body interacts with its own fields, which then exert a force on it, which we call the \emph{self-force}. Part of this force is intimately related to the emission of waves: an accelerated body, or a body moving in curved spacetime, emits radiation that propels it and carries off part of its kinetic energy. This propulsive force is called the \emph{radiation-reaction force}; the work done by it on the particle is equal to the energy carried off by the radiation \cite{work1,work2}. In flat spacetime, the radiation-reaction force and the self-force are identical, while in curved spacetime, the self-force incorporates conservative effects that are not accounted for by simple radiation-reaction. 

The electromagnetic self-force on an accelerated point-charge in flat spacetime has been known for many years \cite{Jackson}. Its effects were studied by Abraham and Lorentz, and Dirac first derived an expression for it within the context of special relativity in 1938~\cite{Dirac}; Ref.~\cite{Lorentz_Dirac_review} provides a pedagogical review. Dirac's result was generalized to curved spacetime by DeWitt and Brehme in 1960~\cite{DeWitt_Brehme} (as corrected by Hobbs~\cite{Hobbs}). Much later, Quinn derived an analogous result for scalar charges~\cite{Quinn}. All of these derivations work in a fixed spacetime; in other words, they assume that neither the charge nor the body's mass influences the spacetime, such that the metric perturbation can be ignored. Recent work by Futamase et al.~\cite{Futamase_EM} has removed this restriction.

A formal expression for the gravitational self-force---the force due to the perturbation $h$---was first derived in 1996 by Mino, Sasaki, and Tanaka \cite{Mino_Sasaki_Tanaka} and Quinn and Wald \cite{Quinn_Wald}; the resulting equation of motion is now known as the MiSaTaQuWa equation. Since then, numerous other derivations have been proffered; see Refs.~\cite{Mino_matching, Detweiler_Whiting, Eran_force, Galley_Hu, Galley_QFT, Fukumoto, Gal'tsov, Gralla_Wald,my_paper} and the reviews \cite{Eric_review, Detweiler_review} for examples.

I will forgo a discussion of the scalar and electromagnetic results. However, all of the aforementioned derivations, including those of the scalar and electromagnetic force, proceed via a careful analysis of the field in the neighbourhood of the particle. For a point-mass, the physically relevant, retarded solution to the wave equation \eqref{linear_wave_eqn} is given by 
\begin{align}
\bar h_{\mu\nu} &= 4\int G_{\mu\nu\mu'\nu'}T^{\mu'\nu'}\sqrt{|g'|}d^4x'\nonumber\\
\quad &= 4m\int_\gamma G_{\mu\nu\mu'\nu'}u^{\mu'}u^{\nu'}dt',\label{point_particle_solution}
\end{align}
where $G_{\mu\nu\mu'\nu'}$ is the retarded Green's function for the wave-operator $E_{\mu\nu}$, discussed in Appendix~\ref{Greens_functions}. The retarded field $h$, obtained by trace-reversal of the above equation, can be split into two pieces: a singular, Coulomb-type field $h^S$, which diverges as $1/r$ at the position of the particle, and a regular field $h^R$, defined as $h^R\equiv h-h^S$. In essence, $h^S$ is spherically symetric about the particle, so it does not contribute to the force, and the acceleration of the particle is determined entirely by the regular field:
\begin{equation}\label{first_self_force_expression}
a^\mu=-\tfrac{1}{2}(g^{\mu\nu}+u^\mu u^\nu)(2h^R_{\nu\alpha;\beta}-h^R_{\alpha\beta;\nu})u^\alpha u^\beta.
\end{equation}
One can most easily motivate this expression by assuming a \emph{generalized equivalence principle}~\cite{Futamase_review}: any asymptotically small, uncharged, nonspinning mass moves on a geodesic of the \emph{smooth} part of the geometry in its neighbourhood. After expanding the geodesic equation ${}^{g+h^R}\del{u}u^\alpha=0$, one arrives at Eq.~\eqref{first_self_force_expression}. Although early derivations essentially assumed this generalized equivalence principle, more recent derivations \cite{Gralla_Wald,my_paper} have derived it by showing that Eq.~\eqref{first_self_force_expression} follows directly from the Einstein equation in the neighbourhood of the particle; it has also been derived directly from general definitions of linear momentum for extended bodies \cite{Futamase_review, Harte_2009}. Note that if we had instead assumed that the small mass moves on a geodesic of the full spacetime $g+h+\order{\e^2}$, as we might guess to be a generalized equivalence principle, then $h^R$ would be replaced by $h$ in the equation of motion \eqref{first_self_force_expression}, and the equation would be divergent; in that sense, Eq.~\eqref{first_self_force_expression} is a \emph{regularized} equation of motion.

One can find the regularized force by writing the expression Eq.~\eqref{first_self_force_expression} in terms of the full field $h$, averaging over a sphere around the particle, and then evaluating the equation on the worldline \cite{Quinn_Wald, Eric_review}. Alternatively, one can attempt to find a meaningful decomposition of $h$ into $h^S$ and $h^R$. The earliest derivations, following methods used for the electromagnetic self-force \cite{DeWitt_Brehme}, split the field into a \emph{direct piece} and a \emph{tail piece}. In flat spacetime, the Green's function $G_{\mu\nu\mu'\nu'}(x,x')$ is given by a delta function with support on the past light cone of $x$; hence, the field at $x$ is fully determined by information on the worldline at the retarded time $u(x)$. (Here $u(x)$ is the proper time on the worldline at the point of intersection with the past light cone of $x$.) However, spacetime curvature allows waves to propagate within, rather than just on, the light cone, which means that the field at $x$ is determined not only by the intersection of the worldline with the past light cone, but also by its entire past history within the light cone. The singular, direct piece of $h$ is the contribution from propagation on the light cone; the regular, tail piece of $h$ is the contribution from within the light cone. Explicitly, the tail is given by
\begin{equation}
\tail_{\mu\nu} = 4m\int_{-\infty}^{u^-} \bar G_{\mu\nu\mu'\nu'}u^{\mu'}u^{\nu'}dt',
\end{equation}
where the upper limit of integration $u^-\equiv u-0^+$ cuts off the integral just prior to the light cone, avoiding the divergence of the Green's function there. Substituting $\tail$ for $h^R$ in Eq.~\eqref{first_self_force_expression} yields the correct equation of motion.

Note, however, that $\tail$ has two undesirable properties. First, it is not differentiable on the worldline: A derivative of $\tail_{\mu\nu}$ with respect to $x^\rho$ will include the term $4m\bar G_{\mu\nu\mu'\nu'}u^{\mu'}u^{\nu'}\partial_\rho u\big|_{x'=z(u)}$, which arises from the derivative of the upper limit of integration. If $x$ is on the worldline, where $z(u)=x$, then the term is evaluated at $x'=x$, and $\partial_\rho u$ is therefore not defined.\footnote{Consider, for example, the flat-spacetime expression $u=t-t'-\sqrt{(x_a-x'_a)(x^a-x'^a)}$, where $x=(t,x^a)$, and $z(u)=(t',x'^a)$ is the retarded point on the worldline. Then $\pdiff{u}{x^a}=n_a$, where $n_a$ is a unit vector pointing from $x'^a$ to $x^a$; since this vector is not defined at $x=z(u)$, neither is $\partial_a u$.} Hence, the tensor $\tail_{\mu\nu;\rho}$ is not defined on the worldline. However, the non-differentiability is limited to derivatives in directions away from the worldline. In addition, $u^\mu u^\nu G_{\mu\nu\mu'\nu'}u^{\mu'}u^{\nu'}$ vanishes at $x=x'$, so the non-differentiability is further limited to derivatives of components of $\tail$ perpendicular to the worldline. Therefore, derivatives along the worldline, and derivatives of the projection $\tail_{\mu\nu}u^\mu u^\nu$, are defined. If $\tail$ is substituted for $h^R$ in Eq.~\eqref{first_self_force_expression}, with a bit of work one can rewrite the equation such that it contains only these well-behaved derivatives.

The second, more substantial undesirable property of $\tail$ is that it is not a solution to the linearized Einstein equation, and so $g+\tail$ is not an approximation to any smooth spacetime satisfying Einstein's equation. Hence, taking it to be $h^R$ in Eq.~\eqref{first_self_force_expression} does not fit comfortably with the generalized equivalence principle invoked above.

An alternative, more meaningful decomposition of $h$ was suggested by Detweiler and Whiting \cite{Detweiler_Whiting}, who were inspired by Dirac's method of flat-spacetime regularization. In flat spacetime, one can write the singular piece of the electromagnetic four-potential as $A^S_\mu = \tfrac{1}{2}(A^{\text{ret}}_\mu +A^{\text{adv}}_\mu)$, where $A^{\text{ret}}_\mu$ and $A^{\text{adv}}_\mu$ are, respectively, the retarded and advanced solutions to Maxwell's equations. Because the retarded and advanced solution share the same singularity structure, taking the average of the two simply yields the singular part of the retarded solution. After subtracting this singular piece from the retarded solution, one arrives at a regular field $A^R_\mu=A^{\text{ret}}_\mu-A^S_\mu=\tfrac{1}{2}(A^{\text{ret}}_\mu-A^{\text{adv}}_\mu)$. The significance of this decomposition is that both $A^S$ and $A^R$ are solutions to Maxwell's equations: $A^S$ is a solution to the inhomogenous equations with a point-charge source, while $A^R$ is a solution to the vacuum equations. These fields can be constructed from singular and regular Green's functions $G^S=\tfrac{1}{2}(G^{\text{ret}}+G^{\text{adv}})$ and $G^R=\tfrac{1}{2}(G^{\text{ret}}-G^{\text{adv}})$.

Detweiler and Whiting formulated an analogous decomposition in curved spacetime by writing, in the case of gravity, $h^S_{\mu\nu} = \tfrac{1}{2}(h^{\text{ret}}_{\mu\nu} + h^{\text{adv}}_{\mu\nu} + h^H_{\mu\nu})$ and $h^R_{\mu\nu} = \tfrac{1}{2}(h^{\text{ret}}_{\mu\nu} - h^{\text{adv}}_{\mu\nu} - h^H_{\mu\nu})$, where $h^H_{\mu\nu}$ is a homogenous solution. In the Lorenz gauge, these fields can be constructed from the singular and regular Green's functions $G^S=\tfrac{1}{2}(G^{\text{ret}}+G^{\text{adv}}+H)$ and $G^R=\tfrac{1}{2}(G^{\text{ret}}-G^{\text{adv}}-H)$, where $H$ is a homogenous function suitably chosen to ensure the causality of the regular field; refer to Ref.~\cite{Eric_review} or Appendices~\ref{Greens_functions} and \ref{point_particle_soln} for details. As in Dirac's flat-spacetime regularization, the singular field is sourced by a point particle, and the regular field is source-free. Stated another way, in this decomposition $h^S$ is the bound field carried with the body---it is the field of the body itself, warped slightly by the curvature of the background spacetime. The regular field $h^R$ is a free radiation field that is originally emitted from the body, but propagates independently of it---an observer in the neighbourhood of the body cannot distinguish its effects from those of the background spacetime~\cite{Detweiler_Whiting, Detweiler_review}. The \emph{Detweiler-Whiting Axiom}, a particular generalized equivalence principle, states that up to order-$\e^2$ errors, an asymptotically small non-spinning body moves on a geodesic of the metric $g+h^R$, which satisfies the vacuum Einstein equation up to order-$\e^2$ errors. Recently, Harte has derived this axiom from general laws of motion for extended bodies~\cite{Harte, Harte_EM, Harte_2009}, though his result in the gravitational case is limited to linear metric perturbations, and in the scalar and electromagnetic cases, to fixed background spacetimes.

From this point on, $h^R$ will always refer to the Detweiler-Whiting field. Beyond satisfying a wave equation, $h^R$ also has the advantage of being differentiable everywhere, even on the worldline. Hence, it can be directly substituted into Eq.~\eqref{first_self_force_expression}. However, on the worldline, the derivatives of $h^R$ differ from certain derivatives of $\tail$ by terms proportional to the Riemann tensor of the background metric (see Appendix~\ref{point_particle_soln}). The Riemann tensor terms exactly cancel in Eq.~\eqref{first_self_force_expression}, such that, assuming the validity of the Detweiler-Whiting Axiom, the self-force can be written as
\begin{equation}\label{self_force_Lorenz}
a^\mu=-\tfrac{1}{2}(g^{\mu\nu}+u^\mu u^\nu)(2\tail_{\nu\alpha\beta}-\tail_{\alpha\beta\nu})u^\alpha u^\beta,
\end{equation}
where I have defined
\begin{equation}
\tail_{\mu\nu\rho} = 4m\int_{-\infty}^{u^-} \del{\rho}\bar G_{\mu\nu\mu'\nu'}u^{\mu'}u^{\nu'}dt',
\end{equation}
which is a well-defined tensor on the worldline. This is the MiSaTaQuWa equation. It is equivalent to the result of substituting $\tail_{\mu\nu}$ for $h^R_{\mu\nu}$ in Eq.~\eqref{first_self_force_expression}, so long as that result is appropriately rearranged to remove undefined derivatives of $\tail_{\mu\nu}$. Note, however, that this equation is less general than Eq.~\eqref{first_self_force_expression}, because it relies on the form of the metric perturbation in the Lorenz gauge; in an alternative gauge, the regular field $h^R$ would have a different form, and Eq.~\eqref{self_force_Lorenz} would no longer hold true. One of the current challenges in the field is finding a unique, equally meaningful definition of $h^S$ and $h^R$ in alternative gauges, in which the Green's function decomposition described above may not be possible.

As a final note in this section, I raise the issue of reduction of order. If one explicitly calculates the self-force given by Eq.~\eqref{first_self_force_expression}, one finds, in addition to the tail terms, an antidamping term $-\tfrac{11}{3}m\dot a^\mu$. Such a term was first discovered by Havas \cite{damping} (as corrected by Havas and Goldberg \cite{damping2}). Because of it, the equation is third-order in time, and hence exhibits physically inadmissible solutions. The usual procedure of eliminating this problem, as is familiar from the case of the Abraham-Lorentz-Dirac force~\cite{Jackson,Lorentz_Dirac_review,order_reduction}, is via a reduction of order: one notes that the right-hand side of the equation should be order $m$, and so one substitutes the leading-order equation of motion into the term $m\dot a^\mu$. In this case, the leading-order equation of motion is $a^\mu=0$, which means the order-reduction entirely eliminates the offending term from the equation. As we shall see in Ch.~\ref{point_particle}, this a posteriori corrective measure is not required if one performs a systematic expansion of the Einstein equation.

\section{Inconsistencies in the self-force}\label{inconsistencies}
\subsection{Incompatible equations of motion}
By assuming a generalized equivalence principle in the above discussion, I have done precisely what cannot be done in GR: assumed an equation of motion independent of the Einstein equation. In order for the solution to the wave equation to also be a solution to the first-order Einstein equation, it must satisfy the gauge condition, which now reads 
\begin{align}
0 &= L_\mu[h]\nonumber\\
  &= 4\int \nabla^\nu G_{\mu\nu\mu'\nu'} T^{\mu'\nu'}[\gamma]dV'\nonumber\\
  &= 4\int G_{\mu\mu'}\del{\nu'}T^{\mu'\nu'}[\gamma]dV'\nonumber\\
  &= 4m\int_\gamma G_{\mu\nu'}a^{\nu'}dt',
\end{align}
where $G_{\mu\mu'}$ is the retarded Green's function for the vector wave equation \eqref{vector Green}. The third line follows from the second after one invokes the identity $\nabla^{\nu}G_{\mu\nu\mu'\nu'}=-G_{\mu(\mu';\nu')}$ (derived in Appendix~\ref{Greens_functions}) and integrates by parts. The final line follows from a straightforward calculation of $\del{\nu'}T^{\mu'\nu'}$ (see, e.g., Ref.~\cite{Eric_review} for that calculation). We see from this sequence of equations that, as noted previously, imposing the gauge condition is equivalent to imposing the conservation equation $T^{\mu\nu}{}_{;\nu}=0$, which is equivalent to imposing the first-order Bianchi identity $\delta G^{\mu\nu}{}_{;\nu} = 0$. And the consequence of any of these conditions is that $a^{\nu}=0$: that is, $\gamma$ must be a geodesic in the background spacetime. This obviously contradicts the equation of motion \eqref{first_self_force_expression}.

A contradiction of this form alerts us to a fundamental difficulty in defining a ``corrected" worldline. At its most fundamental level, the self-force problem consists of finding a pair $(\gamma,h_{\mu\nu})$ that meaningfully and accurately represents the motion and the metric perturbation of an asymptotically small body. This problem is far from trivial. At each order in perturbation theory, the equation of motion, and hence the worldline, is fixed by the Bianchi identity; using any other worldline means that a given $n$th-order perturbation $\hmn{\mu\nu}{\emph{n}}$ is not a solution to the $n$th-order Einstein equation. But at each order, the worldline determined by the Bianchi identity differs from that at every other order. It seems clear that the higher-order equations of motion are corrections to the lower-order ones, but there is no obvious way to self-consistently incorporate those corrections.

As in any problem involving a small parameter, two options present themselves: first, one can assume a regular power-series expansion of every function in the problem, which leads directly to a succession of equations that can be solved exactly, order by order; or second, one can be satisfied with the construction of an approximate solution to the exact equation, however that solution may be arrived at. If the first approach is adopted, then the linearized Bianchi identity fixes the worldline to be a geodesic. In order to incorporate the effects of the self-force, one must introduce additional degrees of freedom: this is accomplished by noting that not just the metric, but the worldline itself carries $\e$-dependence, which must be expanded in a power series. The higher-order terms in this expansion consist of ``deviation vectors" describing infinitesimal corrections to the leading-order geodesic worldline; the effects of the self-force are then realized in evolution equations for these deviation vectors, rather than in an equation of motion for the worldline itself. In this scheme, the first-order perturbation is sourced by a geodesic; the perturbation then determines the first-order deviation vector; the second-order perturbation is determined by the first-order perturbation and deviation vector; and so on. This was the approach taken by Gralla and Wald in a recent derivation of the self-force \cite{Gralla_Wald}. But such an interpretation is meaningful only for a brief time: Since the particle will eventually plunge into the large body, the ``small" correction to the geodesic will eventually grow large. At that point, the entire expansion in powers of $\e$ will have broken down. In other words, the straightforward power series expansion of the Einstein equation is a valid approximation to the actual Einstein equation only on short timescales.

However, in studies of the problem of motion in GR, this first approach is atypical; instead, the second approach is the one typically adopted. In the self-force problem, this has been realized in the procedure of \emph{gauge relaxation} \cite{Mino_Sasaki_Tanaka,Quinn_Wald}, which essentially consists of allowing the Lorenz gauge condition to be slightly violated by the first-order metric perturbation, effectively sidestepping the requirement that $\gamma$ must be a geodesic. In other words, rather than solving the linearized Einstein equation \eqref{linear_EFE}, one solves the equations
\begin{align}
E_{\mu\nu}[h] &= -16\pi T_{\mu\nu}[\gamma],\\
L_{\mu}[h] &=\order{\e^2}.
\end{align}
These equations are useful because the wave equation can be solved for an arbitrary worldline $\gamma$. Solving them is equivalent to solving the approximate Einstein equation
\begin{equation}\label{approx_linear_EFE}
\delta G_{\mu\nu}[h] = 8\pi T_{\mu\nu}[g,\psi]+\order{\e^2}.
\end{equation}
Hence, this procedure yields an approximate solution to the exact Einstein equation, as desired, and it leads to a single worldline obeying a self-consistent equation of motion. It is also similar to successful methods of post-Newtonian theory, in which, prior to any expansion, the exact Einstein equation is written in a ``relaxed" form that can be solved for an arbitrary source. However, the gauge-relaxation used in the self-force problem lacks the systematic nature of the post-Newtonian method, in the sense that the relaxed linear equation has not been shown to follow from a systematic expansion of the exact Einstein equation, and the solution to the relaxed linear problem has not been related to a solution to the exact problem.

One of the goals of this dissertation is to provide such a systematic justification of the gauge-relaxation procedure. Specifically, I seek to split the exact Einstein equation into a sequence of equations that can be solved exactly without expanding the worldline. In order to determine the reliability of this sequence of equations, one must prove that its solution asymptotically approximates a solution to the exact equation. In lieu of obtaining the rigorous error estimates required to achieve such a goal, I seek only a qualitative description of how the solution to the sequence of equations could be obtained by expanding a solution of the exact EFE; this is sufficient to put the expansion at the same level of rigor as a regular Taylor series.

These goals require two key steps: First, a point particle source is mathematically well-defined only for the first-order, linearized equation.\footnote{At least this is true within classical distribution theory \cite{linear_distributions}, since the Einstein tensor of a point particle would contain products of delta distributions and hence be too singular to be treated as a distribution. However, more general methods based on Colombeau algebras, which allow for multiplication of distributions, have been devised to overcome this problem \cite{nonlinear_distributions}.} Hence, an approximation scheme going beyond first order must abandon the point particle approximation. Second, since the worldline in the self-force problem is usually defined via the point particle source, we require a means of defining a (self-consistent) worldline for the extended body.

I remind the reader that these theoretical goals are closely related to experimental ones: in order to extract the parameters of an EMRI from a gravitational wave signal, one requires a template that relates the signal to the motion of the body over a large portion of the inspiral. Such a template must be based on an approximation scheme that is uniform on a domain of size $\sim 1/\e$; in other words, the errors in the approximation must remain small over a long timespan. This can be accomplished only in a scheme that self-consistently incorporates the corrected motion.

\subsection{Singular perturbation theory in General Relativity}
The problems discussed in the previous subsection are closely related to the fact that the perturbation due to a small body is a \emph{singular perturbation}. That is, a regular power-series expansion fails to provide a uniform approximation to an exact solution in the region of interest. This non-uniformity is signaled by two types of errors: First, near the particle, the metric perturbation behaves as $\e/r$; at second order in $\e$, it will behave as $\e^2/r^2$. If the body is anything other than a black hole, the error in the approximation (i.e., the difference between the exact and approximate metrics) will then diverge at $r=0$; and even in the case of a black hole, the expansion will fail near the body, since the second-order Einstein tensor will be ill-defined. Hence, the approximation breaks down due to rapid changes, on the lengthscale $\sim\e$, near the body. The second type of error is that which grows secularly (i.e., accumulates over time): if we use the first-order solution with a point particle source, which requires the particle to move on a geodesic, then the deviation between the exact and approximate position of the body will eventually grow large, and the expansion will once again have broken down. This failure occurs on the radiation-reaction timescale $t_{rr}\equiv1/\e$; in an EMRI, this is the time required for the particle's energy and angular momentum to undergo an order-$1$ change, since their rate of change is proportional to the self-force, itself of order $\e$.

We see from this that we require an expansion that is accurate on the body's length scale $\sim\e$, on the background length scale $\sim\e^0$, and on the radiation-reaction timescale $\sim 1/\e$. Such an expansion cannot be constructed from a regular power series; however, it can be constructed using the methods of singular perturbation theory. In general, for any function $f(x,\e)$, where $x^\mu$ are some coordinates, there are two types of expansions to consider: a \emph{regular expansion}, of the form
\begin{equation}
f(x,\e)=\sum_{n=0}^N\e^nf\coeff{\emph{n}}(x)+\order{\e^{N+1}},
\end{equation}
which is a Taylor expansion at fixed coordinate values; and a \emph{general expansion}, of the form
\begin{equation}\label{singular_expansion}
f(x,\e)=\sum_{n=0}^N\e^nf\coeff{\emph{n}}(x,\e) +\order{\e^{N+1}},
\end{equation}
where the coefficients $f\coeff{\emph{n}}(x,\e)$ are allowed to depend on $\e$, but are nevertheless of order 1, in the sense that there exist positive constants $k$ and $\e_0$ such that $|f\coeff{\emph{n}}(x,\e)|\le k$ for $0\le\e\le\e_0$, but $\displaystyle\lim_{\e\to0}f\coeff{\emph{n}}(x,\e) \not\equiv 0$ (unless $f\coeff{\emph{n}}(x,\e)$ is identically zero). Put simply, the goal of a general expansion is to expand only \emph{part} of a function's $\e$-dependence, while holding fixed some specific $\e$-dependence that captures one or more of the function's essential features.

Several types of general expansions can be considered. First, there are expansions that can be patched together from a finite number of regular expansions; I will call such an expansion a \emph{composite expansion}. For example, in the self-force problem, if we use a rescaled radial coordinate $\tilde r\equiv r/\e$ near the body, then a regular expansion at fixed coordinate values could be accurate on the scale $r\sim\e$, where an expansion at fixed $r$ fails; this \emph{inner expansion} can then be combined with an \emph{outer expansion} valid for $r\sim 1$, yielding a composite expansion that is uniformly accurate both near and far from the body. Composite expansions are most typically used in the method of matched asymptotic expansions, in which regular expansions are constructed in distinct regions, and then any unknown functions in the expansions are determined by comparing the two in a region of mutual validity. This method was first used in fluid dynamics to analyze the behavior of low-viscosity fluid near a boundary. In the context of General Relativity, since the pioneering work of Burke~\cite{Burke}, who studied the effect of radiation-reaction on a post-Newtonian binary, and D'Eath~\cite{DEath_paper,DEath}, who studied the motion of black holes, matched asymptotic expansions have mostly been utilized for two purposes: determining waveforms by matching wave-zone expansions to near-zone expansions (see, e.g., the review \cite{Blanchet_review}), and determining equations of motion by matching an inner expansion near a body to an outer expansion in a larger region (see, e.g., \cite{Kates_motion,Thorne_Hartle,PN_matching}). The latter method was one of the first means of deriving the MiSaTaQuWa equation \cite{Mino_Sasaki_Tanaka}.

Composite expansions are suitable only when the different lengthscales dominate in different regions---e.g., the metric varies on the short lengthscale $\sim\e$ near the small body, while it varies on the background lengthscale $\sim\e^0$ everywhere else. In contrast to this, another type of general expansion, called a \emph{multiscale expansion}, is suitable in situations where multiple lengthscales are relevant everywhere in the region of interest. Multiscale expansions begin directly with the generalized form \eqref{singular_expansion}, where the coefficients $f\coeff{n}(x,\e)$ depend on some specific function of $\e$---for example, in a \emph{two-timescale expansion}, $f\coeff{n}$ is assumed to depend on a time coordinate $t$ and a \emph{slow-time} coordinate $\tilde t=\e t$, which allows the expansion to capture both short-term and long-term effects. Recently, two-timescale expansions have frequently been used in analyzing the self-force problem \cite{other_paper, Mino_Price, Hinderer_Flanagan}.

In this dissertation, I will utilize both of these types of expansions, along with a still-more general type of general expansion: rather than allowing a functional dependence on an $\e$-dependent function from spacetime to $\mathbb{R}$, such as a slow-time coordinate, I will allow a functional dependence on a function from $\mathbb{R}$ \emph{to} spacetime. This will allow me to consider a metric perturbation that is a functional of an $\e$-dependent worldline. In addition to making use of these expansions, I will provide a moderately formal introduction to the methods of singular perturbation theory in the context of General Relativity. Such an explication is timely, since singular perturbation methods are being used increasingly frequently in the field, while very little has been done to explain their formal structure. Thus far, Kates \cite{Kates_structure} has provided the only such exposition. Unfortunately, his work provides only a broad outline of singular perturbation theory; he provides no detailed discussions of particular methods. My presentation will thus focus on precise statements pertaining to particular methods, with the aim of clarifying previous results. In particular, I will discuss the limitations of matched asymptotic expansions as they have been used to derive equations of motion.

\subsection{A self-consistent treatment of the self-force}
As stated above, my principal goal in utilizing singular perturbation techniques is to construct a self-consistent approximation scheme in the self-force problem. My scheme makes use of two general expansions: an inner expansion accurate near the small body, and an outer expansion accurate in the external background spacetime. The essential feature of these expansions, which distinguishes them from previous approaches, is that they treat the worldline of the body as fixed. In other words, the metric is expanded as $\exact{g}=g+\e\hmn{}{1}[\gamma]+\e^2\hmn{}{2}[\gamma]+...$; similarly, the acceleration, treated as a function of time on the fixed worldline, is expanded as $a_\mu(t,\e)=\an{0}_\mu(t)+\e\an{1}_\mu(t;\gamma)+...$. While the notion of a fixed worldline was taken for granted in some earlier derivations, it has never before been considered explicitly. This expansion, because it deals only with a single worldline, is consistent with what would result from an evolution in time that began with (1) some arbitrary initial data and (2) a system of evolution equations that involve only local values of the position, momentum, and metric perturbation of the small body at each value of time.

With the worldline held fixed, I decompose the exact Einstein equation into a sequence of exactly solvable perturbative equations. To accomplish this, I begin by surrounding the body with a worldtube, and I seek a solution outside the tube; the radius $\rad$ of the tube is chosen such that it lies in a \emph{buffer region} defined by the condition $\e\ll r\ll 1$, where $r$ is a radial distance from the body (measured in units of some global, external length scale, such as the large mass $M$ in an EMRI). I assume that the Lorenz gauge can be imposed in the domain of interest and up to the order of interest, such that the Einstein equation in the outer limit is split into a weakly-nonlinear, quasi-hyperbolic equation and a gauge condition. The quasi-hyperbolic equation can be split into a sequence of wave equations that can be solved exactly for an arbitrary worldline. Similarly, the Lorenz gauge condition, coupled with the assumed expansion of the acceleration, yields another sequence of equations that can be solved to determine successively better approximations for the worldline's equation of motion. We shall see that the assumed expansion of the acceleration is necessary to find a sequence of exactly solvable equations; furthermore, it automatically enforces well-behaved, ``order-reduced" equations of motion.

Within this formalism, I derive the self-force by solving the Einstein equation to second order in $\e$ in the buffer region, where the metric can be conveniently expanded in powers of both $\e$ and $r$. The buffer-region expansion yields an expression for the force in terms of a regular part of the metric perturbation, which is identified with the Detweiler-Whiting regular field. Because the buffer region does not provide sufficient boundary conditions to determine this regular field, I also construct a global representation of the metric perturbation in the outer expansion. Following D'Eath \cite{DEath_paper, DEath}, I write a formal solution to the wave equation by expressing it in an integro-differential form, whereby the value of the metric perturbation at any point in the exterior region is related to an integral over the worldtube around the body. The buffer-region expansion then serves to provide boundary data on the tube, determining the global form of the metric perturbation. Hence, both the perturbation and the equation of motion are determined.

This method can be systematically extended to any order, allowing one to determine higher-order perturbations as well as higher-order equations of motion. Some calculations at second order have already been performed \cite{Eran_field, Eran_force, Lousto_second_order}. In particular, Rosenthal has provided a calculation of the second-order self-force. His method uses a variety of techniques, including matching, an integral representation, and a consideration of all possible well-behaved forms of the self-force. However, he calculates the second-order force in a gauge in which the first-order force identically vanishes. This is consistent only on short timescales; in addition, it is inconvenient for practical purposes. The methods presented here have the potential to self-consistently determine the second-order force within the more convenient Lorenz gauge.

\section{Adiabatic approximations}
Unfortunately, even once a reliable expression for the self-force is available, there remain considerable challenges to surmount. First among these is that a straightforward numerical integration of the coupled linearized Einstein equation and the equation of motion for the small body is computationally difficult: for example, because the simulation must be accurate on the extremely long timescale $\sim 1/\e$, numerical errors must be finely controlled; in addition, if the retarded field is evolved, then calculating the self-force at each time-step involves calculating the singular field and subtracting it from the retarded field \cite{regularization2, regularization3, regularization4, regularization5, regularization6, regularization7}, which is a computationally expensive procedure.

Another challenge is related to data analysis: in order to determine if a signal is present in the data stream of a gravitational wave detector, the data must be compared to a collection of template waveforms, using a method known as matched filtering \cite{parameter_estimation2}. Since the parameter space for an EMRI is extremely large, one needs a large number of templates that can be rapidly correlated with the signal. It is unfeasible to achieve this with a direct numerical integration of the linear EFE and equation of motion. Instead, one requires a simplified approximation that is accurate enough to detect a signal in the data stream, but need not be accurate enough to determine the parameters of the physical system that generated the signal---once a signal is detected, more detailed templates can be used to extract parameters from it.

The \emph{adiabatic approximation} refers to one class of potentially useful simplified approximations. The basic assumption in these approximations is that the secular effects of the self-force occur only on a time scale that is much longer than the orbital period $P$. In an EMRI, this assumption is valid during the early stage of inspiral, which I will refer to as the adiabatic stage; it will break down in the final moments, when the orbit transitions to a quasi-radial infall called the plunge \cite{plunge}. From the adiabaticity assumption, numerous approximations have been formulated: for example, (i) since the particle's orbit deviates only slowly from geodesic motion, the self-force can be calculated from a field sourced by a geodesic; (ii) since the radiation-reaction timescale $t_{rr}$ is much longer than the orbital period, periodic effects of the self-force can be neglected; and (iii), based on  arguments to be discussed momentarily, conservative effects of the self-force can be neglected. Because these three approximations are not generally equivalent, I will refer to them under the distinct titles of \emph{geodesic-source approximation}, \emph{secular approximation}, and \emph{radiative approximation}.

A seminal example of an adiabatic approximation is the Peters-Mathews formalism \cite{Peters_Mathews,Peters}, which determines the long-term evolution of a binary orbit by equating the time-averaged rate of change of the orbital energy $E$ and angular momentum $L$ to, respectively, the flux of gravitational-wave energy and angular momentum at infinity. The essential equations in this formalism are the \emph{energy-balance} equation $\left\langle\diff{E}{t}\right\rangle=-\mathscr{F}$, where $\mathscr{F}$ is the flux of gravitational-wave energy, and the corresponding angular-momentum balance equation. On short timescales, the orbit is assumed to be governed by Newtonian physics; the orbital evolution then consists of a slow transition between different Keplerian elliptical orbits. Hence, the $E$ and $L$ appearing in the balance equations are taken to be those of a Keplerian orbit; this is the analogue of the geodesic-source approximation. Also, the formulas are only meaningful in a time-averaged sense, so periodic changes are neglected; this is the secular approximation. Lastly, since the energy function is calculated for a Newtonian, 0PN orbit, while $\mathscr{F}$ is calculated using the quadrupole formula, which is driven by the 2.5PN radiation-reaction force, 1PN and 2PN conservative post-Newtonian effects in the orbital dynamics are being neglected; this is the radiative approximation. (Here, ``nPN" denotes an order-$(v/c)^{2n}$ correction to Newton's law of gravitation, where $v$ is a typical orbital velocity.)

The Peters-Mathews formalism was used to successfully predict the decreasing orbital period of the Hulse-Taylor pulsar. And higher-PN analogues of it are now utilized in constructing post-Newtonian templates for LIGO data analysis, whether by making direct use of a higher--PN-order balance equation or by making use of the flux to construct averaged dissipative parts of the force \cite{parameter_extraction, PN_comparison,PN_comparison2,PN_comparison3}. (Note that in post-Newtonian literature, the ``adiabatic approximation" refers specifically to the use of the energy balance equation for quasicircular inspirals: the energy function is calculated for a circular orbit, and the angular momentum is ignored, such that the orbital evolution is constrained to slowly transition through a sequence of circular orbits. In this approximation, the energy function is corrected to include conservative post-Newtonian effects; hence, any neglected conservative effects, which might arise due to the restricted phase space in the approximation, should be small. For discussions of the efficacy of this approximation, and alternatives to it, see Refs.~\cite{PN_comparison3,Ajith1,Ajith2,Damour2004}.)

In the hopes of achieving similar success, considerable work has been done to formulate an equivalent approximation in the EMRI problem \cite{Mino, Mino_expansion1, Mino_expansion2, Hughes_adiabatic, Drasco, Drasco_revised, Sago1,Sago2, Nakano,Hughes,Hughes2, Mino_Price, Hinderer_Flanagan}. Bound geodesics in Kerr are specified by the initial position along with three constants of motion---the energy $E$, angular momentum $L$, and Carter constant $C$. Hence, if one could easily calculate the rates of change of these quantities, using a method analogous to the Peters-Mathews formalism, then one could determine an approximation to the long-term orbital evolution of the small body in an EMRI, avoiding the lengthy process of regularization involved in directly integrating the self-forced equation of motion. In the early 1980s, Gal'tsov \cite{work1} showed that the average rates of change of $E$ and $L$, as calculated from balance equations that assume geodesic source motion, agree with the averaged rates of change induced by a self-force constructed from the \emph{radiative} Green's function $G^{\text{rad}}$. The radiative Green's function is defined as $G^{\text{rad}}\equiv\tfrac{1}{2}(G^{\text{ret}}-G^{\text{adv}})$---in other words, it is equal to the flat-spacetime regular Green's function, but it differs from the curved spacetime regular Green's function by the homogenous field $H$.\footnote{Note that this nomenclature differs from that of Ref.~\cite{Eric_review}, in which ``radiative Green's function" refers to the regular Green's function $G^R$.} More recently, Mino \cite{Mino} extended Gal'tsov's result by showing that the true self-force and the radiative self-force cause the same averaged rates of change of all three constants of motion. Since then, the radiative Green's function has been used to derive explicit expressions for the rate of change of $C$ in terms of the particle's orbit and wave-amplitudes at infinity \cite{Sago1,Sago2,Nakano}, and the radiative approximation has been concretely implemented \cite{Hughes,Hughes2}.

Because of the potential usefulness of this type of approximation, determining its potential limitations is of paramount importance. Allow me to briefly outline its implications in the self-force problem. First, in the Lorenz gauge, the geodesic-source approximation is equivalent to writing the force as
\begin{equation}\label{geodesic_force}
a^\mu(t)=-\tfrac{1}{2}(g^{\mu\nu}+u^\mu u^\nu)\left(2\tail_{\nu\alpha\beta}[\gamma\coeff{0}]-\tail_{\alpha\beta\nu}[\gamma\coeff{0}]\right)u^\alpha u^\beta,
\end{equation}
where $\gamma\coeff{0}$ is a geodesic worldline that lies tangential to the true worldline at time $t$. This force is calculated from the field sourced by a fictitious, geodesic past history, rather than from the retarded field sourced by the true past history. We can see that the geodesic-source approximation is closely related to a regular, rather than self-consistent, expansion of the Einstein equation: in a regular expansion, the first-order perturbation is sourced by a geodesic, and the deviation vector pointing toward the corrected worldline is determined by $\tail[\gamma\coeff{0}]$. As the deviation vector grows large, this expansion breaks down. But suppose that rather than using a single reference geodesic, we instead move to a new reference geodesic $\gamma\coeff{0}(t)$ at each instant $t$. Then the error in the force might remain small, so long as $\tail_{\alpha\beta\gamma}[\gamma\coeff{0}(t)] \approx \tail_{\alpha\beta\gamma}[\gamma]$. During the adiabatic portion of the inspiral, the two histories are presumably very similar for a long time in the past, so this approximation is presumably fairly accurate.\footnote{It has also been argued that such an equation of motion is accurate in all contexts \cite{quasilocal}.}

The secular approximation consists of averaging the effects of the force over one orbital period. Note that this is not the same as averaging the acceleration over one period. Suppose we wish to calculate the average rate of change of the orbital energy. This is given by $\left\langle\diff{E}{t}\right\rangle=\langle f a^t\rangle$, where $f$ is some function of the coordinates on the worldline. (See, for example, Appendix~\ref{Killing_osculating}.) In order to remove the oscillations in $\diff{E}{t}$ using this formula, one must know the oscillations in both $f$ and $a^t$; we cannot equate $\langle f a^t\rangle$ to $\langle f\rangle\langle a^t\rangle$. Hence, in order to determine the long term behavior of some quantity, we often require information about the oscillatory behavior of some other quantities. Contrapositively, if one calculates the average rates of change of $E$, $L$, and $C$ using information at asymptotic infinity, then this is not equivalent to removing oscillatory behavior in the self-force.

The radiative approximation neglects conservative effects of the self-force. It may not be obvious how this relates to using the radiative Green's function. Since the metric perturbation depends linearly on the Green's function, and the self-force depends linearly on the perturbation, the radiative Green's function defines a radiative field $h^{\text{rad}}\equiv\tfrac{1}{2}(h^{\text{ret}}-h^{\text{adv}})$ and a radiative force $f^{\text{rad}}\equiv\tfrac{1}{2}(f^{\text{ret}}-f^{\text{adv}})$. One can see that $f^{\text{rad}}$ is a dissipative force from its asymmetry: Under the interchange $x\leftrightarrow x'$, the retarded and advanced Green's functions change as $G^{\text{ret}}(x,x')\to G^{\text{adv}}(x',x)$ and $G^{\text{adv}}(x,x')\to G^{\text{ret}}(x',x)$, such that $G^{\text{rad}}(x,x')\to-G^{\text{rad}}(x',x)$. Since the radiative force is built from the (derivatives of the) Green's function at $x=x'$, under the interchange of past and future we have $f^{\text{rad}}\to-f^{\text{rad}}$. This means that the radiative self-force, and the complete self-force in flat spacetime, has only a dissipative part. In the regular Green's function $G^R$, the homogeneous field $H$ breaks this asymmetry, meaning that the true self-force in curved spacetime has a conservative part along with the dissipative part. And the radiative self-force neglects that conservative part.

In this dissertation, I will address the geodesic-source approximation only in my arguments in favor of the self-consistent equation of motion. However, I will describe in detail the subtleties of the secular approximation, and the limitations of the radiative approximation. The relevance of conservative effects has been analyzed in numerous recent publications \cite{Burko_conservative,our_paper, osculating_paper, other_paper, Hinderer_Flanagan, parameter_estimation_conservative_effects}. As was shown in Refs.~\cite{our_paper, osculating_paper, other_paper} (the results of which I will recapitulate in this dissertation), neglecting the conservative effects of the self-force generically leads to long-term errors in the phase of an orbit and of the gravitational wave it produces. These phasing errors are due to both orbital precession and a direct shift in orbital frequency. The direct shift in frequency can be understood by considering a conservative force acting on a circular orbit: the force will be radial, altering the centripetal acceleration, so at a given radius, the frequency will be altered. In early studies \cite{Burko_conservative,our_paper}, the orbital precession was emphasized, but as will be shown in Ch.~\ref{adiabatic}, the direct shift in frequency typically has a much larger secular effect. Despite these errors, a radiative approximation may still suffice for gravitational-wave detection \cite{Hinderer_Flanagan}; for circular orbits, which have minimal conservative effects, radiative approximations may suffice even for parameter-estimation \cite{parameter_estimation_conservative_effects}. However, at this point in time, the results of these analyses remain inconclusive, since they all rely on extrapolations from post-Newtonian results for the conservative part of the self-force. I refer the reader to Ref.~\cite{Hinderer_Flanagan} for the most comprehensive study of these issues.

\section{Relativistic celestial mechanics: a method of osculating orbits}\label{osculating_intro}
In order to characterize the effects of the self-force---such as the effects neglected by an adiabatic approximation---one requires a useful method of analyzing and characterizing the accelerated orbits. Unfortunately, since the discovery of the MiSaTaQuWa equation, research in this area has focused on calculating the force appropriate for a geodesic-source approximation: the force that \emph{would} be exerted on a particle moving on a given geodesic \cite{regularization1, regularization2, regularization3, regularization4, regularization5, regularization6, regularization7}. Because of the technical complexity and computational demand of the problem, no calculation has yet been performed in which the particle is allowed to travel on its self-accelerated orbit, neither within the geodesic-source approximation nor within a fully self-consistent approximation. For my purposes, I take it as a given that such a calculation can be performed; my interest is in devising a method of analyzing the resulting orbit. The result will hence be useful not solely for the self-force problem, but for arbitrary sources of acceleration.

Analysis of orbital motion in GR has historically focused on geodesic motion or post-Newtonian motion. For example, bound geodesics in Kerr have been written in closed form in terms of $E$, $L$, and $C$, in terms of generalizations of eccentricity and other Keplerian parameters \cite{Kerr_parametrization}, and in terms of action-angle variables \cite{action_angle}. Similarly, for post-Newtonian binaries, the conservative portion of the equations of motion has been solved analytically, and the solutions have been usefully parametrized in terms of generalized Keplerian-type parameters \cite{Damour,Damour2004}. However, comparatively little work has been done to characterize accelerated orbits in curved spacetimes. 

In this dissertation, I present a method of characterizing accelerated orbits in any spacetime in which the geodesic equation is integrable. My method is a relativistic extension of the most traditional method of Newtonian celestial mechanics, known variously as the \emph{method of osculating orbits}, the method of variation of constants, or the method of variation of orbital elements (see, e.g., Refs.~\cite{Taff, Beutler}). In this method, the true worldline $z(\lambda)$ is taken to lie tangent to a geodesic $z_G(\lambda)$ at each value of the orbital parameter $\lambda$, such that the true orbit moves smoothly from one geodesic to the next. The instantaneously tangential geodesics are referred to as osculating (meaning ``kissing") orbits. So long as the geodesic equation is integrable, a geodesic can be written in terms of a set of constants $I^A$, called \textit{orbital elements}, and the transition between osculating orbits corresponds to changes in these elements; thus, the method of osculating orbits amounts to parameterizing the true worldline as an evolving geodesic with dynamical orbital elements $I^A(\lambda)$.

Besides its historical relevance, this method has several practical advantages. First, because the orbital elements are constant on a geodesic, the method clearly separates perturbative from non-perturbative effects. (``Perturbative" in this context merely means ``caused by the acceleration"; the ``perturbation" need not be small.) Second, although the orbital elements are equivalent to the set of initial conditions, they are typically chosen so as to provide direct geometric information about the orbit. If the perturbing force is very weak, then the perturbed orbit will lie very close to a geodesic for a long period of time, and changes in the orbital elements will characterize changes in the geometry of the orbit. Thus, although my method is exact, it is most useful in the context of small perturbations. Third, the orbital elements divide into two classes. The first class consists of \textit{principal} orbital elements; these are equivalent to constants of motion such as energy and angular momentum, and they determine the geodesic on which the particle is moving. The second class consists of \textit{positional} orbital elements, which determine the particle's initial position on the selected geodesic, as well as the geodesic's spatial orientation. Generally speaking, long-term changes in the principal orbital elements are produced by dissipative terms in the perturbing force, while long-term changes in the positional elements are produced by conservative terms. (Although Mino has given prescriptions for finding these long-term changes using only the radiative self-force \cite{Mino, Mino_expansion1}, his prescriptions are highly ambiguous in practice \cite{my_thesis}.) Thus, this division into two classes allows one to easily separate conservative from dissipative effects of the perturbing force. Lastly, because it explicitly determines the position and velocity of a tangential geodesic at each instant, the method of osculating orbits explicitly provides the information necessary to implement a geodesic-source approximation in the gravitational self-force problem. These factors make this method an excellent means of testing the utility of adiabatic approximations.

For simplicity, I derive explicit evolution equations for the orbital elements only in a restricted case: I specialize to orbits in Schwarzschild, to orbits bounded between a minimum and a maximum radius (the method is not suitable for the final plunge in an EMRI), and to forces acting within the plane of the orbit. Within these restrictions the force is arbitrary, and in particular, it is not necessarily small. Furthermore, the method could be easily extended to accommodate non-planar motion, or even to orbits in Kerr (using the parametrization of geodesics presented in Ref.~\cite{Kerr_parametrization}).

Prior to my work, there have been at least two notable generalizations of the method of osculating orbits from Newtonian to relativistic mechanics: the adaptation of the method by Damour et~al. to post-Newtonian binary systems \cite{Damour, Damour2004}, and the formulation proposed by Mino for orbits around a Kerr black hole \cite{Mino_expansion1}. The formulation by Damour et~al. is complete and easy to implement, but it is limited to the post-Newtonian regime. Mino's formulation is valid for arbitrary bound orbits in Kerr, but it relies on a rather complicated (though still practical \cite{Drasco_private}) Fourier expansion. I go beyond those presentations by providing a completely general derivation of the evolution equations for orbital elements in a curved spacetime. In addition, when I restrict my results to orbits in Schwarzschild, I utilize a particularly simple parametrization that is both easy to implement and carries intuitive geometric meaning.
   
\subsection{Test cases}
In this dissertation, I implement the method of osculating orbits in two simple test cases. In both cases I focus primarily on emphasizing the limitations of secular and radiative approximations. The first test case I consider consists of a particle, subject to its self-fore, in a weak central gravitational potential; the potential is identical to that of Newtonian gravity, and I hence use the Newtonian limit of my evolution equations for the orbital elements. In this context, I present a two-timescale expansion of the evolution equations, which allows a general analysis of how the conservative and dissipative parts of the self-force contribute to the orbital evolution. In appendix \ref{multiscale_EFE}, I implement the expansion for the particular case of a charged particle subject to its electromagnetic self-force.

The second test case is a system of two compact bodies of mass $m_1$ and $m_2\gg m_1$ in the post-Newtonian regime. To analyze this system of equations with my method of osculating orbits, I use the hybrid equations of motion devised by Kidder, Will, and Wiseman~\cite{Kidder}. These equations take the schematic form  
\begin{equation}
\ddiff{x^a}{t}=-\frac{M}{r^2}\left( 1 + \text{\sc{Schw}}+ \mu \text{\sc{PF}} \right). 
\end{equation}
The spatial separation vector $x^a = x^a_1 - x^a_2$ connects the two bodies, and $M = m_1 + m_2 $ and $\mu = m_1 m_2/M$ are respectively the total mass and reduced mass of the system. The terms on the right-hand side are as follows: the first term is the usual Newtonian gravitational force; the Schwarzschild term $\mbox{\sc{Schw}}$ contains the exact relativistic corrections to Newton's law in a Schwarzschild spacetime of mass $M$, so that $\ddiff{x^a}{t}=-\frac{M}{r^2}\left(1+\text{\sc{Schw}}\right)$ is the exact geodesic equation in that spacetime; the perturbing force $\mu \mbox{\sc{PF}}$ comprises the terms in the post-Newtonian expansion that depend  explicitly on the reduced mass of the system. Since the non-PN terms contained in $\text{\sc{Schw}}$ are of 3PN order and higher, the hybrid equations agree with a post-Newtonian expansion at 2.5PN order. However, they differ from the usual post-Newtonian equations in that they become exact in the test-mass limit $\mu \to 0$. This allows me to apply my method to the post-Newtonian system by taking my osculating orbits to be geodesics in the fictitious Schwarzschild spacetime of mass $M$, and by deriving the perturbing force from $\mu\text{\sc{PF}}$. Unlike in the electromagnetic case, in this case I integrate the evolution equations numerically. My results emphasize not only the long-term impact of the conservative force, but also the long-term impact of the choice of initial conditions.

\section{Organization of this dissertation}
The dissertation contains three main parts: an overview of the foundations of the self-force problem; a calculation of the self-consistent solution to the problem; and an analysis of accelerated motion and the adiabatic approximation to it.

The first part begins in Ch.~\ref{approximations} with a discussion of singular perturbation techniques. I first review those techniques as they are traditionally used: in a fixed coordinate system, with fixed boundary conditions. I then discuss their generalization to GR. My presentation follows that of Kates \cite{Kates_structure}, but with more precise delineations of various types of general expansions. I particularly emphasize the conclusions that can be drawn from the utilization of these expansions; in the case of matched asymptotic expansions, for example, the conclusions will typically be much weaker in GR than in theories with a fixed geometry. I conclude the chapter with a discussion of a fixed-worldline expansion.

The first part continues in Chapters~\ref{point_particle} and \ref{extended_body} with a more thorough explication of the self-force problem within the context of singular perturbation theory. This explication serves two purposes: first, to review both the foundations of the problem and the various derivations in the literature. Much of this review overlaps with previous discussions by Mino \cite{Mino_expansion1, Mino_expansion2, Mino_Price}, Hinderer and Flanagan \cite{Hinderer_Flanagan}, and Gralla and Wald \cite{Gralla_Wald}. However, my presentation differs significantly from those discussions, and it serves to motivate and provide the necessary context for my own approach. The second purpose of the explication is to introduce the notion of a self-consistent expansion in which the metric perturbation is first written as a functional of a worldline and then expanded while holding the worldline fixed. Chapter~\ref{point_particle} presents this expansion in the context of a point particle; Chapter~\ref{extended_body} generalizes it to asymptotically small bodies. After laying that groundwork, the first part of the dissertation concludes in Sec.~\ref{outline} with an outline of the sequences of perturbation equations that must be solved in my self-consistent expansion.

The second part of the dissertation, comprising Chs.~\ref{matching}--\ref{perturbation calculation}, discusses the solution to those equations. In Ch.~\ref{matching}, I discuss a solution using the method of matched asymptotic expansions. I find that very strong conditions must be imposed, on both the form of the inner expansion and its relationship to the outer one, in order to derive useful results. Because this method yields a relatively weak conclusion, I then move on to a less restrictive approach, which makes minimal assumptions about the inner expansion. Chapter~\ref{buffer_region} presents my derivation of the first-order gravitational self-force in terms of the regular field $h^R$. At the end of that chapter, I discuss the interpretation of the field and the acceleration, and I analyze the gauge-dependence of the acceleration. In Sec.~\ref{perturbation calculation}, I calculate the global metric perturbation induced by the body, which determines $h^R$ in terms of tail integrals and recovers the usual MiSaTaQuWa equation.

The third part of the dissertation, comprising Chs.~\ref{osculating} and \ref{adiabatic}, presents my method of osculating orbits and the limitations of adiabatic approximations. Chapter~\ref{osculating} presents the general method of osculating orbits, as well as its application to bound planar orbits in Schwarzschild spacetime. Chapter~\ref{adiabatic} utilizes the method to characterize the limitations of adiabatic approximations in the two test cases: a charged particle in a weakly curved spacetime, and a post-Newtonian binary.

Chapter~\ref{conclusion} concludes the dissertation with a summary of the various approximation techniques discussed throughout the preceding chapters. This includes a comparison to some alternative methods, and a discussion of what is required for derivations of higher order, and globally accurate, approximations.

Many of my calculations utilize a variety of methods that are standard in the self-force literature. For the most part, I confine the description of these methods to appendices. Various calculations, particularly those of unseemly length and those that do not play a large role in the body of the dissertation, are also relegated to appendices. These appendices will be referred to as necessary.

			\chapter{Perturbation theory}\label{approximations}

Perturbation theory is a venerable field of study in GR. In fact, because of the complexity of the field equations, most physically relevant analytical results in GR rely on perturbing away from a known solution. However, most of the foundational work in this area has focused only on descriptions of regular perturbation problems. The underlying formalism of regular perturbation theory has been studied extensively \cite{Geroch_limits, Stewart_limits, Bruni_limits}, and it has been shown that any regular asymptotic expansion of the field equations yields a perturbative solution that approximates at least one exact solution (at least locally) \cite{perturbation_theory1,perturbation_theory2,Rendall_review}. Unfortunately, numerous perturbation problems are singular. Indeed, what may be the single most successful area of research in GR, post-Newtonian theory, centers on a singular perturbation problem. As one would expect, the foundations of that particular singular problem have been studied extensively \cite{Rendall,PN_existence_review}, and it is now known that there exist a large class of exact solutions possessing Newtonian and post-Newtonian expansions \cite{PN_existence,PN_existence1,PN_existence2,PN_existence3}. But general discussions of common singular perturbation techniques are lacking.

Given this situation, in this chapter I provide an overview of the foundations of singular perturbation theory in GR, emphasizing the techniques relevant for the self-force problem. I begin with a review of singular perturbation methods in applied mathematics. More detailed overviews of the subject can be found in numerous textbooks (e.g., Refs.~\cite{Verhulst, Holmes, Kevorkian_Cole, Lagerstrom,Eckhaus}). Among these, the text by Kevorkian and Cole \cite{Kevorkian_Cole} covers the broadest range of topics, and the text by Eckhaus \cite{Eckhaus} provides the most rigorous treatment.

\section{Traditional perturbation theory}
%%%%%%%%%%%%
I begin by defining some useful notation. First, I define the following order symbols: for $x\in\mathbb R^n$,
\begin{itemize}
\item $f(x,\e)=\order{\zeta(\e)}$ if there exist positive constants $k$ and $\e^*$ such that $|f(x,\e)|\le k|\zeta(\e)|$ for fixed $x$ and $0\le\e\le\e^*$.
\item $f(x,\e)=o(\zeta(\e))$ if $\displaystyle\lim_{\e\to0}\frac{f(x,\e)}{\zeta(\e)}=0$ at fixed $x$.
\item $f(x,\e)=O_s(\zeta(\e))$ if $f(x,\e)=O(\zeta(\e))$ and $f(x,\e)\ne o(\zeta(\e))$.
\end{itemize}
For example, $5\e+2\e^{3/2}=\order{\e}=2\e^{3/2}$, $5\e+2\e^{3/2}\ne o(\e)=2\e^{3/2}$, and $5\e+2\e^{3/2}=O_s(\e)\ne 2\e^{3/2}$.

In general, we are concerned with the asymptotic behavior of functions, rather than the behavior of functions evaluated at particular locations. This means we need a norm appropriate for a function. Also, a central issue in perturbation theory is whether or not an approximation is uniformly accurate in a region of interest, where uniformity is defined as follows:
\begin{itemize}
\item $f(x,\e)=\order{\zeta(\e)}$ \emph{uniformly} in a region $D$ if there exist positive constants $k$ and $\e^*$ such that $||f(x,\e)||_D\le k|\zeta(\e)|$ for $0\le\e\le\e^*$, where $||\cdot||_D=\sup_{x\in D}|\cdot|$.
\end{itemize}
Analogous definitions hold for $o$ and $O_s$. These definitions provide a more useful measure of the asymptotic behavior of a function.

Finally, I define several important asymptotic quantities: relative to an asymptotic sequence $\lbrace\zeta_n(\e)\rbrace$, where $\zeta_{n+1}(\e)=o(\zeta_n(\e))$,
\begin{itemize}
\item $f(x,\e)$ is an $N$th-order \emph{asymptotic approximation} of $\exact{f}(x,\e)=O_s(1)$ if $f(x,\e)-\exact{f}(x,\e)=o(\zeta_N(\e))$,\footnote{In the case $\exact{f}(x,\e)=O_s(\zeta_k(\e))$, functions would be rescaled by $\zeta_k(\e)$ before making comparisons.}
\item $f(x,\e)$ is an $N$th-order \emph{asymptotic solution} to a differential equation $\exact{D}[\exact{f}(x,\e)]=0$ if $\exact{D}[f(x,\e)]=o(\zeta_N(\e))$,
\item $f(x,\e)=\displaystyle\sum_{n=0}^N \zeta_n(\e) f\coeff{\emph{n}}(x,\e)$, where $f\coeff{\emph{n}}(x,\e)=O_s(1)$, is an $N$th-order \emph{asymptotic series}. If $f\coeff{\emph{n}}$ is independent of $\e$, then the series is said to be \emph{regular} (sometimes called Poincar\'e-type); if not, then it is \emph{general}. If, in addition, $f$ is an asymptotic approximation to $\exact{f}$, then it is an $N$th-order \emph{asymptotic expansion} of $\exact{f}$.
\end{itemize}
The most common asymptotic sequence is $\lbrace\e^n\rbrace$, which I will use almost exclusively in this dissertation. Note that we are typically uninterested in whether or not an asymptotic series converges as $N\to\infty$. In fact, even if $f$ is an asymptotic series that both converges and asymptotically approximates $\exact{f}$, the function that it converges to might not be $\exact{f}$.

In any given perturbation calculation, one almost always calculates an asymptotic solution to an equation. Determining whether or not an asymptotic solution is also an asymptotic approximation to an exact solution is typically far more difficult. It is, however an essential step in proving the reliability of an expansion, because an asymptotic solution may not be an asymptotic approximation to an exact solution (more pathologically, an asymptotic approximation may not be an asymptotic solution \cite{Verhulst}).

General asymptotic expansions are a powerful tool for solving singular perturbation problems, which are defined by the failure of a regular expansion to provide a uniform approximation. This failure is often signaled by a change of character in the governing differential equation when $\e\to0$: for example, a hyperbolic equation might degenerate into a parabolic equation. The inaccuracy of a regular expansion is also frequently signaled by its failure to satisfy a given boundary condition, or by the expansion growing without bound in a system that we have reason to believe should be bounded. Typically, the underlying origin of the failure is the presence of distinct length scales, one of which appears only for $\e>0$. Often, this means that the exact solution to a problem is singular at $\e=0$; hence, in singular perturbation problems, we assume that $\e\in(0,\e^*]$, which allows us to take the limit $\e\to 0$, but which prevents us from setting $\e=0$. In this chapter, I consider two types of systems: first, systems in which the exact solution undergoes a rapid change near a submanifold; second, systems in which rapid changes occur throughout the region of interest. In the first type of system, the method of matched asymptotic expansions can be used to construct a uniform general expansion; in the second type, the method of multiple scales can be used.

Before proceeding, I define two final bits of notation. $\trans_*$ and $\trans^*$ denote, respectively, the push-forward and pull-back corresponding to a map $\trans$. So, for example, if $f$ is a function of coordinates $x$, and $\tilde x=\trans(x)$, then $\trans_* f$ is the function rewritten in terms of $\tilde x$. $\expand^N_\e\exact{f}$ denotes the $N$th-order regular asymptotic expansion of $\exact{f}$ in the limit of small $\e$, holding fixed the coordinates of which $\exact{f}$ is a function. 

\subsection{Matched asymptotic expansions}
\begin{figure}
\begin{center}
\includegraphics{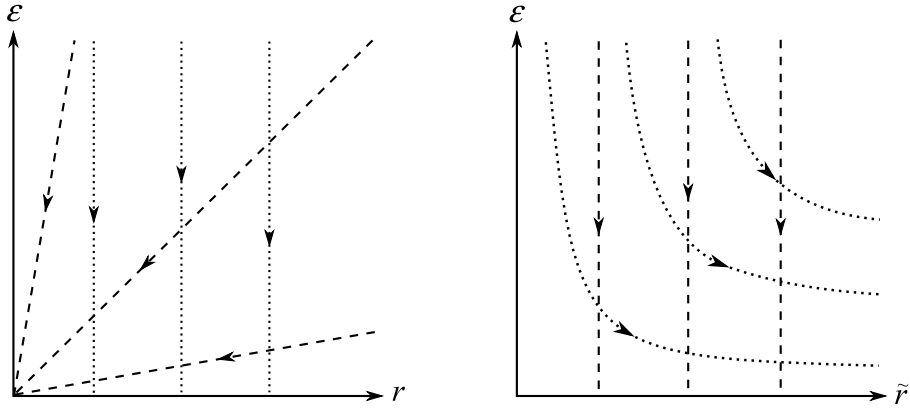}
\end{center}
\caption[Inner and outer limits]{Left: inner limit (dashed curves) and outer limit (dotted curves) in the $(r,\e)$-plane. Right: the same limits in the $(\tilde r, \e)$-plane. The inner limit is defined by $\e\to0$, $\tilde r=r/\e$ fixed; the outer limit, by $\e\to0$, $r$ fixed. From the perspective of the inner limit, the outer limit sends all points to infinity ($\tilde r\to\infty$). From the perspective of the the outer limit, the inner limit sends all points to zero ($r\to 0$).}
\label{limits}
\end{figure}

Matched asymptotic expansions are typically used to solve boundary value problems in which the solution exhibits rapid change in a very small region (or a finite number of such regions). The regions of rapid change are referred to as boundary layers. Frequently, this rapid change prevents a regular expansion from satisfying a given boundary condition, though a ``boundary layer" can sometimes arise away from any boundary. The usual means of solving these problems is to make use of two regular expansions: an inner expansion $f_{\text{in}}$ that is expected to be valid in the boundary layer, and an outer expansion $f_{\text{out}}$ that is expected to be valid outside of it. Suppose we have a one-dimensional problem with coordinate $r$, and that the boundary layer is at $r=r_b$ and has a thickness $\sim\delta(\e)$. Then the outer expansion is simply a regular series at fixed $r$, and the inner expansion is a regular series at fixed values of the rescaled coordinate $\tilde r \equiv (r-r_b)/\delta(\e)$; this can be written as $\trans_\e:r\mapsto\tilde r$. The inner expansion allows us to capture changes over the lengthscale $\delta(\e)$, since $\tilde r$ is of order unity when the original coordinate $r$ is of order $\delta(\e)$. Note that if we treat the problem on a two-dimensional plane with coordinates $(r,\e)$, then the inner and outer expansions can be visualized as expansions along flow lines defined by $r=$constant and $r/\e=$constant, as shown in Fig.~\ref{limits}.

For simplicity, suppose that $r_b=0$ and that boundary data is given at $r=0$ and $r=1$. In this case, the outer expansion typically fails to satisfy the boundary condition at $r=0$, but it can be made to satisfy the condition at $r=1$; conversely, the inner expansion can satisfy only the condition at $r=0$. This leaves each of the expansions underdetermined. The basic idea of matched asymptotic expansions is to fully determine the inner and outer expansions by insisting that they agree in some region of mutual validity. Suppose that $f_{\rm in}$ is an $N$th-order asymptotic approximation of $\exact{f}$ in a region $D_{\text{in}}$, and $f_{\text{out}}$ is an $N$th-order asymptotic approximation in a region $D_{\rm out}$. Then by definition, $\exact{f}-f_{\text{in}}=o(\zeta_N(\e))$ in $D_{\text{in}}$ and $\exact{f}-f_{\text{out}}=o(\zeta_N(\e))$ in $D_{\text{out}}$. Subtracting the second equation from the first, we have the \emph{overlap matching condition}:
\begin{equation}
f_{\text{out}}-f_{\text{in}} = o(\zeta_N(\e)) \text{ in } D_{\text{out}}\cap D_{\text{in}}.
\end{equation}

Note that this condition relies on the existence of the \emph{overlap region} $D_{\text{out}}\cap D_{\text{in}}$. If that region is empty, then the condition is vacuous. And it may not be obvious that such a region ever exists, since the inner expansion trivially appears to be valid only for $\tilde r\sim 1$, and the outer expansion only for $r\sim 1$. However, if $f_{\text{out}}(r)$ is a uniform asymptotic approximation to $\exact{f}(r)$ on an interval $[a,1]$ for constant $a$, then it is also a uniform approximation on the extended interval $[\zeta_i(\e),1]$ for some $\zeta_i(\e)=o(1)$; similarly, if $f_{\text{in}}(\tilde r)$ is a uniform approximation to $\exact{f}(\tilde r)$ on $[0,b]$, then it is a uniform approximation on the extended interval $[0,1/\zeta_j(\e)]$ for some $\zeta_j(\e)=o(1)$ \cite{Lagerstrom, Eckhaus}. If we have access to the exact function $\exact{f}$, then we can explicitly determine the overlap of these extended regions. But in a typical application, without access to an exact solution, we must make use of the \emph{overlap hypothesis}, which states that the overlap region exists. In order to implement the overlap matching condition, one then simply assumes that the constructed asymptotic series of a given order are asymptotic approximations of the same order, and one then takes the overlap region to be the region in which $f_{\text{out}}-f_{\text{in}} = o(\zeta_N(\e))$.

In this thesis, I will not make direct use of the overlap matching condition. Instead, I will use a second, simpler matching condition, which I will refer to as the \emph{coefficient-matching condition}:
\begin{equation}\label{matching_equation}
\expand^k_{\e}\trans^*\expand^m_{\e}\trans_*\exact{f} = \expand^k_{\e}\trans^*\expand^m_{\e}\trans_*\expand^m_{\e}\exact{f}.
\end{equation}
In this matching condition, we match results term-by-term in the expansions. On the left-hand side we have the inner expansion ($\expand^m_{\e}\trans_*\exact{f}$) written as a function of $r$ (via $\trans^*$) and then expanded in the outer limit; on the right-hand side, we have the outer expansion ($\expand^m_{\e}\exact{f}$) written as a function of $\tilde r$ (via $\trans_*$) and expanded in the inner limit, and then rewritten as a function of $r$ and re-expanded. (The right-hand side requires an extra expansion in order to remove terms that would appear as higher-order terms in the inner expansion; this will be illustrated in an example in the following section.) We can write this schematically as
\begin{equation}
\expand_\e f_{\text{in}}(r)=\expand_r f_{\text{out}},
\end{equation}
meaning that when the inner expansion is re-expanded for small $\e$ at fixed $r$, and the outer expansion is re-expanded for small $r$ at fixed $\e$, the two results must agree term-by-term. We can then, for example, equate coefficients of $\e^n r^m$ on the left- and right-hand sides.

If we define the \emph{buffer region} by the inequalities $\e\ll r\ll 1$,\footnote{In the applications of matched asymptotic expansions in GR, the meanings that I have assigned to the terms ``overlap region" and ``buffer region" are often conflated, and the terms are often used interchangeably. For the sake of clarity, I distinguish between the two.} this equation states that the inner and outer expansions must agree term-by-term when they are both expanded in the buffer region; in other words, if the exact solution is expanded first for small $\e$ at fixed $r/\e$ (yielding an inner expansion), and then expanded at fixed $r$ (or in other words, for $r\gg\e$), it must agree, term-by-term, with the result of expanding first for small $\e$ at fixed $r$ and then expanding for $r\ll 1$.

From the perspective of the inner limit, the buffer region lies at asymptotic infinity ($\tilde r\to\infty$); from the perspective of the outer expansion, it lies asymptotically close to $r=0$. From this we can intuit a still simpler matching condition, which I will call the \emph{asymptotic matching condition}:
\begin{equation}
\lim_{\tilde r\to\infty}f\coeff{0}_{\text{in}}=\lim_{r\to0}f\coeff{0}_{\text{out}},
\end{equation}
where $f\coeff{0}_{\text{in}}$ and $f\coeff{0}_{\text{out}}$ are the leading-order terms in, respectively, the inner and outer expansions.

The three matching conditions I have discussed are obviously related. In fact, one can derive the asymptotic matching condition and (a condition similar to) the coefficient-matching condition from the overlap hypothesis. However, one should realize that the overlap hypothesis is merely sufficient to arrive at those two matching conditions: it is not necessary. Functions exist that do not satisfy the overlap hypothesis but nevertheless satisfy the coefficient-matching condition, for example \cite{Eckhaus}.

Once a matching condition has been used to fully determine the inner and outer expansions, one can construct a \emph{composite expansion} that is uniformly accurate on the full domain of the problem. This expansion consists of the sum of the inner and outer expansions, minus the terms that are common to both in the buffer region. Explicitly,
\begin{equation}\label{composite}
f_{\text{comp}} = \expand_\e^m\exact{f} + \trans^*\expand_\e^m\trans_*\exact{f} - \trans^*\expand_\e^m\trans_*\expand_\e^m\exact{f},
\end{equation}
which we can write schematically as
\begin{equation}
f_{\text{comp}} = f_{\text{out}}+f_{\text{in}}-\expand_r f_{\text{out}}.
\end{equation}
Note that this is a general asymptotic expansion of the form $\sum_n\zeta_n(\e)F\coeff{n}(r,r/\delta(\e))$.

\subsubsection{Example}
To illustrate the procedure of matched asymptotic expansions, I consider the following boundary value problem:
\begin{equation}\label{matchDE}
\e\ddiff{\exact{f}}{r}+\diff{\exact{f}}{r}+\exact{f}=0, \qquad \exact{f}(0)=0,\ \exact{f}(1)=1.
\end{equation}
I assume that $\exact{f}$ possesses an outer expansion $f_{\text{out}}=f_{\text{out}}\coeff{0}(r)+\e f_{\text{out}}\coeff{1}(r)+...$. Substituting this expansion into Eq.~\eqref{matchDE} and equating coefficients of each power of $\e$ to zero, we find $\diff{f_{\text{out}}\coeff{0}}{r}+f_{\text{out}}\coeff{0}=0$ and $\diff{f_{\text{out}}\coeff{1}}{r}+f_{\text{out}}\coeff{1}=-\ddiff{f_{\text{out}}\coeff{0}}{r}$; the boundary conditions are $f_{\text{out}}\coeff{0}(0)=f_{\text{out}}\coeff{1}(0)=0$, $f_{\text{out}}\coeff{0}(1)=1$, and $f_{\text{out}}\coeff{1}(1)=0$. The general solution to the zeroth-order differential equation is $f_{\text{out}}\coeff{0}(r)=C\coeff{0}e^{-r}$. This solution can satisfy the boundary condition at $r=1$ (by setting $C\coeff{0}=e$), but it cannot satisfy the condition at $r=0$. Hence, we guess that there is a boundary layer at $r=0$, and we choose only to satisfy the boundary condition at $r=1$. Doing the same for $f_{\text{out}}\coeff{1}$, we find $f_{\text{out}}\coeff{1}=(1-r)e^{1-r}$, yielding the first-order outer expansion
\begin{equation}
f_{\text{out}}= e^{1-r}+\e(1-r)e^{1-r}+...
\end{equation}

Now, in order to construct our inner expansion, we require a choice of rescaled coordinate $\tilde r$. Suppose that we choose $\tilde r=r/\e^p$. Substituting this into Eq.~\eqref{matchDE} and taking the limit $\e\to 0$, we find that if $0<p<1$, then the leading-order differential equation is $\diff{f_{\text{in}}\coeff{0}}{\tilde r}=0$. If $p>1$, then the equation becomes $\ddiff{f_{\text{in}}\coeff{0}}{\tilde r}=0$. And if $p=1$, then it becomes $\ddiff{f_{\text{in}}\coeff{0}}{\tilde r}+\diff{f_{\text{in}}\coeff{0}}{\tilde r}=0$. This is called a distinguished limit (also known as a significant degeneration) of the equation, because it contains within it all the terms appearing in the other two limiting equations. We can intuit that a distinguished limit will yield an approximation with maximal information. And although there is no guarantee that a coordinate leading to a distinguished limit is the ideal choice, it has proven to be the most reliable one.

So, proceeding with the rescaled variable $\tilde r=r/\e$, I rewrite Eq.~\eqref{matchDE} as
\begin{equation}\label{matchDE2}
\ddiff{\exact{f}}{\tilde r}+\diff{\exact{f}}{\tilde r}+\e\exact{f}=0, \qquad \exact{f}(0)=0,
\end{equation}
where, technically, $\exact{f}$ stands in for $\trans_*\exact{f}$. I assume that $\exact{f}(\tilde r)$ possesses an inner expansion $f_{\text{in}}(\tilde r,\e)=f_{\text{in}}\coeff{0}(\tilde r)+\e f_{\text{in}}\coeff{1}(\tilde r)+...$. After substituting this into Eq.~\eqref{matchDE2} and solving order-by-order, we find
\begin{equation}
f_{\text{in}}= D\coeff{0}(1-e^{-\tilde r})+\e\left[D\coeff{1}(1-e^{-\tilde r})-D\coeff{0}\tilde r(1+e^{-\tilde r})\right]+...
\end{equation}

We can now make use of one of the matching conditions to determine the integration constants $D\coeff{0}$ and $D\coeff{1}$. Since $\lim_{\tilde r\to\infty}f_{\text{in}}(\tilde r)\coeff{0}=D\coeff{0}$ and $\lim_{r\to0}f_{\text{out}}\coeff{0}=e$, the asymptotic matching condition implies $D\coeff{0}=e$. In order to determine $D\coeff{1}$, I next make use of the coefficient-matching condition. Rewriting $f_{\text{in}}$ as a function of $r$ and expanding to order $\e$, we find
\begin{equation}\label{fin buffer}
f_{\text{in}}(r)=e+\e D\coeff{1}-er+...
\end{equation}
Note that $e^{-r/\e}=o(\e^n)$ for all $n>0$, so it vanishes in this expansion. Dropping the ellipses, this yields the left-hand side of Eq.~\eqref{matching_equation}:
\begin{equation}
\expand^1_{\e}\trans^*\expand^1_{\e}\trans_*\exact{f}= e+\e D\coeff{1}-er.
\end{equation}
Next, expanding $f_{\text{out}}$ to linear order in $r$, we find
\begin{equation}
f_{\text{out}} = e(1-r)+\e(1-2r)e+...
\end{equation}
This expansion contains an order-$\e r$ term, which is smaller than any term in Eq.~\eqref{fin buffer}; such a term would be matched by a term from $f\coeff{2}_{\text{in}}$, and so we can neglect it here. The extra expansion on the right-hand side of Eq.~\eqref{matching_equation} serves to remove such terms, and so we have
\begin{equation}
\expand^1_{\e}\trans^*\expand^1_{\e}\trans_*\expand^1_{\e}\exact{f} = e(1-r)+e\e.
\end{equation}
Hence, the matching condition now determines that $D\coeff{1}=e$, and we have fully determined the inner expansion:
\begin{equation}
f_{\text{in}}= e(1-e^{-\tilde r})+\e e\left[(1-e^{-\tilde r})-\tilde r(1+e^{-\tilde r})\right]+...
\end{equation}

Using these results, we can construct the uniformly accurate composite expansion
\begin{align}
f_{\text{comp}} &= \left\lbrace e^{1-r}+\e(1-r)e^{1-r}\right\rbrace+\left\lbrace e(1-e^{-r/e})+\e e\left[(1-e^{-r/e})-r/\e(1+e^{-r/\e})\right]\right\rbrace \nonumber\\
&\quad - \left\lbrace e(1-r)+e\e \right \rbrace \nonumber\\
& = e^{1-r}-(1+r)e^{1-r/\e}+\e\left[(1-r)e^{1-r}-e^{1-r/\e}\right],
\end{align}
where the first equality should be compared to Eq.~\eqref{composite}.

\begin{figure}
\begin{center}
\includegraphics[width=2.9 in, height = 2.9 in]{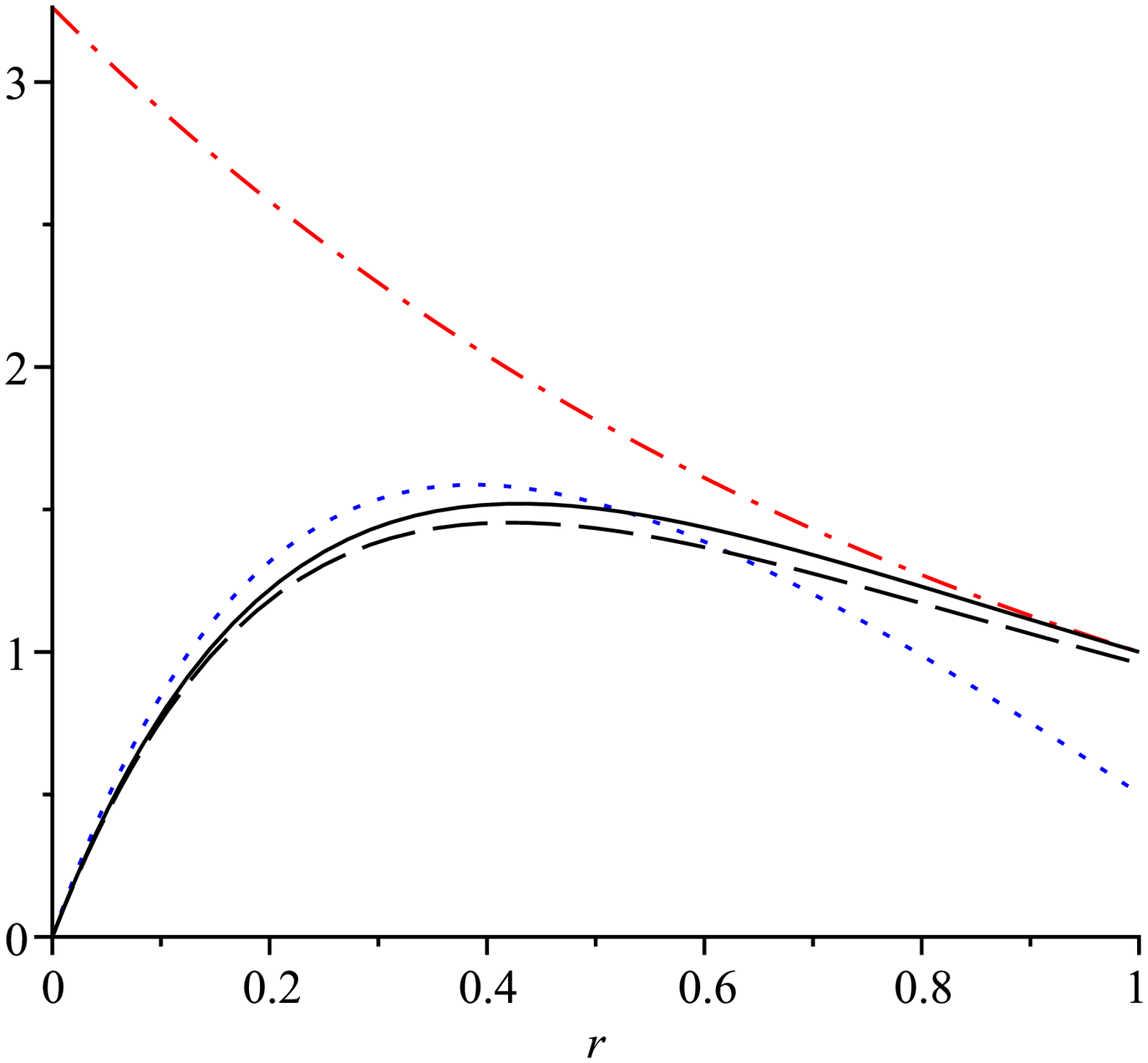}
\includegraphics[width=2.9 in, height = 2.9 in]{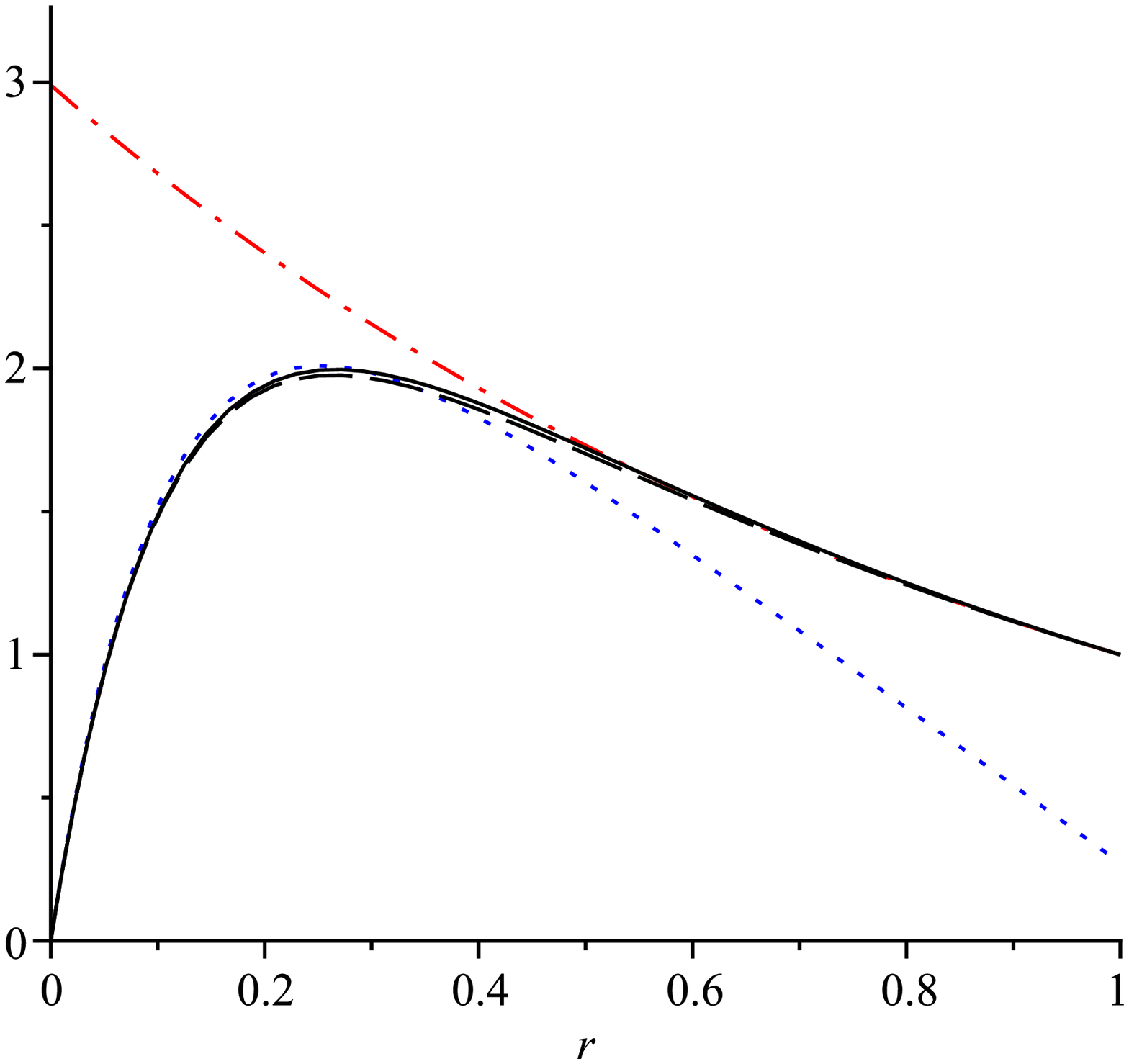}
\end{center}
\caption[An example of matched asymptotic expansions]{Comparisons of the exact solution $\exact{f}$ (the solid black curve), the inner solution $f_{\text{in}}$ (dotted blue), the outer solution $f_{\text{out}}$ (dot-dashed red), and the composite solution $f_{\text{comp}}$ (dashed black). The left plot displays the solutions for $\e=0.2$; the right, for $\e=0.1$.}
\label{matching_example}
\end{figure}

In this case, we can compare our results to the exact solution to Eq.~\eqref{matchDE}, which is given by
\begin{equation}
\exact{f}=\frac{\exp\left[-(1-\sqrt{1-4\e})\frac{r}{2\e}\right]-\exp\left[-(1+\sqrt{1-4\e})\frac{r}{2\e}\right]}{\exp\left[-(1-\sqrt{1-4\e})\frac{1}{2\e}\right]-\exp\left[-(1+\sqrt{1-4\e})\frac{1}{2\e}\right]}.
\end{equation}
Note that this function does not exist at $\e=0$. Hence, the regular series $f_{\text{out}}$ is not a Taylor series expansion of $\exact{f}$. However, the limit $\lim_{\e\to0}\exact{f}$ does exist, and $f_{\text{out}}$ is given by $\sum\frac{\e^n}{n!}\lim_{\e\to 0}\frac{\partial^n\exact{f}}{\partial\e^n}$.
One can straightforwardly check that $f_{\text{in}}$ (when written as a function of $r$) is a uniform, first-order asymptotic approximation on an extended domain $D_{\text{in}}=\lbrace r\,:\, 0\le r\ll 1\rbrace$, and  $f_{\text{out}}$ is a uniform, first-order approximation on $D_{\text{out}}=\lbrace r\,:\,|\e\ln\e|\ll r\le 1\rbrace$. So at first order, the overlap region exists, and it is given by $|\e\ln\e|\ll r\ll 1$ (which is notably smaller than the buffer region). One can also verify that $f_{\text{comp}}$ is a uniform approximation on the whole interval $[0,1]$. Figure~\ref{matching_example} shows a graphical comparison of the exact, inner, outer, and composite solutions.

\subsection{The method of multiple scales}\label{multiple scales}
In some systems, rather than a rapid change occurring near a submanifold, rapid changes occur over the entire region of interest (in other words, the ``boundary layers" are dense in the region). When studying these systems, one cannot make use of regular expansions in separate regions and then combine them to form a uniform approximation. Instead, in order to arrive at a uniform approximation, one must assume a general expansion from the start.

Suppose for simplicity that the rapid changes occur on a scale $\sim 1$ and the slow changes occur on a scale $\sim 1/\e$,\footnote{One could rescale the variables such that the rapid changes occur on the scale $\sim\e$ and the slow changes on the scale $\sim 1$, to accord with the description of a region dense with boundary layers.} and that we seek an approximation to $\exact{f(t,\e)}$ that is uniform on the time-interval $[0,1/\e]$. Then we proceed by introducing a fast time variable $\phi=\phi(t,\e)$ and a slow time variable $\t=\t(t,\e)$ satisfying $\pdiff{\phi}{t}=\omega(t,\e)$ and $\pdiff{\t}{t}=\e\tilde\omega(t,\e)$, where $\omega$ and $\tilde\omega$ are uniformly $O_s(1)$; changes in $\phi$ are of the same order as changes in $t$, while $\t$ changes appreciably only after $t$ changes by a very large amount. (In the simplest case, we have $\phi=t$ and $\t=\e t$.) We invert the transformation in order to write the frequencies as functions of the slow time alone: $\omega(\t,\e)$ and $\tilde\omega(\t,\e)$. I next note that while setting $\omega=1$ will lead to large errors on a timescale $1/\e$ (consider, for example, attempting to approximate $\cos(t+\e t)$ by $\cos t$), setting $\tilde\omega=1$ will lead to large errors only on extremely long timescales outside our range of interest. Hence, I will take the slow time to be given by $\t=\e t$. The remaining frequency, $\omega$, must be determined over the course of the calculation. To make such a goal feasible, I assume $\omega$ possesses a regular expansion $\sum_{n\ge0}\zeta_n(\e)\omega\coeff{\emph{n}}(\t)$.

I next assume that $\exact{f}(t,\e)$ can be written as a function $F(\phi,\t,\e)$, and that $F$ possesses a regular expansion: that is,
\begin{equation}\label{two-time expansion}
\exact{f}(t,\e)=F(\phi,\t,\e)=\sum_{n}\zeta_n(\e)F\coeff{\emph{n}}(\phi,\t).
\end{equation}
Suppose $\exact{f}$ is to satisfy some differential equation $\exact{D}[\exact{f}]=0$. After making the substitution $\exact{f}(t,\e)=F(\phi,\t,\e)$, we use the chain rule to convert derivatives with respect to $t$ into the sum of partial derivatives $\frac{d}{dt}=\omega(\t,\e)\frac{\partial}{\partial\phi} + \e\frac{\partial}{\partial\t}$. We then arrive at a partial differential equation in terms of $\phi$ and $\t$. Now, the essential step in a multiscale expansion consists of treating $\phi$ and $\t$ as independent variables at this point; that is, $F$ is taken to be a solution to the PDE for arbitrary values of $\phi$ and $\t$. Given this assumption, after substituting the expansions for $F$ and $\omega$, we can solve the equation by setting the coefficient of each $\zeta_n$ to zero. If I did not assume that $F$ solves the equation for arbitrary $\phi$ and $\t$, then the $\e$-dependence scattered throughout $\phi(t,\e)$ and $\t(t,\e)$ would prevent us from solving the equation in this manner.

\begin{figure}
\begin{center}
\includegraphics{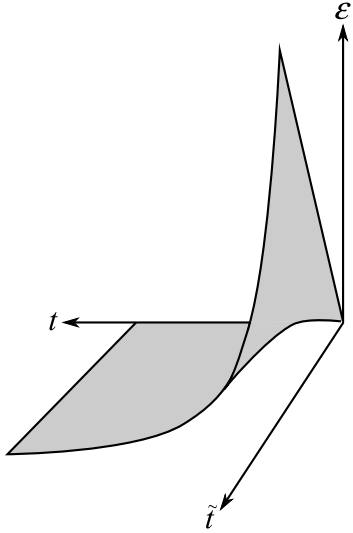}
\end{center}
\caption[The extended manifold of a multiscale expansion.]{In a multiscale expansion, we work on a manifold of larger dimension than that on which the original problem is posed. The solution is eventually evaluated on the submanifold defined by $\t=\e t$, shown here in grey.}
\label{submanifold}
\end{figure}

Treating $\phi$ and $\t$ as independent is equivalent to working on an enlarged manifold with coordinates $(\phi,\t,\e)$. The solution manifold on which $\exact{f}$ lives is a submanifold defined by $\phi=\phi(t,\e)$ and $\t=\t(t,\e)$. (See Fig.~\ref{submanifold} for an illustration in the simple case where $\phi=t$ and $\t=\e t$.) Determining $\omega$ can be viewed as a step in determining this submanifold; in fact, we can note that the transformation from the extrinsic coordinates $x^\alpha=(\phi,\t,\e)$ to the intrinsic coordinates $y^a=(t,\e)$ defines a set of basis vectors $e^\alpha_a$ on the submanifold, given by $e^\alpha_t=(\omega,\e,0)$ and $e^\alpha_\e=(\partial_\e\phi,t,1)$.

Because we are provided with sufficient boundary data for an ODE, rather than for a PDE, we must place some constraints on the function $F$. The most commonly imposed constraint is the \emph{non-secularity} condition, which says that integration constants must be chosen such that any secularly growing term vanishes. Other possible constraints include the demand that each coefficient $F\coeff{\emph{n}}(\phi,\t)$ is a periodic function of $\phi$, and the demand that each coefficient be unique. Obviously, one must apply such constraints judiciously and systematically.

Of course, this procedure relies on a host of assumptions. There is no guarantee that the exact solution $\exact{f}$ possesses an asymptotic expansion of the form \eqref{two-time expansion}; and even if it does, there is no guarantee that the terms in the expansion will necessarily solve the equation for arbitrary $\phi$ and $\t$. However, this method is extremely successful in practice. If one instead assumes a regular expansion, then the dependence of $\exact{f}$ on $\t$ will be expanded in powers of $\e t$. These powers of $t$ will eventually grow large, such that terms initially supposed to be high order become as large as the lower-order terms, preventing the expansion from providing a uniform approximation. In many cases, one can avoid this secular growth by using a rigorous method of averaging, which removes the rapid time dependence and recovers only the leading-order slow time-dependence. However, if one requires the fast time dependence as well, then the two-time method offers the most powerful means of doing so.

Because multiscale expansions do not form an integral part of this dissertation, I relegate further discussion of them to Appendix~\ref{multiscale_EFE}, which contains an illustrative example. In that example, it is demonstrated that even if the assumptions of the multiscale expansion fail at some order, the lower order solution can still yield a uniform asymptotic approximation. For further information, see Refs.~\cite{Kevorkian_Cole,Hinderer_Flanagan}.

\subsection{Singular versus regular perturbation theory}
We should now take note of the essential differences between regular perturbation techniques and singular perturbation techniques. When a regular expansion of an exact solution $\exact{f}$ is substituted into a differential equation $\exact{D}[\exact{f}]=0$, the coefficients in the expansion are guaranteed to solve a hierarchy of differential equations, simply by setting the coefficients of each power of $\e$ to zero. Hence, when constructing a regular series solution to a differential equation, one can determine each term in the solution solely from the given differential equation (and its attendant boundary conditions). But a \emph{general} expansion of an exact solution is not guaranteed to satisfy any such hierarchy, because the coefficients in the expansions depend on $\e$. Hence, when constructing a general series solution to a differential equation, one must impose some extra conditions upon it---e.g., satisfying the overlap hypothesis in the method of matched asymptotic expansions, or satisfying a PDE rather than an ODE in the method of multiple scales---which are not guaranteed to be satisfied given only the form of the general expansion.

This means that proving general properties of solutions is much more difficult using singular perturbation theory. In regular perturbation theory, one can construct proofs of the form ``given an exact solution to such and such a boundary value problem, if it possesses a regular asymptotic expansion then that expansion has such and such behavior"; in singular perturbation theory, we must append further hypotheses to this statement. However, if we seek a very strong statement about the solution to a problem, we must in any case go beyond the form of such a proof: we must also prove that the exact solution actually \emph{does} possess the assumed expansion. This is a difficult feat regardless of whether the assumed expansion is regular or general. While it is usually easier in the case of regular expansions, techniques do exist for handling singular perturbation problems (see Ref.~\cite{Eckhaus} for examples). Furthermore, general expansions provide asymptotic solutions where regular series cannot, and they provide \emph{uniform} asymptotic solutions. Hence, in most cases of interest, their advantages far outweigh any disadvantages.

%%%%%%%%%%%
\section{Perturbation theory in General Relativity}
%%%%%%%%%%%
In General Relativity we typically do not begin with a predetermined manifold with predetermined boundary conditions that uniquely determine an exact solution. Instead, the manifold is (mostly) determined by the leading order ``background" solution to the Einstein equation. Within that manifold, we define boundary conditions that uniquely determine the perturbations. This somewhat complicates the problem, but it also makes the assumptions of singular perturbation theory more reasonable: since we do not seek an approximation to a \emph{unique} exact solution to a given boundary value problem, but only an approximation to \emph{some} exact solution to the EFE, it is eminently reasonable to impose the supplementary conditions required to construct general expansions.

\subsection{Regular perturbation theory}
Before describing singular perturbation theory in GR, I will briefly review regular perturbation theory. In its most geometric description, the formalism begins with a 5D manifold $\mathcal{N}$ carrying a 5D metric $\exact{g'}^{\mu\nu}$ of signature $(0,-,+,+,+)$ and a non-negative scalar field $\e:\mathcal{N}\to\mathbb{R}$. The manifold is foliated by 4D submanifolds $\man_\e$ defined by $d\e=0$, such that $\mathcal{N}\sim \man_{\e}\times\mathbb{R}$. When restricted to act on dual vectors tangent to $\man_\e$, $\exact{g'}^{\mu\nu}$ can be inverted, inducing a 4D Lorentzian metric $\exact{g}_\e$. Each member of the family of metrics $\lbrace\exact{g}_\e\rbrace$ is taken to be an exact solution of Einstein's equation at fixed $\e$. A regular expansion of the pair $(\man_\e,\exact{g}_\e)$ is an expansion around a known ``base" pair $(\man_0,g_0)$. This expansion is performed by first defining a one-to-one relationship between points on $\man_0$ and $\man_\e$ via a diffeomorphism $\varphi_\e:\man_0\to\man_\e$, called an identification map. This map induces a flow on $\mathcal{N}$ with a tangent vector field $X$ that is non-vanishing and nowhere tangent to the submanifolds $\man_\e$. (Note that one could instead begin with a vector field and derive from it an identification map, but beginning with the identification map will be more useful in formalizing general expansions.)

In this context, the regular expansion $\exact{g}_\e=g+\sum_{n\ge 1}\e^n\hmn{}{\emph{n}}$ is given by an expansion along the flow induced by $X$:
\begin{align}
\phi_{\e}^*(\exact{g}_\e) &= e^{\e\!\Lie{X}}\exact{g}\big|_{\man_0},
\end{align}
where $\phi_{\e}^*\exact{g}_\e$ is the pull-back of $\exact{g}_\e$ onto the base manifold $\man_0$, and $\Lie{X}$ is the Lie derivative along the vector $X$. The background metric and the perturbations of it have clear geometrical interpretations: the background metric $g$ is the restriction of $\exact{g}_\e$ to the submanifold defined by $\e=0$; the first-order perturbation $\e \hmn{}{1}\equiv \e(\Lie{X}\exact{g})\big|_{\e=0}$ is the product of the ``distance" $\e$ along a flow line and the rate of change of $\exact{g}$ in the direction of the flow; and so on.

A choice of gauge corresponds to a choice of identification map $\map_\e$. A different choice, say $\psi_\e$, leads to a different tangent vector field $Y$, which in turn leads to a change
\begin{equation}
\psi_{\e}^*(\exact{g}_\e)-\phi_{\e}^*(\exact{g}_\e) = (e^{\e\!\Lie{Y}}-e^{\e\!\Lie{X}})\exact{g}\big|_{\e=0}.
\end{equation}
By expanding the exponentials, one finds that this induces changes $\hmn{}{\emph{n}}\to \hmn{}{\emph{n}}+\Delta \hmn{}{\emph{n}}$. At first and second order, the changes are given explicitly by
\begin{align}
\Delta \hmn{}{1} & = \Lie{\xi\dcoeff{1}}g, \label{gauge_trans1}\\
\Delta \hmn{}{2} & = \tfrac{1}{2}(\Lie{\xi\dcoeff{2}}+\Lie{\xi\dcoeff{1}}^2)g +\Lie{\xi_{(1)}}\hmn{}{1},\label{gauge_trans2}
\end{align}
where $\xi\dcoeff{1}\equiv(Y-X)\big|_{\e=0}$ and $\xi\dcoeff{2}\equiv[X,Y]\big|_{\e=0}$ are vector fields in the tangent bundle of $\man_0$. Note that $\xi\dcoeff{1}$ and $\xi\dcoeff{2}$ are linearly independent, so they can be chosen independently. In terms of coordinates, they correspond to the near-identity transformation
\begin{align}
x^\alpha \to x'^\alpha &= x^\alpha-\e\xi^\alpha_{(1)} +\tfrac{1}{2}\e^2\left(\xi^\alpha_{(1),\beta}\xi^\beta_{(1)}-\xi^\alpha_{(2)}\right) +\order{\e^3},
\end{align}
where the components on the right hand side are in the original coordinates $x^\alpha$. We say that the vectors $\xi\dcoeff{\emph{n}}$ are the generators of the gauge transformation. (See Ref.~\cite{Bruni_limits} for the precise meaning of this phrase.)

\subsection{Singular perturbation theory}
Although singular perturbation techniques have been utilized in many calculations in GR, the only formal description of them was provided by Kates \cite{Kates_structure}. I will review his description in this section, before extending it in the following sections. A singular perturbation problem is characterized by the limit $\e\to 0$ being singular: $\exact{g}_\e$ may not exist at $\e=0$, the topology or dimension of $\man_\e$ may change between $\e=0$ and $\e>0$, etc. This means that the 5D manifold $\mathcal{N}$ does not in general contain a ``base'' manifold $\man_0$; instead, it is given by $\mathcal{N}\sim\man_\e\times(0,\e^*]$. Hence, one cannot build an approximation on the limiting manifold by finding derivatives of the exact metric at $\e=0$. Instead, one works on a ``model manifold'' $\man_M$, on which one constructs a family of approximate solutions
\begin{equation}
g_M(\e)=g(\e)+\sum_{n\ge 1}\e^n h\coeff{n}(\e).
\end{equation}
The topology of the model manifold is taken to be compatible with the leading-order metric $g(\e)$. If there exists an identification map $\map_\e:\man_M\to\man_\e$, which maps a region $\mathcal{U}_M\subset\man_M$ to a region $\mathcal{U}_\e\subset\man_\e$, such that $g_M(\e)$ uniformly approximates $\map^*_\e \exact{g}_\e$ in the region $\mathcal{U}_M$ as $\e\to 0$, then $g_M(\e)$ is a uniform asymptotic approximation (as measured in some suitable norm) to the exact solution in the region $\mathcal{U}_\e$. Once again, the identification map induces a family of curves in the 5D manifold $\mathcal{N}$, but these curves will not, in general, continue smoothly to a base manifold $\man_0$.

As an example, consider a post-Newtonian expansion, which is singular~\cite{Futamase_Schutz, Futamase_particle1, Futamase_review, Rendall}. The Newtonian limit is given by $\e=v/c\to 0$, where $v$ is the supremum of the velocities in the system. In this limit, the light cones of the spacetime fold out into spatial surfaces, and the time-components of the metric blow up---alternatively, if we consider the inverse metric, we see that its time components vanish, such that it degenerates into a 3D spatial metric. Hence, we can infer that the manifold defined by $\e=0$ corresponds to the 3D spatial manifold of Newtonian theory.\footnote{The singular nature of the Newtonian limit is also signaled by the fact that hyperbolic wave equations become elliptic Poisson equations as the speed of gravity's propagation becomes infinite.} In this case, the model manifold and background metric are taken to be those of Minkowski spacetime.

In the next two sections of this chapter, I will formulate matched asymptotic expansions and multiscale expansions within this framework. The description of matched asymptotic expansions follows that given by Kates \cite{Kates_structure}, which built on the work of D'Eath \cite{DEath, DEath_paper}; however, I will more carefully formulate the matching conditions that result from an overlap hypothesis. My discussions of multiscale expansions is original to this work.

In the final section of this chapter, I formulate the self-force problem as a free-boundary value problem. I then discuss means of solving the problem.

\subsection{Matched asymptotic expansions}\label{matching_conditions}
Though most of the description in this section carries over to a more general situation, I will restrict it to the pertinent case of a family of exact solutions $\exact{g}_\e$ containing a body of mass $\sim\e$, on a family of manifolds $\man_\e$. In this section, I will only sketch the formalism of inner and outer limits for this system; in Ch.~\ref{extended_body}, I provide further discussion of concrete applications of these limits.

Suppose that we are given two coordinate systems on $\man_\e$: a local coordinate system $X^\alpha=(T,R,\Theta^A)$ that is centered (in some approximate sense) on the small body, and a global coordinate system $x^\alpha$. For example, in an EMRI, the global coordinates might be the Boyer-Lindquist coordinates of the supermassive Kerr black hole (though we could consider the case in which both coordinate systems are centered on the small body); the local coordinates might be Schwarzschild-type coordinates for the small body. The local coordinates cover some region $D_I$ around (and possibly inside) the body, while the global coordinates cover a larger region $D_E$ outside the body. Assume, without loss of generality, that the two coordinate systems have overlapping domains, and that they are related by a map $\phi_\e:x^\alpha\mapsto X^\alpha$ in the region $D=D_I\cap D_E$.

A regular outer expansion $g_E(x,\e)=g(x)+\e\hmn{}{1}(x)+...$ is constructed by taking the limit $\e\to 0$ at fixed $x^\alpha$. In this limit, the body shrinks toward zero size as all other distances remain roughly constant. For simplicity, I assume that this limit continues to a base manifold $\man_0$. However, the limit certainly does not exist on a remnant curve $\gamma\coeff{0}$ corresponding to the ``position'' of the small body---for example, if one takes a regular limit of the Schwarzschild metric in Schwarzschild coordinates, then there is no limit defined at coordinate values corresponding to $r=0$. Hence, I take the model manifold in the outer expansion to be $\man_E=\man_0\cup\gamma\coeff{0}$, and I take the external background metric to be $g=\exact{g}_0$. Of course, this construction is not essential; the model manifold need not be defined by setting $\e=0$ in this way. But at the very least, for the outer expansion to be regular, we require $g=\lim_{\e\to0}\exact{g}$.

A regular inner expansion $g_I(\tilde X,\e)=g_B(\tilde X)+\e H\coeff{1}(\tilde X)+...$ is constructed by taking the limit $\e\to 0$ at fixed values of the scaled coordinates $\tilde X^\alpha = \trans(X^\alpha)= ((T-T_0)/\e,R/\e,\Theta^A)$. This limit is naturally singular: it follows flow lines that converge at a single point defined by $(T=T_0,R=0)$ in $\man_E$. Explicitly, since the metric written in these coordinates has the form $\exact{g}\sim \e^2f_{\alpha\beta}(\tilde X)d\tilde X^\alpha d\tilde X^\beta$, all distances vanish at $\e=0$. As discussed by D'Eath~\cite{DEath} (see also Ref.~\cite{Gralla_Wald}), to make the limit regular, one must use the conformally rescaled metric $\exact{\tilde g}_\e\equiv \e^{-2}\exact{g}_\e$. This rescaling effectively ``blows up" the distances in  spacetime, such that as $\e\to0$, the size of the small body remains constant while all other distances are sent to infinity; thus, the inner limit serves to ``zoom in" on a small region around the body. The background spacetime defined by $\e=0$ is then defined by the metric $g_B$ of the isolated small body, and the approximation is built on a model manifold $\man_I$ with the topology of that spacetime.\footnote{Note that $\man_I$ generically differs from $\man_E$. Consider, for example, the case of a small black hole orbiting a large black hole. The manifold $\man_I$ possesses a singularity at the ``position" of the small black hole but is otherwise smooth, while the manifold $\man_E$ possesses a singularity at the ``position" of the large black hole but possesses a smooth worldline where the small black hole should be.}

\begin{figure}[tb]
\begin{center}
\includegraphics[scale=1.2]{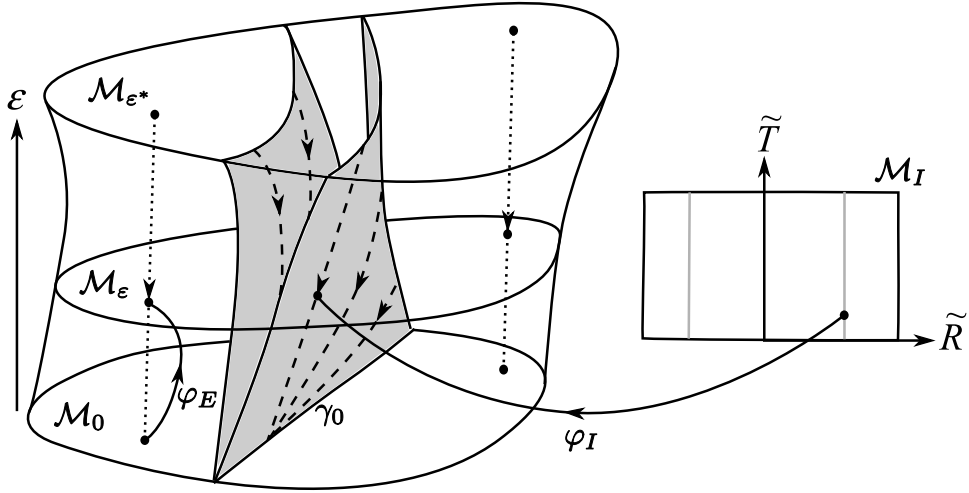}
\end{center}
\caption[Regular expansion of a family of spacetimes]{Regular inner limit (dashed curves) and outer limit (dotted curves) on the manifold $\mathcal{N}\sim\man_\e\times(0,\e^*]$. The inner limit is generated by the map $\map_I$ from the manifold $\man_I$ on which the interior background metric of the isolated small body lives; these curves terminate at a point $(T=T_0,R=0)$ on $\gamma\coeff{0}$. The outer limit is generated by the map $\map_E$ from the manifold $\man_E$ on which the external background metric lives. The external manifold in this case is taken to be equal to $\man_0\cup\gamma\coeff{0}$. The gray region is a surface of constant $\tilde R$, which converges to the $\e$-independent worldline $\gamma\coeff{0}$.}
\label{regular_limit}
\end{figure}

The outer and inner expansions are related to the exact solution via identification maps $\map_E:\man_E\to\man_\e$ and $\map_I:\man_I\to\man_\e$, which respectively identify points on $\man_E$ and $\man_I$ with points on $\man_\e$. (See Fig.~\ref{regular_limit}.) These two identification maps induce a map  $\phi:\man_E\to\man_I$, given by $\phi=\map_I^{-1}\circ\map_E$, which has the identical coordinate description as the original transformation $\phi_\e$ between the global and local coordinates. Gauge transformations in the outer and inner expansions are generated by vector fields $\xi^\alpha(x)$ and $\tilde\xi^\alpha(\tilde X)$, which take their respective values in the tangent bundles of $\man_E$ and $\man_I$. Note that a gauge transformation in the outer expansion generically corresponds to a finite coordinate transformation in the inner expansion, due to the rescaling of the coordinates.

A uniform composite expansion is formed on a model manifold $\man\sim\man_\e$ by cutting out a portion of $\man_E$ and stitching part of $\man_I$ into the excised region. The local and global coordinates each cover a patch of $\man$, identified with the patches $D_I$ and $D_E$ on $\man_\e$ via the maps $\map_I$ and $\map_E$. The uniform metric is constructed on this manifold by adding together the inner and outer approximations in each coordinate system, then removing any ``double-counted'' terms that appear in both metrics.

How would one go about constructing such a uniform approximation if one did not have access to an exact solution? Just as in traditional perturbation theory, one would construct two separate asymptotic solutions to Einstein's equation, but now in possibly two different coordinate systems and on possibly two different manifolds. If one assumes that the two asymptotic solutions are approximations of a single exact solution, and if there exists an overlap region on the manifold $\man$ in which both approximations are valid to the same order, then they must agree in that overlap region. In this case, ``agreement" is defined by the existence of the unique map $\phi$ that relates the two expansions. As usual, we will not worry about a specific overlap region, but instead expand the two solutions in the buffer region.

However, before performing that expansion in the buffer region, one must write $g_I$ and $g_E$ in the same coordinate system. Let us choose this system to be the local coordinates $X^\alpha$. Then, adapting Eq.~\eqref{matching_equation}, the matching condition reads
\begin{equation}\label{strong_matching_condition}
\expand^k_\e\e^2\trans^*\expand^m_\e\e^{-2}\trans_*\exact{g}(X) = \expand^k_\e\e^2\trans^*\expand^m_\e\e^{-2}\trans_*\phi_*\expand^m_\e\exact{g}(x).
\end{equation}
On the left, we begin with the exact metric in the local coordinates $X^\alpha$. It is then expanded to $m$th-order in an inner expansion, by transforming into scaled coordinates $\tilde X^\alpha$ via $\trans$ (along with an appropriate conformal rescaling) and expanding. Next, it is expanded in the buffer region by re-expressing it in the unscaled local coordinates and expanding to $k$th order; this is equivalent to an expansion of the inner solution for $R\gg\e$. On the right-hand side of the equation, we begin with the exact metric in the global coordinates $x^\alpha$. It is expanded to $m$th order in those coordinates, yielding an outer expansion. It is then transformed to the starting point of the left-hand side, by transforming to the local coordinates via $\phi$, then to the scaled local coordinates via $\trans$, then re-expanding to $m$th order to yield an inner expansion. Finally, it is expanded in the buffer region by transforming back to the unscaled local coordinates and re-expanding. The content of this equation is that the expansion in the buffer region must be the same whether it is obtained by first performing an outer expansion or by first performing an inner expansion. Schematically, we can write
\begin{equation}
\expand_\e g_I(X) = \expand_R\phi_*g_E,
\end{equation}
which states that if the inner and outer expansions are written in the same coordinate system, then they must yield the same expansion in the buffer region.

But this is decidedly \emph{not} the matching condition that has been used in practice. Instead, what has been done in practice is the reverse: first, expand the two solutions in the buffer region, and only afterward find the coordinate transformation between them. This is accomplished by setting up a second local coordinate system $Y^{\alpha}=\phi_\gamma(x^\alpha)=(t,r,\theta^A)$ centered on a worldline $\gamma$ in $\man_E$; for example, these might be Fermi normal coordinates, and in the case of regular expansions, they would be centered on $\gamma\coeff{0}$. The outer expansion is then written in these local coordinates and expanded for small $r$, under the presumption that $r\sim R$. After performing this expansion (and the expansion of $g_I$ in the buffer region), one seeks a unique transformation $\phi_{\text{buf}}:Y^\alpha\mapsto X^\alpha$ that maps the buffer-region expansion of $g_E$ into the buffer-region expansion of $g_I$.\footnote{One can see that if everything is correct, the various transformations must be related as $\phi=\phi_{\text{buf}}\circ\phi_{\gamma}$.} Schematically, we can write
\begin{equation}\label{weak_matching_condition}
\expand_\e g_I(X) = \phi_{\text{buf}*}\expand_r \phi_{\gamma*}g_E.
\end{equation}
On the left, the inner expansion $g_I$ is expanded in the buffer region in the local unscaled coordinates $X^\alpha$. On the right, the outer expansion $g_E$ is transformed to the local coordinates $Y^\alpha$ via $\phi_\gamma$, then it is expanded in the buffer region (i.e., for small $r$). Hence, the two buffer-region expansions are written in two different coordinate systems: the inner expansion in the coordinates $X^\alpha$, and the outer expansion in the coordinates $Y^\alpha$. So, in order to make a comparison, as the final step on the right-hand side, the buffer-region expansion of $g_E$ is transformed to the coordinates $X^\alpha$ via $\phi_{\text{buf}*}$. In short, Eq.~\eqref{weak_matching_condition} states that if $g_I$ and $g_E$ are expanded in the buffer region, then the resulting expansions must be related by a coordinate transformation. 

I will call Eq.~\eqref{strong_matching_condition} the \emph{strong matching condition} and Eq.~\eqref{weak_matching_condition} the \emph{weak matching condition}. The weak condition follows from the strong condition, but not vice versa, and one can easily imagine situations in which the weak condition would be satisfied while the strong condition would not. In the weak matching condition, because the metric is already expanded for small $r$ before $\phi_{\text{buf}}$ is determined, $\phi_{\text{buf}}$ will itself be written as an expansion. Thus, the weak matching condition only requires an asymptotic approximation of $\phi_{\text{buf}}$ (or, equivalently, of $\phi=\phi_{\text{buf}}\circ\phi_{\gamma}$). Of course, one can only ever determine an asymptotic approximation---but in the strong matching condition, the approximation is for small $\e$, rather than for both small $\e$ and small $r$. This essentially reduces $\phi_{\text{buf}}$ to a gauge transformation in the buffer-region expansion defined by $R$ (or $r$) and $\e$ both being small. As mentioned above, a gauge transformation in the outer expansion corresponds to a finite coordinate transformation in the inner expansion, and vice versa. Hence, any choice of gauge on $\man_E$ must be compatible with the choice of background coordinates on $\man_I$ (and vice versa). The two matching conditions insist on this compatibility to differing extents.

One should note that though the description in this section makes use of two regular expansions, as in traditional matched asymptotic expansions, the same general description holds for two general expansions. The only differences are that $g\ne\lim_{\e\to 0}\exact{g}$ and that there is no need to conformally scale the metric to arrive at the inner expansion. This is particularly important if one wishes to allow the internal metric to vary on its ``natural" timescale $\tilde T\sim 1$ (i.e., the timescale determined by the mass of the small object). If the metric near the body varies on this timescale, then in the unscaled coordinate time $T=\e\tilde T$, the metric will have a functional dependence on the combination $T/\e$, which will be singular in the limit $\e\to0$. Thus, if both the expansions are to be regular, the internal metric can vary only on the external time $T$, corresponding to an internal slow evolution depending only on $\e\tilde T$. In other words, regularity requires that the internal solution varies quasistatically (see D'Eath's discussion~\cite{DEath}). Of course, for $\e>0$, one could construct a general inner expansion that is identical to the regular inner expansion by rescaling $R$ only, instead of both $T$ and $R$, and then simply assuming the inner expansion varies quasistatically; using this method, a global-in-time expansion can be constructed, and the metric is never conformally rescaled. Also, by using this method, one can remove the quasistatic assumption entirely.

Finally, before moving to the next singular perturbation technique, I will note that just as in traditional singular perturbation theory, there is a distinction between what I have called the overlap region and the buffer region. The buffer region corresponds simply to $\e\ll R\ll 1$. In order for us to express the outer solution in terms of the field $R$, the buffer region must lie within the region $D$, where both the local and global coordinate systems apply, but the size of the region is independent of the order of accuracy of the inner and outer solutions. As I will discuss in Chs.~\ref{extended_body} and \ref{buffer_region}, one can extract considerable information about the metric---and in particular, equations of motion for the small body---by working entirely within the buffer region \cite{Kates_motion, Thorne_Hartle, Gralla_Wald}, without ever constructing explicit inner and outer solutions or making use of an overlap hypothesis.

\subsection{The method of multiple scales}\label{multiple_scales_GR}
In the method of multiple scales, changes on both short and fast time scales occur throughout the spacetime. Thus, one cannot construct a uniform asymptotic approximation based on combining only two limit processes. If we consider a two-timescale expansion, with a fast time $t$ and a slow time $\t=\e t$, there are only two limits that can be easily envisioned: the slow-time limit $\e\to 0$ at fixed $\t$, which follows a congruence of curves in $\mathcal{N}$ that tend toward $t\to\infty$ as $\e\to 0$; or the fast-time limit $\e\to 0$ at fixed $t$, which follows a congruence of curves that tend toward $\t\to 0$. However, in a multiscale expansion, both quantities are to be kept fixed. As discussed above, this is accomplished by treating them as independent variables.

Consider the case of an expansion that holds fixed both a set of coordinates $x^\alpha$ and some scalar field $\zeta(x,\e)$ satisfying $\pdiff{\zeta}{x^\alpha}=o(1)$:
\begin{equation}
\exact{g}(x,\e)=g(x,\zeta)+\sum_{n\ge 1}\e^n \hmn{}{n}(x,\zeta)
\end{equation}
In the simplest case, $\zeta$ is equal to the product of $\e$ and one of the coordinates. When substituting this multiscale expansion into Einstein's equation, one would treat $\zeta$ and $x^\alpha$ as independent coordinates on an extended, 5D manifold $\widetilde{\man}_\e$; these 5D manifolds are stacked atop one another to form a 6D manifold $\widetilde{\mathcal{N}}\sim \man_\e\times\mathbb{R}^2$. The limit $\e\to 0$ is taken at fixed values of both $\zeta$ and $x$, and the actual solution is obtained by restricting the expansion to the submanifold defined by $\zeta=\zeta(x,\e)$.

As in traditional perturbation theory, one might require a fast-time variable $\phi$ that differs from the given coordinate time. Indeed, one might use any coordinates one likes on $\widetilde{\man}_\e$. However, I will forgo any further analysis of the general formulation of multiscale expansions in GR. To provide some flavor of the expansions, in Appendix~\ref{multiscale_EFE} I define gauge transformations and sketch a multiscale expansion of the EFE for the simple case with coordinates $(x^\alpha,\zeta)$. The reader is referred to Ref.~\cite{Kevorkian_Cole} for further details of multiscale expansions in PDEs.

Recently, Hinderer and Flanagan \cite{Hinderer_Flanagan} have constructed a two-timescale formalism tailored to EMRIs. In their method, all dynamical variables (i.e., the metric and the phase space variables of the worldline) are submitted to two-timescale expansions; this expansion captures both the fast dynamics of orbital motion and the slow dynamics of the particle's inspiral and the gravitational backreaction on the background spacetime. Since the metric and the worldline are related by the EFE, it is assumed that the metric can be written as a function of the phase space variables of the worldline. On each timeslice, the limit $\e\to0$ is then taken with the phase space variables held fixed. Specifically, the true worldline is specified by a set of action-angle variables $(J(\t,\e),\varphi(\t,\e))$ and a slow time variable $\t$. Expanding for $\e\to0$ with $\varphi$ and $\t$ held fixed results in a sequence of fast-time and slow-time equations. In the fast-time equations, wherein $\t$ (and therefore $J$) is treated as a constant, the metric is a function of $\varphi$ only; in other words, it is a functional of the geodesic that is instantaneously tangential to the true worldline. From this it follows that the leading-order fast-time equation yields a metric perturbation sourced by that geodesic, as in regular perturbation theory. However, that is only at fixed $\t$---the true worldline and metric perturbation emerge by allowing the variables to vary with $\t$, with a $\t$-dependence determined from the slow-time equations.

Although exceedingly useful for EMRIs, this procedure relies on the background metric being stationary at fixed $\t$, such that it has no fast time dependence, and on the geodesic motion in that background being integrable, such that the metric can be written in terms of the action-angle variables.

\subsection{Fixing the worldline}\label{fixed_worldline_formulation}
In Hinderer and Flanagan's formalism, the metric is written as a function of the phase space variables on the worldline, and then both the metric and those variables are submitted to a two-timescale expansion. The formalism I will now describe is a generalization of this: the metric is written as a functional of the worldline, and then the metric is expanded with that worldline held fixed. In order to motivate this approach, I will first provide a formulation of the exact problem to which we seek an approximate solution.

We wish to determine the mean motion of a small, spatially bounded matter distribution. (For the moment, I neglect the case of a black hole.) In principle, we have some matter field equations to go along with the EFE for this blob of matter. As governed by the field equations, the boundary of the blob traces out some surface in spacetime. In the interior of the boundary, the matter density is finite, and in the exterior it vanishes. To determine the motion of the body, we seek the equation for the generators of this boundary. This is a free-boundary value problem \cite{free_boundary_problems}, in which some boundary values are specified on a boundary that is free to move. In the context of bodies in GR, this problem has received some study  \cite{free_boundary_GR,free_boundary_GR2}, but it is still far from understood, and it must certainly be tackled numerically.

To make progress with an approximation scheme, I reformulate the problem. I surround the body by a tube $\Gamma$ embedded in the buffer region, such that for $\e\to0$, the radius of the tube vanishes. For the moment, consider $\Gamma$ to be defined by constant radius $R=\rad(\e)$ in the local coordinates $X^\alpha$. I assume that the body is fairly widely separated from all other matter sources, such that outside of $\Gamma$ there is a large vacuum region $\Omega$. I also assume that $\Gamma$ is in vacuum; since it lies in the buffer region around the small body, this means that I must restrict my approximation to a small body that is sufficiently compact to not fill the entire buffer region. Now, since the tube is close to the small body (relative to all external length scales), the metric on the tube is primarily determined by the small body's structure. In other words, the information about the body has now been transplanted into boundary conditions on the tube. Recall that the buffer region corresponds to $\tilde R\to\infty$. Hence, on the tube, we can construct a multipole expansion of the body's field, with the form $\sum\tilde R^{-n}$. I assume that the local coordinates $X^\alpha$ are mass-centered, such that the mass dipole term in this expansion vanishes. (See Ref.~\cite{STF_3} and references therein for discussion of multipole expansions in GR; see Refs.~\cite{STF_3, Thorne_Hartle} for discussion of mass-centered coordinates in the buffer region; see, e.g., Ref.~\cite{center_of_mass} for further discussion of definitions of center of mass.) This, then, is another free-boundary value problem: we must determine the equations of motion of the generators of the tube, given the boundary values of the metric on it, and in particular, given that the body lies at the ``center" of it.  With this formulation, we can also determine the motion of a black hole, rather than just a matter distribution.

Now suppose that I want to represent the motion of the body through the external spacetime $(g,\man_E)$, rather than through the exact spacetime. As we can see from Fig.~\ref{regular_limit}, this is easily accomplished by using the regular limit and taking the motion to be represented by the remnant worldline $\gamma\coeff{0}$. However, as discussed in Ch.~1, on long timescales this will provide a very poor representation of the motion.

Let us consider this from another direction. Assume that we were given the exact solution $\exact{g}_\e$ on $\man_\e$, along with the coordinate transformation $\phi_\e$ between the local coordinates $X^\alpha$ and the global coordinates $x^\alpha$ in the buffer region. At fixed $R=\rad(\e)=o(1)$, we could write this transformation as $x^\alpha=\phi^{-1}_\e(T,\rad,\Theta^A)$. In the limit of small $\e$, $\rad$ becomes small as well, meaning that this transformation can be expanded as $x=\phi^{-1}_\e(T,0,\Theta_0^A)+o(1)$, where $\Theta_0^A$ is an arbitrary choice of angles. This transformation thus defines a curve $z^\alpha(T,\e) \equiv \phi^{-1}_\e(T,0,\Theta_0^A)$ in the external manifold $\man_E$. Since the small body is centered ``at" $R=0$, this curve defines a meaningful long-term representative worldline $\gamma$. If we expand $\phi_\e(T,0,\Theta_0^A)$ for small $\e$, then it will not provide a uniform transformation between the inner and outer coordinates; it will contain secularly growing errors of the form $\e t$. So, instead, in order to construct a uniform asymptotic solution, when constructing the external approximation, one must hold $\gamma$ fixed. Determining $\gamma$ then amounts to determining the ``location" at which the (mass-centered) inner expansion is to be performed.

Since we will never be seeking $\phi$ directly, and in case the inverse of $\phi_\e$ does not exist at $R=0$, allow me to present the final reformulation of the problem. Define a tube $\Gamma_E[\gamma]\subset\man_E$ such that it is a surface of constant radius $r$ in Fermi normal coordinates centered on a worldline $\gamma\subset\man_E$. Using the map $\map_E$ from the regular expansion, this defines a tube $\Gamma=\phi_E(\Gamma_E)$. Now, the problem is the following: what equation of motion must $\gamma$ satisfy in order for $\Gamma$ to be mass-centered, in the sense that the mass dipole of the inner expansion vanishes when mapped to $\man_\e$ via $\map_I$? Note that the worldline is a curve in the external manifold $\man_E$. It should not be thought of as a curve in the manifold $\man_\e$ on which the exact metric $\exact{g}_\e$ lives; in fact, if the small body is a black hole, then there is obviously no such curve. 

\begin{figure}[tb]
\begin{center}
\includegraphics[scale=1.2]{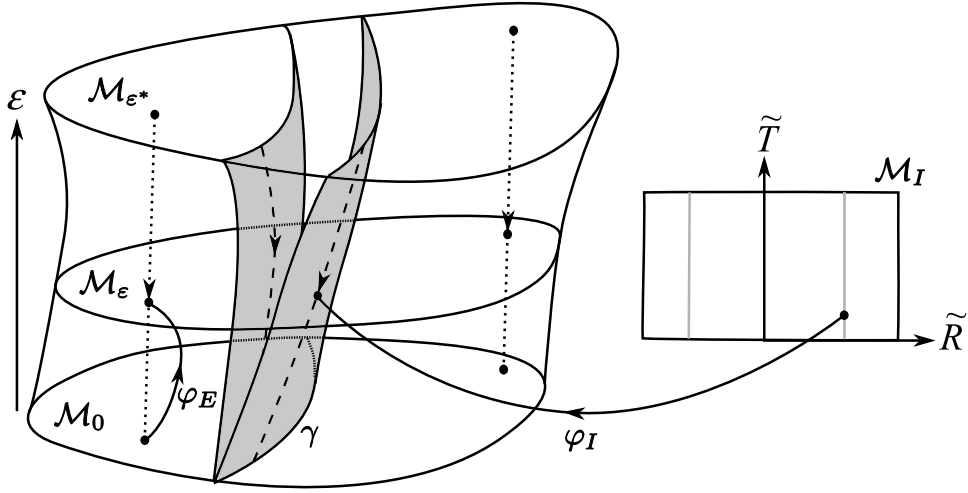}
\end{center}
\caption[Fixed-worldline expansion of a family of spacetimes]{Fixed-worldline expansion of a family of spacetimes. The dotted lines correspond to the outer limit, which lets the body shrink to zero size but keeps its motion fixed. The dashed lines correspond to the inner limit, which keeps the size of the body fixed. Here I display the singular inner limit, which does not rescale the inner time coordinate; hence, dashed lines originating at different times will terminate at different points on $\gamma$. The worldline lies in the manifold $\man_E=\man_0\cup\gamma\coeff{0}$, but it does not correspond to the remnant curve $\gamma\coeff{0}$ defined by the regular limit; instead, it is allowed to have $\e$-dependence, and it is determined by the particular value of $\e$ at which an approximate solution is sought.} 
\label{singular_limit}
\end{figure}

In order to determine the equation of motion of the worldline, I consider a family of metrics $g_E(x,\e;\gamma)$ parametrized by $\gamma$, such that when $\gamma$ is given by the correct equation of motion, we have $g_E(x,\e;\gamma(\e))=\map_E^*\exact{g}_\e(x)$. The metric in the outer limit is thus taken to be the general expansion
\begin{equation}
g_E(x,\e;\gamma) = g(x)+\sum_{n>0}\e^n\hmn{E}{\emph{n}}(x;\gamma).
\end{equation}
Solving Einstein's equations will determine the worldline $\gamma$ for which the inner expansion is mass-centered. I will call this a fixed-worldline expansion, in which the perturbations produced by the body are constructed about a fixed worldline determined by the particular value of $\e$ at which one seeks an approximation. Refer to Figs.~\ref{regular_limit} and \ref{singular_limit} for a graphical comparison between this expansion and a regular one.

Recall that in a multiscale expansion, the expanded equations are solved by assuming that they are valid for arbitrary values of the slow-time variable $\t$, not only on the true solution manifold defined by $\t=\e t$. Similarly, in the fixed-worldline expansion, one method of solving the expanded EFE will consist of assuming that it is valid for arbitrary worldlines; the true solution is found by choosing the true worldline.\footnote{Note that, unlike in the multiscale expansion, where the leading-order term depends on $\t$, in the present expansion, the external background does not depend on $\gamma$. This will be discussed further in Ch.~\ref{extended_body}.} Solving the EFE with an arbitrary worldline seems to require reformulating it in a ``relaxed" form before expanding it, such that, for example, the linearized equation does not immediately determine $\gamma$ to be a geodesic. In the following chapters, I will adopt a certain gauge choice in order to relax the EFE in this way. In the future, other means of solving the EFE in a fixed-worldline expansion should be considered, along with other choices of gauge.

What is the gauge freedom in this expansion? The outer expansion is defined not only by holding $x^\alpha$ fixed, but also by demanding that the mass dipole of the body vanishes when calculated in coordinates centered on $\gamma$. If we perform a gauge transformation generated by a vector $\xi\coeff{1}{}^\alpha(x;\gamma)$, then the mass dipole will no longer vanish in those coordinates. Hence, a new worldline $\gamma'$ must be constructed, such that in coordinates centered on that new worldline, the mass dipole vanishes. In other words, in the outer expansion we have the usual gauge freedom of regular perturbation theory, so long as the worldline is appropriately transformed as well. The transformation law for the worldline is well known \cite{self_force_gauge}; it will be worked out again in Sec.~\ref{comments on force}, in which a gauge-invariant expression for the force will be derived. The only new feature is that $\e$-dependence can be incorporated into the transformation, because the gauge vectors can be functionals of the old worldline; this allows, for example, a ``first-order" gauge vector that is constructed from the tail integral $\tail[\gamma]$. However, to maintain the form of the expansion, we must also insist that $\xi\coeff{\emph{n}}=O_s(1)$. Of course, in addition to this gauge freedom, one can still perform global, $\e$-independent background coordinate transformations.

Despite the fact that the worldline is $\e$-dependent, fixing it does not affect the structure of the Einstein equation in a global coordinate system, since it does not affect the covariant derivative. However, in a local coordinate system, such as Fermi normal coordinates, the coordinate values at a given point $\mathcal{P}$ are defined by their relationship with $\gamma$; they are effectively tethered to it, such that a displacement at $\mathcal{P}$ induces a displacement along $\gamma$. Thus, in a local coordinate system, the fact that the worldline has $\e$-dependence \emph{does} affect the Einstein equation.

All of these issues will be discussed further in the following chapters. In Ch.~\ref{point_particle}, I illustrate the construction of a fixed-worldline expansion for a point particle, emphasizing especially how it differs from a regular expansion. In Ch.~\ref{extended_body}, I return to the case of an extended body and provide a concrete formulation of my fixed-worldline approximation scheme. Chapters~\ref{matching}--\ref{perturbation calculation} present two means of solving the equations in the scheme: via the method of matched asymptotic expansions, and via a calculation in the buffer region that makes minimal assumptions about the solutions in the inner and outer expansions.

			\chapter{The motion of a point particle}
\label{point_particle}
%%%%%%%%%%%
Before tackling the physically and mathematically meaningful problem of an asymptotically small body, I will analyze the simpler problem of an exact point particle. The analysis is, of course, purely formal, because most of the equations fail to make sense in terms of distributions; I will alert the reader to this failure at key points. However, the analysis will serve as an illustration, allowing me to flesh out the differences between a regular expansion and a general expansion with a fixed worldline, thereby setting the stage for my approach in the succeeding chapters.

So, assume for the moment that the exact Einstein equation, $\exact{G}_{\mu\nu}=8\pi\exact{T}_{\mu\nu}$, can be made sense of with the point particle source
\begin{equation}
\exact{T}^{\mu\nu}[\exact{g},\gamma]=\int_\gamma m\exact{u}^\mu\exact{u}^\nu\delta(x,\exact{z}(\exact{t}))d\exact{t},
\end{equation}
where $\gamma$ is the worldline of the particle, $\exact{z}^\alpha(\exact{t})$ are the coordinates on $\gamma$, $\exact{u}^\mu=\diff{\exact{z}^\mu}{\exact{t}}$ is its four-velocity, $\exact{t}$ is proper time with respect to $\exact{g}$ on $\gamma$, and $\delta(x,x')=\delta^4(x^\mu-x'^\mu)/\sqrt{|\exact{g}|}$ is a covariant delta function in the spacetime of the exact solution $\exact{g}_{\mu\nu}$, with $|\exact{g}|$ denoting the absolute value of the determinant of $\exact{g}_{\mu\nu}$. In order to solve the EFE, I will split $\exact{g}$ into a background plus perturbation, $\exact{g}=g+h$.

Now, as discussed in the linearized case in Sec.~\ref{inconsistencies}, the equation of motion of the point particle is entirely determined by the EFE, via the Bianchi identity ${}^{\exact{g}\!}\del{\nu}\exact{G}^{\mu\nu}=0$, which implies the conservation equation ${}^{\exact{g}\!}\del{\nu}\exact{T}^{\mu\nu}=0$. Just as in the linearized case, one can straightforwardly calculate the divergence of $\exact{T}^{\mu\nu}$ to find
\begin{equation}\label{exact_conservation}
{}^{\exact{g}\!}\del{\nu}\exact{T}^{\mu\nu}=\int_\gamma m\exact{a}^\mu\delta(x,\exact{z}(\exact{t}))d\exact{t},
\end{equation}
where $\exact{a}^\mu\equiv{}^{\exact{g}\!}\del{\exact{u}}\exact{u}^\mu$ is the acceleration of the worldline in the full spacetime described by $\exact{g}$. From this it follows that the exact equation of motion is $\exact{a}^\mu=0$; that is, in the full spacetime, $\gamma$ is a geodesic. However, we seek the equation of motion in the background spacetime, not in the full spacetime. Specifically, we seek an expression for $a^\mu\equiv\del{u}u^\mu$, where $u^\mu\equiv \displaystyle\diff{\exact{z}^\mu}{t}$. Although $a^\mu$ is still defined by the exact worldline $\exact{z}^\mu$, it will differ from $\exact{a}^\mu$ for two reasons: one, the Christoffel symbols in the two spacetimes differ; and two, the proper times differ. Following the notation of Appendix~\ref{general_expansions}, I define the tensor $C^\alpha{}_{\beta\gamma}={}^{\exact{g}}\Gamma^\alpha_{\beta\gamma}-\Gamma^\alpha_{\beta\gamma}$, given explicitly by Eq.~\eqref{Cabc}. I also define $t$ to be the proper time with respect to $g$ on the worldline. A brief calculation then determines
\begin{align}
\exact{a}^\alpha &= \ddiff{\exact{z}^\alpha}{\exact{t}}+{}^{\exact{g}}\Gamma^\alpha_{\beta\gamma}\diff{\exact{z}^\beta}{\exact{t}}\diff{\exact{z}^\gamma}{\exact{t}}\\
&= \left(\diff{\exact{t}}{t}\right)^{-2}\left(a^\alpha-\exact{k}u^\alpha+C^\alpha{}_{\beta\gamma}u^\beta u^\gamma\right),
\end{align}
where $\exact{k}\equiv\left(\diff{\exact{t}}{t}\right)^{-1}\ddiff{\exact{t}}{t}$; from the definition of proper time, we have $d\exact{t}=\displaystyle\sqrt{-\exact{g}_{\mu\nu}u^\mu u^\nu}dt$, which implies
\begin{equation}
\exact{k} = \frac{1}{\displaystyle\sqrt{-\exact{g}_{\mu\nu}u^\mu u^\nu}}\diff{}{t}\sqrt{-\exact{g}_{\mu\nu}u^\mu u^\nu}.
\end{equation}
Using $\exact{a}^\mu=0$, we now have the exact equation of motion in the background spacetime:
\begin{equation}\label{exact_eq_motion}
a^\alpha = \exact{k}u^\alpha-C^\alpha{}_{\beta\gamma}u^\beta u^\gamma.
\end{equation}

In the naive approach to the problem, as presented in Sec.~\ref{intro_to_self_force}, the exact EFE and equation of motion are expanded independently without considering how they are related. If we expand the metric as $\exact{g}=g+\e\hmn{}{1}+\order{\e^2}$ and substitute that expansion into the exact equation of motion \eqref{exact_eq_motion}, we quickly arrive at 
\begin{equation}\label{eq_motion}
a^\mu = -\tfrac{1}{2}\e(g^{\mu\nu}+u^\mu u^\nu)(2\hmn{\nu\rho;\sigma}{1}-\hmn{\rho\sigma;\nu}{1})u^\rho u^\sigma+\order{\e^2}.
\end{equation}
Since the metric of a point particle diverges at its position, this equation is ill-defined, but it can be regularized, leading to Eq.~\eqref{first_self_force_expression}. (Note also that because the metric is singular on $\gamma$, so too are the Christoffel symbols, meaning that neither Eq.~\eqref{exact_conservation} nor \eqref{exact_eq_motion} is definable in terms of distributions.)

However, if we substitute the expansion of the metric into the EFE, then a brief calculation leads to the first-order equation $\delta G[\e\hmn{}{1}]=8\pi T[g,\gamma]$, where $T$ is the stress-energy tensor of a point particle in the background spacetime $g$. As discussed in Sec.~\ref{inconsistencies}, the linearized Bianchi identity implies that $T$ must be conserved, which then implies that the worldline must be a geodesic in the background spacetime. This obviously contradicts the equation of motion \eqref{eq_motion}. Historically, this inconsistency has been removed by an a posteriori gauge-relaxation, which was discussed in detail in Sec.~\ref{inconsistencies}, and which I recapitulate here: in this procedure, the linearized EFE is written in the Lorenz gauge, such that it takes on the form of a wave equation $E[\e\hbarmn{}{1}]=-16\pi T[g,\gamma]$, but the Lorenz gauge condition $L_\mu[\e\hmn{}{1}]=0$ is replaced with the milder condition $L_\mu[\e\hmn{}{1}]=O(\e^2)$. The wave equation can be solved with an arbitrary worldline, so it no longer contradicts the equation of motion; and because the errors in the gauge condition are small, the solution to the wave equation is also an approximate solution to the EFE. In essence, this procedure replaces the first-order EFE with the less stringent equation $\delta G[\e\hmn{}{1}]=8\pi T[g,\gamma]+\order{\e^2}$, which allows the stress-energy tensor $T$ to be \emph{not quite} conserved, thereby avoiding the conclusion that $\gamma$ is a geodesic in the background spacetime.

Although this removes the inconsistency, it is ad hoc, having no evident relationship with a systematic expansion of the EFE. In the remainder of this chapter, I present two approximation schemes that more systematically overcome the contradiction between the equation of motion and the linearized EFE. In these schemes, the role of the worldline is more carefully considered, every perturbation equation is solved exactly, and no inconsistencies arise. I first present a regular expansion, which involves an expansion of the worldline and hence fails on long timescales; I then present a general expansion involving a fixed worldline, which (i) overcomes the limitations of the regular expansion, and (ii) offers a more systematic version of the gauge-relaxation procedure.

\section{Regular expansion}\label{regular_expansion_point_particle}
Although a regular expansion in powers of $\e$ might be at the backs of most researchers' minds when they derive an expression for the self-force, only one extant derivation of the MiSaTaQuWa equation \cite{Gralla_Wald} has explicitly sought to remain within the framework of such an expansion. The regular expansion  begins by expanding the metric as 
\begin{equation}
\exact{g}_{\mu\nu}(x,\e)=g(x)+\e\hmn{\mu\nu}{1}(x)+\e^2\hmn{\mu\nu}{2}(x)+\order{\e^3},
\end{equation}
and the Einstein tensor as 
\begin{align}
\exact{G}^{\mu\nu}[\exact{g}]&=G^{\mu\nu}+\e\delta G^{\mu\nu}[\hmn{}{1}]+\e^2\delta G^{\mu\nu}[\hmn{}{2}]+\e^2\delta^2G^{\mu\nu}[\hmn{}{1}] +\order{\e^3},
\end{align}
where $G$ is the Einstein tensor of the background metric $g$, $\delta G^{\mu\nu}[h]$ is linear in $h$ and its derivatives, and $\delta^2 G^{\mu\nu}[h]$ is quadratic in them. Similarly, by expanding the $\sqrt{|\exact{g}|}$ that appears in it, and converting from the proper time in $\exact{g}$ to the proper time in $g$, the stress-energy tensor can be expanded as
\begin{equation}
\exact{T}^{\mu\nu}[\exact{g},\gamma]=\e T^{\mu\nu}[\gamma]+\e^2\delta T^{\mu\nu}[\hmn{}{1},\gamma]+\order{\e^3},
\end{equation}
where the factor of $\e$ is pulled out of $T$ for convenience.

However, given that $\gamma$ satisfies the exact equation \eqref{exact_eq_motion}, it will evidently depend on $\e$, so the above expansion is not yet regular. To make it regular, we must expand the worldline as $\gamma=\gamma\coeff{0}+\e\gamma\coeff{1}+\order{\e^2}$. With the coordinates of the worldline defined by $\exact{z}^\alpha(\tau,\e)$, this expansion takes the form
\begin{equation}
\exact{z}^\alpha(\tau,\e)=\zn{0}^\alpha(\tau)+\e\zn{1}^{\alpha}(\tau)+\order{\e^2},
\end{equation}
where $\tau$ will indicate proper time on the leading-order worldline $\gamma\coeff{0}$, which has the coordinate-form $\zn{0}^\alpha(\tau)$. To make this expansion most meaningful, we can insist that at some time $\tau=\tau_0$ the exact curve $\exact{z}^\alpha$ is tangential to the leading-order curve $\zn{0}^\alpha$; the corrections $\zn{\emph{n}}^\alpha$, $n>0$, then determine the deviation of the exact curve from the geodesic as time progresses away from $\tau=\tau_0$. Since the different terms in the expansion cannot map to different points in a curved spacetime, the ``corrections" are in fact vectors defined on the leading-order worldline. Thus, in a regular expansion, the leading-order approximation $\zn{0}^\mu$ is the only worldline that appears in the background spacetime. The corrections point from this worldline to the true worldline $\exact{z}^\alpha(\tau)$, in the same sense that a geodesic deviation vector points from one geodesic to another, neighbouring one. So in this expansion, the analogue of the MiSaTaQuWa equation will not be an equation for the acceleration of a worldline; instead, it will be an equation for the acceleration of the deviation vector $\zn{1}^\mu$. This acceleration will naturally include a term identical to that of the geodesic deviation equation \cite{Gralla_Wald}, due to the drift of the the true worldline $\gamma$ away from the reference worldline $\gamma\coeff{0}$.

(Note that in these expansions, $g$ is regular, $\hmn{}{1}$ has a delta-function singularity on the worldline, but higher-order terms must solve nonlinear equations, making them ill-defined as distributions. Similarly, $\zn{0}$ is smooth, $\zn{1}$ is regularizable, but higher-order deviation vectors are too singular to be made sense of in terms of distributions.)

By using this expansion of the worldline, we can construct a regular expansion of the stress-energy tensor,
\begin{align}
\exact{T}^{\mu\nu}(\gamma)&=\e T^{\mu\nu}[\gamma\coeff{0}]+\e^2\delta T^{\mu\nu}[\hmn{}{1},\gamma\coeff{0}] +\e^2\tilde{\delta}T_{\mu\nu}[\gamma\coeff{0},\gamma\coeff{1}]+\order{\e^3},
\end{align}
where $\delta T^{\mu\nu}$ is linear in $\hmn{}{1}$, and $\tilde\delta T^{\mu\nu}$ is linear in $\zn{1}$. Substituting this expansion into the Einstein equation, we arrive at a sequence of field equations, written schematically as
\begin{align}
G^{\mu\nu} &= 0,\\
\delta G^{\mu\nu}[\hmn{}{1}] & = 8\pi T^{\mu\nu}[\gamma\coeff{0}], \\
\delta G^{\mu\nu}[\hmn{}{2}] & = 8\pi\delta T^{\mu\nu}[\hmn{}{1},\gamma\coeff{0}]+8\pi\tilde\delta T^{\mu\nu}[\gamma\coeff{0},\gamma\coeff{1}]-\delta^2 G^{\mu\nu}[\hmn{}{1}],\\
&\ \ \vdots\nonumber
\end{align}
These equations can be solved order-by-order for the background metric $g$, the worldline $\zn{0}$, and the perturbations $\hmn{\mu\nu}{{\it n}}$ and $\zn{n}$. My method of solving them is to impose the Lorenz gauge condition on each term in the metric perturbation:
\begin{equation}
L_\mu[\hmn{}{\emph{n}}] = 0.
\end{equation}
With this condition, in each of the above equations, the linearized Einstein tensor becomes the wave-operator $E_{\mu\nu}$, leading to the sequence of wave equations
\begin{align}
E^{\mu\nu}[\hbarmn{}{1}] & = -16\pi T^{\mu\nu}[\gamma\coeff{0}], \label{reg_expansion1}\\
E^{\mu\nu}[\hbarmn{}{2}] & = -16\pi\delta T^{\mu\nu}[\hmn{}{1},\gamma\coeff{0}]-16\pi\tilde\delta T^{\mu\nu}[\gamma\coeff{0},\gamma\coeff{1}]+2\delta^2 G^{\mu\nu}[\hmn{}{1}],\label{reg_expansion2}\\
&\ \ \vdots\nonumber
\end{align}
together with the background EFE, $G^{\mu\nu}= 0$. The $n$th-order wave equation can be readily solved for $\hmn{}{\emph{n}}$ as a functional of $\zn{0}$,..., $\zn{\emph{n}-1}$. Applying the gauge condition to $\hmn{}{\emph{n}}$ ensures that it solves the true $n$th-order EFE, rather than just the wave equation; and because the EFE fully determines the worldline, this means that imposing the gauge condition will do likewise. More precisely, writing the $n$th-order wave equation in the compact form
\begin{equation}
E^{\mu\nu}[\hmn{}{\emph{n}}]=S\dcoeff{\emph{n}}^{\mu\nu},
\end{equation}
its solution is
\begin{equation}
\hbarmn{\mu\nu}{\emph{n}} = -\frac{1}{4\pi}\int G_{\mu\nu\mu'\nu'}S\dcoeff{\emph{n}}^{\mu'\nu'}dV',
\end{equation}
where $S\dcoeff{\emph{n}}^{\mu\nu}$ is a functional of $\zn{0}$,...,$\zn{\emph{n}-1}$. And after making use of Eq.~\eqref{Green1} and integrating by parts, the gauge condition reads
\begin{equation}
0 = L_\mu[\hmn{}{\emph{n}}] = -\frac{1}{4\pi}\int G_{\mu\mu'}\del{\nu'}S\dcoeff{\emph{n}}^{\mu'\nu'}dV',
\end{equation}
where $G_{\mu\mu'}$ is the Green's function for the relativistic vector wave equation \eqref{vector Green}. From this result, we see that, just as was the case in the first-order problem, imposing the conservation of the source $S\dcoeff{\emph{n}}^{\mu\nu}$ is equivalent to imposing the gauge condition on $\hmn{}{\emph{n}}$, and either one will determine $\zn{\emph{n}-1}$. (Alternatively, in this case one could determine every $\zn{\emph{n}}$ by inserting the expansions of the metric and worldline directly into the exact equation of motion \eqref{exact_eq_motion}. However, in general we do not have access to the exact equation of motion.)

Explicitly, the first-order wave equation is integrated to find
\begin{equation}\label{1st order regular}
\hmn{\mu\nu}{1}=4\int_{\gamma\coeff{0}} \bar G_{\mu\nu\mu'\nu'}\un{0}^{\mu'}\un{0}^{\nu'}d\tau',
\end{equation}
where $\un{0}^\mu\equiv\displaystyle\frac{d\zn{0}^\mu}{d\tau}$ is the four-velocity on the leading-order worldline. The gauge condition then reads
\begin{align}
0 &= L_\mu[\hmn{}{1}] = 4m\int_{\gamma\coeff{0}} G_\mu{}^{\nu'}\an{0}_{\nu'}d\tau',
\end{align}
where $\an{0}_{\mu'}$ is the acceleration of $\gamma\coeff{0}$ in $g$. Hence, $\gamma\coeff{0}$ must be a geodesic in the background spacetime.

But $\gamma\coeff{0}$ does not describe the true worldline of the particle: the effect of radiation at first order is incorporated into the correction $\zn{1}^\mu$. Integrating the second-order wave equation, we find
\begin{align}
\hmn{\mu\nu}{2} &= 2\int_{\gamma\coeff{0}} \bar G_{\mu\nu\mu'\nu'}\un{0}^{\mu'}\un{0}^{\nu'}\left(\un{0}^{\rho'}\un{0}^{\sigma'} -g^{\rho'\sigma'}\right)\hmn{\rho'\sigma'}{1}d\tau' \nonumber\\
&\quad +4\int_{\gamma\coeff{0}}\bar G_{\mu\nu\mu'\nu'}\left(2\un{0}^{\mu'}\un{1}^{\nu'} +\un{0}^{\mu'}\un{0}^{\nu'}\un{0}{}_{\gamma'}\un{1}^{\gamma'}\right)d\tau' \nonumber\\
&\quad +4\int_{\gamma\coeff{0}}\bar G_{\mu\nu\mu'\nu';\rho'}\un{0}^{\mu'}\un{0}^{\nu'}\zn{1}^{\rho'}d\tau' -\frac{1}{2\pi}\int \bar G_{\mu\nu\mu'\nu'}\delta^2 G^{\mu'\nu'}dV',
\end{align}
where $\un{1}^\mu\equiv\un{0}^\nu\del{\nu}\zn{1}^\mu$. The first line in this solution arises from $\delta T$, while the second line and the first term in the third arise from $\tilde\delta T$. Imposing the gauge condition $L_\mu[\hmn{}{2}]=0$, making use of Eq.~\eqref{Green1}, integrating by parts, and then making use of the Ricci identity and the second-order Bianchi identity (given by $\del{\nu}\delta^2 G^{\mu\nu}=-\delta\Gamma^\mu_{\beta\gamma}\delta G^{\beta\gamma}-\delta\Gamma^\beta_{\beta\gamma}\delta G^{\mu\gamma}$, where $\delta\Gamma$ is the linear correction to the background Christoffel symbol), we arrive at
\begin{align}\label{force regular}
(g^\mu_\nu+\un{0}^\mu\un{0}{}_\nu)\ddot{z}\dcoeff{1}^\nu-R^\mu{}_{\nu\rho\sigma}\un{0}^\nu\un{0}^\rho\zn{1}^\sigma &= -\tfrac{1}{2}(g^{\mu\nu}+\un{0}^\mu \un{0}^\nu)(2\hmn{\nu\rho;\sigma}{1} -\hmn{\rho\sigma;\nu}{1})\un{0}^\rho\un{0}^\sigma,
\end{align}
where $\ddot{z}_{\scriptscriptstyle{(1)}}^\mu\equiv \un{0}^\rho\del{\rho}\left(\un{0}^\nu\del{\nu}\zn{1}^\mu\right)$.

Equation \eqref{force regular} is the regular-expansion analogue of the MiSaTaQuWa equation. Its solution describes the spatial deviation of the true worldline away from the reference geodesic $\gamma\coeff{0}$. (The equation is spatial in the sense that it has no component along $\un{0}$.) Its right-hand side has precisely the form of Eq.~\eqref{eq_motion}. But its left-hand side is explicitly altered because it is an equation for a deviation vector, and it hence includes a term proportional to the Riemann tensor.

Note that the equation is gauge-invariant in the following sense: Under a gauge transformation generated by a vector $\xi^\alpha$, the deviation vector and metric perturbation change as $\zn{1}^\mu\to\zn{1}^\mu-\xi^\alpha$ and $\hmn{\mu\nu}{1}\to\hmn{\mu\nu}{1}+2\xi_{(\mu;\nu)}$. If we apply these transformations to Eq.~\eqref{force regular}, we find that they have an identical effect on both the left- and right-hand sides: the terms $-(g^\mu_\nu+\un{0}^\mu \un{0}{}_\nu)\ddot{\xi}^\nu+R^\mu{}_{\nu\rho\sigma}\un{0}^\nu\un{0}^\rho\xi^\sigma$ appear. Therefore the equation holds in any gauge. Of course, this gauge transformation has no effect on the leading-order, $\e$-independent worldline $\gamma\coeff{0}$; it alters only the deviation vectors.

While the basic idea of this approach is valid and rigorous, it is unsatisfactory because of its limited realm of validity. For example, in a typical EMRI orbit, the radial coordinate $z^r$ on the particle's leading-order worldline will be of order $\e^0$ for all time, while the deviation vector $\zn{1}^r$ will grow as $\an{1}\cdot(\tau-\tau_0)^2$. This means that the expansion of $\gamma$ is valid only on timescales $\tau\sim\mathcal{R}$: after a dephasing time $\tau\sim\mathcal{R}/\sqrt{\e}$, the ``correction" $\e\zn{1}$ will be of the same order as the leading-order term $\zn{0}$. In other words, the expansion is not uniform in time. And once we commit ourselves to a nonuniform expansion, we must restrict the entire problem to a bounded time-interval $[\tau_i,\tau_f]$. Within this fixed interval, the expansion is valid in the sense that we can guarantee it will be accurate to any given numerical value by making $\e$ sufficiently small; on an unbounded, or a generically $\e$-dependent interval, this statement would not hold true. The restriction to this bounded region has several important consequences. Most obviously, as previously stated, we are specifically interested in large changes that occur on the time-interval $\sim \mathcal{R}/\e$---such as the particle's slow inspiral in an EMRI. Thus, the entire expansion scheme fails on the timescale of interest.

The restriction to a bounded time-interval also restricts the formalism in an important way: since the expansion of the Einstein equation is valid only on a bounded region, the solution to it cannot necessarily be expressed in terms of an unbounded past history. In other words, the boundedness of the domain effectively forces us to cast the problem in an initial value formulation from the beginning. This means that we cannot express the force purely in terms of the usual tail integral; as soon as one writes down the solution as an integral over the entire past history, one assumes that one's expansion is globally valid, rather than just locally valid.\footnote{This point seems to have been missed in Ref.~\cite{Gralla_Wald}.} We can easily see this from the following argument: The correction terms $\zn{n}$ grow large not only for times far in the future of $\tau_0$, but also for times far in the past of $\tau_0$. Hence, at any time $\tau$, the difference between the tail as calculated on $\gamma\coeff{0}$ and the tail as calculated on $\gamma\coeff{0}+\e\gamma\coeff{1}$ will differ by a significant amount, given by $\left|\tail[\gamma\coeff{0}] -\tail[\gamma\coeff{0} +\e\gamma\coeff{1}]\right|\sim \e^2\int_{-\infty}^\tau \zn{1}^{\alpha'}(\tau')\partial_{\alpha'} G(x,z(\tau'))d\tau'$; since $\zn{1}$ grows with $(\tau-\tau_0)^2$, the difference between the two tails appears to be potentially infinite. It is quite likely that the decay of the retarded Green's function would ameliorate this divergence in any case of interest. But there is no obvious reason for extending the domain of the solution beyond the domain of validity of the expansion.

Hence, at each order, the integral over the source must be cut off at the initial time $\tau=\tau_i$, and the remainder of the tail must be replaced by Cauchy data on that initial timeslice. A consequence of this is that the self-force is not naturally expressed in terms of a tail integral over an infinite past history; instead, it is more naturally expressed in terms of a purely local regular field, defined as the retarded field minus a certain local, singular part, in the manner of Detweiler and Whiting. Besides making the solution valid, this also has the advantage of expressing the force in terms of local quantities, with no reference to the past history of the particle; this is useful for a numerical integration in the time domain---and for developing a two-timescale method, as I will discuss momentarily.

\section{General expansion}\label{singular_expansion_point_particle}
Given the limitations of a regular expansion, let us now consider a general expansion. In effect, this expansion will provide a systematic justification of the gauge-relaxation procedure discussed in Sec.~\ref{inconsistencies}. Recall that our basic goal is to find a pair $(\gamma,h)$ satisfying Einstein's equation. In a regular expansion, both the worldline and the metric perturbation are expanded in the limit of small $\e$. In the general expansion we shall now consider, the worldline is held fixed while all other $\e$-dependence is expanded. To find the terms in this expansion, I seek to expand both the exact Einstein equation and the exact equation of motion such that they can be solved with this fixed worldline. With that goal in mind, I decompose the metric as
\begin{equation}\label{worldline_decomposition}
\exact{g}_{\mu\nu}(x,\e)=g_{\mu\nu}(x)+h_{\mu\nu}(x,\e;\gamma).
\end{equation}
I assume that the perturbation can be expanded while holding fixed the functional dependence on $\gamma$:
\begin{equation}\label{worldline_expansion}
h_{\mu\nu}(x;\gamma)=\sum_{n=1}^N\e^n\hmn{\mu\nu}{{\it n}}(x;\gamma)+\order{\e^{N+1}},
\end{equation}
where each term $\hmn{\mu\nu}{{\it n}}$ is a functional of the true worldline $\gamma$ but is nevertheless of order $O_s(1)$. Note that the approximation scheme fails---becoming both inaccurate and internally inconsistent---if any of these coefficients are found to grow larger than order unity. Substituting Eqs.~\eqref{worldline_decomposition} and \eqref{worldline_expansion} into the Einstein equation, we arrive at
\begin{align}\label{not wave-like}
G^{\mu\nu}&+\e\delta G^{\mu\nu}\big[\hmn{}{1}\big]+\e^2\delta G^{\mu\nu}\big[\hmn{}{2}\big]+\e^2\delta^2G^{\mu\nu}\big[\hmn{}{1}\big]+...\nonumber\\
 &= 8\pi\e T^{\mu\nu}[\gamma]+8\pi\e^2\delta T^{\mu\nu}[\hmn{}{1},\gamma]+...
\end{align}

Now, because of the $\e$-dependence in $\gamma$, coefficients of explicit powers of $\e$ are not necessarily equal in the above equation. And if we did attempt to solve the equation order-by-order by equating coefficients in that way, we would arrive at the usual, undesirable conclusion that $\gamma$ must be a geodesic. Hence, we must first reformulate the exact equation. To that end, I adopt the Lorenz gauge for the entire perturbation $h$, rather than for any individual term in its expansion:
\begin{equation}\label{general Lorenz gauge}
L_{\mu}\big[h\big]=0.
\end{equation}
One should note that this choice of gauge can differ significantly from the choice in the regular expansion, where the gauge condition was imposed separately on each individual term in the expansion of $h$. For the moment, I merely assume that Eq.~\eqref{general Lorenz gauge} can always be imposed; I will discuss the validity of that assumption at the end of this section. With the gauge chosen, and after a trivial rearrangement, the Einstein equation is transformed into the weakly nonlinear wave equation
\begin{align}\label{wave-like}
\e E^{\mu\nu}&\big[\hmn{}{1}\big]+\e^2 E^{\mu\nu}\big[\hmn{}{2}\big]+...\nonumber\\
&= 2G^{\mu\nu}-16\pi\e T^{\mu\nu}[\gamma]-16\pi\e^2\delta T^{\mu\nu}[\hmn{}{1},\gamma]+2\e^2\delta^2G^{\mu\nu}\big[\hmn{}{1}\big]+...
\end{align}
Unlike Eq.~\eqref{not wave-like}, in this equation one \emph{can} equate coefficients of powers of $\e$ without determining $\gamma$ in the process. The equation itself is essentially identical to the relaxed Einstein equation, which forms the basis of most post-Minkowski expansions \cite{relaxed_EFE1,relaxed_EFE2}. Both equations are ``relaxed" in the sense that they can be solved without specifying the motion of the source, which is determined only afterward by imposing the gauge condition. Also, in both cases, nonlinearities are treated as source terms for a hyperbolic wave operator, which means that corrections to the null cones are incorporated into the perturbations, rather than into the characteristics of the wave equation.

I now assume that Eq.~\eqref{wave-like} is solved for arbitrary $\gamma$, which implies that coefficients of explicit powers of $\e$ must be equal. This yields the sequence of equations
\begin{align}
G^{\mu\nu} &= 0,\\
E^{\mu\nu}\big[\hbarmn{}{1}\big] & = -16\pi T^{\mu\nu}[\gamma], \label{wave_first}\\
E^{\mu\nu}\big[\hbarmn{}{2}\big] & = -16\pi\delta T^{\mu\nu}\big[\hmn{}{1},\gamma\big]+2\delta^2 G^{\mu\nu}\big[\hmn{}{1}\big],\label{wave_second}\\
&\ \ \vdots\nonumber
\end{align}
which differ from Eqs.~\eqref{reg_expansion1} and \eqref{reg_expansion2} only in that the stress-energy tensor's dependence on $\gamma$ has not been expanded, because the worldline itself has not been. Each of the wave equations can be solved in the same manner as were the corresponding equations in the regular expansions: writing the $n$th-order wave equation in the compact form
\begin{equation}
E^{\mu\nu}\big[\hbarmn{}{\emph{n}}\big]=S\dcoeff{\emph{n}}^{1\mu\nu}[\gamma],
\end{equation}
its solution is
\begin{equation}\label{compact_solutions}
\hbarmn{\mu\nu}{\emph{n}} = -\frac{1}{4\pi}\int G_{\mu\nu\mu'\nu'}S\dcoeff{\emph{n}}^{1\mu'\nu'}[\gamma]dV',
\end{equation}
where, unlike in the regular expansion, the solution is now a functional of the exact worldline $\gamma$. (Note that the superscript 1 is a label, not an index.)

All that remains is to determine that exact worldline. Referring to the exact equation of motion~\eqref{exact_eq_motion}, we see that if we substitute the metric expansion into it, it reads
\begin{equation}\label{expanded_eq_motion}
a^\mu = -\tfrac{1}{2}\e(g^{\mu\nu}+u^\mu u^\nu)(2\hmn{\nu\rho;\sigma}{1}[\gamma]-\hmn{\rho\sigma;\nu}{1}[\gamma])u^\rho u^\sigma+\order{\e^2},
\end{equation}
where $u^\mu=\diff{\exact{z}^\mu}{t}$ and $t$ is proper time with respect to the background metric on $\gamma$. Like the EFE in relaxed form, this equation can be solved with an arbitrary worldline, given an expansion of the acceleration, 
\begin{equation}\label{acceleration_expansion}
a_\mu(t,\e) = \an{0}_\mu(t)+\e\an{1}_\mu(t;\gamma)+\order{\e^2}.
\end{equation}
This is an expansion of a function of time along the fixed worldline; that function will eventually be identified with the acceleration of the worldline, but only once we have obtained an approximation that we deem sufficiently accurate. Before that final step, the worldline itself is left arbitrary. (Note that the equation of motion will cease to be regularizable after $\an{1}$, meaning that the equation of motion of the fixed worldline is itself ill-defined for an exact point particle. This differs from the case of a regular expansion, where the worldline was well defined as a geodesic, but the deviation vectors on it became ill-defined.)

Each term in Eq.~\eqref{acceleration_expansion} can be found from Eq.~\eqref{expanded_eq_motion}. However, since we do not in general have access to such an expanded equation of motion, it is preferable to make use of the Lorenz gauge condition, as was done in the regular expansion. Substituting the expansions of $h_{\mu\nu}$ and $a^\mu$ into the exact gauge condition $L_\mu[h]=0$ and solving with arbitrary $\gamma$, we arrive at the sequence of equations
\begin{align}
L\coeff{0}_\mu\big[\hmn{}{1}\big] &=0, \label{gauge_expansion 1}\\
L\coeff{1}_\mu\big[\hmn{}{1}\big] &= -L\coeff{0}_\mu\big[\hmn{}{2}\big],
 \label{gauge_expansion 2}\\
&\ \ \vdots\nonumber
\end{align}
where $L\coeff{0}[f]\equiv L[f]\big|_{a=\an{0}}$, $L\coeff{1}[f]$ is linear in $\an{1}$, $L\coeff{2}[f]$ is linear in $\an{2}$ and quadratic in $\an{1}$, and so on. More generally, for $n>0$ the equations read
\begin{equation}\label{compact_gauge}
L\coeff{\emph{n}}_\mu\big[\hmn{}{1}\big]=-\sum_{m=1}^n L_\mu\coeff{\emph{n-m}}\big[\hmn{}{\emph{m}+1}\big].
\end{equation}
In these expressions, $L[f]$ is first calculated on an arbitrary worldline, and then to find $L\coeff{\emph{n}}$, the expansion of the acceleration is inserted---while still holding $\gamma$ and $u^\mu$ fixed.

Recall that in the case of the regular expansion, imposing the gauge condition $L[\hmn{}{\emph{n}}]=0$ was equivalent to guaranteeing the conservation of the source $S\dcoeff{\emph{n}}$ in the $n$th-order wave equation. Analogously, in the case of the present general expansion, imposing the gauge condition $L[h]=0$ is equivalent to guaranteeing the conservation of the source in the \emph{exact} wave equation \eqref{wave-like}---that source being $\sum_n \e^nS^1\dcoeff{\emph{n}}=-16\pi\e T-16\pi\e^2\delta T+2\e^2\delta^2G+...$. And the expanded form of the gauge condition, Eq.~\eqref{compact_gauge}, together with the solutions \eqref{compact_solutions}, implies the conservation equation
\begin{equation}
\left(\del{\nu}S\dcoeff{1}^{1\mu\nu}\right)\coeff{\emph{n}}=-\sum_{m=1}^n\left(\del{\nu}S\dcoeff{\emph{m}+1}^{1\mu\nu}\right)\coeff{\emph{n-m}},
\end{equation}
where, in analogy with the notation for $L\coeff{\emph{n}}$, I have defined these quantities such that $\left(\del{\nu} S\dcoeff{\emph{m}}^{1\mu\nu}\right)\coeff{0}=\left(\del{\nu} S\dcoeff{\emph{m}}^{1\mu\nu}\right)\big|_{a=\an{0}}$, $\left(\del{\nu} S\dcoeff{\emph{m}}^{1\mu\nu}\right)\coeff{1}$ is linear in $\an{1}$, and so on.

Allow me to make the algorithm more explicit. The solution to the first-order wave equation, \eqref{wave_first}, is given by
\begin{equation}
\hbarmn{\mu\nu}{1}[\gamma] = 4m\int_\gamma G_{\mu\nu\mu'\nu'} u^{\mu'}u^{\nu'}dt'.\label{1st_funct}
\end{equation}
Note that this is the ``usual" solution obtained by solving the linearized wave equation, as in Sec.~\ref{intro_to_self_force}. The acceleration of the true worldline is determined from Eq.~\eqref{gauge_expansion 1}:
\begin{align}
0 &= L\coeff{0}_\mu\big[\hmn{}{1}\big]\nonumber\\
 & = 4m\int_\gamma G_{\mu}{}^{\mu'}\an{0}_{\mu'}dt',
\end{align}
which implies that $\an{0}_\mu=0$. It does not imply that $a_\mu=0$, however.

Proceeding to second order, the solution to Eq.~\eqref{wave_second} is
\begin{align}
\hbarmn{\mu\nu}{2}[\gamma] &= 2\int_{\gamma} G_{\mu\nu\mu'\nu'}u^{\mu'}u^{\nu'}(u^{\rho'}u^{\sigma'} -g^{\rho'\sigma'})\hmn{\rho'\sigma'}{1}dt' \nonumber\\
&\quad-\frac{1}{2\pi}\int G_{\mu\nu\mu'\nu'}\delta^2 G^{\mu'\nu'}dV'.\label{2nd_funct}
\end{align}
Imposing the gauge condition \eqref{gauge_expansion 2}, making use of Eq.~\eqref{Green1} and the second-order Bianchi identity, and integrating by parts determines the acceleration to order $\e$:
\begin{equation}
\an{1}_\mu = -\tfrac{1}{2}\!\!\left(g_{\alpha}{}^{\beta}\!+\!u_\alpha u^\beta\right)\!\!\left(2\hmn{\beta\gamma;\delta}{1} -\hmn{\delta\gamma;\beta}{1}\right)\!u^\gamma u^\delta\Big|_{a=0}.
\end{equation}
Note that the right-hand side of this equation is evaluated on the worldline, and once evaluated, it contains a term proportional to $-m\dot{a}_\mu$, corresponding to the antidamping phenomenon discovered by Havas \cite{damping} (as corrected by Havas and Goldberg \cite{damping2}). However, my assumed expansion of the acceleration has forced the right-hand side to be evaluated for $a=\an{0}=0$, which serves to automatically yield an ``order-reduced" equation with no higher-order derivatives. Also note that the expansion of the acceleration was necessary to split the gauge condition (or conservation equation) into a sequence of exactly solvable equations. Hence, we can see that the requirement of constructing exact solutions to the perturbation equations eliminates the equations of motion with non-physical solutions that have plagued prior self-consistent approaches. In particular, this method differs from that of the traditional gauge-relaxation procedure. The gauge condition \eqref{gauge_expansion 1} is similar to the relaxed gauge condition $L_\mu[\e\hmn{}{1}]=O(\e^2)$ that has been used historically---but that gauge condition allows the ill-behaved equation of motion and hence requires the posteriori corrective measure of order-reduction. And of course, the gauge-relaxation procedure itself arises as an a posteriori corrective measure, while Eq.~\eqref{gauge_expansion 1} arises as part of a systematic expansion.

However, other than the issue of order-reduction, my method yields an equation of motion that agrees with the expansion given in Eq.~\eqref{eq_motion}. In both cases, the equation of motion applies to the actual worldline $\gamma$, not to a correction to a reference geodesic. Combined with the first-order perturbation given in Eq.~\eqref{1st_funct}, the equation of motion defines a self-consistent solution to the Einstein equation, up to errors of order $\e^2$ on a timescale $\mathcal{R}/\e$; combined with the sum of the first- and second-order perturbations, it defines a solution accurate up to errors of order $\e^3$ on a timescale of order $\mathcal{R}$.

One should note two more important facts about the results just derived. First, from these results, one can easily derive those of the regular expansion, given in Eqs.~\eqref{force regular} and \eqref{1st order regular}, by expanding the worldline and following the usual steps involved in deriving the geodesic deviation equation. Second, while the Lorenz gauge is especially useful for finding the metric perturbation in the general expansion, it is not essential for finding the equation of motion, which could have been found from the conservation of the source. In other words, the equation of motion is gauge-invariant. This is not to say that the value of the acceleration in two different gauges will be the same; rather, a gauge transformation alters both the metric perturbations and the acceleration, such that the relationship between them is unaltered. However, unlike the regular expansion of the previous section, where the leading-order worldline was unaltered, in the present general expansion, the entire worldline is shifted to a new one with a new acceleration.

Beyond these specifics, one should also note the broad similarity between this general expansion and a post-Minkowksian expansion (in particular, the fast-motion approximation \cite{relaxed_EFE1}): the split of the Einstein equation into a wave equation and a gauge condition, the iterative solution to the wave equation in terms of an arbitrary worldline, and use of the gauge condition to determine the acceleration on the worldline. In some sense, then, my approximation scheme serves to elevate the gauge-relaxation procedure used in the self-force problem to the level of systematicness as post-Newtonian theory. Given these commonalities and the many successes of the post-Minkowskian expansion, one might hope that the general expansion suggested here will be equally successful in more general contexts. Note, however, that the character of the solutions given in Eqs.~\eqref{1st_funct} and \eqref{2nd_funct} is significantly different in a curved background than in a flat one, since curvature creates caustics in null cones and allows gravitational perturbations to propagate within, not just on, those cones. These complications suggest that the integrals in Eqs.~\eqref{1st_funct} and \eqref{2nd_funct} might much more easily display secular growth in a curved spacetime. Since the expansion is consistent only in the absence of such secular behaviour, it may be valid only in certain spacetimes and with certain initial conditions.

Perhaps a more significant difference between the above expansion and a post-Min\-kow\-skian one is the choice of gauge. The harmonic gauge used in post-Min\-kow\-ski expansions can be imposed as an exact coordinate condition ${}^{\exact{g}}\Box x^\alpha=0$ on the manifold of the exact solution; as long as the exact solution admits these coordinates, the gauge condition $\partial_\mu h^{\mu\nu}=0$ is automatically imposed on the entire metric perturbation, rather than on any particular term in its expansion. The Lorenz gauge used here, on the other hand, can be imposed only after decomposing the metric into a background plus perturbation, and it is typically formulated only in terms of a first-order perturbation. Up until this point, I have simply assumed that my choice of gauge can always be adopted. Allow me to now justify that assumption to some extent. If we begin with the metric in an arbitrary gauge, then the gauge vectors $\e\xi\dcoeff{1}[\gamma]$, $\e^2\xi\dcoeff{2}[\gamma]$, etc., induce the transformation
\begin{align}
h\to h'&= h+\Delta h\nonumber\\
&=h+\e\Lie{\xi\dcoeff{1}}g+\tfrac{1}{2}\e^2(\Lie{\xi\dcoeff{2}}+\Lie{\xi\dcoeff{1}}^2)g+\e^2\Lie{\xi\dcoeff{1}}\hmn{}{1}+\ldots
\end{align}
If $h'$ is to satisfy the gauge condition $L_\mu[h']$, then $\xi$ must satisfy $L_\mu[\Delta h]=-L_\mu[h]$. After a trivial calculation, this equation becomes
\begin{equation}
\sum_{n>0}\frac{\e^n}{n!}\Box\xi\dcoeff{\emph{n}}^\alpha=-\e L^{\alpha}\big[\hmn{}{1}\big]-\e^2 L^{\alpha}\big[\hmn{}{2}\big]-\e^2 L^\alpha\big[\tfrac{1}{2}\Lie{\xi\dcoeff{1}}^2g+\Lie{\xi\dcoeff{1}}\hmn{}{1}\big]+\order{\e^3}.
\end{equation}
Assuming that this equation is solved for arbitrary $\gamma$, we can equate coefficients of powers of $\e$, leading to a sequence of wave equations of the form
\begin{equation}
\Box\xi\dcoeff{\emph{n}}^\alpha = S^{2\alpha}\dcoeff{\emph{n}},
\end{equation}
where $S^{2\alpha}\dcoeff{\emph{n}}$ is a functional of $\xi\dcoeff{1},...,\xi\dcoeff{\emph{n}-1}$ and $\hmn{}{1},...,\hmn{}{\emph{n}}$. These wave equations have the solution
\begin{equation}
\xi\dcoeff{\emph{n}}^\alpha=-\frac{1}{4\pi}\int G^\alpha{}_{\alpha'}S^{2\alpha'}\dcoeff{\emph{n}}dV'.
\end{equation}
Hence, it seems that the Lorenz gauge can be adopted to any desired order. In any case, regardless of any caveats, and independent of the analogy with post-Minkowskian theory, the general expansion discussed here has the concrete advantage of offering a systematic justification of the self-consistent solution \eqref{1st_funct} and higher order corrections to it.

Other arguments have been made in favor of using the self-consistent solution \eqref{1st_funct} rather than the regular solution \eqref{1st order regular}. The simplest argument is one based on  adiabaticity: because the acceleration is very small, the true worldline deviates only very slowly from a geodesic, so the self-consistent solution can be ``patched together" from a collection of regular solutions. This argument has been made frequently in the past, most recently by Gralla and Wald~\cite{Gralla_Wald}. While it is intuitively reasonable, one must keep in mind its most basic assumption, which is that the (covariant derivative of the) tail integral as calculated over a geodesic $\gamma\coeff{0}$ is nearly identical to the tail integral as calculated over the true worldline $\gamma$. This geodesic-source approximation is a very strong one, since the tail integral potentially contains highly nonlocal contributions \cite{quasilocal, quasilocal2, quasilocal3}. As discussed in the introduction, it is probably true only in a very particular set of situations, such as, for example, an EMRI system in the adiabatic limit \cite{Hughes_adiabatic}, in which the geodesic motion is periodic and the particle executes a large number of orbits before deviating noticeably from the geodesic.\footnote{In fact, the typical derivation of the MiSaTaQuWa equation, which begins with a source moving on a geodesic but ends with a self-force, has sometimes been called an adiabatic approximation \cite{Mino_expansion2}.} Obviously, we would like the self-force to be valid in more general regimes---for example, in the final moments of plunge in the EMRI orbit.

Hinderer and Flanagan's two-timescale expansion, discussed in Sec.~\ref{multiple_scales_GR}, provides a more systematic meth\-od of ``patching together'' regular expansions. At each value of the ``slow time," one can perform a regular expansion, from which the self-force can be derived as discussed above, using the actual field and the position and momentum of the particle as initial data---this is one reason why the force as derived in a regular expansion should be expressed in terms of the actual field, rather than the tail integral over the entire past history of a geodesic. By letting the slow time evolve continuously, a series of regular expansions are automatically patched together to arrive at a self-consistent evolution.

Another expansion has been devised by Mino~\cite{Mino_expansion1, Mino_expansion2}. He begins with an expansion similar to the one presented here, but he then performs a second expansion of each $\hmn{}{n}$, in such a way that each term in the expansion of $\hmn{}{n}$ depends only on information from the instantaneously tangential worldline governed by the $(n-1)$th-order self-force; that is, each term in the expansion of the leading-order perturbation $\hmn{}{1}$ depends only on the geodesic instantaneously tangential to the true worldline, $\hmn{}{2}$ depends only on the worldline governed by the first-order self-force, and so on.

The methods developed in this dissertation  are intended to complement the above approaches. It is hoped that they will be valid in more general contexts, though more detailed studies would be required to bear out that hope.

			\chapter{The motion of an extended body}\label{extended_body}
Since the general expansion presented in the previous section is based on an exact point particle source, it is ill-behaved beyond first order. As such, we must now consider methods of accounting for the extension of an asymptotically small body. Specifically, we must consider how to formulate an asymptotic expansion in which a representative worldline for the small body is held fixed.

Perhaps the most obvious approach is to work with a body of arbitrary size and then take the limit as that size becomes small. Such a method has been used by Harte \cite{Harte_extended_bodies, Harte, Harte_EM, Harte_2009} in deriving self-force expressions, following the earlier work of Dixon \cite{Dixon1,Dixon2,Dixon}. Working with an extended body of arbitrary size is of course rather difficult, because one cannot necessarily disentangle the body's internal field from the external fields \cite{Ehlers_Rudolph}; it is only in the limit of small size that one can meaningfully speak of a body moving through an external spacetime. Hence, in this dissertation, I will be interested only in approaches that treat the body as asymptotically small from the start. The simplest means of doing so is to treat the body as an \emph{effective} point particle at leading order, with finite size effects introduced as higher-order effective fields, as done by Galley and Hu \cite{Galley_Hu}. However, while this approach is computationally efficient, allowing one to perform high-order calculations with (relative) ease, it requires one to introduce methods such as dimensional regularization and mass renormalization in order to arrive at meaningful results. Because of these undesirable requirements, I will not consider such a method here.

\section{Point particle limits}\label{point-particle limits}
In order to move from an exactly pointlike body to an asymptotically small one, we must consider a family of metrics $g(\e)$ containing a body whose mass scales as $\e$ in the limit $\e\to0$. (That is, $m\sim\e\mathcal{R}$.) If each member of the family is to contain a body of the same type, then the size of the body must also approach zero with $\e$. The precise scaling of the size with $\e$ is determined by the type of body, but this precise scaling is not generally relevant.\footnote{However, as discussed previously, my calculations require the existence of a vacuum buffer region around the body. If the body is not sufficiently compact, then it will extend throughout the buffer region, and my calculation will not apply. Likewise, my calculation fails when a body becomes tidally disrupted.} What \emph{is} relevant is the ``gravitational size"---the length scale relevant to the metric outside the body---and this size always scales linearly with the mass. If the body is compact, as is a neutron star or a black hole, then its gravitational size is also its actual linear size.

Point particle limits such as this have been used to derive equations of motion many times in the past, including in derivations of geodesic motion at leading order \cite{Infeld,Geroch_particle1,Geroch_particle2} and in constructing post-Newtonian limits \cite{Futamase_particle1, Futamase_particle2, Futamase_review}. In general, deriving corrections to geodesic motion requires considering two types of point particle limits: the inner and outer limits discussed in Ch.~\ref{approximations}. The inner limit can be expected to be valid for $r\ll \mathcal{R}$, where $r$ is some measure of radial distance from the body; and the outer limit can be expected to be valid for $r\gg\e\mathcal{R}$.

These two limits can be utilized in multiple ways. For example, the outer limit can be used to examine the effect of the small body on the external spacetime, while the inner limit can be used to study the effect of the external spacetime on the metric of the small body \cite{Manasse, Thorne_Hartle, Eric_tidal}. What is of interest in this dissertation is how the two limits mesh in the buffer region, since the metric in that region will determine the motion. To understand this, note that the buffer region, at fixed time, is approximately flat, since it is simultaneously in the asymptotic far zone of the body (because $r\gg m$) and in a small local patch in the external spacetime (because $r\ll\mathcal{R}$). Thus, in the buffer region, the linear momentum of the small body, or some other measure of motion, can be defined. Speaking roughly, an equation for the derivative of this linear momentum will then provide an equation of motion for the body.

We can consider two basic methods of deriving equations of motion in the buffer region. The first method is that of matched asymptotic expansions, which requires explicit construction of the inner and outer expansions before comparing them in the buffer region. The second method foregoes an explicit calculation of an approximation in either the inner or outer limit (or both), instead working entirely in the buffer region and using some local definition of the motion of the body. Although both make use of inner and outer expansions, the two methods are logically and practically distinct. However, both methods have sometimes been referred to as the method of matched asymptotic expansions (e.g., in Ref.~\cite{Kates_motion}).

D'Eath was the first to apply these methods to the problem of motion in General Relativity. He used matched asymptotic expansions to show that at leading order, a rotating black hole moves on a geodesic of the external spacetime \cite{DEath, DEath_paper}. Since D'Eath's pioneering work, these methods have been used in many contexts: to show that the leading-order equation of motion for an arbitrarily structured body is that of a geodesic \cite{Kates_motion}; to show that the leading-order equation of motion for a charged body is the Lorentz force law \cite{Kates_Lorenz_force}; to derive post-Newtonian equations of motion \cite{Kates_PN, Futamase_particle1, Futamase_particle2, Futamase_review, PN_matching}; to derive general laws of motion due to the coupling of the body's multipoles with those of the external spacetime \cite{Thorne_Hartle}; and most pertinently, to derive the gravitational self-force \cite{Mino_Sasaki_Tanaka, Eric_review, Eric_matching, Detweiler_review, Fukumoto, Gralla_Wald}. These derivations of the gravitational self-force will be the subject of the remainder of this section.

Let us first consider the earliest such derivation, performed by Mino, Sasaki, and Tanaka \cite{Mino_Sasaki_Tanaka}, and in slightly different manners by Poisson \cite{Eric_review, Eric_matching} and Detweiler \cite{Detweiler_review}. These derivations take the small body to be a Schwarzschild black hole (with the hope that more general bodies would obey the same equation of motion), such that in the inner limit the exact metric $\exact{g}$ can be approximated by $\exact{g}=g_B(\tilde R)+H(\tilde R)+\order{\e^2}$, where the internal background metric $g_B$ is the metric of the isolated black hole, $H(\tilde R)$ consists of tidal perturbations, and $\tilde R$ is the scaled radial coordinate discussed in Ch.~\ref{approximations}. In the outer limit, the metric is written as $\exact{g}=g+h[\gamma]+\order{\e^2}$, where $g$ is an arbitrary vacuum metric and $h[\gamma]$ is the perturbation due to a point particle traveling on a worldline $\gamma$. Expanding the external metric in normal coordinates centered on the worldline, expanding the internal metric for $r\gg m$, and insisting that the results of these expansions are identical, then determines an equation of motion for $\gamma$. Mino, Sasaki, and Tanaka \cite{Mino_matching} later used a similar method to determine an equation of motion for a small Kerr black hole; they followed Thorne and Hartle's \cite{Thorne_Hartle} approach of defining the spin and angular momentum of the body as an integral over a closed spatial surface in the buffer region, and they derived an equation of motion by combining these definitions with the assumed point particle perturbation in the external spacetime.

Allow me to more precisely state the underlying logic of these derivations, which one might hope to be as follows: Suppose there exists a metric $\exact{g}$ such that (i) in a region $D_I$, $\exact{g}$ is well approximated by the metric of a tidally perturbed black hole, (ii) in a region $D_E$, $\exact{g}$ is well approximated by the metric of some vacuum spacetime as perturbed by a point-like source moving on a worldline  $\gamma \subset \man_E$, and (iii) the regions $D_I$ and $D_E$ overlap. Then the worldline $\gamma$ is governed by the MiSaTaQuWa equation. Alternatively, a weaker formulation might be stated as follows: the approximate solutions to the Einstein equation given by $g+h[\gamma]$ and $g_B+H$, as defined above, can be combined to form a global approximate solution if and only if $\gamma$ is governed by the MiSaTaQuWa equation. As I will discuss in Ch.~\ref{matching}, the method of matching used in derivations of the self-force actually provides a significantly weaker result than either of the above two statements: it yields a unique result for the acceleration only when further assumptions are made. This follows from the fact that the coordinate transformation between the inner and outer solutions is unique only when it is strongly restricted.

We can see several problems with this approach. First, it suffers from the same problem described in the previous chapter: since the point particle solution solves the linearized Einstein equation only if the point particle travels on a geodesic, a non-systematic gauge relaxation must be invoked. Second, it does not offer any way to go beyond first order, since it provides no means of determining the external perturbations (though see Refs.~\cite{Eran_field, Eran_force} for an extension to second order). These two problems are resolved in the fixed-worldline approach. But a third problem is that the result of the matching calculation is extremely weak, since it requires many assumptions in order to determine an acceleration. Of course, it is still a marked improvement over the earliest point particle derivations, since it derives the self-force from the consistency of the field equation and makes no questionable assumptions about the behavior of singular quantities.

More recent derivations using inner and outer limits have been performed by Fukumoto, Futamase, and Itoh \cite{Fukumoto} and Gralla and Wald \cite{Gralla_Wald}. These derivations work entirely in the buffer region, rather than using matching; they do not assume that the external perturbation is that of a point particle at leading order; and they do not restrict the small body to be a Schwarzschild black hole. Fukumoto et al., following the work of Futamase \cite{Futamase_particle1, Futamase_particle2, Futamase_review} and Thorne and Hartle \cite{Thorne_Hartle}, defined the linear momentum of the body as an integral in the buffer region and derived the acceleration by simply differentiating it. While this derivation is quite simple relative to most others, it contains at least one questionable aspect: it relies on an assumed relationship between the body's linear momentum, as defined in the buffer region, and the four-velocity of the body's worldline (justified by an analogy with post-Newtonian results).

Gralla and Wald explicitly restricted themselves to a regular expansion, defining the acceleration of the body via a regular expansion of the body's worldline, roughly as described in the previous section. The only questionable aspect of their derivation is that it writes the solution to the first-order Einstein equation as an integral over the past history of the leading-order, geodesic worldline $\gamma\coeff{0}$, and it expresses the force in terms of the tail integral $\tail[\gamma\coeff{0}]$. As discussed above, this is not obviously justified: in order to remain consistently within the domain of validity of a regular expansion, the tail should be cut off at some finite past time and complemented with an integral over an initial data surface. The practical drawback of this derivation is that it is obviously limited to short timescales, as discussed previously. While a regular expansion such as this can be used to derive the self-force and then incorporated into a two-timescale expansion, my goal here is to provide a self-consistent approach in which the worldline is never treated as a geodesic.

\section{Definitions of the worldline}
While the derivations described above are increasingly satisfactory, none of them have satisfactorily defined the worldline of the asymptotically small body. To see this, we must examine the various definitions of this worldline.

Most of these definitions are in terms of the outer limit \cite{Infeld, DEath, DEath_paper, Geroch_particle1, Geroch_particle2, Gralla_Wald}. At each value of $\e$, the body can be surrounded by a worldtube $\Gamma$ of radius $\rad=o(1)$; the worldline of the body is then defined as the limit $\Gamma_{\e=0}$ of these worldtubes, which defines a curve in the limiting spacetime $\exact{g}(\e=0)$, as portrayed in Fig.~\ref{regular_limit}. Kates generalized this to the case when the limit $\e\to0$ is singular \cite{Kates_motion}.

These definitions are extremely problematic, because the worldline seems to naturally emerge only when the body is exactly point-like, which is the case only in the limiting spacetime defined by $\e\to0$. But in this spacetime, the worldline must be $\e$-independent; any $\e$-dependence would automatically be ``pure gauge," such that the self-force could be set to zero over the entire domain of the regular expansion. How, then, can the worldline be defined such that it can accurately and meaningfully reflect the motion of the body for $\e>0$? If we reject the use of small ``corrections" to a worldline, represented by a deviation vector, how can we find the worldline that the deviation vector ``points" to?

One possible definition has been suggested by Futamase \cite{Futamase_particle1, Futamase_particle2, Futamase_review, Fukumoto}. Rather than defining the worldline as the limit of a family of worldtubes of radius $\sim\e$, he defines the worldline as the curve that remains within every such tube as $\e\to0$. One can easily show that if such a curve exists, then it is unique. However, one can just as easily show that, in general, such a curve exists only for a short time: for any two values of $\e$, say $\e_1$ and $\e_2$, the interiors of the worldtubes $\Gamma_{\e_1}$ and $\Gamma_{\e_2}$ will intersect only for a brief time period $t\lesssim\mathcal{R}$. Hence, this definition does not improve upon the previous one. It also has the disadvantage that it cannot apply to small black holes.

Yet another approach is to define the worldline implicitly. Consider how this is realized in matched asymptotic expansions (see Ref.~\cite{PN_matching} for the clearest example), in which the worldline is defined roughly as follows: if the perturbed metric of the external spacetime near a worldline $\gamma$ is equivalent (up to diffeomorphism) to the perturbed metric of the small body, then $\gamma$ is said to be the worldline of the body. In practice, this means that the worldline is defined operationally by the fact that, in the buffer region, the external metric in some coordinates centered on $\gamma$ is identical to the internal metric in some mass-centered coordinates. More precisely, suppose we are given an inner expansion $g_I=g_B+H$ in coordinates $X^\mu$ on $\man_I$, and an outer expansion $g_E=g+h$ in coordinates $x^\mu$ on the manifold $\man_E$. We can always transform $g_E$ into a coordinate system (e.g., Fermi coordinates) centered on a worldline $\gamma\subset\man_E$, via a map $\phi_\gamma:\man_E\to\man_E$---this map is defined by the choice of worldline $\gamma$. The worldline is then defined to be that of the small body if there exists a \emph{unique} map $\phi_{\text{buf}}:\man_E\to\man_I$ in the buffer region such that for $\phi=\phi_{\text{buf}}\circ\phi_\gamma$, we have $\phi(x^\mu)=X^\mu$ and $\phi_* g_E(x)=g_I(X)$, where an equal sign indicates equality at the lowest common order.\footnote{I refer the reader back to Sec.~\ref{matching_conditions}, where these maps were introduced, and to Sec.~\ref{fixed_worldline_formulation}, where this definition of the worldline was first implied.}

So long as we restrict our attention to approximate solutions of the Einstein equations, this operational definition seems to be valid. It will undoubtedly result in a metric that solves the Einstein equation to some specified order in some large region of spacetime. However, at first glance it might seem unlikely that such an approximation might arise from the expansion of an exact solution to the EFE, since a worldline seems to arise naturally only at $\e=0$; it is not apparent how the $\e$-dependent worldline could arise in an exact solution. One simple means of motivating this is to suppose first that we know $g_I(X)$. Then $g_E(x)$ is given by the $\gamma$-dependent mapping $\phi^*_\gamma\circ\phi^*_{\text{buf}}g_I$, meaning that the metric in the global coordinates can obviously be written as a functional of $\gamma$. In this sense, $\gamma$ is simply a means of parametrizing the metric (c.f. the discussion in Ref.~\cite{Racine_Flanagan}).

We can also consider this from a different angle. As shown by Sciama et al.~\cite{Sciama}, any exact solution of the Einstein equation can be written in an integral formulation. Consider a bounded vacuum region $\Omega\subset\man_\e$ with a boundary $\partial\Omega$. At any point $x$ in the interior of $\Omega$, the metric will satisfy
\begin{equation}\label{Sciama formal solution}
\exact{g}^{\alpha\beta}(x) = \int\limits_{\partial\Omega_\e}{}^\exact{g}\del{\sigma'} \exact{G}^{\alpha\beta\nu'}{}_{\nu'}(x,x')dS^{\sigma'},
\end{equation}
where $\exact{G}^{\alpha\beta}{}_{\mu'\nu'}(x,x')$ is a Green's function for the operator 
\begin{equation}
\exact{D}_{\mu\nu\rho\sigma} = \tfrac{1}{2}\exact{g}^{\alpha\beta}\exact{g}_{\mu(\rho}\exact{g}_{\sigma)\nu} {}^\exact{g}\del{\alpha}{}^\exact{g}\del{\beta} +\exact{R}_{\mu(\rho\sigma)\nu}.
\end{equation}
(The proof of the integral identity in Ref.~\cite{Sciama} is restricted to a convex normal neighbourhood of $x$, but for the sake of argument, assume that it is valid even if $\Omega$ extends beyond that neighbourhood.) Assume that in some four-dimensional region $\mathcal{U}$ around the body, a scalar field $R$ provides a measure of distance from the body. The region $\mathcal{U}$ need not include the body itself, but should have the topology of $\mathcal{S}\times[T_1,T_2]$, where $\mathcal{S}$ is a spatial shell around the body and $[T_1,T_2]$ is a timelike interval. Now suppose that the ``inner" boundary of $\Omega$ is a timelike worldtube $\Gamma\subset\mathcal{U}$ of fixed radius $R\equiv\rad=o(1)$ around the body. The surface $\Gamma$ is parametrized by two angles $\theta^A$ ($A=1,2$) and by a time $T$. Thus, $\Gamma$ is generated by a collection of timelike curves $\gamma_{(\rad,\theta)}:T\mapsto x^\alpha(T,\rad,\theta^A)$. Note that since $\rad$ is arbitrary within some interval, we can use it interchangeably with $R$, implying that $(t,\rad,\theta^A)$ defines a local coordinate system $X^\mu$ near the small body. The collection of maps $\gamma_{(\rad,\theta)}$ thus defines the coordinate transformation $\phi$ between these local coordinates and the global coordinates $x^\mu$.

The metric can be written as 
\begin{align}
\exact{g}^{\alpha\beta} &= \int\limits_{S^2}\int\limits_{\gamma_{(\rad,\Theta)}}{}^\exact{g}\del{\sigma'} \exact{G}^{\alpha\beta\nu'}{}_{\nu'}\sqrt{|\exact{g}'|}n^{\sigma'}dtd\theta^1d\theta^2 +\int\limits_{\partial\Omega -\Gamma_\e}{}^\exact{g}\del{\sigma'}\exact{G}^{\alpha\beta\nu'}{}_{\nu'}dS^{\sigma'}.
\end{align}
For small values of $\e$, the radius $\rad$ of the tube is also small, so each of the curves $\gamma_{(\rad,\theta)}$ can be expanded about $\rad=0$. If $\gamma_{(\rad,\theta)}$ is sufficiently well behaved, this expansion is valid even if $x^\alpha(T,\rad=0)$ does not describe a timelike curve (or any curve) in $\man_\e$. However, note that the integrand in the above integral will generically diverge at $\rad=0$, since the small body's contributution to the metric will contain terms diverging as $\rad^{-n}$; thus, the integrand itself cannot be naively expanded in powers of $\rad$ without carefully expanding it in powers of $\e$ at the same time. (Alternatively, one could perform such an expansion and introduce some regularization method afterward.) Nevertheless, in the limit of small $\e$, the metric outside of the tube will naturally be expressed as a functional of a single worldline $\gamma: T\mapsto x^\alpha(T,\rad=0)$. This curve is made unique by demanding that the mass dipole of the body vanishes, up to some desired order, when calculated in the local coordinates $(T,\rad,\theta^A)$.

Based on these plausibility arguments, we can reasonably believe that an exact metric for a small body could naturally be expressed as a functional of an $\e$-dependent curve that represents the motion of the body. Hence, we can reasonably believe that a general expansion in which the metric perturbations are treated as functionals of a fixed, $\e$-dependent worldline, can approximate an exact metric. Actually proving that the expansion in this dissertation approximates an exact solution would presumably require a monumental effort. However, as discussed above, the worldline of the body is uniquely defined in an operational sense, and the metric that depends on it will provide a (hopefully uniform) approximate solution to the Einstein equation, whether or not it provides an approximation to an exact solution. 

Thus, the metric in the outer limit will be taken to be
\begin{equation}
\exact{g}(x,\e) = g(x)+h(x,\e;\gamma)=g(x)+\sum_{n>0}\e^n\hmn{}{\emph{n}}(x;\gamma),
\end{equation}
as first presented in Ch.~\ref{approximations}. Based on the formal solution \eqref{Sciama formal solution}, we can now see why $g(x)$ does not depend on $\gamma$ in this expansion. Taking part of the boundary to be a spatial surface $\Sigma$ that intersects the timelike worldtube $\Gamma$ and assuming that at leading order the interior (in $\man_E$) of the worldtube is smooth, it follows that for the leading-order solution outside the tube, the integral over the tube can be replaced by an integral over a spacelike ``cap" that joins smoothly with $\Sigma$; hence, the background metric will not depend on $\gamma$. I remind the reader that while the background metric in this case agrees with the background metric in a regular expansion, the terms $\hmn{}{\emph{n}}$ will \emph{not} be equal to derivatives of $\exact{g}(\e)$ with respect to $\e$ at $\e=0$, since they depend on an $\e$-dependent worldline.

With this additional, hopefully illuminating discussion now behind us, I refer the reader again to Figs.~\ref{regular_limit} and \ref{singular_limit} for a schematic comparison between the approximations constructed with and without a fixed worldline.

\section[Outline of approximation scheme]{Outline for the construction of a uniform and self-consist\-ent approximation scheme}\label{outline}
The foregoing discussions have made clear that every derivation of the gravitational self-force has at least one questionable aspect. Some of these questionable aspects are fundamental---e.g., a reliance on an exact point particle source or an assumed form for the force---while others are relatively innocuous. However, at this point in time, nearly a dozen derivations have arrived at roughly the same expression for the force. Thus, there can be little doubt, if there ever was, that the equation for the self-force is essentially correct.

The derivation presented in Chs.~\ref{buffer_region} and \ref{perturbation calculation} is not intended to remedy the first condition mentioned above: it is not without questionable aspects of its own (though these are relatively few). What the derivation \emph{is} intended to do is utilize singular perturbation theory to construct a self-consistent approximation scheme. In some sense, my analysis of singular versus regular perturbation theory has already clarified an important issue: though all earlier calculations ended with an expression for the force in terms of a tail integral, they left some ambiguity as to whether the integral is to be evaluated over the true past history of the body, or over a fictitious geodesic past. Straightforward analysis suggested that the integral \emph{must} be over a geodesic, even if, contradictorily, the motion is accelerated---but we have seen in the previous chapter why that is a faulty conclusion: it is based on regular perturbation theory, which is valid only for short times, and which hence should never have included an integral over the entire past.

Beyond providing clarifications such as that, the larger goal of my approach is to provide an approximation scheme that is potentially uniform over times $t\lesssim\mathcal{R}/\e$. I will now outline the structure of my scheme. I consider a family of metrics $\exact{g}(\e)$, where $\e>0$, in a large vacuum region $\Omega$ outside the body. Eventually, the parameter $\e$ will be identified with the mass $m_0$ of the body at some initial time. I will ensure, by construction, that the expansion is an asymptotic solution to the Einstein equation; I will hope, based on the plausibility arguments offered in the previous section, that the expansion is also an asymptotic approximation to an exact solution. As a technicality, I assume that all quantities have been rescaled by the infimum of the external length scales in $\Omega$, such that we can meaningfully speak of the mass of the body, or a radial coordinate near the body, being small or large relative to unity.

I choose $\Omega$ to lie outside a worldtube $\Gamma$ surrounding the body. The tube's radius $\rad$ is chosen to satisfy $\e\ll\rad\ll 1$; in other words, $\Gamma$ is chosen to be embedded in the buffer region where both inner and outer expansions are valid. Hence, from the point of view of the outer expansion, the radius of the worldtube is asymptotically small ($\rad\ll 1$), and its interior forms part of a smooth manifold $\man_E$, on which the external background metric is defined. The worldline lies in $\man_E$, at the center of this smooth interior. But from the point of view of the inner expansion, the radius of the tube is asymptotically large ($\rad\gg \e$), and its interior is a subset of a manifold $\man_I$, on which the internal background metric is defined, and in which there is potentially a black hole and no meaningful worldline. The worldline $\gamma$ at the center of the worldtube's interior---in $\man_E$---is defined to be the body's worldline if the body also lies at the center of the worldtube's interior---in $\man_I$---in the sense that its mass dipole vanishes on the worldtube. Using the worldtube to divide the spacetime into an inner region and an outer region in this way serves to ``cut out" the singularities that would appear in the metric perturbation in the outer limit, were it extended into the interior of the worldtube.

Although I am interested in the solution outside the tube, I will require some information from the metric in the inner limit. I assume the existence of some local polar coordinates $X^\alpha=(T,R,\Theta^A)$, such that the metric can be expanded for $\e\to 0$ while holding fixed $\tilde R\equiv R/\e$, $\Theta^A$, and $T$. This leads to the ansatz
\begin{align}\label{internal ansatz}
\exact{g}(X,\e) =g_I(T,\tilde R,\Theta^A)= g_B(T,\tilde R,\Theta^A)+ H(T,\tilde R,\Theta^A,\e),
\end{align}
where $H$ at fixed $(T,\tilde R,\Theta^A)$ is a perturbation beginning at order $\e$. The leading-order term $g_B(T,\tilde R,\Theta^A)$ at fixed $T$ is the metric of the small body if it were isolated. For example, if the body is a small Schwarzschild black hole of ADM mass $\e m(T)$, then in Schwarzschild coordinates $g_B(T,\tilde R,\Theta^A)$ is given by
\begin{align}
ds^2 &= -\left(1-2m(T)/\tilde R\right)dT^2 +\left(1-2m(T)/\tilde R\right)^{-1}\e^2d\tilde{R}^2 \nonumber\\
&\quad + \e^2\tilde{R}^2\left(d\Theta^2+\sin^2\Theta d\Phi^2\right).
\end{align}
Since the metric becomes one-dimensional at $\e=0$, the limit $\e\to0$ is singular. As discussed in Ch.~\ref{approximations}, the limit can be made regular by rescaling time as well, such that $\tilde T=(T-T_0)/\e$, and then rescaling the entire metric by a conformal factor $1/\e^2$. This is equivalent to using the above general expansion and assuming that the metric $g_B$ and its perturbations are quasistatic (evolving only on timescales $\sim 1$). Both are equivalent to assuming that the exact metric contains no high-frequency oscillations occurring on the body's natural timescale $\sim\e$. In other words, the body is assumed to be in equilibrium.

Given these assumptions, the vacuum EFE $\exact{G}=0$ can be expanded as
\begin{equation}
0=\exact{G} = G_I[g_B]+\delta G_I[H]+\delta^2 G_I[H]+...,
\end{equation}
where each term is further expanded as
\begin{align}
G_I[g_B] &= \e^{-2}\left(G\coeff{0}_I[g_B]+\e G\coeff{1}_I[g_B]+\e^2 G\coeff{2}_I[g_B]\right),\\
\delta^k G_I[g_B] &= \e^{-2}\left(\delta^kG\coeff{0}_I[H]+\e \delta^kG\coeff{1}_I[H]+\e^2 \delta^kG\coeff{2}_I[H]\right).
\end{align}
The overall factors of $\e^{-2}$ result from $\tilde R=R/\e$ and the fact that the Einstein tensor scales as the metric divided by two powers of length. The correction terms contain derivatives with respect to $T$, which are each suppressed by a factor of $\e$; specifically, $G_I\coeff{\emph{n}}$ and $\delta^kG_I\coeff{\emph{n}}$ consist of the terms in $G_I$ and $\delta^k G_I$ that contain $n$ derivatives with respect to $T$. Now, suppose $H$ possesses an expansion
\begin{equation}
H(T,\tilde R,\Theta^A,\e)=\sum_{n=1}^{N_I}\e^n H\coeff{\emph{n}}(T,\tilde R,\Theta^A).
\end{equation}
Substituting this expansion of $H$ into the above expansion of the EFE, and then solving order-by-order in powers of $\e$, leads to the sequence
\begin{align}
G\coeff{0}_I{}^{\mu\nu}[g_B] &= 0,\label{inner_eqn0}\\
\delta G\coeff{0}_I{}^{\mu\nu}[H\coeff{1}] &= -G\coeff{1}_I{}^{\mu\nu}[g_B],\label{inner_eqn1}\\
\delta G\coeff{0}_I{}^{\mu\nu}[H\coeff{2}] &= -\delta^2 G\coeff{0}_I{}^{\mu\nu}[H\coeff{1}] - \delta G\coeff{1}_I{}^{\mu\nu}[H\coeff{1}]-G\coeff{2}_I{}^{\mu\nu}[g_B],\label{inner_eqn2}\\
&\vdots\nonumber
\end{align}
Note that there is only one timescale here, so these equations automatically follow from the assumed form of the expansion of the metric; there is none of the potential failings of a two-timescale expansion. In the following chapter and Appendix~\ref{perturbed_BH}, I discuss a particular solution to this sequence of equations.

In the outer limit, I follow the procedure outlined in Sec.~\ref{singular_expansion_point_particle}. I expand in the limit $\e\to 0$ while holding fixed some global coordinates $x^\alpha$ as well as the worldline $\gamma$. This leads to the ansatz
\begin{equation}\label{external ansatz}
\exact{g}(x,\e) =g_E(x,\e;\gamma)= g(x)+h(x,\e;\gamma),
\end{equation}
where
\begin{equation}
h(x,\e;\gamma)=\sum_{n=1}^{N_E}\e^n\hmn{E}{\emph{n}}(x;\gamma) +\order{\e^{N_E+1}}.
\end{equation}
In order to solve the Einstein equation with a fixed worldline, I assume that the Lorenz gauge can be imposed everywhere in $\Omega$ on the entirety of $h$, such that $L_\mu[h]=0$.\footnote{Note that this is a stronger assumption than in the point particle case, because if the metric is given in some other gauge, the gauge vector(s) transforming to the Lorenz gauge must satisfy not only some weakly nonlinear wave equation, but also some suitable boundary conditions on the worldtube $\Gamma$. However, in practice I will be satisfied by the existence of an approximate solution to the Einstein equation that approximately satisfies the gauge condition up to errors of order $\e^3$.} With this gauge condition, the vacuum Einstein equation $\exact{R}_{\mu\nu}=0$ is reduced to a weakly nonlinear wave equation that can be expanded and solved at fixed $\gamma$, leading to the sequence of wave equations
\begin{align}
E_{\mu\nu}[\hmn{E}{1}] &=0, \label{h_E1 eqn}\\
E_{\mu\nu}[\hmn{E}{2}] &=2\delta^2 R_{\mu\nu}[\hmn{E}{1}], \label{h_E2 eqn},\\
&\vdots\nonumber
\end{align}
where $E_{\mu\nu}$ is the wave operator defined in Eq.~\eqref{wave op def}. I discuss the formal solution to these equations in Sec.~\ref{perturbation calculation}.

For simplicity, I assume that each term in the expansion of the metric perturbation minimally violates the Lorenz gauge, in the sense that if a solution truncated at some finite order violates the Lorenz gauge, then that violation is solely due to the acceleration. Again solving at fixed $\gamma$, this assumption leads to the equations
\begin{align}
L\coeff{0}_\mu\big[\hmn{E}{1}\big] &=0,\label{h_E1 gauge}\\
L\coeff{1}_\mu\big[\hmn{E}{1}\big] &= -L\coeff{0}_\mu\big[\hmn{E}{2}\big], \label{h_E2 gauge}
\end{align}
which follow from an assumed expansion of the acceleration:
\begin{equation}
a_i(t,\e) = \an{0}_i(t)+\e\an{1}_i(t;\gamma)+\order{\e^2}.\label{a expansion}
\end{equation}
I remind the reader that $L_\mu$ is the gauge operator defined in Eq.~\eqref{gauge op def}, $L\coeff{0}[f]\equiv L[f]\big|_{a=\an{0}}$, and $L\coeff{1}[f]$ consists of the terms in $L[f]$ that are linear in $\an{1}$.

In the next chapter, I discuss the solution of these equations using the method of matched asymptotic expansions. I will argue that the method, at least as it has been utilized in calculations of the self-force, provides unduly weak results. In the subsequent two chapters, I will present an alternative method, which follows the approach of Kates \cite{Kates_motion} and Gralla and Wald \cite{Gralla_Wald} by working first in the buffer region to determine the equation of motion, then using the results from the buffer region to generate a global solution. The calculation of the equation of motion in the buffer region is presented in Ch.~\ref{buffer_region}; the calculation of the global metric perturbation is presented in Ch.~\ref{perturbation calculation}.

			%%%%%%%%%%%
\chapter{Calculation of the self-force from matched asymptotic expansions}\label{matching}
%%%%%%%%%%%
In this chapter, I consider the most intuitive means of solving the systems of equations presented in the previous chapter: the method of matched asymptotic expansions. As outlined in Ch.~\ref{approximations}, in this method the perturbation equations in the inner and outer expansions are solved independently, and then any free functions are identified by insisting that the two metrics agree in the buffer region around the body. Following the tradition of the field, in matching the two metrics, I will make use of the weak matching condition, rather than the strong condition.

My presentation of the matching procedure will roughly follow that of Refs.~\cite{Eric_review,Eric_matching}, though most of my conclusions will apply as well to the earlier calculation performed by Mino, Sasaki, and Tanaka \cite{Mino_Sasaki_Tanaka}. However, my goal is not simply to review those earlier calculations, but to pinpoint their underlying assumptions. First among these assumptions is a very strong restriction on the relationship between the inner and outer solutions: essentially, the two solutions must be assumed to differ only by generically ``small" coordinate transformations in the buffer region. This restriction is required because the weak matching condition, which has always been used in matched-expansion derivations of the self-force, is found to be \emph{too} weak to yield unique results. The required restriction amounts to introducing a ``refined" matching condition midway between the weak and strong conditions.

Second among the underlying assumptions is the restrictive choice of inner solution, which effectively already removes many of the integration constants that would normally be fixed by a matching procedure. As discussed in Ch.~\ref{approximations}, in traditional matched asymptotic expansions the leading-order inner and outer solutions are determined entirely by boundary conditions, while in the matched expansions used in the self-force problem, the leading-order solutions must be chosen based on some desired physical properties; only after the leading-order solutions are chosen can boundary conditions be imposed. In the self-force problem, the leading-order outer solution is taken to be some desired vacuum metric. For the EMRI problem, the desired metric is that of a Kerr black hole. Typically, for simplicity, the leading-order inner solution is taken to be that of a Schwarzschild black hole, though one could instead choose, for example, that of a neutron star~\cite{Love_numbers1,Love_numbers2}.

However, in derivations of the self-force, the inner and outer solutions have been even further restricted: the form of the perturbations have also been largely selected, rather than determined by matching. For example, the inner perturbations have been taken to be of a particular form presumed to correspond to the influence of tidal fields on the small black hole. And the outer perturbation has been taken to be that of a point particle. In this chapter, I will make use of these assumed forms for the inner and outer solutions. As will be discussed in Chapter~\ref{perturbation calculation}, the assumption of a point particle perturbation can be removed, because the point particle solution follows directly from the assumed existence of an inner expansion. In the meantime, in this chapter, I simply take it as an assumption. On the other hand, I will never justify the assumed form of the tidally perturbed black hole metric. Instead, I will point out the ways in which this metric restricts the generality of the inner solution.

My analysis begins with a discussion of the outer expansion. Section~\ref{internal solution} then describes the metric in the inner expansion. In Sec.~\ref{coordinate transformation}, I present the coordinate transformation between them, focusing on the restrictions that must be imposed on the transformation to yield a unique result. I conclude the chapter with a discussion of the method.

Beginning in this chapter, I require two important computational techniques: near-coincidence expansions and STF (symmetric trace-free) decompositions. The former is reviewed in Appendix~\ref{local_expansions}; the latter, in Appendix~\ref{STF tensors}. I also make extensive use of the Fermi and retarded coordinate systems discussed in Appendix~\ref{coordinates} and the Green's functions defined in Appendix~\ref{Greens_functions}. Readers unfamiliar with these techniques and quantities should peruse those appendices as necessary.

\section{External solution}
I require an expansion of the background metric $g$ and the first-order external perturbation $\hmn{E}{1}$ in the buffer region. To find these expansions, I adopt Fermi coordinates $(t,x^a)$ centered on $\gamma$ and then expand in powers of the geodesic distance $r\equiv\sqrt{\delta_{ij}x^ix^j}$. The construction of the coordinate system is sketched in Appendix~\ref{coordinates}; refer to Ref.~\cite{Eric_review} for a detailed description.

Although the solution to the wave equation is more naturally expressed in terms of retarded coordinates \cite{Eric_review}\footnote{Again, the construction of these coordinates is sketched in Appendix~\ref{coordinates}, and detailed in Ref.~\cite{Eric_review}}, in the calculations in the following two chapters, Fermi coordinates are more advantageous; for example, the solution to the wave equation with a point particle source is expressed as an integral over the worldline up to a retarded time $u$, but in Ch.~\ref{perturbation calculation}, the solution to the wave equation will be expressed as an integral over a worldtube, which will be evaluated just as easily in Fermi coordinates as in retarded coordinates. Thus, in those later calculations, the simpler form of the background metric in Fermi coordinates outweighs the advantages of retarded coordinates, and I adopt them in this chapter as well for consistency.

I will be interested only in components in the Cartesian-type coordinates $(t,x^a)$, but I will express these components in terms of $r$ and two angles $\theta^A$, which are defined in the usual way in terms of $x^a$. I also introduce the unit one-form $n_\alpha\equiv \partial_\alpha r$, which depends only on the angles $\theta^A$. I will use the  multi-index notation $n^L\equiv n^{i_1}...n^{i_\ell}\equiv n^{i_1...i_\ell}$. Angular brackets around indices denote the STF combination of the enclosed indices; a caret over a tensor denotes the STF part of that tensor. Finally, I define the coordinate one-forms $t_\alpha\equiv\partial_\alpha t$ and $x^a_\alpha\equiv \partial_\alpha x^a$.

In Fermi coordinates, the components of the background metric are given by the standard results \eqref{Fermi_tt}--\eqref{Fermi_ab}. Rather than using those results directly, I express them in terms of the tidal fields $\etide_{ab}\equiv R_{a0b0}$ and $\btide_{ab}\equiv\tfrac{1}{2}\epsilon_a{}^{cd}R_{0bcd}$. I then decompose each component of the metric into irreducible STF form, as outlined in Appendix~\ref{STF tensors}. The resulting expression for the metric is
\begin{align}
g_{tt} &= -1-2ra_in^i-\tfrac{1}{3}r^2 a_i a^i-a_{\langle i}a_{j\rangle}\nhat^{ij} -r^2\etide_{ij}\nhat^{ij} +O(r^3), \label{Fermi background tt}\\
g_{ta} &= \tfrac{2}{3}r^2\epsilon_{aik}\btide^k_j\nhat^{ij}+O(r^3),\label{Fermi background ta}\\
g_{ab} &= \delta_{ab} -\tfrac{1}{9}r^2\delta_{ab}\etide_{ij}\nhat^{ij} -\tfrac{1}{9}r^2\etide_{ab} +\tfrac{2}{3}r^2\etide_{i\langle a}\nhat^i_{b\rangle}+O(r^3).\label{Fermi background ab}
\end{align}
Here the tidal fields are functions on the worldline, and are therefore functions of $t$ (and potentially $\gamma$) only.

One should note that the coordinate transformation $x^\alpha(t,x^a)$ between Fermi coordinates and the global coordinates is $\e$-dependent, since Fermi coordinates are tethered to an $\e$-dependent worldline. If one were using a regular expansion, then this coordinate transformation would devolve into a background coordinate transformation to a Fermi coordinate system centered on a geodesic worldline, combined with a gauge transformation to account for the $\e$-dependence. But in the present general expansion, the transformation is purely a background transformation, because the $\e$-dependence in the transformation is reducible to the $\e$-dependence in the fixed worldline.

The transformation hence induces not only new $\e$-dependence into the perturbations $\hmn{E}{\emph{n}}$, but also $\e$-dependence in the background metric $g$. (Despite its $\e$-dependence, $g$ is the background metric of the outer expansion, and I will use it to raise and lower indices on $h$.) This new $\e$-dependence takes two forms: a functional dependence on $z^\alpha(t)=x^\alpha(t,x^a=0)$, the coordinate form of the worldline written in the global coordinates $x^\alpha$; and a dependence on the acceleration vector $a^\alpha(t)$ on that worldline. For example, the first type of dependence appears in the components of the Riemann tensor (or tidal fields) in Fermi coordinates, which are related to the components in the global coordinates via the relationship $R_{IJKL}(t)=R_{\alpha\beta\gamma\delta}(z^\mu(t))e^\alpha_I e^\beta_J e^\gamma_K e^\delta_L$. The second type of $\e$-dependence consists of factors of the acceleration $a^\mu(t)$, which has the assumed expansion $a_i(t,\e) = \an{0}_i(t)+\e\an{1}_i(t;\gamma)+\order{\e^2}$.

Hence, in the buffer region we can opt to work with the quantities $g$ and $h_E$, which are defined with $a$ fixed, or we can opt to re-expand these quantities by substituting into them the expansion of $a$. (In either case, we would still hold fixed the functional dependence on $z^\mu$.) Substituting the expansion of $a$ in Fermi coordinates yields the \emph{buffer-region expansions}
\begin{align}
g_{\mu\nu} & = g_{\mu\nu}\coeff{0}(t,x^a;\gamma)+\e g_{\mu\nu}\coeff{1}(t,x^a;\gamma)+\order{\e^2},\label{buffer_expansion g}\\
\hmn{E\alpha\beta}{\emph{n}} &= \hmn{\alpha\beta}{\emph{n}}(t,x^a;\gamma) +\order{\e}\label{buffer_expansion h},
\end{align}
where $g_{\mu\nu}\coeff{0}\equiv g_{\mu\nu}\big|_{a=\an{0}}$, $g_{\mu\nu}\coeff{1}$ is linear in $\an{1}_i$, and $\hmn{\mu\nu}{\emph{n}}\equiv\hmn{E\mu\nu}{\emph{n}}\big|_{a=\an{0}}$. Because the inner expansion does not hold the acceleration fixed, for the sake of matching, in this chapter I will use the buffer-region quantities $g\coeff{0}$, $g\coeff{1}$, and $\hmn{}{1}$; in the following two chapters, I will make use of buffer-region quantities as well the original quantities $g$ and $\hmn{E}{\emph{n}}$.

In order to determine $\hmn{}{1}$, I rewrite the wave equation \eqref{h_E1 eqn} as
\begin{equation}
E_{\alpha\beta}[\hmn{E}{1}] = -16\pi (T_{\alpha\beta}-\tfrac{1}{2}g_{\alpha\beta}g^{\mu\nu}T_{\mu\nu}),
\end{equation}
where $T_{\alpha\beta}$ is the stress-energy tensor of a point particle, given in Eq.~\eqref{stress_energy_point_particle}. Note that there is no contradiction between this equation and Eq.~\eqref{h_E1 eqn}, since the latter applies only in the vacuum region $\Omega$, where $T_{\alpha\beta}$ vanishes pointwise. As discussed in the first two chapters of this dissertation, the solution to this wave equation can be expressed in terms of an integral over the worldline $\gamma$. Near the worldline, the solution can then be expanded in powers of $r$, using the methods of Appendix~\ref{local_expansions}; that calculation is presented in Appendix~\ref{point_particle_soln}. The result is the following:  
\begin{align}\label{external metric buffer expansion}
\hmn{Ett}{1} &= \frac{2m}{r}+\A{}{1,0}+3ma_in^i+r\left[4ma_ia^i+\A{i}{1,1}n^i+m\left(\tfrac{3}{4}a_{\langle i}a_{j\rangle} +\tfrac{5}{3}\etide_{ij}\right)\nhat^{ij}\right]\nonumber\\
&\quad+O(r^2),\\
\hmn{Eta}{1} &= \C{a}{1,0}+r\big(\B{}{1,1}n_a-2m\dot a_a+\C{ai}{1,1}n^i+\epsilon_{ai}{}^j\D{j}{1,1}n^i +\tfrac{2}{3}m\epsilon_{aij}\btide^j_k\nhat^{ik}\big)\nonumber\\
&\quad+O(r^2), \\
\hmn{Eab}{1} &= \frac{2m}{r}\delta_{ab}+(\K{}{1,0}-ma_in^i)\delta_{ab}+\H{ab}{1,0} +r\Big\lbrace\delta_{ab}\big[\tfrac{4}{3}ma_ia^i+\K{i}{1,1}n^i \nonumber\\
&\quad+\left(\tfrac{3}{4}ma_{\langle i}a_{j\rangle}-\tfrac{5}{9}m\etide_{ij}\right)\nhat^{ij}\big] +\tfrac{4}{3}m\etide^i_{\langle a}\nhat_{b\rangle i} +4ma_{\langle a}a_{b\rangle}-\tfrac{38}{9}m\etide_{ab} \nonumber\\
&\quad+\H{abi}{1,1}n^i +\epsilon_i{}^j{}_{(a}\I{b)j}{1,1}n^i+\F{\langle a}{1,1}n^{}_{b\rangle}\Big\rbrace+O(r^2),
\end{align}
where the uppercase script quantities are STF Cartesian tensors that are functions of time alone; they are constructed from the tail integral, the acceleration, and $\etide$, and their exact form is specified in Table~\ref{STF wrt tail}. The naming convention for those tensors follows that in Eqs.~\eqref{generic_STF tt}--\eqref{generic_STF ab}. $\hmn{}{1}$ is given by setting $a_i=\an{0}_i$ in these expressions.

The buffer-region expansion of the full metric in the outer limit can now be written as
\begin{equation}
g_{E\alpha\beta}=g\coeff{0}_{\alpha\beta}+\e g\coeff{1}_{\alpha\beta}+\e\hmn{\alpha\beta}{1}+\order{r^3,\e r^2,\e^2},
\end{equation}
where $g_{\alpha\beta}$ is given in Fermi coordinates in Eqs.~\eqref{Fermi background tt}--\eqref{Fermi background ab}.

\section{Internal solution}\label{internal solution}
I assume that the internal solution is that of a perturbed Schwarzschild black hole, and I adopt retarded Eddington-Finkelstein coordinates $(U,X^a)$ adapted to that spacetime. The background metric $g_B$ is then given by
\begin{align}
g_{B} &= -f(U,\tilde R)dUdU -2\widetilde\Omega_adUdX^a+(\delta_{ab}-\widetilde\Omega_{ab})dX^a dX^b,
\end{align}
where $f(U,\tilde R)=1-2M(U)/\tilde R$, and $\widetilde\Omega_a\equiv X^a/R$ is a function of two angles $\Theta^A$. (Note that in this equation, I have written the metric in non-rescaled coordinates, but I have written the components of the metric in terms of the scaled coordinate $\tilde R$.) Here $M(U)$ is the Bondi mass of the spacetime, divided by the mass at $U=0$. The mass is allowed to depend on $U$ because $g_B$ is required only to solve Eq.~\eqref{inner_eqn0}, which contains no time-derivatives. Next, I expand the components of the metric perturbation $H$ as
\begin{equation}
H_{\mu\nu}(U,\tilde R, \Theta^A,\e) = \e H_{\mu\nu}\coeff{1}(U,\tilde R,\Theta^A)+\e^2 H_{\mu\nu}\coeff{2}(U,\tilde R,\Theta^A)+...
\end{equation}
As a boundary condition on these perturbations, I require that they remain regular on the event horizon. In addition, I adopt the light cone gauge~\cite{light_cone_gauge}, defined in retarded polar coordinates by the condition $H^{(n)}_{UR}=H^{(n)}_{RR}=H^{(n)}_{RA}=0$. In this gauge, $U$ and $R$ maintain their geometrical meaning even in the perturbed spacetime: $U$ is constant on each outgoing light cone, and $R$ is an affine parameter on outgoing light rays. I assume that this gauge condition can always be imposed.

The first- and second-order perturbations, along with the time-dependence of $g_B$, must satisfy the vacuum Einstein equations \eqref{inner_eqn1}--\eqref{inner_eqn2}. In Appendix~\ref{perturbed_BH}, I show that $\diff{M}{U}=0$. This implies that Eq.~\eqref{inner_eqn1} becomes $\delta G\coeff{0}_I[H\coeff{1}]=0$, the linearized vacuum EFE for static perturbations. The solutions to this equation have been thoroughly studied~\cite{Zerilli, Regge_Wheeler, Wald_perturbations, Eric_perturbations}. Because of the spherical symmetry of the background spacetime, the equation can be most easily solved by decomposing $H\coeff{1}$ into spherical harmonics: the various harmonics decouple in the linearized Ricci tensor, such that they can be solved independently. In addition, for $\ell>0$, the harmonics can be decomposed into even- and odd-parity sectors, which also decouple. It is known that that the gauge-invariant content of the monopole terms in the solution correspond to a constant shift of the black hole's mass parameter; odd-parity dipole terms correspond to a shift to a nonzero, constant spin; and even-parity dipole perturbations correspond to a shift in center of mass, which can always be removed via a coordinate transformation. In addition, it is known that for all $\ell$, the solutions behave as $\tilde R^\ell$ for $\tilde R\gg 1$.

In the derivation provided by Mino, Sasaki, and Tanaka \cite{Mino_Sasaki_Tanaka}, the monopole and dipole terms in $H\coeff{1}$ were set to zero, on the basis that they correspond to either pure gauge or to mere redefinitions of mass and angular momentum. However, this step is not justified, since the ``constant" shifts in the black hole's parameters are actually functions of time, with a time-dependence to be determined by the higher-order perturbation equations. Also, the fact that the even-parity dipole term corresponds to a shift in center of mass does not mean that it can be trivially ignored; this will be discussed further in the following sections. For $\ell>1$, Mino, Sasaki, and Tanaka took the terms to necessarily behave as $\tilde R^\ell$ in the buffer region. This means that $H\coeff{\emph{n}}$ cannot contain terms of $\ell>n$: since $\e^n\tilde R^\ell=\e^{n-\ell}R^\ell$, if $\ell>n$ then such a term would correspond to negative powers of $\e$ in the outer expansion. Hence, $H\coeff{1}$ can contain only monopole and dipole terms, and since these are set to zero, $H\coeff{1}$ itself must be zero. It then follows that Eq.~\eqref{inner_eqn2} becomes another linearized vacuum EFE for static perturbations, $\delta G\coeff{0}_I[H\coeff{2}]=0$. The solutions to this equation must be purely quadrupolar, since monopole and dipole terms are set to vanish and $H\coeff{2}$ cannot contain terms of $\ell>2$. However, even if the monopole and dipole terms are set to zero, this reasoning remains specious, because solutions with $\ell>n$ \emph{can} exist: though the asymptotically dominant terms behave as $\tilde R^\ell$, subdominant terms can grow less rapidly with $\ell$, as is shown explicitly in Appendix~\ref{perturbed_BH}.

In the derivation provided by Poisson Ref.~\cite{Eric_review,Eric_matching}, all of the above steps were taken, but the quadrupole terms were then further constrained. Rather than finding a general inner solution and then restricting it by imposing a matching condition, Poisson simplified the possible forms of the metric by first imposing a form of the asymptotic matching condition. (Refer back to Ch. 2 for the definition of this condition.) Specifically, he demanded that for $\tilde R\gg 1$, the metric must asymptotically approach that of a vacuum spacetime in retarded coordinates centered on a geodesic; that metric is given explicitly in Eqs.~\eqref{ret_coords1}--\eqref{ret_coords5}. This demand motivated the following ansatz for the internal metric in polar coordinates:
\begin{align}
g_{IUU} &= -f\left[1+\e^2\R^2e_1(\R)\etide^*(U)\right]+\order{\e^3},\\
g_{IUR} & = -1, \\
g_{IUA} & = R\left[\tfrac{2}{3}\e^2\R^2\left(e_2(\tilde R)\etide^*_A+b_2(\tilde R)\btide^*_A\right)+\order{\e^3}\right], \\
g_{IRR} & = g_{IRA} = 0, \\
g_{IAB} & = R^2\left[\Omega_{AB}-\tfrac{1}{3}\e^2\R^2\left(e_3(\R)\etide^*_{AB}+b_3(\R)\btide^*_{AB}\right)+\order{\e^3}\right],
\end{align}
where $e_1$, $e_2$, $e_3$, $b_2$, and $b_3$ are undetermined functions constrained to approach 1 for $\tilde R\gg 1$, and the quantities $\ein^*$, $\bin^*_A$, etc., are constructed from tidal fields $\ein_{ab}$ and $\bin_{ab}$, as displayed in Eqs.~\eqref{Estar}--\eqref{Bstar_AB}. Note that even with the constraint on the asymptotic behavior of the internal metric, the above ansatz is more restrictive than it need be: generally, the free functions of $\R$ could also be functions of $U$, with a $U$-dependence to be determined by the higher-order EFE.

One might wonder why the internal metric is constrained to approach that of a vacuum metric in coordinates centered on a geodesic, rather than being constrained to approach a vacuum metric in coordinates centered on an arbitrarily accelerating worldline, to agree with the form of the external background metric in retarded coordinates. The reason is that the terms linear in the acceleration in the external background metric are even-parity dipole terms, which have been set to zero to ensure that the coordinates are mass-centered. I will return to the relevance of these terms in the following sections, but I will note now that this assumed form already suggests that the body must be moving on a geodesic of some spacetime. That spacetime will turn out to be $g+h^R$, rather than $g$. See Ref.~\cite{Detweiler_review} for further discussion of this point.

Substituting the ansatz into the linearized EFE and imposing regularity at the event horizon determines the free functions. After transforming the resulting metric back into Cartesian-type coordinates, one finds
\begin{align}
g_{IUU} &= -f-\e^2 f^2\R^2\ein^*+O(\e^3),\\
g_{IUa} &= -\Omega_a+\tfrac{2}{3}\e^2\R^2 f(\ein^*_a+\bin^*_a)+O(\e^3),\\
g_{Iab} &= \delta_{ab}-\Omega_{ab}-\tfrac{1}{3}\e^2\R^2\left(1-\frac{2 M^2}{\R^2}\right)\ein^*_{ab}-\tfrac{1}{3}\e^2\R^2\bin^*_{ab}+O(\e^3),
\end{align}
where $\ein_a^*=\ein_A^*\Omega^A_a$, $\bin_a^*=\bin_A^*\Omega^A_a$, $\ein_{ab}^*=\ein_{AB}^*\Omega^A_a\Omega^B_b$, and $\bin_{ab}^*=\bin_{AB}^*\Omega^A_a\Omega^B_b$. Note that there is no a priori relationship between the mass $\e M$ of the internal spacetime and the mass $\e m$ of the point particle perturbation in the external spacetime. Similarly, although the inner solution was specifically constructed to asymptotically approach the form of an external metric in the buffer region, there is no priori relationship between $\ein_{ab}$ and $\etide_{ab}$ or between $\bin_{ab}$ and $\btide_{ab}$. These relationships are to be determined in the matching procedure.

To expand the metric in the buffer region, we rewrite $\tilde R$ as $R/\e$ and then re-expand in powers of $\e$; this corresponds to an expansion for $R\gg \e$. In order to agree with the external metric, which is constructed in Fermi coordinates and in the Lorenz gauge, we must also transform from retarded coordinates and the lightcone gauge into Fermi-like harmonic coordinates $(T,X^a)$; and the result must be decomposed into its irreducible STF pieces. That calculation is shown in Appendix~\ref{perturbed_BH}. The final result is
\begin{align}\label{internal metric buffer expansion}
g_{ITT} & = -1+\e\frac{2M}{R}+\tfrac{5}{3}\e MR\ein_{ij}\hat N^{ij}-R^2\ein_{ij}\hat N^{ij}+\order{\e^2,\e R^2, R^3}, \\
g_{ITa} & = 2\e MR\ein_{ai}N^i+\tfrac{2}{3}\e MR\epsilon_{aij}\bin^j_k\hat N^{ik}+\tfrac{2}{3}R^2\epsilon_{aik}\bin^k_j\hat N^{ij}+\order{\e^2,\e R^2, R^3}, \\
g_{Iab} & = \delta_{ab}\left(1+\e\frac{2M}{R}-\tfrac{5}{9}\e MR\ein_{ij}\hat N^{ij}-\tfrac{1}{9}R^2\ein_{ij}\hat N^{ij}\right) +\tfrac{64}{21}\e MR\ein_{i\langle a}\hat N_{b\rangle}{}^i \nonumber\\
&\quad-\tfrac{46}{45}\e MR\ein_{ab}-\tfrac{1}{9}R^2\ein_{ab}+\tfrac{2}{3}\e MR\ein_{ij}\hat N_{ab}{}^{ij}+\tfrac{2}{3}R^2\ein_{i\langle a}\hat N^i_{b\rangle}\nonumber\\
&\quad -\tfrac{4}{3}\e MR\epsilon_{jk(a}\bin_{b)}^kN^j+\order{\e^2,\e R^2, R^3},
\end{align}
where $N^i=X^i/R$. This is the metric that I will use in the matching procedure, even though, as pointed out above, it has already been heavily restricted. I refer the reader to Appendix~\ref{perturbed_BH} for a more thorough discussion of the derivation of this metric and of its generality.

\section{The matching procedure}\label{coordinate transformation}
\subsection{Zeroth-order matching}
I now consider the relationship between the two metrics. Beginning with the zeroth-order weak matching condition, we have the metric in the outer expansion given by
\begin{align}
g_{Ett} &= -1-2r\an{0}_in^i-\tfrac{1}{3}r^2 \an{0}_i \an{0}{}^i-r^2\an{0}_{\langle i}\an{0}_{j\rangle}\nhat^{ij}-r^2\etide_{ij}\nhat^{ij} +O(\e, r^3),\\
g_{Eta} &= \tfrac{2}{3}r^2\epsilon_{aik}\btide^k_j\nhat^{ij}+O(\e, r^3),\\
g_{Eab} &= \delta_{ab} -\tfrac{1}{9}r^2\delta_{ab}\etide_{ij}\nhat^{ij} -\tfrac{1}{9}r^2\etide_{ab}+\tfrac{2}{3}r^2\etide_{i\langle a}\nhat^i_{b\rangle}+O(\e, r^3).
\end{align}
while the metric in the inner expansion is given by
\begin{align}
g_{ITT} &= -1-R^2\ein_{ij}\hat N^{ij} +O(\e, R^3),\label{gITT_buffer}\\
g_{ITa} &= \tfrac{2}{3}R^2\epsilon_{aik}\bin^k_j\hat N^{ij}+O(\e,R^3),\\
g_{Iab} &= \delta_{ab}-\tfrac{1}{9}R^2\delta_{ab}\ein_{ij}\hat N^{ij}-\tfrac{1}{9}R^2\ein_{ab}+\tfrac{2}{3}R^2\ein_{i\langle a}\hat N^i_{b\rangle}+O(\e,R^3)\label{gIab_buffer}.
\end{align}
It seems that we may immediately identify these two metrics and conclude that $T=t$, $X^a=x^a$, $\ein_{ab}=\etide_{ab}$, and most importantly, $\an{0}_\mu=0$. However, the matching condition does not require that these two metrics be identical, since they may be in different coordinate systems; the matching condition requires only that these two metrics be related by a diffeomorphism. But this condition places no restriction at all on the acceleration of the worldline: The form of the inner metric is that of an arbitrary background written in Fermi coordinates centered on a geodesic worldline. The form of the outer metric is that of a known background written in Fermi coordinates centered on a possibly accelerated worldline. Regardless of the value of the acceleration, if the geodesic is embedded in the external spacetime, then these two solutions are obviously related by a diffeomorphism, since the geodesic can be transformed to the accelerated worldline.

Evidently, some information has been lost here. I assumed from the beginning that the inner and outer expansions were performed ``around" the same worldline. In the inner expansion, the ``location" of the body is encoded into the coordinate system by the condition that the body's mass dipole vanishes in that coordinate system; in the outer expansion, the ``location" of the body is encoded in the worldline sourcing the perturbation. If we use the weak matching condition, in which we expand the metric before finding the coordinate transformation between the inner and outer expansions, then this information is lost.

However, one might wonder if this ambiguity might be removed by supplementing the weak matching condition with some other condition. One such condition appears obvious: the coordinate transformation between the inner and outer expansions in the buffer region must be ``small"--that is, it must vanish in the limit $\e\to0$. This removes the possibility of transforming from an arbitrary geodesic to an arbitrarily accelerated worldline. In the buffer region, $r\to0$ as $\e\to0$, so this allows transformations that have no explicit $\e$-dependence, but which do have explicit $r$-dependence. I trust the reader to convince himself that under such a transformation, we must have $R=r$, $T=t$, the tidal fields appearing in the inner metric are identical (up to $\order{\e}$ corrections) with those constructed from the Riemann tensor in the outer solution---and the leading-order term in the acceleration must vanish: $\an{0}=0$. The two coordinate systems may, of course, be related by rotations, but these are insignificant.

Hence, we can adopt a stronger, refined matching condition: the inner and outer expansions in the buffer region must be equal up to a unique small coordinate transformation. Unfortunately, this refined condition is still insufficient. The reason is that the inner expansion \emph{could} have included acceleration-type terms. In fact, we can always include such terms by transforming the metric into an accelerating frame. Suppose we begin with the Schwarzschild metric in Kerr-Schild form,
\begin{equation}\label{Kerr-Schild}
g_{B\mu\nu} = \eta_{\mu\nu} + \frac{\e M}{\bar R}\ell_\mu \ell_\nu,
\end{equation}
where $\eta=\text{diag}(-1,1,1,1)$ is the Minkowski metric, $\ell_\mu=\left(1,\frac{\bar X}{\bar R},\frac{\bar Z}{\bar R},\frac{\bar Z}{\bar R}\right)$ is a null vector, $\bar R=\displaystyle\sqrt{\bar X^2+\bar Y^2+\bar Z^2}$, and the (unscaled) coordinates are $(\bar T,\bar X,\bar Y,\bar Z)$. Now, by  using the flat-spacetime transformation from an inertial frame to an accelerated one, we can transform the metric to a new set of accelerated retarded coordinates $(U', R',\Theta'^A)$. For simplicity, assume that the acceleration is in the $\bar Z$-direction. Then the transformation is given by
\begin{align}
\bar T & = T_0(U')+R'\left[\cos\Theta'\sinh q(U')+\cosh q(U')\right],\\
\bar X & = R' \sin\Theta'\cos\Phi',\\
\bar Y & = R' \sin\Theta'\sin\Phi',\\
\bar Z & = Z_0(U')+R'\left[\cos\Theta'\cosh q(U')+\sinh q(U')\right],
\end{align}
where $T_0=\int\cosh q(U')dU'$, $Z_0=\int\sinh q(U')dU'$, and $q(U')=\int \alpha(U')dU'$, where $\alpha(U')$ is the magnitude of the acceleration. Under this transformation, $g_B$ maintains the form in Eq.~\eqref{Kerr-Schild}; $\eta_{\mu\nu}$ becomes the metric of flat spacetime in retarded coordinates, given in Eqs.~\eqref{ret_coords1}--\eqref{ret_coords5}, while $\ell_\mu$ takes on a more complicated (and unenlightening) form. Note that in flat spacetime, this transformation translates the spatial origin from $\bar Z=0$ to $\bar Z = Z_0(U')$. And in the spacetime of $g_B$, the same interpretation applies at large distances from the black hole---that is, in the buffer region. In other words, the new coordinates are not mass-centered: the center of mass is accelerating away from the origin.

Although the metric takes on an inconveniently complicated form in this non-mass--centered coordinate system, in principle one could use it in constructing the inner expansion $g_I$. If one did so, then when $g_I$ was expanded in the buffer region, it would become $g_{I\mu\nu}=\eta_{\mu\nu}+O(\e,R^2)$, as we can infer immediately from the form of Eq.~\eqref{Kerr-Schild}. But in this expansion, $\eta_{\mu\nu}$ is the metric of flat spacetime centered on an accelerating worldline, not on a geodesic. Therefore, if we transform the metric to Fermi-type coordinates $(T,X^a)$, we arrive at
\begin{align}
g_{ITT} &= -1-2R\alpha\cos\Theta+O(\e,R^2),\\
g_{ITa} &= O(\e,R^2),\\
g_{Iab} &= \delta_{ab}+O(\e,R^2).
\end{align}
This metric agrees with the one in the outer expansion, regardless of the value of the acceleration. We may identify $\alpha\cos\Theta$ with $\an{0}_in^i$, and the matching procedure, even with the refined matching condition, provides no information whatsoever about the worldline.

We can readily see why the matching procedure has failed: we have not insisted on any relationship between the inner and outer expansions. In order for matching to be successful, we must insist that the ``position" of the black hole in the inner expansion can be identified with the position of the worldline in the outer expansion. To make this identification mathematically precise, I insist that the two expansions are to be expanded and matched in the buffer region only when the outer expansion is evaluated in a coordinate system centered on the worldline and the inner expansion is evaluated in a mass-centered coordinate system. If this condition is imposed, then the accelerating coordinate system $(U',R',\Theta^A)$ is inadmissible, since it is not mass-centered. Therefore, we can discount it and others like it---and we can once again, now more confidently, conclude that the acceleration of the worldline must vanish in the limit $\e\to0$. Such a condition serves to implicitly define the worldline, and it is necessary for the matching procedure to be well defined and to yield unambiguous results.

Based on the above analysis of the zeroth-order matching procedure, I suggest the following matching condition: if the inner expansion is written in a mass-centered coordinate system and the outer expansion is written in a worldline-centered coordinate system, then the two expansions must be equal up to a small coordinate transformation when expanded in the limit of small (outer) radial coordinate distances. (Here ``outer" radial coordinate means a coordinate that is formally of order $1$ in the outer expansion and of order $1/\e$ in the inner expansion.) Making use of this condition allows us to determine the acceleration of the worldline at zeroth order. However, as we shall see in the next subsection, it requires still more restrictions.

\subsection{First-order matching}
Comparing the expression for the external solution with that for the internal solution, we find that the $1/r$ terms agree if and only if we make the identification $m=M$. In order for the other terms to be made to agree, there must exist a coordinate transformation, from the external coordinates to the internal coordinates, that induces a gauge transformation $g_E\to g_E+\e\delta g_E+\order{\e^2}$, where
\begin{align}\label{required_trans}
\delta g_{Ett} & = -\A{}{1,0}+r(2\an{1}_i-\A{i}{1,1})n^i+\order{r^2}, \\
\delta g_{Eta} & = -\C{a}{1,0}-\tfrac{1}{6}r\partial_t(\A{}{1,0}+3\K{}{1,0})n_a-r\C{ai}{1,1}n^i  -r\epsilon_{ai}{}^j\D{j}{1,1}n^i+2mr\etide_{ai}n^i\nonumber\\
&\quad+\order{r^2},\\
\delta g_{Eab} & = -\delta_{ab}\K{}{1,0}-\H{ab}{1,0}-r\delta_{ab}\K{i}{1,1}n^i -\tfrac{3}{10}r\left(\K{\langle a}{1,1}-\A{\langle a}{1,1}+2\partial_t\C{\langle a}{1,1}\right)n_{b\rangle}\nonumber\\
&\quad-r\H{abi}{1,1}n^i-r\epsilon_i{}^j{}_{(a}\I{b)j}{1,1}n^i +\tfrac{12}{7}mr\etide_{i\langle a}\nhat_{b\rangle}{}^i+\tfrac{16}{5}mr\etide_{ab} +\tfrac{2}{3}mr\etide_{ij}\nhat_{ab}{}^{ij}\nonumber\\
&\quad -\tfrac{4}{3}mr\epsilon_{jk(a}\btide^k_{b)}n^j+\order{r^2}.
\end{align}
I remind the reader that $a_i$ is to be set to $\an{0}_i=0$ in the explicit expressions for the uppercase script tensors.

This transformation is generated by a vector field satisfying $\delta g_{E\alpha\beta}=2\xi_{(\alpha;\beta)}$. I assume the field can be expanded as
\begin{equation}
\begin{array}{lr}
\xi_t = \sum_{n\ge 0}r^n\xi_t\coeff{\emph{n}}, & \xi_a = \sum_{n\ge 0}r^n\xi_a\coeff{\emph{n}},
\end{array}
\end{equation}
where the coefficients $\xi_t\coeff{n}$ and $\xi_a\coeff{n}$ are decomposed as
\begin{align}
\xi_t\coeff{\emph{n}} & = \sum_{\ell\ge 0}\Xi_L\coeff{\emph{n}}\nhat^L,\\
\xi_a\coeff{\emph{n}} & = \sum_{\ell\ge 1}\left(\Upsilon_{aL-1}\coeff{\emph{n}}\nhat_{L-1} +\epsilon_{ab}{}^c\nhat^{bL-1}\Lambda_{cL-1}\coeff{\emph{n}}\right)+\sum_{\ell\ge 0}\Psi_L\coeff{\emph{n}}\nhat_a{}^L.
\end{align}
The Cartesian tensors $\Xi_L$, $\Upsilon_L$, $\Lambda_L$, and $\Psi_L$ are STF in $L$, and they depend only on time.

Calculating $2\xi_{(\alpha;\beta)}$ from the above expansion is straightforward. Demanding that the result of this calculation agrees with Eq.~\eqref{required_trans} at each order in $r$ then determines a sequence of equations for $\xi\coeff{\emph{n}}$. No $\order{1/r}$ terms appear in Eq.~\eqref{required_trans}, so from the $\order{1/r}$ terms in $2\xi_{(\alpha;\beta)}$ we find that $\partial_a\xi_t\coeff{0}=0$ and $\partial_{(a}\xi_{b)}\coeff{0}=0$. From this we determine that $\xi_{\alpha}\coeff{0}$ must be independent of angle: $\xi_t\coeff{0} = \Xi\coeff{0}$ and
$\xi_a\coeff{0} = \Upsilon_a\coeff{0}$.

From the $\order{r^0}$ terms, we find
\begin{align}
\partial_t\Xi\coeff{0} &= -\tfrac{1}{2}\A{}{1,0},\\
(n_a+r\partial_a)\xi_t\coeff{1} &= -\partial_t\xi_a\coeff{0}-\C{a}{1,0},\\
(n_{(a}+r\partial_{(a})\xi_{b)}\coeff{1} &= -\tfrac{1}{2}\delta_{ab}\K{}{1,0}-\tfrac{1}{2}\H{ab}{1,0},
\end{align}
from the $tt$-, $ta$-, and $ab$-component, respectively. The first of these equations determines that $\Xi\coeff{0}=-\tfrac{1}{2}\int\!\A{}{1,0}dt$, the second determines that $\Xi_a\coeff{1}=-\C{a}{1,0}-\partial_t\Upsilon_a\coeff{0}$, and the last determines that $\Psi\coeff{1}=-\tfrac{1}{2}\K{}{1,0}$, $\Upsilon_{ab}\coeff{1}=-\tfrac{1}{2}\H{ab}{1,0}$, and $\Lambda_c\coeff{1}$ is arbitrary. All other terms in $\xi_\alpha\coeff{1}$ vanish.

Finally, from the $\order{r}$ terms, we find:
\begin{align}
\partial_t\xi_t\coeff{1} & = \etide^j_in^i\Upsilon_j\coeff{0}\! +\an{1}_in^i\! -\tfrac{1}{2}n^i\A{i}{1,1}\!\!,\\
(2n_a+r\partial_a)\xi_t\coeff{2} & = -\partial_t\xi_a\coeff{1}+2\etide_{ai}n^i\Xi\coeff{0} -\tfrac{1}{6}\partial_t(\A{}{1,0}+3\K{}{1,0})n_a\nonumber\\
&\quad -\C{ai}{1,1}n^i-\epsilon_{ai}{}^j\D{j}{1,1}n^i-2m\etide_{ai}n^i -2R_{0i}{}^j{}_an^i\Upsilon_j\coeff{0},\\
2(2n_{(a}+r\partial_{(a})\xi_{b)}\coeff{2} & = \tfrac{4}{3}R_{0(ab)i}n^i\Xi\coeff{0} -\tfrac{4}{3}R^j{}_{(ab)i}n^i\Upsilon_j\coeff{0} -\delta_{ab}\K{i}{1,1}n^i+\tfrac{12}{7}m\etide_{i\langle a}\nhat_{b\rangle}{}^i\nonumber\\
&\quad +\tfrac{16}{5}m\etide_{ab}-\H{abi}{1,1}n^i -\!\tfrac{3}{10}\big(\K{\langle a}{1,1}\!\!-\A{\langle a}{1,1}\!+\!2\partial_t\C{\langle a}{1,1}\big)n_{b\rangle}\nonumber\\
&\quad +\tfrac{2}{3}m\etide_{ij}\nhat_{ab}{}^{ij} -\tfrac{4}{3}m\epsilon_{jk(a}\btide^k_{b)}n^j -\epsilon_i{}^j{}_{(a}\I{b)j}{1,1}n^i.
\end{align}
Again, these equations follow from the $tt$-, $ta$-, and $ab$-component, respectively. The first of them yields the equation of motion
\begin{equation}\label{matching equation of motion}
\partial_t^2\Upsilon_i\coeff{0}+\etide^j_i\Upsilon_j\coeff{0} = \tfrac{1}{2}\A{i}{1,0}-\partial_t\C{i}{1,0}-\an{1}_i,
\end{equation}
the second of them yields
\begin{align}
\Xi\coeff{2} & = -\tfrac{1}{12}\partial_t\A{}{1,0}, \\
\Xi\coeff{2}_{ab} & = \tfrac{5}{16}\partial_t\H{ab}{1,0}+\tfrac{5}{4}\etide_{ab}\Xi\coeff{0} -\tfrac{5}{8}\C{ab}{1,1}-\tfrac{5}{4}m\etide_{ab} +\tfrac{5}{4}\epsilon^j{}_{i\langle a}\btide_{b\rangle}^i\Upsilon_j\coeff{0},\\
\partial_t\Lambda_c\coeff{1} & = -\D{c}{1,1} +\tfrac{1}{2}\epsilon_c{}^{pq}\epsilon^{i}{}_{jp}\btide^j_q\Upsilon_i\coeff{0},
\end{align}
and the last of them yields (after some algebra)
\begin{align}
\Upsilon_a\coeff{2} & = -\tfrac{1}{2}\etide^j_a\Upsilon_j\coeff{0}+\tfrac{3}{16}\A{a}{1,1} -\tfrac{3}{8}\partial_t\C{a}{1,0},\\
\Upsilon_{ab}\coeff{2} & = \tfrac{6}{5}m\etide_{ab},\\
\Upsilon_{abc}\coeff{2} & = -\tfrac{1}{4}\H{abc}{1,1},\\
\Psi_a\coeff{2} & = -\tfrac{9}{20}(\K{a}{1,1}+\tfrac{1}{4}\A{a}{1,1} -\tfrac{1}{2}\partial_t\C{a}{1,0})+\tfrac{1}{2}\etide^j_a\Upsilon_j\coeff{0},\\
\Psi_{ab}\coeff{2} & = \tfrac{1}{3}m\etide_{ab},\\
\Lambda_{ab}\coeff{2} & = -\tfrac{1}{2}\I{ab}{1,1}-\tfrac{2}{3}m\btide_{ab} -\tfrac{2}{3}\btide_{ab}\Xi\coeff{0}+\tfrac{2}{3}\epsilon_i{}^j_{(a}\etide^i_{b)}\Upsilon_j\coeff{0}.
\end{align}
All other terms vanish.

In summary, the first three terms in the expansion of the gauge vector field are given by
\begin{align}
\xi_t\coeff{0} &= -\tfrac{1}{2}\int\!\!\A{}{1,0}dt,\\
\xi_a\coeff{0} &= \Upsilon_a\coeff{0},
\end{align}
where $\Upsilon_a\coeff{0}$ is a function of time satisfying the equation of motion \eqref{matching equation of motion},
\begin{align}
\xi_t\coeff{1} &= (\C{i}{1,0}-\partial_t\Upsilon_i\coeff{0})n^i,\\
\xi_a\coeff{1} &= \epsilon_a{}^{ij}n_i\left(\int\!\!\D{j}{1,0}dt +\tfrac{1}{2}\epsilon_j{}^{pq}\epsilon^\ell{}_{kp} \int\!\!\btide^k_q\Upsilon_\ell\coeff{0}dt\right)+\tfrac{1}{2}\K{}{1,0}n_a+\tfrac{1}{2}\H{ai}{1,0}n^i,
\end{align}
and
\begin{align}
\xi_t\coeff{2} &= \tfrac{5}{8}\big(-\tfrac{1}{2}\partial_t\H{ij}{1,0} +2\etide_{ij}\Xi\coeff{0}  +\C{ij}{1,1}+2m\etide_{ij} +2\epsilon^k{}_{p\langle i}\btide^c_{j\rangle}\Upsilon_k\coeff{0}\big)\nhat^{ij}\nonumber\\
&\quad +\tfrac{1}{12}\partial_t\A{}{1,0},\\
\xi_a\coeff{2} &= \left[\tfrac{1}{2}\Upsilon_j\coeff{0}\etide^j_i +\tfrac{9}{20}(\K{i}{1,1}+\tfrac{1}{4}\A{i}{1,1} -\tfrac{1}{2}\partial_t\C{i}{1,0})\right]\nhat_a^i +\tfrac{1}{3}m\etide_{ij}\nhat_{ab}{}^{ij} \nonumber\\
&\quad-\tfrac{1}{2}(\Upsilon_j\coeff{0}\etide^j_a+\tfrac{3}{8}\A{a}{1,1} -\tfrac{3}{4}\partial_t\C{a}{1,0})-\tfrac{6}{5}m\etide_{ai}n^i+\tfrac{1}{4}\H{abi}{1,1}n^i \nonumber\\
&\quad+\epsilon_{aij}\nhat^{ik}\Big(\tfrac{2}{3}\epsilon^{cd}{}_{(j} \etide_{k)c}\Upsilon_d\coeff{0} +\tfrac{1}{2}\I{jk}{1,1}+\tfrac{2}{3}m\btide_{jk} -\tfrac{1}{3}\btide_{jk}\int\!\!\A{}{1,0}dt\Big).
\end{align}
This is the most general transformation that succeeds in making the exterior solution identical to the interior solution, up to order $\e r$. It has one free function of time: $\Upsilon_a\coeff{0}$.

Despite the refinement of the matching condition formulated in the zeroth-order matching procedure, this coordinate transformation has failed to uniquely identify the acceleration of the worldline. Instead, it determines an equation for $\Upsilon_a\coeff{0}$, given by Eq.~\eqref{matching equation of motion}. Consider the meaning of this equation. In the internal solution, the mass dipole and all dipole perturbations have been set to zero, and an acceleration term in the buffer region corresponds to a dipole perturbation. Equation~\eqref{matching equation of motion} thus tells us that for any given acceleration $\an{1}_i$, we can perform a small, angle- and $r$-independent spatial translation (in the buffer region) that ensures the dipole perturbation vanishes.

In order to arrive at the correct equation, $\an{1}_i=\tfrac{1}{2}\A{i}{1,0}-\partial_t\C{i}{1,0}$, one must further restrict the matching condition. Recall that the coordinate transformation must be small, which implies that $\Upsilon_i\coeff{0}$ must remain of order unity. If the right-hand side of Eq.~\eqref{matching equation of motion} does not vanish, then $\Upsilon_i\coeff{0}$ will generically grow large; more precisely, on a timescale such as $\sim1/\e$, which becomes unbounded in the limit $\e\to 0$, $\Upsilon_i\coeff{0}$ will generically become larger than order unity. However, it will not \emph{necessarily} grow large (e.g., if the right-hand side of Eq.~\eqref{matching equation of motion} is purely oscillatory). Thus, we cannot conclude that the right-hand side must vanish based on the refined matching condition of the previous section. Instead, I propose a final version of the \emph{refined matching condition}: \emph{if the inner expansion is written in a mass-centered coordinate system and the outer expansion is written in a worldline-centered coordinate system, then the two expansions must be equal up to a} necessarily \emph{small coordinate transformation when the inner expansion is expanded in the limit of small (outer) radial coordinate distances.} In other words, the coordinate transformation must not only be small, but must necessarily remain so on all timescales of interest.

With this final refinement, we can conclude the following: if on an unbounded timescale, (i) the exact metric possesses inner and outer expansions, (ii) there exists a local coordinate system in which the metric in the inner expansion is given by that of a tidally perturbed black hole, up to errors of order $\e^3$, (iii) there exists a global coordinate system in which the metric in the outer expansion is that of the external background $g$ plus a point-particle solution to the wave equation~\eqref{h_E1 eqn}, up to errors of order $\e^2$, and (iii) the exact solution satisfies the refined matching condition presented above, then the worldline defining the point particle perturbation has an acceleration given by
\begin{equation}
\an{1}_i = \tfrac{1}{2}\A{i}{1,0}-\partial_t\C{i}{1,0},
\end{equation}
where $\A{i}{1,0}$ and $\C{i}{1,0}$ are obtained by setting $a_i=\an{0}_i=0$ in Table~\ref{STF wrt tail}. This is the MiSaTaQuWa equation.

It may be that some or all of these assumptions can be removed, even within the context of matched asymptotic expansions. For example, in Chapter~\ref{perturbation calculation}, I will show that if the inner and outer expansions exist, then the solution to the wave equation~\eqref{h_E1 eqn} is that with a point particle source. Similarly, the inner metric could correspond to a body other than a black hole. If it were taken to be a tidally perturbed, otherwise spherically symmetric neutron star, for example, then the equation of motion would be unaffected: outside the star, the metric would be altered only by the presence of induced tidal moments, which scale as $\e^{2\ell+1}$ and hence would not appear in the first-order matching procedure \cite{Love_numbers1,Love_numbers2}. It is also possible that that matching the inner and outer solutions to higher order in $r$ would show that $\Upsilon_a\coeff{0}$ must vanish, and that the refined matching condition is needlessly strong; however, there is no obvious indication of the order at which this would occur.

\section{Interpretation and commentary}\label{commentary}
Let us interpret the above calculation. First, note that a large part of the transformation consists of removing tail terms. This can be understood as follows: In the Fermi coordinates centered on the worldline in $g$, the spacetime appears to be that of a singular monopole perturbation $h^S$, plus a regular homogeneous perturbation $h^R$, plus the field of the smooth background metric $g$ expanded about some worldline. But in the coordinates $X^\alpha$, at a large distance $R\gg \e$ from the body, the spacetime appears to be simply a singular monopole perturbation atop some smooth background field that is expanded around a geodesic. Therefore, transforming between these coordinates can be understood as transforming from the Fermi coordinates of $g$ into the Fermi coordinates of $g+h^R$, where $h^R$ is the Detweiler-Whiting regular field. Reference~\cite{Detweiler_review} contains further discussion of this point.

For example, the angle- and $r$-independent monopole term $\A{}{1,0}=\tail_{tt}$ is removed because proper time must be measured in $g+h^R$, rather than in $g$. Similarly, the dipole terms in the perturbation are removed because the body is nonspinning and non-accelerating in $g+h^R$,  rather than in $g$. And if we proceeded to order $\e r^2$ in the matching procedure, we would find that the tidal fields appearing in the inner expansion are those of $g+h^R$, rather than those of $g$.

These general points are shared with the derivation in Ref.~\cite{Eric_review}. Note, however, that my results differ considerably from those of Refs.~\cite{Mino_Sasaki_Tanaka, Eric_review}. The first difference is that in those earlier calculations it was found that the tetrad on the worldline is not parallel-propagated in the external background spacetime. This followed from the fact that the spin dipole perturbation in the internal solution had been set to zero via a choice of gauge; effectively, the Fermi tetrad was required to rotate with the perturbed gravitational field, to set the total spin to zero. However, this is not necessary: transforming from the external, nonrotating frame, to the internal, rotating frame (or to a nonrotating frame in $g+h^R$), simply requires a gauge transformation (specifically generated by $\Lambda\coeff{1}_c$). There is no reason to require the external, background Fermi tetrad in $g$ to spin.

More importantly, my analysis has shown that the weak matching condition that is normally utilized is actually too weak to yield unique results. In order to arrive at an equation of motion, I have had to formulate a refined matching condition in a somewhat ad hoc manner. But even if that refinement is accepted, all of these derivations rely on another strong assumption: the form of the inner expansion, which fixes not only the background metric, but also the form the perturbations. By assuming that the metric in the inner expansion appears to that of a singular monopole perturbation atop some smooth background field that is expanded around a geodesic, the matching derivations of the self-force seem to implicitly assume a generalized equivalence principle: they assume that in vacuum, the black hole, as viewed from a distance, moves on a geodesic of some smooth spacetime. Given that such a geodesic exists, the matching procedure provides a means of determining \emph{which} smooth spacetime the geodesic lies in---but it does not prove the existence of the geodesic.

If the inner expansion is to be sufficiently general for the matching procedure to \emph{derive} the generalized equivalence principle, rather than assume it, then one must use a less restricted inner expansion. For example, at linear order, no acceleration-like dipole term (i.e. one behaving as $\sim r$ in the buffer region) can arise in the inner expansion without also introducing a mass dipole. However, in the inner expansion an acceleration term $\e r\an{1}$ corresponds to a second-order perturbation $\e^2\tilde R \an{1}$, so in order to maintain that no such term can arise in mass-centered coordinates, one must solve the second-order EFE in generality. In addition, if one uses the refined matching condition or some variant of it, one must prove that an acceleration term cannot arise from a generically small coordinate transformation in the buffer region. It is not immediately clear that at second order, being in mass-centered coordinates implies that any acceleration-like term must vanish.\footnote{Kinnersley's photon rocket is an example of an exact solution to the EFE with acceleration-like dipole terms sourced by radiation \cite{Kinnersley, Bonnor_rocket, Damour_rocket}.} Solving the EFE, in full generality, at second order also requires one to include potentially time-evolving shifts in the mass and spin of the black hole. The time-evolution, if any, of these parameters would be determined at second order. Indeed, a time-dependent correction to the mass will be found in the next chapter.

In that chapter, I will perform a second-order analysis, but of a different sort than the one just suggested. Instead of assuming that the small body is a black hole and trying to solve the second-order EFE in the whole of that black hole's spacetime, I will allow the body to be arbitrarily structured, and I will solve the second-order EFE in the outer expansion, and only in the buffer region.
			\chapter{General expansion in the buffer region}\label{buffer_region}
%%%%%%%%%%%
Given the failings of the method of matched asymptotic expansions, I now make use of a different approach. Rather than solving the internal and external problems separately and then matching the solutions, I work entirely in the buffer region, making minimal assumptions about the forms of the internal and external solutions. By solving the Einstein equation up to second order in $\e$ in the buffer region, I determine the equation of motion of the worldline up to $\order{\e^2}$ errors. In addition to being free of the problems found in the method of matched asymptotic expansions, this method will also be valid for an arbitrarily structured body; the only restriction placed on the body is that it must be sufficiently compact to admit a buffer region free of matter. Although I perform this calculation in the Lorenz gauge, the choice of gauge should be of little significance.

Over the course of this calculation, we will find that the external metric perturbation in the buffer region is expressed as the sum of two solutions: one solution that diverges at $r=0$ and which is entirely determined from a combination of (i) the multipole moments of the internal background metric $g_B$, (ii) the Riemann tensor of the external background $g$, and (iii) the acceleration of the worldline $\gamma$; and a second solution that is regular at $r=0$ and must be determined from the global past history of the body. At leading order, these two solutions are identified as the Detweiler-Whiting singular and regular fields $h^S$ and $h^R$, and the self-force is determined entirely by $h^R$. Along with the self-force, the acceleration of the worldline includes the Papapetrou spin-force \cite{Papapetrou}. This leaves us with the self-force in terms of the metric perturbation induced by the body.

In the next chapter, I proceed to obtain a global, formal solution for the metric perturbation in the Lorenz gauge. Following the method of D'Eath \cite{DEath_paper, DEath}, I write the formal solution to the wave equation in an integral representation, whereby the value of the metric perturbation at any point in the exterior region is related to an integral over the worldtube around the body. Since the tube is chosen to lie in the buffer region, the previously obtained expansion in that region then serves to provide the boundary data on the tube. This approach allows me to determine $h^R$ in the buffer-region expansion by appealing to the consistency of the integral representation of the wave equation. Given the results of the buffer-region expansion as boundary values, evaluating the integral representation at a point just outside the worldtube must return the general solution in the buffer region. This consistency condition determines the unknown functions in terms of a tail integral. With the solution in the buffer region determined, the worldline is also determined; at the same time, since the boundary values are determined, the solution in the external spacetime is also determined. But let us not get ahead of ourselves.

\section{The form of the expansion} \label{buffer_expansion}
Now, I no longer wish to assume that $\hmn{E}{1}$ is the metric perturbation due to a point particle. Instead, I wish to justify that conclusion. So I must first determine the general form of an expansion in powers of $r$ for the metric perturbations $\hmn{E}{\emph{n}}$. To accomplish this, I consider the form of the internal metric $g_B+H$. Again, I no longer wish to assume a form for this internal metric. Instead, I merely assume that in the buffer region there exists a smooth coordinate transformation between the local coordinates $(T,R,\Theta^A)$ and the Fermi coordinates $(t,x^a)$ such that $T\sim t$, $R\sim r$, and $\Theta^A\sim\theta^A$. The buffer region corresponds to asymptotic infinity $r\gg\e$ (or $\tilde r\gg1$) in the internal spacetime. So after re-expressing $\tilde r$ as $r/\e$, the internal background metric can be expanded as
\begin{align}
g_{B\alpha\beta}(t,\tilde r,\theta^A) &= \sum_{n\geq0}\left(\frac{\e}{r}\right)^n g\coeff{\emph{n}}_{B\alpha\beta}(t,\theta^A).
\end{align}
There is no a priori reason to exclude negative values of $n$, since $g_B$ is an unknown function of $\tilde r$. However, since the internal and external solutions must be approximations to the same metric, they must agree with one another. And since the external expansion has no negative powers of $\e$, neither has the internal expansion. Furthermore, since $g+h=g_B+H$, we must have $g_B\coeff{0}=g(x^a=0)$, since these are the only terms independent of both $\e$ and $r$. Thus, noting that $g(x^a=0)=\eta$, where $\eta\equiv\text{diag}(-1,1,1,1)$, I can write
\begin{align}
g_{B\alpha\beta}(t,\tilde r,\theta^A) &= \eta_{\alpha\beta}+\frac{\e}{r} g\coeff{1}_{B\alpha\beta}(t,\theta^A) +\left(\frac{\e}{r} \right)^2g\coeff{2}_{B\alpha\beta}(t,\theta^A)+\order{\e^3/r^3},
\end{align}
implying that the internal background spacetime is asymptotically flat.

I assume that the perturbation $H$ can be similarly expanded in powers of $\e$ at fixed $\tilde r$,
\begin{align}
H_{\alpha\beta}(t,\tilde r,\theta^A,\e) &= \e H\coeff{1}_{\alpha\beta}(t,\tilde r,\theta^A;\gamma) +\e^2 H\coeff{2}_{\alpha\beta}(t,\tilde r,\theta^A;\gamma)+\order{\e^3},
\end{align}
and that each coefficient can be expanded in powers of $1/\tilde r=\e/r$ to yield
\begin{align}\label{buffer ansatz}
\e H\coeff{1}_{\alpha\beta}(\tilde r) &= rH\coeff{0,1}_{\alpha\beta}+\e H\coeff{1,0}_{\alpha\beta}+ \frac{\e^2}{r}H\coeff{2,-1}_{\alpha\beta}+\order{\e^3/r^2},\\
\e^2 H\coeff{2}_{\alpha\beta}(\tilde r) & = r^2 H\coeff{0,2}_{\alpha\beta} +\e rH\coeff{1,1}_{\alpha\beta}+\e^2 H\coeff{2,0}_{\alpha\beta}+\e^2\ln r H\coeff{2,0,ln}_{\alpha\beta}+\order{\e^3/r},\\
\e^3 H\coeff{3}_{\alpha\beta}(\tilde r) & = \order{\e^3,\e^2r, \e r^2,r^3},
\end{align}
where $H\coeff{\emph{n,m}}$, the coefficient of $\e^n$ and $r^m$, is a function of $t$ and $\theta^A$ (and potentially a functional of $\gamma$). Again, the form of this expansion is constrained by the fact that no negative powers of $\e$ can appear in the buffer region.\footnote{One might think that terms with negative powers of $\e$ could be allowed in the expansion of $g_B$ if they are exactly canceled by terms in the expansion of $H$, but the differing powers of $r$ in the two expansion makes this impossible.} Note that explicit powers of $r$ appear because $\e\tilde r=r$. Also note that I allow for a logarithmic term at second order in $\e$; this term arises because the retarded time in the internal background includes a logarithmic correction of the form $\e\ln r$ (e.g., $t-r\to t-r^*$ in Schwarzschild coordinates). Since I seek solutions to a wave equation, this correction to the characteristic curves induces a corresponding correction to the first-order perturbations. 

The expansion of $H$ may or may not hold the acceleration fixed. (In the previous chapter, it did not.) Regardless of this choice, the general form of the expansion remains valid: incorporating the expansion of the acceleration would merely shuffle terms from one coefficient to another. And since the internal metric $g_B+H$ must equal the external metric $g+h$, the general form of the above expansions of the $g_B$ and $H$ completely determines the general form of the external perturbations:
\begin{align}
\hmn{E\alpha\beta}{1} & = \frac{1}{r}\hmn{E\alpha\beta}{1,-1} +\hmn{E\alpha\beta}{1,0} +r\hmn{\alpha\beta}{1,1}+\order{r^2}, \label{h_E1 expansion}\\
\hmn{E\alpha\beta}{2} & = \frac{1}{r^2}\hmn{E\alpha\beta}{2,-2} +\frac{1}{r}\hmn{E\alpha\beta}{2,-1} +\hmn{E\alpha\beta}{2,0}+\ln r\hmn{E\alpha\beta}{2,0,ln}+\order{r},\label{h_E2 expansion}
\end{align}
where each $\hmn{E}{\emph{n,m}}$ depends only on $t$ and $\theta^A$, along with an implicit functional dependence on $\gamma$. If the internal expansion is performed with $a$ held fixed, then the internal and external quantities are related order-by-order: e.g., $\sum_m H\coeff{0,m}=g$, $\hmn{E}{1,-1}=g_B\coeff{1}$, and $\hmn{E}{1,0}=H\coeff{1,0}$. Since I am not concerned with determining the internal perturbations, the only such relationship of interest is $\hmn{E}{\emph{n,-n}}=g_B\coeff{\emph{n}}$. This equality tells us that the most divergent, $r^{-n}$ piece of the $n$th-order perturbation $\hmn{E}{\emph{n}}$ is defined entirely by the $n$th-order piece of the internal background metric $g_B$, which is the metric of the body if it were isolated.

To obtain a general solution to the Einstein equation, I write each $\hmn{E}{\emph{n,m}}$ as an expansion in terms of irreducible symmetric trace-free pieces:
{\allowdisplaybreaks\begin{align}
\hmn{Ett}{{\it n,m}} &= \sum_{\ell\ge0}\A{L}{{\it n,m}}\nhat^L, \\
\hmn{Eta}{{\it n,m}} &= \sum_{\ell\ge0}\B{L}{{\it n,m}}\nhat_a{}^L +\sum_{\ell\ge1}\left[\C{aL-1}{{\it n,m}}\nhat^{L-1} +\epsilon_{ab}{}^c\D{cL-1}{{\it n,m}}\nhat^{bL-1}\right],\\
\hmn{Eab}{{\it n,m}} &= \delta_{ab}\sum_{\ell\ge0}\K{L}{{\it n,m}}\nhat^L+\sum_{\ell\ge0}\E{L}{{\it n,m}}\nhat_{ab}{}^L +\sum_{\ell\ge1}\!\left[\F{L-1\langle a}{{\it n,m}}\nhat^{}_{b\rangle}{}^{L-1} +\epsilon^{cd}{}_{(a}\nhat_{b)c}{}^{L-1}\G{dL-1}{{\it n,m}}\right] \nonumber\\
&\quad+\sum_{\ell\ge2}\!\left[\H{abL-2}{{\it n,m}}\nhat^{L-2}+\epsilon^{cd}{}_{(a}\I{b)dL-2}{{\it n,m}} \nhat_c{}^{L-2}\right].
\end{align}}
Here a hat indicates that a tensor is STF with respect to $\delta_{ab}$, angular brackets $\langle\rangle$ indicate the STF combination of enclosed indices, parentheses indicate the symmetric combination of enclosed indices, and all the uppercase script symbols are functions of time (and potentially functionals of $\gamma$) and are STF in all their indices. Each term in this expansion is linearly independent of all the other terms. All the quantities on the right-hand side are flat-space Cartesian tensors; their indices can be raised or lowered with $\delta_{ab}$. Refer to Appendix~\ref{STF tensors} for more details about this expansion.

Now, since the wave equations \eqref{h_E1 eqn} and \eqref{h_E2 eqn} are covariant, they must still hold in the new coordinate system, despite the additional $\e$-dependence. Thus, both equations could be solved for arbitrary acceleration in the buffer region. However, due to the length of the calculations involved, I will instead solve the equations
\begin{align}
E_{\alpha\beta}[\hmn{E}{1}] & = 0, \label{h_1 eqn}\\
E\coeff{0}_{\alpha\beta}[\hmn{}{2}] & = 2\delta^2 R\coeff{0}_{\alpha\beta}[\hmn{}{1}]+\order{\e},\label{h_2 eqn}
\end{align}
where $E\coeff{0}[f]\equiv E[f]\big|_{a=\an{0}}$ and $\delta^2 R\coeff{0}[f]\equiv \delta^2R[f]\big|_{a=\an{0}}$.\footnote{In analogy with the notation used for $L\coeff{\emph{n}}$, $E\coeff{1}[f]$ and $\delta^2 R\coeff{1}[f]$ would be linear in $\an{1}$, $E\coeff{2}[f]$ and $\delta^2 R\coeff{2}[f]$ would be linear in $\an{2}$ and quadratic in $\an{1}$, and so on. For a function $f\sim 1$, $L\coeff{\emph{n}}[f]$, $E\coeff{\emph{n}}[f]$, and $\delta^2 R\coeff{\emph{n}}[f]$ correspond to the coefficients of $\e^n$ in expansions in powers of $\e$.} The first equation is identical to Eq.~\eqref{h_E1 eqn}. The second equation follows directly from substituting Eqs.~\eqref{buffer_expansion g} and \eqref{buffer_expansion h} into Eq.~\eqref{h_E2 eqn}; in the buffer region, it captures the dominant behavior of $\hmn{E}{2}$, represented by the approximation $\hmn{}{2}$, but it does not capture its full dependence on acceleration. If one desired a global second-order solution, one would solve Eq.~\eqref{h_E2 eqn}, but for my purpose, which is to determine the first-order acceleration $\an{1}$, Eq.~\eqref{h_2 eqn} will suffice.

Unlike the wave equations, the gauge conditions \eqref{h_E1 gauge} and \eqref{h_E2 gauge} already incorporate the expansion of the acceleration. As such, they are unmodified by the replacement of the second-order wave equation \eqref{h_E2 eqn} with its approximation \eqref{h_2 eqn}. So we can write
\begin{align}
L\coeff{0}_\mu\big[\hmn{E}{1}\big] &=0, \label{h_1 gauge}\\
L\coeff{1}_\mu\big[\hmn{E}{1}\big] &= -L\coeff{0}_\mu\big[\hmn{}{2}\big], \label{h_2 gauge}
\end{align}
where the first equation is identical to Eq.~\eqref{h_E1 gauge}, and the second to Eq.~\eqref{h_E2 gauge}. (The second identity holds because $L\coeff{0}_\mu\big[\hmn{}{2}\big]=L\coeff{0}_\mu\big[\hmn{E}{2}\big]$, since $\hmn{}{2}$ differs from $\hmn{E}{2}$ by $\an{1}$ and higher acceleration terms, which are set to zero in $L\coeff{0}$.) I remind the reader that while this gauge choice is important for finding the external perturbations globally, any other choice would suffice in the buffer region calculation. For example, one could expand $g$ and $h$ in buffer region expansions that incorporate the expansion of the acceleration, enabling one to solve the full Einstein equation order-by-order in $\e$; it might be difficult to make this mesh with a global expansion in the external spacetime, but it would suffice to determine the acceleration. Alternatively, one could construct a two-timescale expansion in the buffer region, which would mesh with a global two-timescale expansion of the Einstein equation in the external spacetime.

As a final, important point, I assume the partial time-derivative of any term in an expansion is of the same order as the term itself. In what follows, the reader may safely assume that all calculations are lengthy unless noted otherwise.

%%%%%%%%%%%
\section{First-order solution in the buffer region}\label{buffer_expansion1}
%%%%%%%%%%%
In principle, solving the first-order Einstein equation in the buffer region is straightforward. One need simply substitute the expansion of $\hmn{E}{1}$, given in Eq.~\eqref{h_E1 expansion}, into the linearized wave equation \eqref{h_1 eqn} and the gauge condition \eqref{h_1 gauge}. Equating powers of $r$ in the resulting expansions then yields a sequence of equations that can be solved for successively higher-order terms in $\hmn{E}{1}$. Solving these equations consists primarily of expressing each quantity in its irreducible STF form, using the decompositions~\eqref{decomposition_1} and \eqref{decomposition_2}; since the terms in this STF decomposition are linearly independent, we can solve each equation term-by-term. This calculation is aided by the fact that $\del{\alpha}=x^a_\alpha\partial_a+\order{r^0}$, so for example, the wave operator $E_{\alpha\beta}$ consists of a flat-space Laplacian $\partial^a\partial_a$ plus corrections of order $1/r$. Appendix B also contains many useful identities, particularly $\partial_\alpha r =n_\alpha$, $n^\alpha\partial_\alpha\nhat^L=0$, and the fact that $\nhat^L$ is an eigenvector of the flat-space Laplacian: i.e., $\partial^a\partial_a\nhat^L =-\frac{\ell(\ell+1)}{r^2}\nhat^L$. Because the calculation consists mostly of simple, albeit lengthy algebra, I will for the most part simply summarize results. 

Of course, the Einstein equation in the buffer region does not completely determine the solution: auxiliary boundary data must also be provided. Since the most singular term, $\hmn{E\alpha\beta}{1,-1}$, is the order-$1/r$ term in the internal background metric $g_B$, it will be fully determined in terms of the mass of the internal spacetime. Some of the subleading terms will also be determined by the mass, while others will remain unknown. The unknowns form the Detweiler-Whiting regular field; they will eventually be expressed in terms of a tail integral in Sec.~\ref{perturbation calculation}.

So, we begin with the the most divergent term in the wave equation: the order-$1/r^3$, flat-space Laplacian term
\begin{equation}
\frac{1}{r}\partial^c\partial_c\hmn{E\alpha\beta}{1,-1} =0.
\end{equation}
The $tt$-component of this equation is
\begin{equation}
0=-\sum_{\ell\ge0}\ell(\ell+1)\A{L}{1,-1}\nhat^L,
\end{equation}
from which we read off that $\A{}{1,-1}$ is arbitrary and $\A{L}{1,-1}$ must vanish for all $\ell\ge1$. The $ta$-component is
\begin{align}
0&=-\sum_{\ell\ge0}(\ell+1)(\ell+2)\B{L}{1,-1}\nhat_a{}^L -\sum_{\ell\ge1}\ell(\ell-1)\C{aL-1}{1,-1}\nhat^{L-1}\nonumber\\
&\quad-\sum_{\ell\ge1}\ell(\ell+1)\epsilon_{abc}\D{cL-1}{1,-1}\nhat_b{}^{L-1},
\end{align}
from which we read off that $\C{a}{1,-1}$ is arbitrary and all other coefficients must vanish. Lastly, the $ab$-component is
\begin{align}
0&=-\delta_{ab}\sum_{\ell\ge0}\ell(\ell+1)\K{L}{1,-1}\nhat^L -\sum_{\ell\ge0}(\ell+2)(\ell+3)\E{L}{1,-1}\nhat_{ab}{}^L\nonumber\\
&\quad-\sum_{\ell\ge1}\ell(\ell+1)\F{L-1\langle a}{1,-1}\nhat^{}_{b\rangle}{}^{L-1} -\sum_{\ell\ge1}(\ell+1)(\ell+2)\epsilon_{cd(a}\nhat_{b)}{}^{cL-1}\G{dL-1}{1,-1} \nonumber\\
&\quad-\sum_{\ell\ge2}(\ell-2)(\ell-1)\H{abL-2}{1,-1}\nhat^{L-2} -\sum_{\ell\ge2}\ell(\ell-1)\epsilon^{}_{cd(a}\I{b)dL-2}{1,-1}\nhat_c{}^{L-2},
\end{align}
from which we read off that $\K{}{1,-1}$ and $\H{ab}{1,-1}$ are arbitrary and all other coefficients must vanish. Thus, we find that the wave equation constrains $\hmn{E}{1,-1}$ to be
\begin{align}
\hmn{E\alpha\beta}{1,-1}& =\A{}{1,-1}t_\alpha t_\beta+ 2\C{a}{1,-1}t_{(\beta}x^a_{\alpha)} +(\delta_{ab}\K{}{1,-1}+\H{ab}{1,-1})x^a_\alpha x^b_\beta.
\end{align}

This is further constrained by the most divergent, $1/r^2$ term in the gauge condition:
\begin{equation}
-\frac{1}{r^2}\hmn{E\alpha c}{1,-1}n^c+\frac{1}{2r^2}n_\alpha\eta^{\mu\nu}\hmn{E\mu\nu}{1,-1}=0.
\end{equation}
From the $t$-component of this equation, we read off $\C{a}{1,-1}=0$; from the $a$-component, $\K{}{1,-1}=\A{}{1,-1}$ and $\H{ab}{1,-1}=0$. Thus, $\hmn{E\alpha\beta}{1,-1}$ depends only on a single function of time, $\A{}{1,-1}$. By the definition of ADM mass, this function (times $\e$) must be twice the mass of the internal background spacetime. Thus, $\hmn{E}{1,-1}$ is fully determined to be
\begin{equation}\label{h1n1}
\hmn{E\alpha\beta}{1,-1}=2m(t)(t_\alpha t_\beta +\delta_{ab}x^a_\alpha x^b_\beta),
\end{equation}
where $m(t)$ is defined to be the mass at time $t$ divided by the initial mass $\e\equiv m_0$. (Alternatively, we could set $\e$ equal to unity at the end of the calculation, in which case $m$ would simply be the mass at time $t$; obviously, the difference between the two approaches is immaterial.)

At the next order, $\hmn{E}{1,0}$, along with the acceleration of the worldline and the time-derivative of the mass, first appears in the Einstein equation. The order-$1/r^2$ term in the wave equation is
\begin{equation}
\partial^c\partial_c\hmn{E\alpha\beta}{1,0}=-\frac{2m}{r^2}a_cn^c(3t_\alpha t_\beta -\delta_{ab}x^a_\alpha x^b_\beta),
\end{equation}
where the terms on the right arise from $\Box$ acting on $\frac{1}{r}\hmn{E}{1,-1}$. This equation constrains $\hmn{E}{1,0}$ to be
\begin{equation}\label{h10}
\begin{split}
\hmn{Ett}{1,0}&=\A{}{1,0}+3ma_cn^c, \\
\hmn{Eta}{1,0}&=\C{a}{1,0}, \\
\hmn{Eab}{1,0}&=\delta_{ab}\left(\K{}{1,0}-ma_cn^c\right)+\H{ab}{1,0}.
\end{split}
\end{equation}
Substituting this result into the order-$1/r$ term in the gauge condition, we find
\begin{equation}
-\frac{4}{r} t_\alpha\partial_t m +\frac{4m}{r}\an{0}_ax^a_\alpha=0.
\end{equation}
Thus, both the leading-order part of the acceleration and the rate of change of the mass of the body vanish:
\begin{equation}
\begin{array}{lcr}
\displaystyle\frac{\partial m}{\partial t}=0\,, && \an{0}_i =0.
\end{array}
\end{equation}

At the next order, $r\hmn{E}{1,1}$, along with squares and derivatives of the acceleration, first appear in the Einstein equation, and the tidal fields of the external background couple to $\frac{1}{r}\hmn{E}{1,-1}$. The order-$1/r$ term in the wave equation becomes
\begin{align}
\left(r\partial^c\partial_c+\frac{2}{r}\right)\hmn{Ett}{1,1} & = -\frac{20m}{3r}\etide_{ij}\nhat^{ij}
-\frac{3m}{r}a_{\langle i}a_{j\rangle}\nhat^{ij} +\frac{8m}{r}a_ia^i, \\
\left(r\partial^c\partial_c+\frac{2}{r}\right)\hmn{Eta}{1,1} & = -\frac{8m}{3r}\epsilon_{aij}\btide^j_k\nhat^{ik}-\frac{4m}{r}\dot a_a,\\
\left(r\partial^c\partial_c+\frac{2}{r}\right)\hmn{Eab}{1,1} & = \frac{20m}{9r}\delta_{ab}\etide_{ij}\nhat^{ij}  -\frac{76m}{9r}\etide_{ab} -\frac{16m}{3r}\etide^i_{\langle a}\nhat_{b\rangle i}
+\frac{8m}{r}a_{\langle a}a_{b\rangle}\nonumber\\
&\quad+\frac{m}{r}\delta_{ab}\!\left(\tfrac{8}{3}a_ia^i\!-3a_{\langle i}a_{j\rangle}\nhat^{ij}\right).
\end{align}
From the $tt$-component, we read off that $\A{i}{1,1}$ is arbitrary, $\A{}{1,1}=4ma_ia^i$, and $\A{ij}{1,1}=\tfrac{5}{3}m\etide_{ij} +\tfrac{3}{4}ma_{\langle i}a_{j\rangle}$; from the $ta$-component, $\B{}{1,1}$, $\C{ij}{1,1}$, and $\D{i}{1,1}$ are arbitrary, $\C{i}{1,1}=-2m\dot a_i$, and $\D{ij}{1,1}=\tfrac{2}{3}m\btide_{ij}$; from the $ab$ component, $\K{i}{1,1}$, $\F{i}{1,1}$, $\H{ijk}{1,1}$, and $\I{ij}{1,1}$ are arbitrary, and $\K{}{1,1}=\tfrac{4}{3}ma_ia^i$, $\K{ij}{1,1}=-\tfrac{5}{9}m\etide_{ij}+\tfrac{3}{4}ma_{\langle i}a_{j\rangle}$, $\F{ij}{1,1}=\tfrac{4}{3}m\etide_{ij}$, and $\H{ij}{1,1}=-\tfrac{38}{9}m\etide_{ij}+4ma_{\langle i}a_{j\rangle}$.

Substituting this into the order-$r^0$ terms in the gauge condition, we find
\begin{align}
0&=(n^i+r\partial^i)\hmn{E\alpha i}{1,1} -\tfrac{1}{2}\eta^{\mu\nu}(n_a-r\partial_a)\hmn{E\mu\nu}{1,1}x^a_\alpha -\partial_t\hmn{E\alpha t}{1,0} -\tfrac{1}{2}\eta^{\mu\nu}\partial_t\hmn{E\mu\nu}{1,0}t_\alpha\nonumber\\
&\quad +\tfrac{4}{3}m\etide_{ij}\nhat^{ij}n_\alpha +\tfrac{2}{3}m\etide_{ai}n^ix^a_\alpha,
\end{align}
where the equation is to be evaluated at $a=\an{0}=0$. From the $t$-component, we read off
\begin{equation}\label{B11}
\B{}{1,1}=\tfrac{1}{6}\partial_t\left(\A{}{1,0}+3\K{}{1,0}\right).
\end{equation}
From the $a$-component,
\begin{equation}\label{F11}
\F{a}{1,1}=\tfrac{3}{10}\left(\K{a}{1,1}-\A{a}{1,1}+\partial_t\C{a}{1,0}\right).
\end{equation}
It is understood that both these equations hold only when evaluated at $a=\an{0}$.

Thus, the order-$r$ component of $\hmn{E}{1}$ is
\begin{equation}\label{h11}
\begin{split}
\hmn{Ett}{1,1}&=4ma_ia^i+\A{i}{1,1}n^i+\tfrac{5}{3}m\etide_{ij}\nhat^{ij}  +\tfrac{3}{4}ma_{\langle i}a_{j\rangle}\nhat^{ij}, \\
\hmn{Eta}{1,1}&=\B{}{1,1}n_a-2m\dot a_a+\C{ai}{1,1}n^i+\epsilon_{ai}{}^j\D{j}{1,1}n^i +\tfrac{2}{3}m\epsilon_{aij}\btide^j_k\nhat^{ik},\\
\hmn{Eab}{1,1}&=\delta_{ab}\big(\tfrac{4}{3}ma_ia^i+\K{i}{1,1}n^i -\tfrac{5}{9}m\etide_{ij}\nhat^{ij} +\tfrac{3}{4}ma_{\langle i}a_{j\rangle}\nhat^{ij}\big) +\tfrac{4}{3}m\etide^i_{\langle a}\nhat_{b\rangle i}
\\
&\quad-\tfrac{38}{9}m\etide_{ab}+4ma_{\langle a}a_{b\rangle}+\H{abi}{1,1}n^i +\epsilon\indices{_i^j_{(a}}\I{b)j}{1,1}n^i+\F{\langle a}{1,1}n^{}_{b\rangle}.
\end{split}
\end{equation}
where $\B{}{1,1}$ and $\F{a}{1,1}$ are constrained to satisfy Eqs.~\eqref{B11} and \eqref{F11}.

To summarize the results of this section, we have $\hmn{E\alpha\beta}{1}=\frac{1}{r}\hmn{E\alpha\beta}{1,-1} +\hmn{E\alpha\beta}{1,0} +r\hmn{E\alpha\beta}{1,1}+\order{r^2}$, where $\hmn{E\alpha\beta}{1,-1}$ is given in Eq.~\eqref{h1n1}, $\hmn{E\alpha\beta}{1,0}$ is given in Eq.~\eqref{h10}, and $\hmn{E\alpha\beta}{1,1}$ is given in Eq.~\eqref{h11}. In addition, we have determined that the ADM mass of the internal background spacetime is time-independent, and that the acceleration of the body's worldline vanishes at leading order.

%%%%%%%%%%%
\section{Second-order solution in the buffer region}\label{buffer_expansion2}
%%%%%%%%%%%
Though the calculations are much lengthier, solving the second-order Einstein equation in the buffer region is essentially no different than solving the first. I seek to solve the approximate wave equation \eqref{h_2 eqn}, along with the gauge condition \eqref{h_2 gauge}, for the second-order perturbation $\hmn{}{2}\equiv\hmn{E}{2}\big|_{a=\an{0}}$; doing so will also, more importantly, determine the acceleration $\an{1}$. In this calculation, the acceleration is set to $a=\an{0}=0$ everywhere except in the left-hand side of the gauge condition, $L\coeff{1}[\hmn{E}{1}]$, which is linear in $\an{1}$.

Substituting the expansion
\begin{align}
\hmn{\alpha\beta}{2}&=\frac{1}{r^2}\hmn{\alpha\beta}{2,-2} +\frac{1}{r}\hmn{\alpha\beta}{2,-1}+\hmn{\alpha\beta}{2,0} +\ln(r)\hmn{\alpha\beta}{2,0,ln} +\order{\e,r}
\end{align}
and the results for $\hmn{E}{1}$ from the previous section into the wave equation and the gauge condition again yields a sequence of equations that can be solved for coefficients of successively higher-order powers (and logarithms) of $r$. Due to its length, the expansion of the second-order Ricci tensor is given in Appendix~\ref{second-order expansions}. Note that since the approximate wave equation \eqref{h_2 eqn} contains an explicit $\order{\e}$ correction, $\hmn{}{2}$ will be determined only up to $\order{\e}$ corrections. For simplicity, I omit these $\order{\e}$ symbols from the equations in this section; note, however, that these corrections do not effect the gauge condition, as discussed above. 

To begin, the most divergent, order-$1/r^4$ term in the wave equation reads
\begin{align}
\frac{1}{r^4}\left(2+r^2\partial^c\partial_c\right)\hmn{\alpha\beta}{2,-2} & = \frac{4m^2}{r^4}\left(7\nhat_{ab} + \tfrac{4}{3}\delta_{ab}\right)x^a_\alpha x^b_\beta -\frac{4m^2}{r^4}t_\alpha t_\beta,
\end{align}
where the right-hand side is the most divergent part of the second-order Ricci tensor, as given in Eq.~\eqref{ddR0n4}. From the $tt$-component of this equation, we read off $\A{}{2,-2}=-2m^2$, and that $\A{a}{2,-2}$ is arbitrary. From the $ta$-component, $\B{}{2,-2}$, $\C{ab}{2,-2}$, and $\D{c}{2,-2}$ are arbitrary. From the $ab$-component, $\K{}{2,-2}=\tfrac{8}{3}m^2$, $\E{}{2,-2}=-7m^2$, and $\K{a}{2,-2}$, $\F{a}{2,-2}$, $\H{abc}{2,-2}$, and $\I{ab}{2,-2}$ are arbitrary.

The most divergent, order-$1/r^3$ terms in the gauge condition similarly involve only $\hmn{}{2,-2}$; they read
\begin{equation}
\frac{1}{r^3}\left(r\partial^b-2n^b\right)\hmn{\alpha b}{2,-2}-\frac{1}{2r^3}\eta^{\mu\nu}x^a_\alpha\left(r\partial_a -2n_a\right)\!\hmn{\mu\nu}{2,-2}=0.
\end{equation}
After substituting the results from the wave equation, the $t$-component of this equation determines that $\C{ab}{2,-2}=0$. The $a$-component determines that $\H{abc}{2,-2}=0$, $\I{ab}{2,-2}=0$, and
\begin{equation}\label{F2n2}
\F{a}{2,-2}=3\K{a}{2,-2}-3\A{a}{2,-2}.
\end{equation}
Thus, the order-$1/r^2$ part of $\hmn{}{2}$ is given by
\begin{equation}
\begin{split}
\hmn{tt}{2,-2} & = -2m^2+\A{i}{2,-2}n^i, \\
\hmn{ta}{2,-2} & = \B{}{2,-2}n_a+\epsilon_{a}{}^{ij}n_i\D{j}{2,-2},\\
\hmn{ab}{2,-2} & = \delta_{ab}\left(\tfrac{8}{3}m^2+\K{i}{2,-2}n^i\right) -7m^2\nhat_{ab}+\F{\langle a}{2,-2}n^{}_{b\rangle},
\end{split}
\end{equation}
where $\F{a}{2,-2}$ is given by Eq. \eqref{F2n2}.

The metric perturbation in this form depends on five free functions of time. However, from calculations in flat spacetime, we know that order-$\e^2/r^2$ terms in the metric perturbation can be written in terms of two free functions: a mass dipole and a spin dipole. We transform the perturbation into this ``canonical" form by performing a gauge transformation (c.f. Ref.~\cite{STF_2}). The transformation is generated by $\xi_\alpha=-\frac{1}{r}\B{}{2,-2}t_\alpha-\frac{1}{2r}\F{a}{2,-2}x^a_\alpha$, the effect of which is to remove $\B{}{2,-2}$ and $\F{a}{2,-2}$ from the metric.  This transformation is a refinement of the Lorenz gauge. (Effects at higher order in $\e$ and $r$ will be automatically incorporated into the higher-order perturbations.) The condition $\F{a}{2,-2}-3\K{a}{2,-2}+3\A{a}{2,-2}=0$ then becomes $\K{a}{2,-2}=\A{a}{2,-2}$. The remaining two functions are related to the ADM momenta of the internal spacetime: 
\begin{equation}
\begin{array}{lcr}
\A{i}{2,-2} =2M_i\,, && \D{i}{2,-2}=2S_i,
\end{array}
\end{equation}
where $M_i$ is such that $\partial_t M_i$ is proportional to the ADM linear momentum of the internal spacetime, and $S_i$ is the ADM angular momentum. $M_i$ is a mass dipole term; it is what would result from a transformation $x^a\to x^a+M^a/m$ applied to the $1/r$ term in $\hmn{E}{1}$. $S_i$ is a spin dipole term. Thus, the order-$1/r^2$ part of $\hmn{}{2}$ reads
\begin{equation}\label{h2n2}
\begin{split}
\hmn{tt}{2,-2} & = -2m^2+2M_in^i, \\
\hmn{ta}{2,-2} & = 2\epsilon_{aij}n^iS^j,\\
\hmn{ab}{2,-2} & = \delta_{ab}\left(\tfrac{8}{3}m^2+2M_in^i\right)-7m^2\nhat_{ab}.
\end{split}
\end{equation}

At the next order, $1/r^3$, because the acceleration is set to zero, $\hmn{}{2,-2}$ does not contribute to $E\coeff{0}[\hmn{}{2}]$, and $\hmn{}{1,-1}$ does not contribute to $\delta^2R\coeff{0}[\hmn{}{1}]$. The wave equation hence reads
\begin{equation}
\frac{1}{r}\partial^c\partial_c\hmn{\alpha\beta}{2,-1}= \frac{2}{r^3}\ddR{\alpha\beta}{0,-3}{\hmn{}{1}},
\end{equation}
where $\ddR{\alpha\beta}{0,-3}{\hmn{}{1}}$ is given in Eqs.~\eqref{ddR0n3_tt}--\eqref{ddR0n3_ab}. The $tt$-component of this equation implies $r^2\partial^c\partial_c\hmn{tt}{2,-1}=6m\H{ij}{1,0}\nhat^{ij}$, from which we read off that $\A{}{2,-1}$ is arbitrary and $\A{ij}{2,-1}=-m\H{ij}{1,0}$. The $ta$-component implies $r^2\partial^c\partial_c\hmn{ta}{2,-1}=6m\C{i}{1,0}\nhat_a^i$, from which we read off $\B{i}{2,-1}=-m\C{i}{1,0}$ and that $\C{a}{2,-1}$ is arbitrary. The $ab$-component implies
\begin{align}
r^2\partial^c\partial_c\hmn{ab}{2,-1}&=6m\left(\A{}{1,0} +\K{}{1,0}\right)\nhat_{ab}-12m\H{i\langle a}{1,0}\nhat_{b\rangle}{}^{i}+2m\delta_{ab}\H{ij}{1,0}\nhat^{ij},
\end{align}
from which we read off that $\K{}{2,-1}$ is arbitrary, $\K{ij}{2,-1}=-\tfrac{1}{3}m\H{ij}{1,0}$, $\E{}{2,-1}=-m\A{}{1,0}-m\K{}{1,0}$, $\F{ab}{2,-1}=2m\H{ab}{1,0}$, and $\H{ab}{2,-1}$ is arbitrary. This restricts $\hmn{}{2,-1}$ to the form
\begin{equation}
\begin{split}
\hmn{tt}{2,-1}&=\A{}{2,-1}-m\H{ij}{1,0}\nhat^{ij}, \\
\hmn{ta}{2,-1}&=-m\C{i}{1,0}\nhat_a^i+\C{a}{2,-1}, \\
\hmn{ab}{2,-1}&=\delta_{ab}\left(\K{}{2,-1} -\tfrac{1}{3}m\H{ij}{1,0}\nhat^{ij}\right) -m\left(\A{}{1,0}+\K{}{1,0}\right)\nhat_{ab}\\
&\quad +2m\H{i\langle a}{1,0}\nhat^{}_{b\rangle}{}^i+\H{ab}{2,-1}.
\end{split}
\end{equation}

We next substitute $\hmn{}{2,-2}$ and $\hmn{}{2,-1}$ into the order-$1/r^2$ terms in the gauge condition. The $t$-component becomes 
\begin{equation}
\frac{1}{r^2}\left(4m\C{i}{1,0}+12\partial_tM_i+3\C{i}{2,-1}\right)n^i=0,
\end{equation}
from which we read off
\begin{equation}
\C{i}{2,-1}=-4\partial_tM_i-\tfrac{4}{3}m\C{i}{1,0}.
\end{equation}
And the $a$-component becomes
\begin{align}
0&=\frac{1}{r^2}\left(-\tfrac{4}{3}m\A{}{1,0}-\tfrac{4}{3}m\K{}{1,0} -\tfrac{1}{2}\A{}{2,-1}+\tfrac{1}{2}\K{}{2,-1}\right)n_a\nonumber\\
&\quad +\left(\tfrac{2}{3}m\H{ai}{1,0}-\H{ai}{2,-1}\right)n^i -2\epsilon_{ija}n^i\partial_tS^j,
\end{align}
from which we read off
\begin{align}
\A{}{2,-1}&=\K{}{2,-1}-\tfrac{8}{3}m\left(\A{}{1,0}+\K{}{1,0}\right),\\
\H{ij}{2,-1}&=\tfrac{2}{3}m\H{ij}{1,0},
\end{align}
and that the angular momentum of the internal background is constant at leading order:
\begin{equation}
\partial_tS^i=0.
\end{equation}

Thus, the order-$1/r$ term in $\hmn{}{2}$ is given by
\begin{equation}\label{h2n1}
\begin{split}
\hmn{tt}{2,-1}&=\K{}{2,-1}-\tfrac{8}{3}m\left(\A{}{1,0}+\K{}{1,0}\right) -m\H{ij}{1,0}\nhat^{ij}, \\
\hmn{ta}{2,-1}&=-m\C{i}{1,0}\nhat_a^i-4\partial_tM_i-\tfrac{4}{3}m\C{i}{2,-1}, \\
\hmn{ab}{2,-1}&=\delta_{ab}\left(\K{}{2,-1} -\tfrac{1}{3}m\H{ij}{1,0}\nhat^{ij}\right) -m\left(\A{}{1,0}+\K{}{1,0}\right)\nhat_{ab}\\
&\quad +2m\H{i\langle a}{1,0}\nhat^{}_{b\rangle}{}^i+\tfrac{2}{3}m\H{ab}{1,0}.
\end{split}
\end{equation}
Note a peculiar feature of this term: the undetermined function $\K{}{2,-1}$ appears in precisely the form of a mass monopole. The value of this function will never be determined (though its time-dependence will be). This ambiguity arises because the mass $m$ that I have defined is the mass of the internal \emph{background} spacetime, which is based on the internal limit process that holds $\e/r$ fixed. A term of the form $\e^2/r$ appears as a perturbation of this background, even when, as in this case, it is part of the mass monopole of the body. This is equivalent to the ambiguity in any expansion in one's choice of small parameter: one could expand in powers of $\e$, or one could expand in powers of $\e+\e^2$, and so on. It is also equivalent to the ambiguity in defining the mass of a non-isolated body; whether the ``mass" of the body is taken to be $m$ or $m+\tfrac{1}{2}\K{}{2,-1}$ is a matter of taste. As we shall discover, the time-dependent part of $\K{}{2,-1}$ is constructed from the tail terms in the first-order metric perturbation. Hence, the ambiguity in the definition of the mass is, at least in part, equivalent to whether or not one chooses to include the free gravitational field induced by the body in what one calls its mass. (In fact, any order-$\e$ incoming radiation, not just that originally produced by the body, will contribute to this effective mass.) In any case, I will define the ``correction" to the mass as $\delta m\equiv\tfrac{1}{2}\K{}{2,-1}$.

We next move to the order-$\ln(r)/r^2$ terms in the wave equation, and the order-$\ln(r)/r$ terms in the gauge condition, which read
\begin{align}
\ln r\partial^c\partial_c\hmn{\alpha\beta}{2,0,ln} &= 0,\\
\ln r \left(\partial^b\hmn{\alpha b}{2,0,ln}-\tfrac{1}{2}\eta^{\mu\nu}x^a_\alpha\partial_a \hmn{\mu\nu}{2,0,ln}\right) & = 0.
\end{align}
From this we determine
\begin{align}
\hmn{\alpha\beta}{2,0,ln}& =\A{}{2,0,ln}t_\alpha t_\beta+ 2\C{a}{2,0,ln}t_{(\beta}x^a_{\alpha)} \nonumber\\
&\quad+(\delta_{ab}\K{}{2,0,ln}+\H{ab}{2,0,ln})x^a_\alpha x^b_\beta.
\end{align}

Finally, we arrive at the order-$1/r^2$ terms in the wave equation. At this order, the body's tidal moments become coupled to those of the external background. The equation reads
\begin{equation}\label{2n2 wave equation}
\partial^c\partial_c\hmn{\alpha\beta}{2,0} +\frac{1}{r^2}\!\!\left(\hmn{\alpha\beta}{2,0,ln}\!\! +\tilde{E}_{\alpha\beta}\right)=\frac{2}{r^2}\ddR{\alpha\beta}{0,-2}{\hmn{}{1}},
\end{equation}
where $\tilde E_{\alpha\beta}$ comprises the contributions from $\hmn{}{2,-2}$ and $\hmn{}{2,-1}$, given in Eqs.~\eqref{E_tt}, \eqref{E_ta}, and \eqref{E_ab}. The contribution from the second-order Ricci tensor is given in Eqs.~\eqref{ddR0n2_tt}--\eqref{ddR0n2_ab}.

Foregoing the details, after some algebra we can read off the solution
\begin{align}\label{h20}
\hmn{tt}{2,0} & = \A{}{2,0}+\A{i}{2,0}n^i+\A{ij}{2,0}\nhat^{ij}+\A{ijk}{2,0}\nhat^{ijk}\\
\hmn{ta}{2,0} & = \B{}{2,0}n_a+\B{ij}{2,0}\nhat_a{}^{ij} +\C{a}{2,0}+\C{ai}{2,0}\nhat_a{}^i\nonumber\\
&\quad+\epsilon_a{}^{bc}\left(\D{c}{2,0}n_b+\D{ci}{2,0}\nhat_b{}^i +\D{cij}{2,0}\nhat_b{}^{ij}\right)\\
\hmn{ab}{2,0} & = \delta_{ab}\left(\K{}{2,0}+\K{i}{2,0}n^i +\K{ijk}{2,0}\nhat^{ijk}\right) +\E{i}{2,0}\nhat_{ab}{}^i+\E{ij}{2,0}\nhat_{ab}{}^{ij}\nonumber\\
&\quad +\F{\langle a}{2,0}\nhat_{b\rangle} +\F{i\langle a}{2,0}\nhat_{b\rangle}{}^i+\F{ij\langle a}{2,0}\nhat_{b\rangle}{}^{ij} +\epsilon^{cd}{}_{(a}\nhat_{b)c}{}^i\G{di}{2,0}\nonumber\\
&\quad+\H{ab}{2,0}+\H{abi}{2,0}n^i+\epsilon^{cd}{}_{(a}\I{b)d}{2,0}n_c,
\end{align}
where each of the STF tensors is listed in Table~\ref{h20_tensors}.

In solving Eq.~\eqref{2n2 wave equation}, we also find that the logarithmic term in the expansion becomes uniquely determined:
\begin{equation}\label{h2ln}
\hmn{\alpha\beta}{2,0,ln} = -\tfrac{16}{15}m^2\etide_{ab}x^a_\alpha x^b_\beta.
\end{equation}
This term arises because the sources in the wave equation \eqref{2n2 wave equation} contain a term $\propto\etide_{ab}$, which cannot be equated to any term in $\partial^c\partial_c\hmn{ab}{2,0}$. Thus, the wave equation cannot be satisfied without including a logarithmic term. Recall that the logarithmic term arises at the order we would expect it to: the first-order perturbation alters the null cone of the spacetime, such that, e.g., $t-r\to t-r-2\e m\ln r$, which naturally introduces a correction $\sim \e^2\ln r$ to the order-$\e$ terms in the solution to the wave equation.

\begin{table}[tb]
\caption[STF tensors in the order-$\e^2r^0$ part of the metric perturbation]{Symmetric trace-free tensors appearing in the order-$\e^2r^0$ part of the metric perturbation in the buffer region around the body. Each tensor is a function of the proper time $t$ on the worldline $\gamma$, and each is STF with respect to the Euclidean metric $\delta_{ij}$.}
\begin{tabular*}{\linewidth}{l}
\hline\hline
$\begin{array}{rcl}
\A{}{2,0} & \phantom{=}& \text{ is arbitrary} \\ 
\A{i}{2,0} &=& -\partial^2_tM_i-\tfrac{4}{5}S^j\btide_{ji}+\tfrac{1}{3}M^j\etide_{ji} -\tfrac{7}{5}m\A{i}{1,1}-\tfrac{3}{5}m\K{i}{1,1} +\tfrac{4}{5}m\partial_t\C{i}{1,0}\\
\A{ij}{2,0} &=& -\tfrac{7}{3}m^2\etide_{ij}\\
\A{ijk}{2,0} &=& -2S_{\langle i}\btide_{jk\rangle} +\tfrac{5}{3}M_{\langle i} \etide_{jk\rangle} -\tfrac{1}{2}m\H{ijk}{1,1} \\
\B{}{2,0} &=& m\partial_t\K{}{1,0} \\
\B{ij}{2,0} &=& \tfrac{1}{9}\left(2M^l\btide^k_{(i} -5S^l\etide^k_{(i}\right)\epsilon_{j)kl}-\tfrac{1}{2}m\C{ij}{1,1} \\
\C{i}{2,0} & \phantom{=}& \text{ is arbitrary} \\
\C{ij}{2,0} &=& 2\left(S^l\etide^k_{(i}-\tfrac{14}{15}M^l\btide^k_{(i}\right)\epsilon_{j)lk}
-m\left(\tfrac{6}{5}\C{ij}{1,1}-\partial_t\H{ij}{1,0}\right)\\
\D{i}{2,0} &=& \tfrac{1}{5}\left(6M^j\btide_{ij} -7S^j\etide_{ij}\right)+2m\D{i}{1,1} \\
\D{ij}{2,0} &=& \tfrac{10}{3}m^2\btide_{ij} \\
\D{ijk}{2,0} &=&\tfrac{1}{3}S_{\langle i}\etide_{jk\rangle} +\tfrac{2}{3}M_{\langle i}\btide_{jk\rangle} \\
\K{}{2,0} &=& 2\delta m \\
\K{i}{2,0} &=& -\partial^2_tM_i-\tfrac{4}{5}S^j\btide_{ij} -\tfrac{5}{9}M^j\etide_{ij}+\tfrac{13}{15}m\A{i}{1,1}\!+\tfrac{9}{5}m\K{i}{1,1}\! -\tfrac{16}{15}m\partial_t\C{i}{1,0}\\
\K{ijk}{2,0} &=& -\tfrac{5}{9}M_{\langle i}\etide_{jk\rangle} +\tfrac{2}{9}S_{\langle i}\btide_{jk\rangle}-\tfrac{1}{6}m\H{ijk}{1,1}\\
\E{i}{2,0} &=& \tfrac{2}{15}M^i\etide_{ij}+\tfrac{1}{5}S^j\btide_{ij} +\tfrac{1}{10}m\partial_t\C{i}{1,0}
-\tfrac{9}{20}m\K{i}{1,1} -\tfrac{11}{20}m\A{i}{1,1}\\
\E{ij}{2,0} &=& \tfrac{7}{5}m^2\etide_{ij}\\
\F{i}{2,0} &=& \tfrac{184}{75}M^j\etide_{ij}+\tfrac{72}{25}S^j\btide_{ij} +\tfrac{46}{25}m\partial_t\C{i}{1,0}
-\tfrac{28}{25}m\A{i}{1,1}+\tfrac{18}{25}m\K{i}{1,1} \\
\F{ij}{2,0} &=& 4m^2\etide_{ij}\\
\F{ijk}{2,0} &=& \tfrac{4}{3}M_{\langle i}\etide_{jk\rangle} -\tfrac{4}{3}S_{\langle i}\btide_{jk\rangle} +m\H{ijk}{1,1}\\
\G{ij}{2,0} &=& -\tfrac{4}{9}\epsilon^{}_{lk(i}\etide_{j)}^kM^l\! -\tfrac{2}{9}\epsilon^{}_{lk(i}\btide_{j)}^kS^l\!+\tfrac{1}{2}m\I{ij}{1,1}\\
\H{ij}{2,0} & \phantom{=}& \text{ is arbitrary} \\
\H{ijk}{2,0} &=& \tfrac{58}{15}M_{\langle i}\etide_{jk\rangle} -\tfrac{28}{15}S_{\langle i}\btide_{jk\rangle}+\tfrac{2}{5}m\H{ijk}{1,1}\\
\I{ij}{2,0} &=& -\tfrac{104}{45}\epsilon^{}_{lk(i}\etide_{j)}^kM^l -\tfrac{112}{45}\epsilon^{}_{lk(i}\btide_{j)}^kS^l+\tfrac{8}{5}m\I{ij}{1,1}
\end{array}$\\
\hline\hline
\end{tabular*}
\label{h20_tensors}
\end{table}

We now move to the final equation in the buffer region: the order-$1/r$ gauge condition. This condition will determine the acceleration $\an{1}$. At this order, $\hmn{E}{1}$ first contributes to Eq.~\eqref{h_2 gauge}:
\begin{equation}
\gauge{\alpha}{1,-1}{\hmn{E}{1}}=\frac{4m}{r}\an{1}_ax^a_\alpha .
\end{equation}
The contribution from $\hmn{}{2}$ is most easily calculated by making use of Eqs.~\eqref{gauge_help1} and \eqref{gauge_help2}. After some algebra, we find that the $t$-component of the gauge condition reduces to
\begin{align}
0&=-\frac{4}{r}\partial_t\delta m +\frac{4m}{3r}\partial_t\A{}{1,0} +\frac{10m}{3r}\partial_t\K{}{1,0},
\end{align}
and the $a$-component reduces to 
\begin{align}\label{accelerations}
0 &=\frac{4}{r}\partial_t^2 M_a+\frac{4m}{r}\an{1}_a+\frac{4}{r}\etide_{ai}M^i+\frac{4}{r}\btide_{ai}S^i -\frac{2m}{r}\A{a}{1,1}+\frac{4m}{r}\partial_t\C{a}{1,0}.
\end{align}
The reader is reminded that these equations are valid only when evaluated at $a(t)=\an{0}(t)=0$, except in the term $\frac{4m}{r}\an{1}_a$ that arose from $\gauge{\alpha}{1}{\hmn{E}{1}}$. In the following subsection, this will allow me to swap partial derivatives with covariant derivatives on the worldline.

The $t$-component determines the rate of change of the mass correction $\delta m$. It can be immediately integrated to find
\begin{align}\label{mdot}
\delta m(t)&=\delta m(0) +\tfrac{1}{6}m\left[2\A{}{1,0}(t)+5\K{}{1,0}(t)\right]-\tfrac{1}{6}m\left[2\A{}{1,0}(0)+5\K{}{1,0}(0)\right].
\end{align}
If one felt so inclined, one could incorporate $\delta m(0)$ into the leading-order mass $m$. The time-dependent terms correspond to the effective mass created by the gravitational waves emitted by the body.

The $a$-component of the gauge condition determines the acceleration of the worldline. Note the most important feature of Eq.~\eqref{accelerations}, which is that it contains two types of accelerations: $\partial_t^2 M_i$ and $\an{1}_i$. The first type is the second time derivative of the body's mass dipole (or the first derivative of its ADM linear momentum), as measured in a frame centered on the worldline $\gamma$. The second type is the covariant acceleration of the worldline relative to the external spacetime. In other words, $\partial_t^2M_i$ corresponds to the acceleration of the body's center of mass relative to the center of the coordinate system, while $a_i$ measures the acceleration of the coordinate system itself. I define the worldline to be that of the body if the mass dipole vanishes for all times, meaning that the body is centered on the worldline for all times. If we start with initial conditions $M_i(0)=0=\partial_t M_i(0)$, then the mass dipole remains zero for all times if and only if the worldline satisfies the equation
\begin{equation}\label{a1}
\an{1}_a=\tfrac{1}{2}\A{a}{1,1}-\partial_t\C{a}{1,0}-\tfrac{1}{m}S_i\btide^i_a.
\end{equation}
This equation of motion contains two types of terms: a Papapetrou spin force, given by $-S_i\btide^i_a$, which arises due to the coupling of the body's spin to the local magnetic-type tidal field of the external spacetime; and a self-force, arising from homogenous terms in the wave equation.

Note that if we had followed the path of Gralla and Wald \cite{Gralla_Wald}, we would have identified $\an{0}$ as the acceleration of the worldline $\gamma\coeff{0}$. This would be the only actual worldline in play; all the corrections to the motion would be vectors defined on it. Hence, when we found $\an{0}=0$, we would have identified the worldline as a geodesic, and there would be no corrections $\an{n}$ for $n>0$. We would then have arrived at the equation of motion
\begin{equation}
\partial^2_tM_a+\etide_{ab}M^b = \tfrac{1}{2}\A{a}{1,1}-\partial_t\C{a}{1,0}-\tfrac{1}{m}S_i\btide^i_a.
\end{equation}
This is precisely the equation of motion derived by Gralla and Wald. It describes the drift of the body away from the reference geodesic $\gamma\coeff{0}$. If the external background is flat, then the mass dipole has a valid meaning as a displacement vector regardless of its magnitude; the second derivative $\partial_t^2M_i$ then provides a perfectly valid definition of the body's acceleration for all times. However, if the external background is curved, then $M_i$ has meaning only if the body is ``close" to the worldline. Thus, $\partial_t^2M_i$ is a meaningful acceleration only for a short time, since it will generically grow large as the body drifts away from the reference worldline. On that short timescale of validity, the deviation vector defined by $M^i$ accurately points from $\gamma\coeff{0}$ to a ``corrected" worldline $\gamma$; that worldline, the approximate equation of motion of which is given in Eq.~\eqref{a1}, accurately tracks the motion of the body. After a short time, when the mass dipole grows large and the regular expansion scheme begins to break down, the deviation vector will no longer correctly point to the corrected worldline.

To summarize the results of this section, the second-order perturbation in the buffer region is given by $h\coeff{2}_{\alpha\beta} =\frac{1}{r^2}\hmn{\alpha\beta}{2,-2} +\frac{1}{r}\hmn{\alpha\beta}{2,-1} +\hmn{\alpha\beta}{2,0} +\ln(r)\hmn{\alpha\beta}{2,0,ln} +\order{\e,r}$, where $\hmn{}{2,-2}$ is given in Eq.~\eqref{h2n2}, $\hmn{}{2,-1}$ in Eq.~\eqref{h2n1}, $\hmn{}{2,0}$ in Eq.~\eqref{h20}, and $\hmn{}{2,0,ln}$ in Eq.~\eqref{h2ln}. At order $\e^2/r^2$, the metric is written in terms of the mass and spin dipoles of the internal background metric $g_B$. The mass dipole is set to zero by an appropriate choice of worldline. At leading order in $\e$, the body's spin is constant along the worldline. At order $\e^2/r$, there arises an effective correction to the body's mass, given by Eq.~\eqref{mdot}. I note that this mass correction is entirely gauge-dependent: it could be removed by redefining the time coordinate on the worldline. The principal result of this section is the order-$\e$ term in the expansion of the body's acceleration, given by Eq.~\eqref{a1}. I remind the reader that the equation of motion would contain an antidamping term \cite{damping, damping2,Quinn_Wald} if I had not assumed that the acceleration possesses an expansion of the form given in Eq.~\eqref{a expansion}, and that such an expansion is necessary to determine a sequence of exactly solvable equations.

\section{A discussion of the force and the field in the buffer region}\label{comments on force}
The foregoing calculation completes the derivation of the gravitational self-force, in the sense that, given the metric perturbation in the neighbourhood of the body, the self-force is uniquely determined by irreducible pieces of that perturbation. Explicitly, the terms that appear in the self-force are given by
\begin{align}
\A{a}{1,1} &= \frac{3}{4\pi}\int n_a\hmn{tt}{1,1}d\Omega,\\
\C{a}{1,0} &= \hmn{ta}{1,0}.
\end{align}
Making use of the fact that $\hmn{\alpha\beta}{1,-1}$ is a monopole, we can write the acceleration in a variety of forms:
\begin{align}
\an{1}_a &= \lim_{r\to0}\left(\frac{3}{4\pi}\int\frac{n_a}{2r}\hmn{tt}{1}d\Omega -\partial_t\hmn{ta}{1}\right)\\
&= \lim_{r\to0}\frac{1}{4\pi}\int\left(\tfrac{1}{2}\partial_a\hmn{tt}{1} -\partial_t\hmn{ta}{1}\right)d\Omega\\
&= \lim_{r\to0}\frac{3}{4\pi}\int\left(\tfrac{1}{2}\partial_i\hmn{tt}{1} -\partial_t\hmn{ti}{1}\right)n^i_a d\Omega. \label{reg force}
\end{align}
One can easily derive these equalities from the STF decomposition of $\hmn{}{1}$ and the integral identities \eqref{n_integral}--\eqref{nhat_integral}. The form of the force in the second line is the method of regularization used by Quinn and Wald~\cite{Quinn_Wald}; the form in the third line is used to derive a gauge-invariant equation of motion, as was was first noted by Gralla~\cite{Gralla_gauge}. I will return to that notion momentarily.

This is all that is needed to incorporate the motion of the body into a dynamical system that can be numerically evolved; at each timestep, one simply needs to calculate the field near the worldline and decompose it into irreducible pieces in order to determine the acceleration of the body. (Obviously, such a procedure is vastly more complicated than what I have just implied \cite{regularization1, regularization2, regularization3, regularization4, regularization5, regularization6, regularization7}.) The remaining difficulty is to actually determine the field at each timestep. In the next chapter, I will write down formal expressions for the metric perturbation, and in particular, I will determine the metric perturbation at the location of the body in terms of a tail integral.

However, before doing so, I will emphasize some important features of the self-force and the field near the body. First, note that the first-order external field $\hmn{E}{1}$ separates into two distinct pieces. There is the singular piece $h^S$, given by
\begin{align}
h^S_{tt} &= \frac{2m}{r}\Big\lbrace 1+\tfrac{3}{2}ra_in^i+2r^2a_ia^i +r^2\left(\tfrac{3}{8}a_{\langle i}a_{j\rangle} +\tfrac{5}{6}\etide_{ij}\right)\nhat^{ij}\Big\rbrace+ \order{r^2} \\
h^S_{ta} &= -2mr\dot a_a +\tfrac{2}{3}m r\epsilon_{aij}\btide^j_k\nhat^{ik}+\order{r^2} \\
h^S_{ab} &= \frac{2m}{r}\Big\lbrace\delta_{ab}\big[1-\tfrac{1}{2}ra_in^i +\tfrac{2}{3}r^2a_ia^i +r^2\left(\tfrac{3}{8}a_{\langle i}a_{j\rangle}-\tfrac{5}{18}\etide_{ij}\right)\nhat^{ij}\big] +2r^2a_{\langle a}a_{b\rangle}\nonumber\\
&\quad -\tfrac{19}{9}r^2\etide_{ab}+\tfrac{2}{3}r^2\etide^i_{\langle a}\nhat_{b\rangle i}\Big\rbrace+\order{r^2}.
\end{align}
This field is a solution to the homogenous wave equation for $r>0$, but it is divergent at $r=0$. It is the generalization of the $1/r$ Newtonian field of the body, as perturbed by the tidal fields of the external spacetime $g$. Following the method used in Sec.~5.3.5 of Ref.~\cite{Eric_review}, one can easily show that this is precisely the Detweiler-Whiting singular field, given by
\begin{equation}
h^S_{\alpha\beta}=4m\int_\gamma \bar G^S_{\alpha\beta\alpha'\beta'}u^{\alpha'}u^{\beta'}dt',
\end{equation}
where $G^S_{\alpha\beta\alpha'\beta'}$ is the singular Green's function (defined in Appendix~\ref{Greens_functions}).

Next, there is the regular field $h^R\equiv \hmn{E}{1}-h^S$, given by
\begin{align}
h^R_{tt} &= \A{}{1,0}+r\A{i}{1,1}n^i+\order{r^2}, \\
h^R_{ta} &= \C{a}{1,0} +r\Big(\B{}{1,1}n_a+\C{ai}{1,1}n^i +\epsilon_{ai}{}^j\D{j}{1,1}n^i\Big)+\order{r^2},\\
h^R_{ab} &= \delta_{ab}\K{}{1,0}+\H{ab}{1,0}+r\Big(\delta_{ab}\K{i}{1,1}n^i+\H{abi}{1,1}n^i +\epsilon\indices{_i^j_{(a}}\I{b)j}{1,1}n^i+\F{\langle a}{1,1}n^{}_{b\rangle}\Big) \nonumber\\
&\quad+\order{r^2}.
\end{align}
This field is a solution to the homogeneous wave equation even at $r=0$. It is a free radiation field in the neighbourhood of the body. And it contains all the free functions in the buffer-region expansion.

Now, the acceleration of the body is given by
\begin{align}
\an{1}_a = \tfrac{1}{2}\partial_ah^R_{tt} -\partial_t h^R_{ta} -\tfrac{1}{m}S_i\btide^i_a,
\end{align}
which we can rewrite as 
\begin{align}
\an{1}{}^\alpha &= -\tfrac{1}{2}\left(g^{\alpha\delta}+u^{\alpha}u^{\delta}\right) \!\left(2h^R_{\delta\beta;\gamma}-h^R_{\beta\gamma;\delta}\right)\!\!\big|_{a=0} u^{\beta}u^{\gamma} +\frac{1}{2m}R^{\alpha}{}_{\beta\gamma\delta}u^\beta S^{\gamma\delta}
\end{align}
where $S^{\gamma\delta}\equiv e_c^\gamma e_d^\delta\epsilon^{cdj}S_j$. In other words, a non-spinning body (for which $S^{\gamma\delta}=0$), moves on a geodesic of a spacetime $g+\e h^R$, where $h^R$ is a free radiation field in the neighbourhood of the body; a local observer would measure the ``background spacetime,'' in which the body is in free fall, to be $g+\e h^R$, rather than $g$. If we performed a transformation into Fermi coordinates in $g+\e h^R$, the metric would contain no acceleration term, and it would take the simple form of a smooth background plus a singular perturbation. These points were first realized by Detweiler and Whiting \cite{Detweiler_Whiting} and since emphasized especially by Detweiler \cite{Detweiler_review}. They are, perhaps, made especially clear in the derivation presented here, which naturally demarcates the singular and regular fields.

Lastly, I comment on the gauge-dependence of the acceleration. First, suppose that we had not chosen a worldline for which the mass dipole vanishes, but instead had chosen some ``nearby" worldline. Then Eq.~\eqref{accelerations} provides the relationship between the acceleration of that worldline, the mass dipole relative to it, and the first-order metric perturbations (I neglect spin for simplicity). Now, the mass dipole is given by $M_i=\frac{3}{8\pi}\lim_{r\to 0}\int r^2\hmn{tt}{2}n_i d\Omega$, which has the covariant form
\begin{equation}
M_{\alpha'} = \frac{3}{8\pi}\lim_{r\to0}\int\! g^\alpha_{\alpha'}n_\alpha r^2\hmn{\mu\nu}{2}u^\mu u^\nu d\Omega,
\end{equation}
where a primed index corresponds to a point on the worldline. Note that the parallel propagator does not interfere with the angle-averaging, because in Fermi coordinates, $g^\alpha_{\beta'}=\delta^\alpha_\beta+O(\e,r^2)$. One can also rewrite the first-order-metric-perturbation terms in Eq.~\eqref{accelerations} using the form given in Eq.~\eqref{reg force}. We then have Eq.~\eqref{accelerations} in the covariant form
\begin{align}\label{gauge-invariant form}
\frac{3}{8\pi}\lim_{r\to0}&\int\! g^\alpha_{\alpha'}\! \left(\!g_{\alpha\beta}\frac{D^2}{d\tau^2}+\etide_{\alpha\beta}\!\right)\!n^\beta r^2\hmn{\mu\nu}{2}u^\mu u^\nu d\Omega\big|_{a=\an{0}} \nonumber\\
&= -m\an{1}_{\alpha'}-\frac{3m}{8\pi}\lim_{r\to0}\int\!g^\alpha_{\alpha'}\left(2\hmn{\beta\mu;\nu}{1}-\hmn{\mu\nu;\beta}{1}\right)u^\mu u^\nu n_\alpha^\beta d\Omega.\big|_{a=\an{0}}
\end{align}

Now consider a gauge transformation generated by $\e\xi\coeff{1}[\gamma]+\tfrac{1}{2}\e^2\xi\coeff{2}[\gamma]+...$, where $\xi\coeff{1}$ is bounded as $r\to0$, and $\xi\coeff{2}$ diverges as $1/r$. More specifically, I assume the expansions $\xi\coeff{1}=\xi\coeff{1,0}(t,\theta^A)+O(r)$ and $\xi\coeff{2}=\frac{1}{r}\xi\coeff{2,-1}(t,\theta^A)+O(1)$.\footnote{The dependence on $\gamma$ appears in the form of dependence on proper time $t$. Each term could in addition depend on the acceleration, but such dependence would not affect the result.} This transformation preserves the presumed form of the outer expansion, both in powers of $\e$ and in powers of $r$. According to Eqs.~\eqref{gauge_trans1}--\eqref{gauge_trans2}, the metric perturbations transform as 
\begin{align}
\hmn{\mu\nu}{1} &\to \hmn{\mu\nu}{1}+2\xi\coeff{1}_{(\mu;\nu)},\\
\hmn{\mu\nu}{2} &\to \hmn{\mu\nu}{2}+\xi\coeff{2}_{(\mu;\nu)}+\hmn{\mu\nu;\rho}{1}\xi\coeff{1}{}^\rho + 2\hmn{\rho(\mu}{1}\xi\coeff{1}{}^\rho{}_{;\nu)}+\xi\coeff{1}{}^\rho\xi\coeff{1}_{(\mu;\nu)\rho} \nonumber\\
&\quad+\xi\coeff{1}{}^\rho{}_{;\mu}\xi\coeff{1}_{\rho;\nu}+\xi\coeff{1}{}^\rho{}_{;(\mu}\xi\coeff{1}_{\nu);\rho}.
\end{align}
Using the results for $\hmn{}{1}$, the effect of this transformation on $\hmn{tt}{2}$ is given by
\begin{equation}
\hmn{tt}{2}\to\hmn{tt}{2}-\frac{2m}{r^2}n^i\xi\coeff{1}_i+O(r^{-1}).
\end{equation}
The order-$1/r^2$ term arises from $\hmn{\mu\nu;\rho}{1}\xi\coeff{1}{}^\rho$ in the gauge transformation. On the right-hand side of Eq.~\eqref{gauge-invariant form}, the metric-perturbation terms transform as
\begin{equation}
(2\hmn{\beta\mu;\nu}{1}-\hmn{\mu\nu;\beta}{1})u^\mu u^\nu n^\beta \to (2\hmn{\beta\mu;\nu}{1}-\hmn{\mu\nu;\beta}{1})u^\mu u^\nu n^\beta +2n_\beta\left(g_\beta^\gamma\frac{D^2}{d\tau^2}+\etide_\beta^\gamma\right)\xi\coeff{1}_\gamma.
\end{equation}
The only remaining term in the equation is $m\an{1}_\alpha$. If we extend the acceleration off the worldline in any smooth manner, then it defines a vector field that transforms as $a^\alpha\to a^\alpha+\e\Lie{\xi\coeff{1}}a^\alpha+...$. Since $\an{0}=0$, this means that $\an{1}\to\an{1}$---it is invariant under a gauge transformation.

From these results, we find that the left- and right-hand sides of Eq.~\eqref{gauge-invariant form} transform in the same way:
\begin{equation}\label{gauge_effect}
{\rm LHS\ or\ RHS} \to {\rm LHS\ or\ RHS} - \frac{3}{4\pi}\lim_{r\to0}\int g^\alpha_{\alpha'} n_\alpha^\beta\left(g_\beta^\gamma\frac{D^2}{d\tau^2}+\etide_\beta^\gamma\right)\xi\coeff{1}_\gamma d\Omega.
\end{equation}
Therefore, Eq.~\eqref{gauge-invariant form} provides a gauge-invariant relationship between the acceleration of a chosen fixed worldline, the mass dipole of the body relative to that worldline, and the first-order metric perturbations. So suppose that we begin in the Lorenz gauge, and we choose the fixed worldline $\gamma$ such that the mass dipole vanishes relative to it. Then in some other gauge, the mass dipole will no longer vanish relative to $\gamma$, and we must adopt a different, nearby fixed worldline $\gamma'$. If the mass dipole is to vanish relative to $\gamma'$, then the acceleration of that new worldline must be given by $a_\alpha=\e\an{1}_\alpha+o(\e)$, where
\begin{equation}
\an{1}_{\alpha'}=-\frac{3m}{8\pi}\lim_{r\to0}\int\! g^\alpha_{\alpha'}(2\hmn{\beta\mu;\nu}{1}-\hmn{\mu\nu;\beta}{1})u^\mu u^\nu n_\alpha^\beta d\Omega.\big|_{a=\an{0}}.
\end{equation}
Hence, this is a covariant and gauge-invariant form of the first-order acceleration. (By that I mean the \emph{equation} is valid in any gauge, not that the value of the acceleration is the same in every gauge; under a gauge transformation, a new fixed worldline is adopted, and the value of the acceleration on the new worldline is related to that on the old worldline according to Eq.~\eqref{gauge_effect}.) An argument of this form was first presented by Gralla~\cite{Gralla_gauge} for the case of a regular expansion of the worldline; it is now extended to the case of a fixed-worldline expansion.

			\chapter{The metric perturbation in the external spacetime}\label{perturbation calculation}
%%%%%%%%%%%%%%%%%%%%%%%%%%%%%%%%%%%%%%%%%%%%%%%%%%%%%%%%%%%%%%%%%%%%%%%%%%%%%%%%%%%%%%%%%%%%%%%%%%%%%%%%%%
%%%%%%%%%%%%%%%%%%%%%%%%%%%%%%%%%%%%%%%%%%%%%%%%%%%%%%%%%%%%%%%%%%%%%%%%%%%%%%%%%%%%%%%%%%%%%%%%%%%%%%%%%%
A solution to the self-force problem consists of a pair $(\gamma,h)$. In the previous chapter, we have determined the equation of motion of $\gamma$; we now require a means of determining the metric perturbation.

%%%%%%%%%%%
\section{Integral formulation of the Einstein equation in the external spacetime}\label{integral_expansion}
%%%%%%%%%%%
\begin{figure}[tb]
\begin{center}
\includegraphics{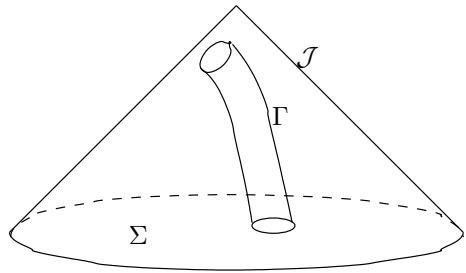}
\end{center}
\caption[The vacuum region $\Omega$]{The spacetime region $\Omega$ is bounded by the union of the spacelike surface $\Sigma$, the timelike worldtube $\Gamma$, and the null surface $\mathcal{J}$.} 
\label{volume}
\end{figure}

On the external manifold $\man_E$, I seek an approximate solution of Einstein's equation in a vacuum region $\bar\Omega\equiv\Omega\cup\partial\Omega$, where $\Omega$ is a bounded, open subset of $\man_E$. I now specify this region to be the future range of dependence of a surface formed by the union of a worldtube $\Gamma$ and a spatial surface $\Sigma$. This implies that the future boundary of $\Omega$ is a null surface $\mathcal{J}$. Refer to Fig.~\ref{volume} for an illustration. The boundary of the domain is hence $\partial\Omega \equiv \Gamma\cup\mathcal{J}\cup\Sigma$. The worldtube $\Gamma$ is defined by a constant Fermi radial coordinate distance $r=\rad$ from the worldline $\gamma\subset\man_E$. Since the tube is an artificial division of spacetime, and it may be located anywhere in the buffer region, any valid solution cannot depend on $\rad$. The spatial surface $\Sigma$ is chosen to intersect $\Gamma$ at the initial time $t=0$.

In $\Omega$, the Lorenz gauge is imposed on the entire perturbation $h$, splitting the Einstein equation into the weakly nonlinear wave equation
\begin{align}\label{wave_tube}
E_{\alpha\beta}\big[h\big] &=2\delta^2R_{\alpha\beta}\big[h\big]+\order{\e^3}
\end{align}
and the gauge condition $L_{\mu}\big[h\big]=0$. Note that if a solution to the wave equation satisfies the gauge condition on $\partial\Omega$, then the wave equation ensures that the gauge condition is satisfied everywhere. And since I have already determined the equation of motion using the expansion in the buffer region, I will hence not be interested in the gauge condition here.

As discussed in Secs.~\ref{singular_expansion_point_particle} and \ref{outline}, I assume the expansion $h_{\alpha\beta}(x,\e;\gamma)=\sum_n\e^n\hmn{E\alpha\beta}{n}(x;\gamma)$ and arrive at the sequence of wave equations
\begin{align}
E_{\alpha\beta}\big[\hmn{E}{1}\big] & = 0, \label{first_order_wave_tube}\\
E_{\alpha\beta}\big[\hmn{E}{2}\big] & = 2\delta^2R_{\alpha\beta}[\hmn{E}{1}]. \label{second_order_wave_tube}
\end{align}
Following D'Eath \cite{DEath,DEath_paper,Eric_review}, I rewrite the wave equations as integro-differential equations by calculating $E[G^{\text{adv}}]h-E[h]G^{\text{adv}}$ (where $G^{\text{adv}}$ represents the advanced Green's function for $E_{\mu\nu}$), integrating both sides of the resulting equation, making use of Stokes' law, and finally simplifying the result using the reciprocality relation $G^{\text {adv}}_{\alpha'\beta'\alpha\beta}(x',x)= G_{\alpha\beta\alpha'\beta'}(x,x')$. The resulting equations are
\begin{align}
\hmn{E\alpha\beta}{1} & = \frac{1}{4\pi}\oint\limits_{\partial\Omega}\! \Big(G_{\alpha\beta}{}^{\gamma'\delta'}\hmn{E\gamma'\delta';\mu'}{1} -\hmn{E\gamma'\delta'}{1} G_{\alpha\beta}{}^{\gamma'\delta'}{}_{;\mu'}\Big)dS^{\mu'}\!,\label{first_order_Kirchoff}\\
\hmn{E\alpha\beta}{2} & = \frac{1}{4\pi}\oint\limits_{\partial\Omega}\! \Big(G_{\alpha\beta}{}^{\gamma'\delta'}\hmn{E\gamma'\delta';\mu'}{2} -\hmn{E\gamma'\delta'}{2} G_{\alpha\beta}{}^{\gamma'\delta'}{}_{;\mu'}\Big)dS^{\mu'}\nonumber\\
&\quad-\frac{1}{2\pi}\int\limits_\Omega G_{\alpha\beta}{}^{\gamma'\delta'} \delta^2R_{\gamma'\delta'}[\hmn{E}{1}]dV'. \label{second_order_Kirchoff}
\end{align}

Alternatively, we might rewrite Eq.~\eqref{wave_tube} directly:
\begin{align}\label{all_orders_Kirchoff}
h_{\alpha\beta} & = \frac{1}{4\pi}\oint\limits_{\partial\Omega}\Big(G_{\alpha\beta}{}^{\gamma'\delta'} \del{\mu'}h_{\gamma'\delta'}-h_{\gamma'\delta'}\del{\mu'} G\indices{_{\alpha\beta}^{\gamma'\delta'}}\Big) dS^{\mu'}\nonumber\\
&\quad-\frac{1}{2\pi}\int_\Omega G_{\alpha\beta}{}^{\gamma'\delta'}\delta^2R_{\gamma'\delta'}\big[h\big]dV'+O(\e^3).
\end{align}
Note that any solution to Eq.~\eqref{wave_tube} in $\Omega$ will also satisfy this integro-differential equation; however, because of the $\e$-dependence of the true worldline $\gamma$, not every solution to Eq.~\eqref{wave_tube} will admit an expansion satisfying the two equations~\eqref{first_order_wave_tube} and \eqref{second_order_wave_tube} (though a solution to the latter is obviously a solution to the former). In that sense, Eq.~\eqref{all_orders_Kirchoff} is more robust than Eqs.~\eqref{first_order_Kirchoff} and \eqref{second_order_Kirchoff}.

In any case, these integral representations all have several important properties in common. First, the integral over the boundary is, in each case, a homogeneous solution to the wave equation, while the integral over the interior is an inhomogeneous solution.\footnote{The integral over the interior will also contain homogeneous solutions. However, these will be $\rad$-dependent, and they will exactly cancel corresponding $\rad$-dependent terms in the boundary integral.} Second, the integral over the boundary can be split into an integral over the worldtube $\Gamma$ and the spatial surface $\Sigma$; the contribution of the null surface $\mathcal{J}$ vanishes by construction. Also note that $x$ must lie in the interior of $\Omega$; an alternative expression must be derived if $x$ lies on the boundary \cite{Greens_functions}. 

Furthermore, the integral representations avoid any divergence in the second-order solution. Comparing Eq.~\eqref{all_orders_Kirchoff} to the analogous expression for a point particle, given in Eqs.~\eqref{1st_funct} and \eqref{2nd_funct}, we see that the point particle source terms have been replaced by an integral over a worldtube surrounding the small body, as we desired. And the volume integral over the interior of $\Omega$ does not diverge in $\Omega$, as it would in Eq.~\eqref{2nd_funct}, because the region of integration excludes the interior of the worldtube. 

Finally, one should note the essential character of these integrals. They provide a type of Kirchoff representation \cite{Friedlander, Sciama, Eric_review} of a solution to the wave equation \eqref{wave_tube}. However, while the integral representation is satisfied by any solution to the associated wave equation, it does not \emph{provide} a solution. That is, one cannot prescribe arbitrary boundary values on $\Gamma$ and then arrive at a solution. The reason is that the worldtube is a timelike boundary, which means that field data on it can propagate forward in time and interfere with the data at a later time. However, by applying the wave operator $E_{\alpha\beta}$ onto equation \eqref{all_orders_Kirchoff}, we see that the Kirchoff representation of $h$ is guaranteed to satisfy the wave equation at each point $x\in\Omega$. In other words, the problem arises not in satisfying the wave equation in a pointwise sense, but in simultaneously satisfying the boundary conditions. However, since the tube is chosen to lie in the buffer region, these boundary conditions can be supplied by the buffer-region expansion. This can presumably be accomplished in a variety of ways, two of which I will discuss presently. Note that since the buffer-region expansion has been made to satisfy the Lorenz gauge to some order in $\rad$, using it as boundary data will enforce the Lorenz gauge in $\Omega$ to the same order.

Now, recall that in almost all the derivations of the gravitational self-force (excluding those in Refs.~\cite{Fukumoto, Gralla_Wald}), the first-order external perturbation was assumed to be that of a point particle. This was justified to some extent by an argument first made by D'Eath \cite{DEath, DEath_paper} and later used by Rosenthal \cite{Eran_field}. The argument is based on the integral Eq.~\eqref{first_order_Kirchoff} and the asymptotically small size of the worldtube. First, note that the directed area element on the worldtube behaves as $\sim\rad^2(-n^{\mu'})$. Also, in constructing the external solution, we formally assume $r\sim 1$ (since the limit is constructed with fixed coordinate values in mind), which means that we can treat the Green's functions and its derivatives as quantities of order unity. Thus, the dominant term in the worldtube integral is determined by the derivative of the $m/r$ term in $\hmn{E}{1}$; using the result from the buffer-region expansion, this yields
\begin{align}
-\rad^2 n^{\mu'}\del{\mu'}\!\!\left[\frac{2m}{r'}(2u_{\alpha'}u_{\beta'} +g_{\alpha'\beta'})\right]\!\!\bigg|_{r'=\rad}
=2m(2u_{\alpha'}u_{\beta'}+g_{\alpha'\beta'})+\order{\rad}.
\end{align}
Hence, the boundary integral can be written as
\begin{align}\label{external tube approx}
\hmn{E\alpha\beta}{1} &= \frac{1}{2\pi}\int\limits_{\Gamma} mG_{\alpha\beta}{}^{\gamma'\delta'}(2u_{\alpha'}u_{\beta'} +g_{\alpha'\beta'})dt'd\Omega' +\hmn{\Sigma\alpha\beta}{1}+\order{\rad},
\end{align}
where $\hmn{\Sigma\alpha\beta}{1}$ is the contribution from the initial data surface $\Sigma$. Expanding the Green's function on the worldtube about the worldline $\gamma$, this becomes
\begin{align}\label{external point particle}
\hmn{E\alpha\beta}{1} &= \int\limits_\gamma 2mG_{\alpha\beta\bar\alpha\bar\beta}(2u^{\bar\alpha}u^{\bar\beta} +g^{\bar\alpha\bar\beta})d\bar t +\hmn{\Sigma\alpha\beta}{1}+\order{\rad},
\end{align}
where the barred coordinates correspond to points on the worldline, and $\bar t$ is proper time, running from $\bar t=0$ to $\bar t\sim 1/\e$. Equation \eqref{external point particle} is the solution to the wave equation with a point particle source---except for the corrections of order $\rad$. It can be put in the more usual form of Eq.~\eqref{1st_funct} by using the identity \eqref{Green3}.

In the original derivation presented by D'Eath \cite{DEath,DEath_paper}, $\rad$ was set to zero with no explicit justification. In Rosenthal's later derivations \cite{Eran_field}, this step was justified based on the notion that we are interested in the limit in which the small body shrinks to a point. However, if the size of the body vanishes, then so too does its mass, in which case there is no perturbation at all; and at second order, setting $\rad$ to zero would create a divergent solution. Hence, discarding the order-$\rad$ corrections based on this argument is not justified. We could also argue that the order-$\rad$ terms must be discarded because the external solution cannot depend on the arbitrary radius of the tube. However, this second argument is also specious: One could just as easily express Eq.~\eqref{external tube approx} as an integral over \emph{any} curve in the interior of $\Gamma$, rather than the central curve $\gamma$. But if one did so, then one would introduce mass dipole terms into the metric, and an explicit calculation of the error terms would show that they do \emph{not} vanish. In some sense, this correctly implies that the choice of worldline at leading order is inconsequential, since any choice within the worldtube results only in the introduction of a mass dipole, which is a second-order term, and the self-force will by definition set the resulting mass dipole to zero. However, this resolution becomes murky when we consider that the size of the tube must be left arbitrary to achieve a valid solution, and the mass dipole in the buffer region calculation is precisely order $\e^2$, rather than order $\e\rad$.

Instead, I present here an alternative argument to justify D'Eath's conclusion: Suppose we take our buffer region expansion of $\hmn{E}{1} $ to be valid everywhere in the interior of $\Gamma$ (in $\man_E$), rather than just in the buffer region. This is a meaningful supposition in a distributional sense, since the $1/r$ singularity in $\hmn{E}{1}$ is locally integrable even at $\gamma$. Note that the extension of the buffer-region expansion is not intended to provide an accurate or meaningful approximation in the interior; it is used only as a means of determining the field in the exterior. I can do this because the field values in $\Omega$ are entirely determined by the field values on $\Gamma$, so using the buffer-region expansion in the interior of $\Gamma$ leaves the field values in $\Omega$ unaltered. Now, given the extension of the buffer-region expansion, it follows from Stokes' law that the integral over $\Gamma$ in Eq.~\eqref{first_order_Kirchoff} can be replaced by a volume integral over the interior of the tube, plus two surface integrals over the ``caps" $\mathcal{J}_{cap}$ and $\Sigma_{cap}$, which fill the ``holes" in $\mathcal{J}$ and $\Sigma$, respectively, where they intersect $\Gamma$. Schematically, we can write Stokes' law as $\int_{\text{Int}(\Gamma)}=\int_{\mathcal{J}_{cap}} +\int_{\Sigma_{cap}}-\int_{\Gamma}$, where $\text{Int}(\Gamma)$ is the interior of $\Gamma$; this is valid as a distributional identity in this case.\footnote{Note that the ``interior" here means the region bounded by $\Gamma\cup\Sigma_{cap}\cup\mathcal{J}_{cap}$. $\text{Int}(\Gamma)$ does not refer to the set of interior points in the point-set defined by $\Gamma$.} The minus sign in front of the integral over $\Gamma$ accounts for the fact that the directed surface element in Eq.~\eqref{first_order_Kirchoff} points \emph{into} the tube. Because $\mathcal{J}_{cap}$ does not lie in the past of any point in $\Omega$, it does not contribute to the perturbation at $x\in\Omega$. Hence, we can rewrite Eq.~\eqref{first_order_Kirchoff} as 
\begin{align}
\hmn{E\alpha\beta}{1} &= -\frac{1}{4\pi}\!\!\!\int\limits_{\text{Int}(\Gamma)}\!\!\! \del{\mu'}\Big(G_{\alpha\beta}{}^{\alpha'\beta'}\nabla^{\mu'}\hmn{E\alpha'\beta'}{1} -\hmn{E\alpha'\beta'}{1}\nabla^{\mu'}G_{\alpha\beta}{}^{\alpha'\beta'}\Big)dV' +\hmn{\bar\Sigma\alpha\beta}{1}\nonumber\\
&=-\frac{1}{4\pi}\!\!\!\int\limits_{\text{Int}(\Gamma)}\!\!\! \Big(G_{\alpha\beta}{}^{\alpha'\beta'}E_{\alpha'\beta'}[\hmn{E}{1}] -\hmn{E\alpha'\beta'}{1}E^{\alpha'\beta'}[G_{\alpha\beta}]\Big)dV' +\hmn{\bar\Sigma\alpha\beta}{1},
\end{align}
where $\hmn{\bar\Sigma\alpha\beta}{1}$ is the contribution from the spatial surface $\bar\Sigma\equiv\Sigma\cup\Sigma_{cap}$, and $E^{\alpha'\beta'}[G_{\alpha\beta}]$ denotes the action of the wave-operator on $G_{\alpha\beta}{}^{\gamma'\delta'}$. Now note that $E^{\alpha'\beta'}[G_{\alpha\beta}]\propto\delta(x,x')$; since $x\notin\text{Int}(\Gamma)$, this term integrates to zero. Next note that $E_{\alpha'\beta'}[\hmn{E}{1}]$ vanishes everywhere except at $\gamma$. This means that the field at $x$ can be written as
\begin{align}
\hmn{E\alpha\beta}{1} &= \frac{-1}{4\pi}\!\lim_{\rad\to 0}\!\!\!\int\limits_{\text{Int}(\Gamma)} \!\!\!\!\!G_{\alpha\beta}{}^{\alpha'\beta'}E_{\alpha'\beta'}[\hmn{E}{1}]dV'+\hmn{\bar\Sigma\alpha\beta}{1}.
\end{align}
Making use of the fact that $E_{\alpha\beta}[\hmn{E}{1}] = \partial^c\partial_c(1/r)\hmn{E\alpha\beta}{1,-1}+\order{r^{-2}}$, along with the identity $\partial^c\partial_c(1/r)=-4\pi\delta^3(x^a)$, where $\delta^3$ is a coordinate delta function in Fermi coordinates, we arrive at the desired result
\begin{equation}
\hmn{E\alpha\beta}{1} = 2m\int_\gamma G_{\alpha\beta\bar\alpha\bar\beta}(2u^{\bar\alpha}u^{\bar\beta} +g^{\bar\alpha\bar\beta})d\bar t+\hmn{\bar\Sigma\alpha\beta}{1}.
\end{equation}
Thus, simply neglecting the $\order{\rad}$ terms in Eq.~\eqref{external point particle} yields the correct result, and in the region $\Omega$, the leading-order perturbation produced by the asymptotically small body is identical to the field produced by a point particle.

Gralla and Wald \cite{Gralla_Wald} have provided an alternative derivation of the same result, using distributional methods to prove that the distributional source for the linearized Einstein equation must be that of a point particle in order for the solution to diverge as $1/r$. One can understand this by considering that the most divergent term in the linearized Einstein tensor is a Laplacian acting on the perturbation, and the Laplacian of $1/r$ is a flat-space delta function; the less divergent corrections are due to the curvature of the background, which distorts the flat-space distribution into a covariant curved-spacetime distribution.

At second order, the above method can be used to simplify Eq.~\eqref{second_order_Kirchoff} by replacing at least part of the integral over $\Gamma$ with an integral over $\gamma$. I will not pursue this simplification here, however. Instead, I will present an alternative means of determining the metric perturbation. This method is based on a direct calculation of the boundary integral in Eq.~\eqref{all_orders_Kirchoff}. As such, it is somewhat similar in spirit to the Direct Integration of the Relaxed Field Equations (DIRE) used by Will et al. in post-Newtonian theory \cite{DIRE}. While the method used above relied on $E_{\mu\nu}[h]$ being well defined as a distribution, a direct integration of the boundary integral can be performed, in principle, regardless of the behavior of $h$ in the buffer region. Hence, it might be used at any order in perturbation theory.

The method of direct integration proceeds as follows. As noted above, the Kirchoff representation of the solution is guaranteed to satisfy the wave equation at all points in $\Omega$, but it provides a valid solution only if, in addition, it agrees with the data on the boundary $\partial\Omega$. Thus, the Kirchoff representation is guaranteed to be a $C^1$ solution in $\bar\Omega$ if it satisfies the consistency conditions
\begin{equation}
\begin{split}
\lim_{x\to x'}h_{\alpha\beta} & = h_{\alpha'\beta'} \\
\lim_{x\to x'}n^\mu\del{\mu}h_{\alpha\beta} & = n^{\mu'}\del{\mu'}h_{\alpha'\beta'}
\end{split}\quad
\text{for } x'\in\Gamma.
\end{equation}
However, these conditions allow $h$ to contain a term such as $(r-\rad)^2\ln(r-\rad)$; both the term itself and its first derivative vanish in the limit $r\to\rad$, but the second derivative does not. Since we seek a solution that is smooth and independent of $\rad$, I demand that $h$ satisfy the following, stronger condition: Since the radius $\rad$ of the tube is small, the boundary data $h'$ can be expressed as an expansion in powers of $\rad$ and $\e$---this is the buffer-region expansion. If $x$ is near the worldtube, then $r\sim\rad$, meaning that $h_{\alpha\beta}$ can similarly be expressed as an expansion in powers of $r$ and $\e$. Recalling that $\expand_s(f(s))$ denotes an expansion of $f$ for small $s$, I write the expansion of the boundary values as $\expand_\e(\expand_\rad(h'))$, and I write the expansion of the integral representation of the solution in $\Omega$ as $\expand_\e(\expand_r(h))$. I demand that these expansions are identical:
\begin{equation}\label{consistency}
\expand_\e(\expand_\rad(h'))\big|_{\rad=r}=\expand_\e(\expand_r(h)).
\end{equation}
Hence, by expanding the integral representation of the perturbation near the worldtube and insisting that the result is consistent with the boundary data provided by the buffer-region expansion, all the free functions in the buffer region expansion will be determined. 

Since the equation of motion depends only on first-order terms, for the purposes of this dissertation I will limit the expansion just described to first order. The expansion is performed only in the buffer region, meaning that it provides an explicit expression for the perturbation only in that region. However, by imposing the consistency condition, the boundary data on $\Gamma$ can be determined to any desired order of accuracy in $\rad$; using this boundary data, the solution in $\Omega$ will then be determined to the same order of accuracy. A similar procedure could be adopted at second order and above. At those orders, the expansion of the boundary integral would yield $\rad$-dependent terms that would be grouped with the volume integral over $\Omega$; this combination would yield an approximation to the inhomogenous part of the solution. The homogenous part of the solution would be dealt with in the same manner as the first-order perturbation.

%%%%%%%%%%%
\section{The boundary integral}\label{integral_expansion1}
%%%%%%%%%%%

\begin{figure}[tb]
\begin{center}
\includegraphics{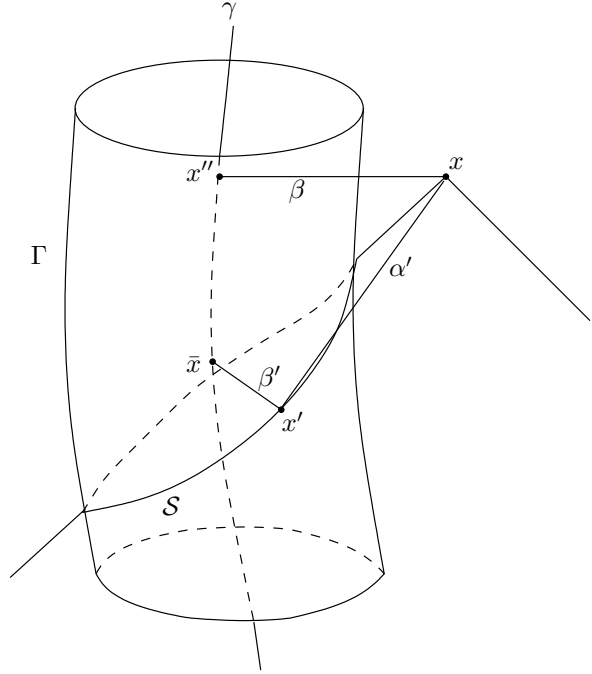}
\end{center}
\caption[Points near the worldtube $\Gamma$]{The two-dimensional hypersurface $\mathcal{S}$ is defined by the intersection of the worldtube $\Gamma$ with the past light cone of the point $x$. $x$ is linked to a point $x'\in\mathcal{S}$ by a null geodesic $\alpha'$. $x$ and $x'$ are separately linked to points $x''=\gamma(t)$ and $\bar{x}=\gamma(t')$ by spacelike geodesics $\beta$ and $\beta'$, each of which is perpendicular to $\gamma$.} 
\label{tube}
\end{figure}

Since the calculation in this section is intended primarily as a proof of principle, rather than calculating $\hmn{E}{1}$ I will calculate its approximation $\hmn{}{1}$; in other words, I will consistently neglect acceleration terms. Hence I take the boundary data on the tube to be defined by $h=\frac{\e}{r}\hmn{}{1,-1}+\e\hmn{}{1,0}+\e r\hmn{}{1,1}$, and the field outside the tube to be the expansion of 
\begin{align}
h_{\alpha\beta}&=\frac{1}{4\pi}\oint\limits_{\partial\Omega}\! \Big(G_{\alpha\beta}{}^{\gamma'\delta'}\hmn{\gamma'\delta';\mu'}{1}
-\hmn{\gamma'\delta'}{1}G\indices{_{\alpha\beta}^{\gamma'\delta'}_{;\mu'}}\Big) dS^{\mu'} +\order{\e^2}
\end{align}
to order $\e r$. Since the volume integral contributes only $\order{\e^2,r^2}$ terms, it is neglected here.

Two parts of the boundary lie within the causal past of $x$: the spatial hypersurface $\Sigma$ and the worldtube $\Gamma$. The contribution to the field from the data on $\Sigma$ is given by
\begin{equation}\label{Sigma contribution}
\hmn{\Sigma\alpha\beta}{1}=\frac{1}{4\pi}\int\limits_{\Sigma}\! \Big(G_{\alpha\beta}{}^{\gamma'\delta'}\hmn{\gamma'\delta';\mu'}{1}
-\hmn{\gamma'\delta'}{1}G\indices{_{\alpha\beta}^{\gamma'\delta'}_{;\mu'}}\Big) dS^{\mu'},
\end{equation}
where the data $\hmn{\gamma'\delta'}{1}$ is constrained to satisfy the Lorenz gauge and merge smoothly with the buffer region expansion. I assume that $\hmn{\Sigma}{1}$ can be expanded in a regular power series in $r$,
\begin{equation}
\hmn{\Sigma\alpha\beta}{1}=\sum_{m\ge 0}r^m\hmn{\Sigma\alpha\beta}{1,\emph{m}},
\end{equation}
and that each $\hmn{\Sigma\alpha\beta}{1,\emph{m}}$ can be decomposed into irreducible STF pieces. Because this data can only contribute to the homogenous, free functions in the buffer region expansion, we can infer the nonzero pieces of $\hmn{\Sigma\alpha\beta}{1}$ from that expansion:
\begin{align}
\hmn{\Sigma tt}{1,0}&=\A{\Sigma}{1,0}, \\
\hmn{\Sigma ta}{1,0}&=\C{\Sigma a}{1,0}, \\
\hmn{\Sigma ab}{1,0}&=\delta_{ab}\K{\Sigma}{1,0}+\H{\Sigma ab}{1,0},\\
\hmn{\Sigma tt}{1,1}&=\A{\Sigma i}{1,1}n^i, \\
\hmn{\Sigma ta}{1,1}&=\B{\Sigma}{1,1}n_a+\C{\Sigma ai}{1,1}n^i +\epsilon_{ai}{}^j\D{\Sigma j}{1,1}n^i,\\
\hmn{\Sigma ab}{1,1}&=\delta_{ab}\K{\Sigma i}{1,1}n^i+\H{\Sigma abi}{1,1}n^i +\epsilon\indices{_i^j_{(a}}\I{\Sigma b)j}{1,1}n^i +\F{\Sigma\langle a}{1,1}n^{}_{b\rangle}.
\end{align}

Now consider the integration over $\Gamma$. In the buffer region, I define $\zeta(\e)$ such that $r\sim\rad\sim\zeta(\e)$. Since the function $\zeta(\e)$ is arbitrary, except that it must vanish in the limit $\e\to0$, we can use $\e$ and $\zeta$ as independent expansion parameters. The volume element on $\Gamma$ is given by $dS_{\mu'}=-n_{\mu'}N(x')\rad^2dt'd\Omega'$, where $N(x)=1+\tfrac{1}{3}\etide_{cd}(t)x^{cd}+\order{\zeta^3,\e}$, and $t'$, $\rad$, and $\theta'^A$ are Fermi coordinates based at $\gamma$. The boundary data is constructed from 
\begin{equation}
\hmn{\gamma'\delta'}{1}=\frac{1}{\rad}\hmn{\gamma'\delta'}{1,-1} +\hmn{\gamma'\delta'}{1,0} +\rad\hmn{\gamma'\delta'}{1,1} +\order{\zeta^2},
\end{equation}
where $\hmn{\gamma'\delta'}{1,-1}$, $\hmn{\gamma'\delta'}{1,0}$, and $\hmn{\gamma'\delta'}{1,1}$ are obtained by setting the acceleration to zero in Eqs.~\eqref{h1n1}, \eqref{h10}, and \eqref{h11}.

The integral over $\Gamma$ can be divided into two regions: the convex normal neighbourhood $\mathcal{N}$ of $x$---consisting of all the points that are connected to $x$ by unique geodesics---and the complement of the convex normal neighbourhood. In $\mathcal{N}$, the Green's function admits the Hadamard decomposition \cite{Eric_review}
\begin{equation}
G_{\alpha\beta}{}^{\gamma'\delta'}= U_{\alpha\beta}{}^{\gamma'\delta'}\delta_+(\sigma) +V_{\alpha\beta}{}^{\gamma'\delta'}\theta_+(-\sigma),
\end{equation}
where $\sigma(x,x')$ is Synge's world function, which is equal to one-half the squared geodesic distance between $x$ and $x'$. Derivatives of this biscalar will be denoted by, e.g., $\sigma_\mu\equiv\sigma_{;\mu}$. The delta function $\delta_+(\sigma(x,x'))$ has support on the past light cone of $x$, while the Heaviside function $\theta_+(-\sigma(x,x'))$ has support within the past light cone.

Substituting these expressions into the boundary integral, we find that it can be broken into several pieces:
\begin{align}
4\pi h_{\alpha\beta} &= \int\limits_{\Gamma\cap\mathcal{N}}\!\! \Big[\h^{\text{tail}}_{\alpha\beta}\theta_+(-\sigma) +\h^{\text{dir}1}_{\alpha\beta}\delta_+(\sigma) +\h^{\mathrm{dir}2}_{\alpha\beta}\delta'_+(\sigma)\Big]Ndt'd\Omega'\nonumber\\
&\quad+\!\!\!\!\!\!\!\!\!\int\limits_{\ (\Gamma\setminus\mathcal{N})\cap\past}\!\!\!\!\!\!\!\!\!\h^{\text{tail}}_{\alpha\beta}Ndt' d\Omega' +\hmn{\Sigma\alpha\beta}{1}+\order{\e^2},
\end{align}
where $\past$ is the past of $x$, and $\delta'$ is the derivative of the delta function. Inside the normal neighbourhood, the terms in the integrand are given by
\begin{align}
\h^{\text{tail}}_{\alpha\beta} &= \Big(\hmn{\gamma'\delta'}{1,-1}-\rad\del{n'}\hmn{\gamma'\delta'}{1,-1}\Big) V\indices{_{\alpha\beta}^{\gamma'\delta'}} +\rad\hmn{\gamma'\delta'}{1,-1} \del{n'}V\indices{_{\alpha\beta}^{\gamma'\delta'}}+\order{\zeta^2},\\
\h^{\text{dir}1}_{\alpha\beta} &= \Big(\hmn{\gamma'\delta'}{1,-1}\!\!-\!\rad\del{n'}\hmn{\gamma'\delta'}{1,-1}\!\! -\!\rad^2\del{n'}\hmn{\gamma'\delta'}{1,0}\!\! -\!\rad^2\hmn{\gamma'\delta'}{1,1}\Big) U\indices{_{\alpha\beta}^{\gamma'\delta'}}\nonumber\\
&\quad+\Big(\rad\hmn{\gamma'\delta'}{1,-1} +\rad^2\hmn{\gamma'\delta'}{1,0}\Big)\del{n'} U\indices{_{\alpha\beta}^{\gamma'\delta'}}-\rad \hmn{\gamma'\delta'}{1,-1}V\indices{_{\alpha\beta}^{\gamma'\delta'}}\sigma_{\mu'}n^{\mu'} +\order{\zeta^3},\\
\h^{\text{dir}2}_{\alpha\beta} &= \Big(\rad\hmn{\gamma'\delta'}{1,-1}+\rad^2\hmn{\gamma'\delta'}{1,0} +\rad^3\hmn{\gamma'\delta'}{1,1}\Big)U\indices{_{\alpha\beta}^{\gamma'\delta'}} \sigma_{\mu'}n^{\mu'}+\order{\zeta^5},
\end{align}
where $\del{n'}\equiv n'^\alpha\del{\alpha'}$. The ``direct" and ``tail" titles should be self-explanatory. Outside the normal neighbourhood, the term in the integrand is
\begin{align}
\h^{\text{tail}}_{\alpha\beta} &= \Big(\hmn{\gamma'\delta'}{1,-1}-\rad\del{n'}\hmn{\gamma'\delta'}{1,-1}\Big) G\indices{_{\alpha\beta}^{\gamma'\delta'}}+\rad\hmn{\gamma'\delta'}{1,-1} \del{n'}G\indices{_{\alpha\beta}^{\gamma'\delta'}}+\order{\zeta^2}.
\end{align}

These expressions are completely general; they can be simplified by making use of the fact that $\del{n'}\hmn{\gamma'\delta'}{1,-1} = \order{\zeta^2,\zeta\e} = \del{n'}\hmn{\gamma'\delta'}{1,0}$. Note that $\h^{\text{tail}}_{\alpha\beta}$, $\h^{\text{dir}1}_{\alpha\beta}$, and $\h^{\text{dir}2}_{\alpha\beta}$ are scalars at $x'$ and rank-two tensors at $x$. Also note that we require the final answer to be accurate up to errors of $\order{\zeta^2}$, in order to determine the free functions in $\hmn{}{1,1}$; each of the above expansions is performed to an order sufficient to meet this requirement, given that $\delta_+(\sigma)\sim 1/\zeta^2$ and $\delta'_+(\sigma)\sim 1/\zeta^4$.

It is convenient to adopt $\sigma$ as an integration variable, which can be done using the transformation $dt'=\displaystyle\frac{d\sigma}{\r}$, where $\r\equiv\sigma_{\alpha'}(x,x')u^{\alpha'}$ can be thought of as a measure of the luminosity distance from $x$ to $x'$. Note that the four-velocity at $x'$ is defined as the tangent to a curve of constant $r$ and $\theta^A$: that is, $u^{\alpha'}=\frac{\partial x^{\alpha'}}{\partial t'}\big|_\Gamma$. Since $t'$ is the proper time on the worldline, rather than the proper time on the generators of the worldtube, this four-velocity is not normalized. (This implies that $\r$ is not an affine parameter on the geodesic connecting $x$ to $x'$.)

After performing this change of variables, we eliminate the $\delta'$ term in the boundary integral by using the identity
\begin{equation}
\int\limits_{\Gamma\cap\mathcal{N}}\!\!\!\h^{\text{dir2}}_{\alpha\beta} N\delta'(\sigma)dt'd\Omega' =-\oint\limits_{\mathcal{S}}\frac{1}{\r} \partial_{t'}\bigg(\frac{N}{\r}\h^{\text{dir2}}_{\alpha\beta}\bigg)d\Omega'.
\end{equation}
Here $\mathcal{S}$ is the intersection of the past light cone of $x$ with the worldtube. For simplicity, I assume that the normal neighbourhood of $x$ is sufficiently large for $\mathcal{S}$ to be well defined and for the intersection $\mathcal{S}\cap\mathcal{N}$ to be closed, such that it has the topology of a sphere. This also requires $x$ to be sufficiently late in time to prevent an intersection of $\mathcal{S}$ with $\Sigma$. (If $x$ is not sufficiently late in time, then $\mathcal{S}$ will be ``cut off" where it intersects $\Sigma$.)

We can now express $h$ as
\begin{align}
h_{\alpha\beta}&=\frac{1}{4\pi}\oint\limits_\mathcal{S} \h^{\text{dir}}_{\alpha\beta}Nd\Omega' +\frac{1}{4\pi}\!\!\!\int\limits_{\Gamma\cap\past}\!\!\! \h^{\text{tail}}_{\alpha\beta}Nd\Omega'dt' +\hmn{\Sigma\alpha\beta}{1}+\order{\e^2},
\end{align}
where
\begin{equation}
\h^{\text{dir}}_{\alpha\beta} = \frac{1}{\r}\h^{\text{dir}1}_{\alpha\beta} -\frac{1}{N\r}\partial_{t'}\bigg(\frac{\h^{\text{dir}2}_{\alpha\beta}}{\r}\bigg) +\order{\zeta^2,\e}.
\end{equation}
The metric perturbation has three types of contributions: the ``direct" type arising from data on the light cone; the ``tail" part arising from the interior of the light cone; and the contribution from the initial data surface.

The ``direct" and ``tail" contributions are calculated explicitly in Appendix~\ref{boundary_integral}. The result for the direct contribution is
\begin{align}\label{direct terms}
\frac{1}{4\pi}\oint\limits_\mathcal{S}\h^{\text{dir}}_{\alpha\beta}Nd\Omega' &= \frac{2m}{r}\left(t_\alpha t_\beta + \delta_{ab}x^a_\alpha x^b_\beta\right) +\tfrac{5}{3}mr\etide_{ij}\nhat^{ij}t_\alpha t_\beta\nonumber\\
&\quad +4mr\left(\etide_{bi}n^i+\tfrac{1}{3}\epsilon_{bij}\btide^j_k\nhat^{ik}\right) t_{(\alpha}x^b_{\beta)}\nonumber\\
&\quad +\tfrac{1}{9}mr\Big[12\etide_{i\langle a}\nhat_{b\rangle}^i-5\delta_{ab}\etide_{ij}\nhat^{ij} +\left(12\rad^2/r^2-2\right)\etide_{ab}\Big]x^a_\alpha x^b_\beta\nonumber\\
&\quad+\order{\zeta^2,\e}.
\end{align}
Note that the $1/r$ term in this result agrees with the $1/r$ term that was used for boundary data.

The result for the tail terms is
\begin{align}\label{tail terms}
\frac{1}{4\pi}\!\!\!\int\limits_{\Gamma\cap\past}\!\!\! \h^{\text{tail}}_{\alpha\beta}Nd\Omega'dt' & =  \Big(\tail_{\Gamma 00}+\tail_{\Gamma 00c}x^c\Big)t_\alpha t_\beta+2\Big(\tail_{\Gamma 0b}+\tail_{\Gamma 0bc}x^c\Big)t_{(\alpha}x^b_{\beta)}\nonumber\\
&\quad+\Big(\tail_{\Gamma ab}+\tail_{\Gamma abc}x^c\Big)x^a_\alpha x^b_\beta  -4m\bigg(r+\frac{\rad^2}{3r}\bigg)\etide_{ab}x^a_\alpha x^b_\beta\nonumber\\
&\quad +\order{\zeta^2,\e}.
\end{align}
Note that the $\rad$-dependent term in this equation exactly cancels the $\rad$-dependent term in Eq.~\eqref{direct terms}. In addition, note that this expansion is identical to the one in Sec.~\ref{buffer_expansion} only after explicit factors of the acceleration are set to zero. This means, in effect, that when comparing individual components of our expansion here to those in our previous expansion in the buffer region, we should replace the covariant derivative in the Fermi-coordinate expression for $\tail_{\Gamma\alpha''\beta''\gamma''}$ with a partial derivative.

%%%%%%%%%%%%%%
\section{Identification of unknown functions}
%%%%%%%%%%%%%
We now combine the results of Eqs.~\eqref{Sigma contribution}, \eqref{direct terms}, and \eqref{tail terms} to arrive at an expansion of the form
\begin{equation}
h_{\alpha\beta}=\frac{1}{r}\hmn{\alpha\beta}{1,-1}+\hmn{\alpha\beta}{1,0} +r\hmn{\alpha\beta}{1,1}+\order{\zeta^2,\e},
\end{equation}
which we will identify with the expansion defined by Eqs.~\eqref{h1n1}, \eqref{h10}, and \eqref{h11}. After defining the tail terms 
\begin{align}
\tail_{IJ} &=\tail_{\Gamma IJ}+\hmn{\Sigma IJ}{1,0}, \\ \tail_{IJc}n^c &=\tail_{\Gamma IJc}n^c+\hmn{\Sigma IJ}{1,1},
\end{align}
and decomposing the results into STF pieces, we find
\begin{align}\label{h1n1 tail}
\hmn{\alpha\beta}{1,-1} &= \frac{2m}{r}\left(t_\alpha t_\beta + \delta_{ab}x^a_\alpha x^b_\beta\right),\\
\hmn{\alpha\beta}{1,0} &= \tail_{00}t_\alpha t_\beta +2\tail_{0b}t_{(\alpha}x^b_{\beta)}  +\left(\tail_{\av{ab}} +\tfrac{1}{3}\delta_{ab}\delta^{ij}\tail_{ij}\right) x^a_{(\alpha}x^b_{\beta)},\label{h10 tail}
\end{align}
and
\begin{align}\label{h11 tail}
\hmn{tt}{1,1} &= \tfrac{5}{3}m\etide_{ij}\nhat^{ij}+\tail_{00i}n^i, \\
\hmn{ta}{1,1} &= 2m\etide_{ai}n^i+\tfrac{2}{3}m\epsilon_{aij}\btide^j_k\nhat^{ik}+\tail_{0\langle ac\rangle}n^c\nonumber\\
&\quad+\tfrac{1}{3}\tail_{0ij}\delta^{ij}n_a +\tfrac{1}{2}\epsilon_{aci}\epsilon^{ijk}\tail_{0jk}n^c,\\ 
\hmn{ab}{1,1} &= \tfrac{4}{3}m\etide_{i\langle a}\nhat_{b\rangle}^i-\tfrac{5}{9}m\delta_{ab}\etide_{ij}\nhat^{ij} -\tfrac{38}{9}m\etide_{ab}\nonumber\\
&\quad+\mathop{\STF}_{ab}\left[\tfrac{2}{3}\epsilon_{iac} \mathop{\STF}_{ib}\left(\tail_{\langle ij\rangle d}\epsilon_b{}^{jd}\right)n^c+\tfrac{3}{5}\delta^{ij}\tail_{\langle ib\rangle j}n_a\right]\nonumber\\
&\quad+\tfrac{1}{3}\delta_{ab}\delta^{ij}\tail_{ijc}n^c+\tail_{\langle abc\rangle}n^c.
\end{align}
After setting explicit factors of the acceleration to zero, this expansion agrees with Equations~\eqref{h1n1}, \eqref{h10}, and \eqref{h11}. By comparing the two sets of equations, we identify all the unknown STF tensors in the buffer region expansion. The results of this identification are listed in Table~\ref{STF wrt tail}. Note that these identifications are modulo the acceleration that appears in the covariant derivative in $\tail_{\Gamma\alpha\beta\gamma}$.

For this solution to agree with the results of the buffer region expansion, it must satisfy the relationships given in Eqs.~\eqref{B11} and \eqref{F11}. In terms of the tail integral, these relationships read 
\begin{align}
\delta^{ij}\tail_{\langle ai\rangle j} &= \tfrac{1}{6}\delta^{ij}\tail_{ija} -\tfrac{1}{2}\tail_{00a} +\partial_t\tail_{0a},\\
\delta^{ij}\tail_{0ij} &=\tfrac{1}{2}\partial_t\Big(\tail_{00} +\delta^{ij}\tail_{ij}\Big),
\end{align}
where it is understood that the equations hold only for $a=0$. By using the Green's functions identities \eqref{Green1}, \eqref{Green2}, and \eqref{Green3}, and neglecting acceleration terms, one can easily show that the tail terms  $\tail_{\Gamma}$ satisfy these relationships. Hence, we must constrain the initial data terms $\hmn{\Sigma}{1}$ to independently satisfy them, which implies 
\begin{align}
\B{\Sigma}{1,1} &= \tfrac{1}{6}\partial_t\left(\A{\Sigma}{1,0} +3\K{\Sigma}{1,0}\right), \\
\F{\Sigma a}{1,1} &= \tfrac{3}{10}\left(\K{\Sigma a}{1,1} -\A{\Sigma a}{1,1} + \partial_t\C{\Sigma a}{1,0}\right).
\end{align}

The reader should take note of two important facts about the metric perturbation derived here. First, as we expected, the expansion displayed above is identical to the expansion of the point particle solution in the neighbourhood of the worldline. Second, and again as we expected, the expansion is completely determined by the most singular, $\e/r$, term in the metric. Although the nonsingular terms are required to maintain consistency at the boundary, one can derive all of them simply by using the $1/r$ term as boundary data.

Now, the principal purpose of the calculation of the boundary integral was to express the equations of motion in terms of the body's past history. The correction to the body's mass, given in Eq.~\eqref{mdot}, can now be written as
\begin{align}\label{mdot_tail}
\delta m(t)=\delta m(0)+\tfrac{1}{3}m\tail_{00} +\tfrac{5}{18}m\delta^{ab}\tail_{ab}.
\end{align}
In covariant form, this is
\begin{equation}
\delta m(t) = \delta m(0)+\tfrac{1}{18}m\left(5g^{\alpha\beta} +11u^{\alpha}u^{\beta}\right)\tail_{\alpha\beta}.
\end{equation}
This is similar, but not identical to, a result found by Mino, Sasaki, and Tanaka~\cite{Mino_Sasaki_Tanaka}. The source of disagreement between the two results is not clear. It is worth noting that both results appear at one order lower than that given by Thorne and Hartle \cite{Thorne_Hartle}, who chose to eliminate the homogeneous field $h^R$.

The leading-order acceleration of the body, given in Eq.~\eqref{a1}, is
\begin{equation}\label{a1_tail}
\an{1}_a=\tfrac{1}{2}\tail_{00a}-\tail_{0a0}-\tfrac{1}{m}S_i\btide^i_a.
\end{equation}
(Here, again, the right-hand side of this equation is to be evaluated at $a=\an{0}=0$.) In covariant form, this result can be written as
\begin{align}
a^{\alpha} &= -\tfrac{1}{2}\e\left(g^{\alpha\delta}+u^{\alpha}u^{\delta}\right) \left(2\tail_{\delta\beta\gamma}-\tail_{\beta\gamma\delta}\right) u^{\beta}u^{\gamma} +\frac{\e}{2m}R^{\alpha}{}_{\beta\gamma\delta}u^\beta S^{\gamma\delta}+\order{\e^2}
\end{align}
where I have again used the definition $S^{\gamma\delta}\equiv e_c^\gamma e_d^\delta\epsilon^{cdj}S_j$. The spin term is the usual Papapetrou spin force. The tail term is the usual MiSaTaQuWa self-force---except that the tail integral is defined as the sum of an integral over the worldline, cut off at $t=0$, and an integral over an initial data surface. Of course, Eqs.~\eqref{mdot_tail} and \eqref{a1_tail} hold only in the Lorenz gauge.

This concludes what might seem to be the most egregiously lengthy derivation of the self-force yet performed. It is hoped, however, that along with the additional length has come additional insight.

			\chapter{The method of osculating orbits}\label{osculating}
The preceding chapters have resulted in a formal equation of motion for a small body moving in an arbitrary background spacetime. In order to actually make use of that equation of motion to perform gravitational wave astronomy, we require a means of characterizing orbits in particular, experimentally relevant spacetimes. More specifically, since the acceleration is small, we require a useful means of characterizing gradual deviations from geodesic motion. The method I will adopt is a traditional method of Newtonian celestial mechanics: the \emph{method of osculating orbits}, described in Sec.~\ref{osculating_intro}.

\begin{figure}[tb]
\begin{center}
\includegraphics{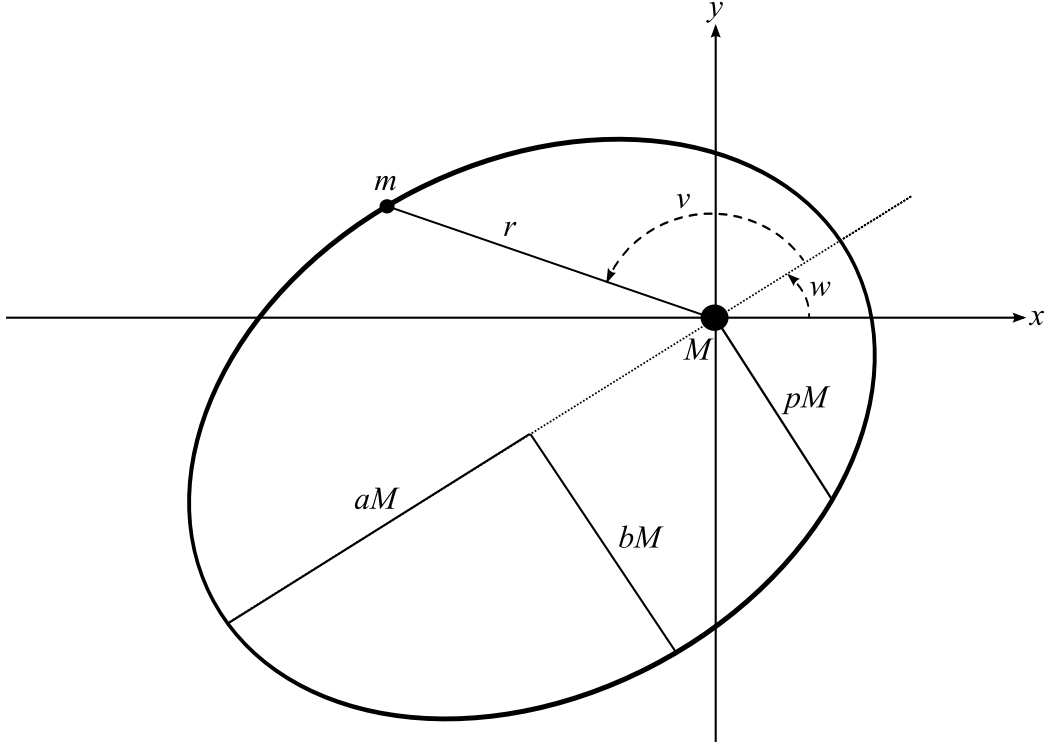}
\end{center}
\caption[An elliptical orbit]{An elliptical orbit of a mass $m$ about a second mass $M$, which lies at a focus of the ellipse. The size and shape of the ellipse are defined by the (dimensionless) semi-latus rectum $p$ and eccentricity $e$ (not shown); alternatively, the size and shape are defined by the (dimensionless) semi-major axis $a$ and semi-minor axis $b$, which are related to $p$ and $e$ by the formulas $e=\sqrt{1-(b/a)^2}$ and $p=a(1-e^2)$. Periapsis, the point of nearest approach, is at a distance $pM/(1+e)$ from $M$; apoapsis, the point of greatest distance, is at a distance $pM/(1-e)$. The argument of periapsis, $w$, defines the orientation of the ellipse relative to the fixed coordinate axes $x$ and $y$. The true anomaly, $v$, defines the angular position relative to the semi-major axis. The position of the mass is specified by the polar coordinates $r$ and $\phi$, where $\phi=v+w$.} 
\label{ellipse}
\end{figure}

Such a method is especially suitable for weakly perturbed orbits, because it allows us to characterize the orbit according to the physically meaningful orbital elements of the geodesic. In Newtonian celestial mechanics restricted to planar motion, the method consists of writing the true orbit as a Keplerian, elliptical orbit with a time-dependent eccentricity $e$, semi-latus rectum $p$, and argument of periapsis $w$, or some equivalent set of orbital elements. (See Fig.~\ref{ellipse}.) Lagrange first derived the correct evolution equations for the orbital elements in the case of a perturbing force that can be written as the gradient of a potential; Gauss later generalized Lagrange's results to the case of an arbitrary perturbing force. In this dissertation, I will begin by discussing the method as applied to an arbitrary worldline in an arbitrary spacetime, assuming only that the geodesics of the spacetime are integrable; I will then specialize to the case of bound, planar geodesics in Schwarzschild spacetime; finally, from my results in Schwarzschild, I re-derive Gauss' perturbation equations. Although the orbits of principal interest for EMRI systems are accelerated orbits in Kerr, rather than accelerated orbits in Schwarzschild, the method of osculating orbits in Schwarzschild has some utility in itself, as will be shown in the next chapter; it also has the benefit of straightforwardly generalizing an historically important method of Newtonian celestial mechanics into the realm of relativistic celestial mechanics.

The notation in this and the following chapter differs from that of preceding chapters. Proper time on the worldline will henceforth be denoted by $\tau$, and the coordinates $(t,r,\theta,\phi)$ will now be global coordinates centered on the large body in an EMRI, rather than on the small body.

\section{The general case}
\label{general case}

I first consider the completely general situation of a point particle 
moving on an arbitrary worldline $\orbit(\lambda)$ parametrized by
$\lambda$. I define the acceleration $a^\alpha$ acting on the particle via the equation of motion  
\begin{equation}\label{affine eq mot}
\ddot z^{\alpha}+\Chr{\alpha}{\beta}{\gamma} 
\dot z^{\beta} \dot z^{\gamma} = a^{\alpha},
\end{equation}
where an overdot indicates an ordinary total derivative with respect to the proper time $\tau$ on the worldline. The normalization condition 
$\dot z^{\alpha}\dot z_{\alpha} = -1$ implies the orthogonality
condition $a^{\alpha}\dot z_{\alpha} = 0$, which will be essential for
later calculations. The relation between $a^{\alpha}$ and the Newtonian
perturbing force is discussed in Sec.~\ref{Newtonian}.   

Using the relations $\dot z^{\alpha} =
\diff{\orbit}{\lambda}\dot\lambda$ and $\ddot z^{\alpha} =
\ddiff{\orbit}{\lambda}\dot\lambda^2+\diff{\orbit}{\lambda}\ddot\lambda$, 
the equation of motion becomes 
\begin{equation}
\ddiff{\orbit}{\lambda}+\Chr{\alpha}{\beta}{\gamma} 
\diff{z^{\beta}}{\lambda}\diff{z^{\gamma}}{\lambda} 
= a^{\alpha}\left(\diff{\tau}{\lambda}\right)^2
+ \kappa(\lambda)\diff{\orbit}{\lambda},
\label{eq mot}
\end{equation}
where $\kappa=-\ddot\lambda/\dot\lambda^2$. The first term on the 
right-hand side is due to the force acting on the particle, while the
second term is present whenever $\lambda$ is a non-affine parameter. For future convenience, I note that $\kappa$ can also be written as
\begin{equation}
\kappa=\left(\diff{\tau}{\lambda}\right)^{-1}\diff{}{\lambda}\diff{\tau}{\lambda}.
\end{equation}

My goal is to transform the equation of motion~\eqref{eq mot} into 
evolution equations for a set of orbital elements $I^A$. That is, I
seek a transformation $\{\orbit,\dot z^{\alpha}\} \to I^A$. Letting 
$\geo(I^A,\lambda)$ be a geodesic with orbital elements $I^A$, the
\textit{osculation condition} states the following:  
\begin{eqnarray}
\orbit(\lambda) & = & \geo(I^A(\lambda),\lambda), 
\label{osc 1} \\
\diff{\orbit}{\lambda}(\lambda) & = & 
\pdiff{\geo}{\lambda}(I^A(\lambda),\lambda), 
\label{osc 2}
\end{eqnarray}
where the partial derivative in the second equation holds $I^A$ 
fixed. These two equations assert that at each value of $\lambda$ we
can find a set of orbital elements $I^A(\lambda)$ such that the
geodesic with those elements has the same position and velocity as the 
accelerated orbit. I can freely make this assertion because the
number of orbital elements is equal to the number of degrees of
freedom on the orbit. 

As a consequence of the osculation condition, all relations that are
obtained using only algebraic manipulations of coordinates and
velocities on a geodesic are also valid on the true orbit. However, it  
is important to note that $\kappa$ is altered by the acceleration of 
the worldline, because it involves second derivatives. Hence, an
expression for $\kappa(\lambda)$ that is valid on an osculating
geodesic will not be valid on the tangential accelerated
orbit. Nevertheless, $\ddot\lambda=0$ for an affine parameter
$\lambda$ on both orbits, so affine parameters remain affine.   

Now, combining the osculation condition with the equations of motion 
generates evolution equations for $I^A$. From Eq.~\eqref{osc 1} we
have that $\diff{\orbit}{\lambda}=\diff{\geo}{\lambda}$, which implies 
$\diff{\orbit}{\lambda} = \pdiff{\geo}{\lambda} 
+ \pdiff{\geo}{I^A}\diff{I^A}{\lambda}$, where the index $A$ is summed
over. Comparing this result with Eq.~\eqref{osc 2}, we find 
\begin{equation}
\pdiff{\geo}{I^A}\diff{I^A}{\lambda} = 0.
\label{diffI 1}
\end{equation}
Furthermore, $\geo$ satisfies the geodesic equation
\begin{equation}
\pddiff{\geo}{\lambda}+\Chr{\alpha}{\beta}{\gamma}
\pdiff{z_G^{\beta}}{\lambda}\pdiff{z_G^{\gamma}}{\lambda} 
= \kappa_G(\lambda,I^A(\lambda))\pdiff{\geo}{\lambda},
\end{equation}
where $\kappa_G(\lambda,I^A)$ is the measure of non-affinity of $\lambda$
on the geodesic, given explicitly by
\begin{equation}
\kappa_G=\left(\diff{\tau}{\lambda}\right)^{-1}\pdiff{}{\lambda}\diff{\tau}{\lambda},
\end{equation}
where $\diff{\tau}{\lambda}$ is treated as a function of $\lambda$ and $I^A(\lambda)$. Subtracting this geodesic equation from the equation
of motion~\eqref{eq mot} and using Eq.~\eqref{osc 2} to remove the
Christoffel terms, we obtain  
\begin{equation}
\ddiff{\orbit}{\lambda} = \pddiff{\geo}{\lambda}
+ a^{\alpha}\left(\diff{\tau}{\lambda}\right)^2 
+ \left[\kappa(\lambda)
- \kappa_G(\lambda)\right]\pdiff{\geo}{\lambda}. 
\end{equation}
But differentiating Eq.~\eqref{osc 2} yields
$\ddiff{\orbit}{\lambda}=\pddiff{\geo}{\lambda} 
+\left(\pdiff{}{I^A}\pdiff{\geo}{\lambda}\right)\diff{I^A}{\lambda}$.
Comparing these results, we find   
\begin{eqnarray}
\left(\pdiff{}{I^A}\pdiff{\geo}{\lambda}\right)\diff{I^A}{\lambda} 
& = & a^{\alpha}\left(\diff{\tau}{\lambda}\right)^2 
+ \left[\kappa(\lambda)-\kappa_G(\lambda)\right]
\pdiff{\geo}{\lambda}. 
\end{eqnarray}
By making use of the equations for $\kappa$ and $\kappa_G$, we can rewrite this as
\begin{eqnarray}
\left(\pdiff{}{I^A}\pdiff{\geo}{\lambda}\right)\diff{I^A}{\lambda} 
& = & a^{\alpha}\left(\diff{\tau}{\lambda}\right)^2 
+ \pdiff{\geo}{\lambda}\left(\diff{\tau}{\lambda}\right)^{-1} \diff{I^A}{\lambda}\pdiff{}{I^A}\diff{\tau}{\lambda}
.\label{diffI 2} 
\end{eqnarray}

Equations \eqref{diffI 1} and \eqref{diffI 2} form a closed system
of first-order differential equations for the orbital elements
$I^A$. Two sources of change in the orbital elements are apparent: a 
direct source due to the perturbing force $a^{\alpha}$, and an
indirect source due to the change in the affinity of the
parametrization of the accelerated orbit. If we use the
affine parameter $\lambda=\tau$, then the equations simplify to  
\begin{eqnarray}
\pdiff{\geo}{I^A}\dot I^A & = & 0,
\label{diffI 1b} \\
\pdiff{\dot z_G^{\alpha}}{I^A}\dot I^A & = & a^{\alpha} 
\label{diffI 2b}.
\end{eqnarray}
These equations can be easily inverted to solve for the derivatives 
$\dot I^A$, which is done in Sec.~\ref{evolution}. If a non-affine
parameter $\lambda$ is required in a specific application, one may
easily find $\diff{I^A}{\lambda}$ by multiplying the above equations
by $\diff{\tau}{\lambda}$, which will also be done in
Sec.~\ref{evolution}; alternatively, one could use Eqs.~\eqref{diffI 1} and \eqref{diffI 2} directly. 

%%%%%%%%%%%%%%%%%%%%%%%%%%%%%%%%%%%%%%%%%%%%%%%%%%%%%%%%%%%%%%%%%%%%%%%

\section{Geodesics in Schwarzschild spacetime}
\label{geodesics}

I now focus on the specific case of bound orbits in Schwarzschild
spacetime. The osculating orbits in this case are bound geodesics, 
for which I use the parametrization presented in the text by
Chandrasekhar~\cite{Chandra} and described in detail in
Ref.~\cite{parametrization}. This parametrization is given in 
Schwarzschild coordinates and can be easily derived as follows. 

Because of the spherical symmetry of the Schwarzschild spacetime, we 
can freely set $\theta=\pi/2$. The geodesic equations in a
Schwarzschild spacetime with mass parameter $M$ can be easily solved
for the remaining coordinates to find 
\begin{eqnarray}
\dot t & = & E/f,
\label{tdot}\\
\dot{r}^2 & = & E^2-U_{\rm eff}, 
\label{rdot}\\
\dot\phi & = & \frac{L}{r^2}, 
\label{phidot}
\end{eqnarray}
where $f = 1-2M/r$, $E$ and $L$ are constants equal to energy and
angular momentum per unit mass, respectively, the effective potential
is $U_{\rm eff}=f(1+L^2/r^2)$, and an overdot represents a derivative
with respect to the proper time $\tau$ on the orbit.  

We are interested in bound orbits that oscillate between a minimal
radius $r_1$ and a maximal radius $r_2$; the locations at which these minimal and maximal radii are achieved are respectively referred to as
periapsis and apoapsis. Adapting the tradition of celestial mechanics, 
I define the (dimensionless) semi-latus rectum $p$ and the
eccentricity $e$ such that the turning points are given by 
\begin{eqnarray}
r_1 & = & \frac{pM}{1+e}, \label{periapsis}\\
r_2 & = & \frac{pM}{1-e}, \label{apoapsis}
\end{eqnarray}
where $0 \leq e < 1$. These two constants describe the geometry of the 
orbit, just as in Keplerian orbits: $p$ is a measure of the radial
extension of the orbit, while $e$ is a measure of its deviation from
circularity. These constants can be related to $E$ and $L$ by letting
$\dot{r}=0$ in Eq.~\eqref{rdot}, which leads to 
\begin{eqnarray}
E^2 & = & \frac{(p-2-2e)(p-2+2e)}{p(p-3-e^2)}, \label{E}\\
L^2 & = & \frac{p^2M^2}{p-3-e^2}. \label{L}
\end{eqnarray}

Continuing to exploit the analogy with Keplerian orbits, I introduce
a non-affine parameter $\chi$ that runs from 0 to $2\pi$ over one radial cycle,
such that $r(\chi)$ takes the elliptical form   
\begin{equation}
r(\chi) = \frac{pM}{1+e\cos(\chi-w)},
\label{r}
\end{equation}
where $w$ is the value of $\chi$ at periapsis, referred to as the
argument of periapsis. The radial component of the velocity is hence 
\begin{equation}
r'(\chi) = \frac{pMe\sin(\chi-w)}
{\bigl[ 1 + e\cos(\chi-w) \bigr]^2},
\label{rprime}
\end{equation}
where a prime henceforth indicates a derivative with respect to 
$\chi$. 

From these results we can relate the parameter $\chi$ to the proper 
time $\tau$ using $\diff{\tau}{\chi}=\frac{r'}{\dot r}$, which yields 
\begin{equation}
\diff{\tau}{\chi} = \frac{p^{3/2}M (p-3-e^2)^{1/2}}
{(p-6-2e\cos v)^{1/2}(1+e\cos v)^2},
\label{chidot}
\end{equation}
where I have introduced the variable 
\begin{equation} 
v \equiv \chi - w
\end{equation} 
for brevity; this quantity is analogous to the true anomaly in Keplerian orbits. Along with Eqs.~\eqref{tdot}, \eqref{phidot}, \eqref{E},
and \eqref{L}, this leads to the following parametrizations for
$t(\chi)$ and $\phi(\chi)$: 
{\allowdisplaybreaks\begin{eqnarray}
\phi(\chi) & = & \Phi + \int^{\chi}_w\phi'(\tilde{\chi})d\tilde{\chi}, 
\label{phi}\\
\phi '(\chi) & = & \sqrt{\frac{p}{p-6-2e\cos v}}, 
\label{phiprime}\\
t(\chi) & = & T + \int^{\chi}_wt'(\tilde{\chi})d\tilde{\chi}, 
\label{t}\\
t'(\chi) & = & \frac{p^2M}{(p-2-2e\cos v)(1+e\cos v)^2}
\sqrt{\frac{(p-2-2e)(p-2+2e)}{p-6-2e\cos v}},
\label{tprime}
\end{eqnarray}}
where I have defined the constants $T$ and $\Phi$ as the values of 
$t$ and $\phi$ at periapsis, respectively. 

My parametrization of bound geodesics consists of Eqs.~\eqref{r},
\eqref{rprime}, and \eqref{phi}--\eqref{tprime}. We see that a
geodesic is uniquely specified by the orbital elements
$I^A=\lbrace p,e,w,T,\Phi \rbrace$. The principal elements $p$ and $e$
determine the spatial shape of the orbit and are equivalent to
specifications of energy and angular momentum; they determine the
choice of geodesic. The positional elements $w$ and
$\Phi$ determine the spatial orientation of the
orbit; the final positional element, $T$, specifies the starting point of the particle on the selected geodesic. All together, the specification of the orbital elements is equivalent to the specification of initial values for the
position and velocity of the particle. We need three initial coordinate values (including time)  
for a planar orbit, and we need two initial components of the velocity (three minus
one, by virtue of the normalization condition on the velocity
vector); this counting matches the number of orbital elements.  

% NEW PAR BEGINS
%
I note that my choice of orbital elements is closely related to
Mino's in Ref.~\cite{Mino_expansion1}. When the orbital motion is restricted 
to the equatorial plane of a Kerr black hole, Mino uses the
principal elements $E$ and $L$ and positional elements that are
identical to my $w$, $T$, and $\Phi$. To use $(p,e)$ instead of
$(E,L)$ is mostly a matter of taste; I believe that the set $(p,e)$
is more useful than $(E,L)$ because it gives a simpler
parametrization, and because $p$ and $e$ are geometrically more  
informative. In the following subsection I will deviate more strongly  
from Mino's parameterization: for reasons that will be explained, I
shall avoid directly evolving the elements $T$ and $\Phi$.  
%
% NEW PAR ENDS

All the equations presented in this section remain valid for a
perturbed orbit, with the exception of Eqs.~\eqref{periapsis} and
\eqref{apoapsis}, which lose their meaning. The alteration that I
shall make to account for the perturbation is that in each equation,
the orbital elements will become functions of $\chi$. 

%%%%%%%%%%%%%%%%%%%%%%%%%%%%%%%%%%%%%%%%%%%%%%%%%%%%%%%%%%%%%%%%%%

\section{Evolution equations}
\label{evolution}

If I restrict the perturbing force to lie in the plane of the orbit, 
and assume that the orbit remains bound, then the geodesics described
in the last section form a sufficient set of osculating orbits. Using
my parametrization of these geodesics, along with the results of the 
general analysis in Sec.~\ref{general case}, we can now find evolution 
equations for the orbital elements. Multiplying both sides of
Eq.~\eqref{diffI 1b} by $\diff{\tau}{\chi}$, we find 
\begin{eqnarray}
\pdiff{r}{p}p' + \pdiff{r}{e}e' + \pdiff{r}{w}w'  & = & 0, 
\label{wdot}\\
\pdiff{t}{p}p' + \pdiff{t}{e}e' + \pdiff{t}{w}w'  + T' & = & 0, 
\label{Tdot}\\
\pdiff{\phi}{p}p' + \pdiff{\phi}{e}e' + \pdiff{\phi}{w}w' 
+ \Phi' & = & 0.
\label{Phidot} 
\end{eqnarray}
Similarly, from Eq.~\eqref{diffI 2b} we find
\begin{eqnarray}
\pdiff{\dot t}{p}p' + \pdiff{\dot t}{e}e' + \pdiff{\dot t}{w}w' 
& = & a^t\tau', 
\label{ft}\\
\pdiff{\dot r}{p}p' + \pdiff{\dot r}{e}e' + \pdiff{\dot r}{w}w' 
& = & a^r\tau', 
\label{fr}\\
\pdiff{\dot \phi}{p}p' + \pdiff{\dot \phi}{e}e' 
+ \pdiff{\dot\phi}{w}w' & = & a^{\phi}\tau'. 
\label{fphi}
\end{eqnarray}

% NEW PAR BEGINS
%
The orthogonality condition $a^{\alpha}\dot z_{\alpha}=0$ allows us to
remove one component of Eq.~\eqref{diffI 2b} from the set of
equations; I use this freedom to remove Eq.~\eqref{ft}. The remaining 
equations decouple into a closed system of ordinary differential
equations for $p$, $e$, and $w$ and two auxiliary equations for $T$
and $\Phi$. We shall find that the evolution equations for $p$, $e$,
and $w$ are simple. The equations for $T$ and $\Phi$, however, are
not: Factors such as $\pdiff{t}{p}$ in Eqs.~\eqref{Tdot} and
\eqref{Phidot} introduce elliptic integrals of the form 
$\int_w^\chi\pdiff{t'}{p}(\tilde\chi)d\tilde\chi$ into the expressions
for $T'$ and $\Phi'$. These integrals would have to be evaluated at 
each time-step in a numerical evolution, and they would create an
excessive computational cost. Additionally, the integrals generally
grow linearly with $\chi$, and this produces terms in $T(\chi)$ and
$\Phi(\chi)$ that grow quadratically with $\chi$, as well as terms
that oscillate with a linearly increasing amplitude. Such terms
greatly confuse both numerical and analytical descriptions, and they
are largely an artefact of my parametrization. (This statement applies
also to Mino's parameterization \cite{Mino_expansion1}.) I note that
similar (though less severe) difficulties arise also in the method of 
osculating orbits in Newtonian celestial mechanics; refer for example
to the discussion on pp. 248--250 in the text by Beutler
\cite{Beutler}. In the Newtonian context, alternative orbital elements 
are typically selected so as to overcome these problems. With no
obvious choice of alternative elements in the relativistic context, I
opt instead to directly evolve the coordinates $t$ and $\phi$ rather
than the elements $T$ and $\Phi$.     
%
% NEW PAR ENDS

The phase space thus consists of $\{p,e,w,t,\phi\}$. This choice of 
phase space does not allow an easy separation of perturbative from
geodesic effects in the evolutions of $t$ and $\phi$, nor does
it allow a clean separation of conservative from dissipative effects. 
But it is overwhelmingly more convenient than the alternative choice 
$\{p,e,w,T,\Phi\}$. If $T$ and $\Phi$ are required in an application, 
they may be found as, e.g., $T=t-\int_w^\chi
t'(\tilde{\chi})d\tilde{\chi}$. This may be necessary if initial 
conditions are required on an osculating orbit, or if one wishes to
fully isolate perturbative effects. 

Solving for $w'$ from Eq.~\eqref{wdot}, and noting that
$\pdiff{r}{w}=-r'$, we find 
\begin{equation}
w' =\frac{1}{r'}\left(\pdiff{r}{p}p' + \pdiff{r}{e}e'\right).
\label{wprime_raw} 
\end{equation}
Substituting this into Eqs.~\eqref{ft} and \eqref{fphi}, we can solve
for $p'$ and $e'$ to find 
\begin{eqnarray}
p' & = & \frac{\mathcal{L}_e(\phi)a^r-\mathcal{L}_e(r)a^{\phi}}
{\mathcal{L}_e(\phi)\mathcal{L}_p(r)-\mathcal{L}_e(r)\mathcal{L}_p(\phi)}
\tau', \\
e' & = & \frac{\mathcal{L}_p(r)a^{\phi}-\mathcal{L}_p(\phi)a^r}
{\mathcal{L}_e(\phi)\mathcal{L}_p(r)-\mathcal{L}_e(r)\mathcal{L}_p(\phi)}
\tau',
\end{eqnarray}
where $\mathcal{L}_a(x)\equiv \pdiff{\dot x}{a}
+ \frac{1}{r'}\pdiff{r}{a}\pdiff{\dot x}{w}$. Explicitly, the
results are 
{\allowdisplaybreaks\begin{eqnarray}
p' & = & \frac{2p^{7/2}M^2(p-3-e^2)(p-6-2e\cos v)^{1/2} 
  (p-3-e^2\cos^2v)}{(p-6+2e)(p-6-2e)(1+e\cos v)^4} a^{\phi}\nonumber\\
&&
- \frac{2p^3Me(p-3-e^2)\sin v}{(p-6+2e)(p-6-2e)(1+e\cos v)^2} a^r,
\quad  
\label{pprime} \\
e' & = & \frac{p^{5/2}M^2(p-3-e^2)}{(p-6+2e)(p-6-2e)}\Bigg\lbrace \frac{e(p^2-10p+12+4e^2)}{(p-6-2e\cos v)^{1/2}(1+e\cos v)^4}\nonumber\\
&& +\frac{(p-6-2e^2)
  \left[(p-6-2e\cos v)e\cos v+2(p-3)\right]\cos v}{(p-6-2e\cos v)^{1/2}
  (1+e\cos v)^4}\Bigg\rbrace a^{\phi}
\nonumber\\
&& + \frac{p^2M(p-3-e^2)(p-6-2e^2)\sin v}{(p-6+2e)
  (p-6-2e)(1+e\cos v)^2} a^r, 
\label{eprime} \\
w' & = & \frac{p^{5/2}M^2(p-3-e^2)\sin v}{e(p-6+2e)(p-6-2e)}\Bigg\lbrace\frac{-4e^3\cos v 
  }{(p-6-2e\cos v)^{1/2}(1+e\cos v)^4} \nonumber\\
&&+ \frac{(p-6)
\left[(p-6-2e\cos v)e\cos v+2(p-3)\right]}{(p-6-2e\cos v)^{1/2}(1+e\cos v)^4} \Bigg\rbrace a^{\phi} 
\nonumber\\
&& -\frac{p^2M(p-3-e^2)\left[(p-6)\cos v+2e\right]}{e(p-6+2e)(p-6-2e)
  (1+e\cos v)^2} a^r.
\label{wprime}
\end{eqnarray}}
These equations could be rewritten in any number of ways, in terms of
alternative linear combinations of $a^t$, $a^r$, and $a^{\phi}$, by
using the orthogonality relation $a_{\alpha}\dot{z}^{\alpha}=0$, which
has the explicit form  
\begin{equation}\label{orthogonality}
f t' a^t - f^{-1} r' a^r - r^2 \phi' a^{\phi} = 0. 
\end{equation}
The result of such a rearrangement might in fact be simpler, but it
may also be ill-behaved from a numerical point of view. One such
alternative combination is given in 
Appendix~\ref{Killing_osculating}.  

My first formulation of the method of osculating orbits is
complete. We have first-order evolution equations for each one of the
dynamical variables in the set $\{p,e,w,t,\phi\}$; the equations for
$t$ and $\phi$ were obtained in the preceding subsection, and for
convenience they are reproduced here: 
\begin{eqnarray} 
t' & = & \frac{p^2M}{(p-2-2e\cos v)(1+e\cos v)^2}
\sqrt{\frac{(p-2-2e)(p-2+2e)}{p-6-2e\cos v}}, 
\\ 
\phi' & = & \sqrt{\frac{p}{p-6-2e\cos v}}. 
\end{eqnarray}
Equations (\ref{pprime}), (\ref{eprime}), and (\ref{wprime}) form a
complete set of equations for $p(\chi)$, $e(\chi)$, and $w(\chi)$;
once these functions are known, $t(\chi)$ and $\phi(\chi)$ can be
obtained from the remaining two equations. We recall that $v = \chi -
w(\chi)$.   

One may note that $w'$ diverges as $e \to 0$. This corresponds to the 
fact that $w$ loses its geometric meaning for circular orbits. To
overcome this difficulty we can again follow celestial mechanics and 
define alternative orbital elements $\alpha=e\sin w$ and $\beta=e\cos
w$. The radial coordinate in terms of these elements is  
\begin{equation}
r=\frac{pM}{1+\Psi+\Omega},
\end{equation}
where $\Psi=\alpha\sin\chi$ and $\Omega=\beta\cos\chi$ are introduced
for the sake of brevity in later expressions. While $\alpha$ and
$\beta$ do not possess a clear geometric meaning, which limits their 
usefulness for generic orbits, they do allow one to analyze
small-eccentricity or quasi-circular orbits. Their evolution equations
can be easily calculated as $\alpha '=e'\sin w+ew'\cos w$ and $\beta
'=e'\cos w-ew'\sin w$. Using the identities $e\cos
v=\alpha\sin\chi+\beta\cos\chi$ and $e\sin
v=\beta\sin\chi-\alpha\cos\chi$ to simplify the results, we find 
\begin{eqnarray}
\beta ' & = & \frac{p^{5/2}M^2(p-3-\alpha^2-\beta^2)}
{\sqrt{p-6-2(\Psi+\Omega)}((p-6)^2-4(\alpha^2+\beta^2))(1+\Psi+\Omega)^4}\nonumber\\
&&\times\Bigg\lbrace-4\alpha\left[\alpha\beta\cos2\chi
+\frac{1}{2}(\alpha^2-\beta^2)\sin2\chi\right]+\beta\left[p^2-10p+12+4(\alpha^2+\beta^2)\right] \nonumber\\ 
&& +\left[2(p-3)+(p-6)(\Psi+\Omega)-2(\Psi\!+\!\Omega)^2\right]
\left[(p-6)\cos\chi-2\beta(\Psi+\Omega)\right]
\Bigg\rbrace a^{\phi} \nonumber\\
&& +\frac{p^2M(p-3-\alpha^2-\beta^2)
\left[(p-6-2\beta^2)\sin\chi+2\alpha(1+\Omega)\right]} 
{((p-6)^2-4(\alpha^2+\beta^2))(1+\Psi+\Omega)^2}a^r,\\
\alpha ' & = & \frac{p^{5/2}M^2(p-3-\alpha^2-\beta^2)}
{\sqrt{p-6-2(\Psi+\Omega)}((p-6)^2-4(\alpha^2+\beta^2))(1+\Psi+\Omega)^4}\nonumber\\
&& \times\Bigg\lbrace 
4\beta\left[\alpha\beta\cos2\chi+\frac{1}{2}(\alpha^2\!
- \!\beta^2)\sin2\chi\right]+\alpha\left[p^2-10p+12+4(\alpha^2\!+\!\beta^2)\right] \nonumber\\
&& +\left[2(p-3)+(p-6)(\Psi\!+\!\Omega)-2(\Psi\!+\!\Omega)^2\right]
\left[(p-6)\sin\chi-2\alpha(\Psi\!+\!\Omega)\right]
\Bigg\rbrace a^{\phi} \nonumber\\
&& -\frac{p^2M(p-3-\alpha^2-\beta^2)
\left[(p-6-2\alpha^2)\cos\chi+2\beta(1+\Psi)\right]} 
{((p-6)^2-4(\alpha^2+\beta^2))(1+\Psi+\Omega)^2}a^r.
\end{eqnarray}
To evolve the full system we must also express $p'$, $t'$, and 
$\phi'$ in terms of $\alpha$ and $\beta$: 
{\allowdisplaybreaks\begin{eqnarray}
p' & = & \frac{2p^{7/2}M^2\sqrt{p-6-2(\Psi+\Omega)}
(p-3-\alpha^2-\beta^2)(p-3-(\Psi+\Omega)^2)}
{[(p-6)^2-4(\alpha^2+\beta^2)](1+\Psi+\Omega)^4}a^{\phi}\nonumber\\ 
&&
-\frac{2p^3M(p-3-\alpha^2-\beta^2)(\beta\sin\chi-\alpha\cos\chi)}
{[(p-6)^2-4(\alpha^2+\beta^2)](1+\Psi+\Omega)^2}a^r,\\ 
\nonumber\\
t'(\chi) & = &
\frac{p^2M\sqrt{(p-2)^2-4(\alpha^2+\beta^2)}}
{(p-2-2(\Psi+\Omega))\sqrt{p-6-2(\Psi+\Omega)}(1+\Psi+\Omega)^2},\\ 
\nonumber\\\phi '(\chi) & = & \sqrt{\frac{p}{p-6-2(\Psi+\Omega)}}.
\end{eqnarray}}
This is my second formulation of the method of osculating orbits. The
first formulation involves shorter equations, but it becomes
ill-behaved when $e$ is small. The second formulation is well
behaved, but it involves longer equations.

\section{Newtonian osculating orbits}
\label{Newtonian}

Since my work extends the standard methods of Newtonian celestial
mechanics, it is worthwhile to show that my equations
reduce to those for perturbed Keplerian orbits in Newtonian
mechanics. In this section I derive the Newtonian limit of my
expressions by expanding in powers $p^{-1}$; since $p^{-1} \propto
r^{-1} \sim v^2$, this is equivalent to a post-Newtonian expansion. I
shall first describe the general relationship between the Newtonian
and relativistic perturbing forces. Next I shall show that my
geodesic parametrization reduces to Keplerian ellipses and that my
evolution equations for the orbital elements $p$, $e$, and $w$ reduce
to Gauss' perturbation equations of celestial mechanics.  

Substituting the Christoffel symbols of the Schwarzschild metric into
the equations of motion~\eqref{affine eq mot} yields the following
equations for the force: 
\begin{eqnarray}
a^r & = & \ddot r +f\frac{M}{r^2}\dot t^2 
- f^{-1}\frac{M}{r^2}\dot r^2+f\dot\phi^2, \\
a^\phi&=&\ddot\phi+2\frac{\dot r\dot\phi}{r}, \\ 
a^t & = & \ddot t + f^{-1}\frac{2M}{r^2}\dot{r}^2, 
\end{eqnarray}
where $f = 1-2M/r$. The time-component of the force can be written in
a more useful form using the orthogonality
relation~\eqref{orthogonality}.  

These expressions for the relativistic force differ nontrivially from
those in the Newtonian case. I define $F$, the Newtonian
perturbing force per unit mass, via Newton's second law: 
\begin{equation}
\ddot{\bm{x}} = \bm{g}+\bm{F},
\end{equation}
where $\bm{x}$ is a 3-vector representing the spatial coordinates of
the particle and $\bm{g}=-\frac{M}{r^2}\hat{\bm{r}}$ is the Newtonian
gravitational acceleration. For convenience I have defined the
Newtonian acceleration as the second derivative of $\bm{x}$ with
respect to proper time rather than coordinate time. I also define the
radial and tangential components of the perturbing force via
\begin{equation} 
\bm{F}\equiv F^r \hat{\bm{r}} + 
F^{\phi} \hat{\bm{\phi}}, 
\end{equation} 
where $\hat{\bm{r}}$ and $\hat{\bm{\phi}}$ form an orthonormal basis
in the orbital plane. Given these definitions, writing $\ddot{\bm{x}}$
in polar coordinates $(r,\phi)$ leads to  
\begin{eqnarray}
F^r & = & \ddot r - r\dot\phi^2+\frac{M}{r^2} \\
F^{\phi} & = & r\ddot\phi +2\dot r\dot\phi.
\end{eqnarray}

Comparing the Newtonian and relativistic expressions for the
perturbing force, we see they are related by the equations 
\begin{eqnarray}
a^r & = & F^r + r\left(1-f\right)\dot\phi^2 +\frac{M}{r^2}\left(f\dot{t}^2+f^{-1}\dot{r}^2-1\right), 
\label{fr relation}\\
a^{\phi} & = & \frac{F^{\phi}}{r}\label{fphi relation}. 
\end{eqnarray}
Thus, $a^r$ differs from $F^r$ by relativistic corrections,
while $a^{\phi}$ differs from $F^{\phi}$ only by a factor of
the orbital radius. 

I next consider my parametrization of geodesics. From
Eqs.~\eqref{phiprime}, \eqref{tprime}, and \eqref{chidot} one
trivially finds the leading-order terms in $\phi'$, $t'$, and
$\dot\chi$ to be  
\begin{equation}
\phi' = 1,\quad
t' = \frac{p^{3/2}M}{[1+e\cos(\chi-w)]^2},\quad
\dot\chi = \frac{[1+e\cos(\chi-w)]^2}{p^{3/2}M}.
\label{chidot expansion}
\end{equation}
Thus, in the Newtonian limit we have $\phi=\chi$ and $t=\tau$ and the
resulting parametrization 
\begin{equation}
r = \frac{pM}{1+e\cos(\phi-w)},\quad \diff{\phi}{t} = \frac{[1+e\cos(\phi-w)]^2}{p^{3/2}M}.\label{phidot Kepler}
\end{equation}

In terms of the orbital elements, we see that $w=\Phi$ in the
Newtonian limit. This corresponds to the loss of one degree
of freedom, as we would expect from the fact that $t$ in Newtonian
physics is a universal parameter rather than a coordinate. We can also 
easily find that the energy and angular momentum per unit mass reduce
to $E=1-\frac{1-e^2}{2p}$ and $L=\sqrt{p}M$, respectively. The first
term in $E$ is the rest energy of the particle, while the second term
is the Newtonian energy $\frac{1}{2}v_iv^i-\frac{M}{r}$. 

With the exception of the inclusion of the rest mass, the above
results are standard Keplerian relationships. They can be solved analytically by introducing the eccentric anomaly $E_{anom}$, defined by the relationships $\cos E_{anom}=\frac{x}{a}=\frac{e+\cos v}{1+e\cos v}$ and $\sin E_{anom}=\frac{y}{b}=\frac{\sqrt{1-e^2}\sin v}{1+e\cos v}$. After substituting this into the equation \eqref{phidot Kepler} for $\diff{\phi}{t}$, one can integrate the equation to find \emph{Kepler's equation} $E_{anom}-e\sin E_{anom}=\frac{2\pi t}{\av{P}}$. Here $\av{P}$ is the orbital period, given by $2\pi M p^{3/2}/(1-e^2)^{3/2}$, and $\frac{2\pi t}{\av{P}}$ is a quantity called the mean anomaly, which measures the fraction of a cycle . Kepler's equation can be solved via a root-finding algorithm, or it can be solved in terms of a convergent sum of Bessel functions. That yields $E_{anom}(t)$, which then straightforwardly leads to $v(t)$ and thence to $\phi(t)$. 

Just as the parametrization of Schwarzschild geodesics reduces to the standard parame\-tri\-za\-tion of Keplerian orbits, my evolution equations for
the orbital elements reduce to the standard evolution equations for perturbed Keplerian
orbits. Substituting Eq.~\eqref{chidot
 expansion} into Eqs.~\eqref{orthogonality}, \eqref{fr relation}, and
\eqref{fphi relation}, we find the leading-order expressions for the
perturbing force: 
\begin{eqnarray}
a^r & = & F^r \\
a^{\phi}&=&\frac{F^{\phi}}{r} \\
a^t & = & \frac{e\sin(\phi-w)}{\sqrt{p}}F^r 
+ \frac{1+e\cos(\phi-w)}{\sqrt{p}}F^{\phi}. \qquad
\end{eqnarray}
These results allow us to expand Eqs.~\eqref{pprime}, \eqref{eprime},
and \eqref{wprime} to find the leading-order expressions for the
orbital elements: 
\begin{eqnarray}
\diff{p}{t} & = & \frac{2p^{3/2}}{1+e\cos(\phi-w)}F^{\phi}, \label{pdot_Newtonian}\\ 
\diff{e}{t} & = & \sqrt{p}\ \frac{e+2\cos(\phi-w)+e\cos^2(\phi-w)}
{1+e\cos(\phi-w)}F^{\phi}+\sqrt{p}\ \sin(\phi-w)F^r, \label{edot_Newtonian}\\
\diff{w}{t} & = & \frac{\sqrt{p}}{e}\ 
\frac{\sin(\phi-w)[2+e\cos(\phi-w)]}{1+e\cos(\phi-w)}F^{\phi} 
\nonumber\\
&& -\frac{\sqrt{p}}{e}\ \cos(\phi-w)F^r.\label{wdot_Newtonian}
\end{eqnarray}
These are Gauss' perturbation equations.

In the following chapter, I will make use of both the fully relativistic evolution equations and their Newtonian limit. Before doing so, for the sake of brevity I define the following quantities: 
\begin{align}
P &\equiv \frac{p^{3/2}M}{(1+e\cos v)^2},\\
f^p_\phi &\equiv \frac{2p^3M}{(1+e\cos v)^3},\label{fp}\\
f^e_r &\equiv \frac{p^2M\sin v}{(1+e\cos v)^2},\\
f^e_\phi &\equiv p^2M\frac{e+2\cos v +e\cos^2v}
{(1+e\cos v)^3},\\
f^w_r &\equiv -\frac{p^2M\cos v}{e(1+e\cos v)^2}, \\
f^w_\phi &\equiv \frac{p^2M\sin v (2+e\cos v)}{e(1+e\cos v)^3},\label{fw}
\end{align}
where $v\equiv\phi-w$ and
\begin{equation}
f^p\equiv f^p_\phi F^\phi, \quad f^e\equiv f^e_rF^r+f^e_\phi F^\phi, \quad f^w\equiv f^w_rF^r+f^w_\phi F^\phi.
\end{equation}
In terms of these quantities, we have the evolution equations
\begin{equation}
\diff{t}{\phi}=P(I^A,\phi),\quad \diff{I^A}{\phi}=f^A(I^B,\phi),\label{dphi}
\end{equation}
where $A\in\lbrace p,e,w\rbrace$.
			\chapter{Adiabatic approximations}\label{adiabatic}
My primary use of the method of osculating orbits will be to study the efficacy of an adiabatic approximation. As discussed in the introduction, an adiabatic approximation attempts to circumvent a complicated procedure of calculating the regular field at each timestep by using information about asymptotic wave amplitudes. This approximation can do any or all of the following: calculate the waves at each instant as if the particle moved on a geodesic, then slowly move between geodesics (the geodesic-source approximation); neglect oscillations, because the information at infinity is related to the local behavior of the particle only in a time-averaged sense (the secular approximation); and neglect conservative effects, because the information at infinity is calculated using the radiative Green's function (the radiative approximation).

These issues were studied in Refs.~\cite{osculating_paper,our_paper,other_paper}; however, the most comprehensive study was more recently performed by Hinderer and Flanagan \cite{Hinderer_Flanagan}. They found that for orbits in Kerr, the three types of sub-approximations identified above are consistent at leading order in a multiscale expansion. (However, it is worth noting that when implementing the geodesic-source approximation, along with the point-particle perturbation they include a perturbation due to evolution of the large black hole's mass and angular momentum---the point-particle perturbation alone would not provide an approximation accurate at first order.) This leading-order approximation is formally accurate up to errors of order $O_s(1)$ in the orbital phase over a radiation-reaction time. Such an approximation might be sufficiently accurate for gravitational-wave detection. However, in order to extract orbital parameters from a waveform, we require an approximation that is accurate up to errors of order $o(1)$ over a radiation-reaction time. A simple adiabatic approximation fails in this regard.

In this chapter, I analyze the types of errors that can arise in an adiabatic approximation. Most obviously, the neglect of conservative effects induces secular errors in the orbital phase. We can identify at least two sources of such errors: first, there is an orbital precession, corresponding to long-term changes in the argument of periapsis $w$. This is akin to the well-known precession of the perihelion of Mercury. The second effect is a direct shift of the orbital period. For example, in Newtonian physics, on a circular orbit the acceleration is given by $a=r\omega^2$, so for constant initial positions, a shift in the radial, conservative force causes a shift in the frequency. In terms of the orbital elements, this means that for given values of the elements, the frequency $\omega$ is different.\footnote{Note that in the post-Newtonian ``adiabatic approximation," which is tailored to circular orbits, this conservative correction to the orbital frequency is actually accounted for, because the energy that appears in the energy-balance equation is calculated based on 3PN conservative dynamics, rather than being based on 0PN dynamics. This is an important difference between the PN adiabatic approximation and the EMRI adiabatic approximation.}

In the first section of this chapter, I provide a general discussion of these issues in the context of a multiscale expansion of the Newtonian osculating orbit equations. My presentation essentially serves to summarize the results of Hinderer and Flanagan \cite{Hinderer_Flanagan} in a simplified setting: I conclude as they do that in order to construct a sufficiently accurate approximation for EMRI inspirals, one requires the full first-order self-force and the radiative part of the second-order force. I also present a concrete application of the multiscale expansion, but to avoid cluttering the discussion, I relegate the application to Appendix~\ref{multiscale_EFE}.

In the second section of this chapter, I apply the method of osculating orbits in Schwarz\-schild to analyze post-Newtonian binary systems. My numerical results for these systems reveal the importance of the conservative correction to the orbital period; the effect of orbital precession, on the other hand, is found to be relatively minor. My results also reveal that the long timespan of an inspiral leads to a large, long-term impact of initial conditions.

\section{Multiscale expansion of Newtonian osculating orbits}\label{multiscale_osculating}
Consider the evolution equations \eqref{phidot Kepler} for $\phi(t,\e)$ and \eqref{pdot_Newtonian}--\eqref{wdot_Newtonian} for the orbital elements $I^A(t,\e)$, where I now consider the restricted set of elements $A=(p,e,w)$. Suppose that the self-force is given by $\bm{F}=\e\bm{F}\dcoeff{1}+\e^2\bm{F}\dcoeff{2}+...$, and that each term has a radiation-reaction/dissipative part $\bm{F}\dcoeff{\emph{n}}{}_{\rm rad}$ and a conservative part $\bm{F}\dcoeff{\emph{n}}{}_{\rm con}$. Using a multiscale expansion, as outlined in Sec.~\ref{multiple scales}, we can determine the generic effects of these forces.

In order to efficiently distinguish between secular and oscillatory terms, I will use the azimuthal angle $\phi$, rather than time $t$, as the independent variable, and solve for $t(\phi,\e)$ and $I^A(\phi,\e)$. In this case, the function $t(\phi,\e)$ implicitly represents the evolution of the orbital phase as a function of time. As presented at the end of Ch.~\ref{osculating}, the evolution equations can be written as  
\begin{align}
\diff{t}{\phi} &= P(\phi,I^B),\label{t_gen}\\
\diff{I^A}{\phi} &= \e f^A\dcoeff{1}(\phi,I^B)+\e^2f^A\dcoeff{2}(\phi,I^B)+...,\label{IA_gen}
\end{align}
where $P(\phi,I^A)=p^{3/2}M/(1+e\cos(\phi-w))^2$, $f^A\dcoeff{1}$ is constructed from $\bm{F}\dcoeff{1}$, and $f^A\dcoeff{2}$ is constructed from $\bm{F}\dcoeff{2}$.

I now introduce the slow variable $\tilde\phi=\e\phi$,\footnote{We could instead use $\phi$ in combination with the slow time $\tilde t=\e t$, or some other combination of variables; but computations with these combinations are made difficult by the fact that $\phi(t)$ contains oscillations even for an unperturbed, Keplerian orbit.} and I assume the multiscale expansions
\begin{align}
t(\phi,\e) &= \e^{-1}t\dcoeff{-1}(\phi,\tilde\phi)+t\dcoeff{0}(\phi,\tilde\phi)+\e t\dcoeff{1}(\phi,\tilde\phi)+...,\\
I^A(\phi,\e) &= I^A\dcoeff{0}(\phi,\tilde\phi)+\e I^A\dcoeff{1}(\phi,\tilde\phi)+\e^2I^A\dcoeff{2}(\phi,\tilde\phi)+...
\end{align}
The expansion of $t$ includes an inverse power of $\e$ because it grows secularly even in the absence of a perturbation, so for $\tilde\phi\sim 1$ we expect $t\sim\e^{-1}$. I assume that each term in these expansions is $2\pi$-periodic in $\phi$: that is, $t\dcoeff{\emph{n}}(\phi+2\pi,\tilde\phi)=t\dcoeff{\emph{n}}(\phi,\tilde\phi)$ and $I^A\dcoeff{\emph{n}}(\phi+2\pi,\tilde\phi)=I^A\dcoeff{\emph{n}}(\phi,\tilde\phi)$, etc. For future brevity, I introduce the notation $\langle f\rangle =\frac{1}{2\pi}\int_0^{2\pi}f(\phi)d\phi$ to denote an average over one period. Since $\phi$ and $\tilde\phi$ are to be treated as independent variables, these integrals hold $\tilde\phi$ fixed. 

Substituting the expansions into the differential equations \eqref{t_gen}--\eqref{IA_gen}, we arrive at
\begin{align}
\e^{-1}\pdiff{t\dcoeff{-1}}{\phi}+& \pdiff{t\dcoeff{-1}}{\tilde\phi}+\pdiff{t\dcoeff{0}}{\phi} +\e\left(\pdiff{t\dcoeff{0}}{\tilde\phi}+\pdiff{t\dcoeff{1}}{\phi}\right)+... \nonumber\\
&= P(\phi,I^A\dcoeff{0})+\e \pdiff{P}{I^A}(\phi,I^B\dcoeff{0})I^A\dcoeff{1}+...,\\
\!\!\!\!\!\!\!\!\pdiff{I^A\dcoeff{0}}{\phi} +&\e\left(\pdiff{I^A\dcoeff{0}}{\tilde\phi}+\pdiff{I^A\dcoeff{1}}{\phi}\right)+\e^2\left(\pdiff{I^A\dcoeff{1}}{\tilde\phi}+\pdiff{I^A\dcoeff{2}}{\phi}\right)+... \nonumber\\
&= \e f^A\dcoeff{1}(\phi,I^B\dcoeff{0})+\e^2\left(f^A\dcoeff{2}(\phi,I^B\dcoeff{0})+\pdiff{f^A\dcoeff{1}}{I^C}(\phi,I^B\dcoeff{0})I^C\dcoeff{1}\right)+...
\end{align}
I now treat $\phi$ and $\tilde\phi$ as independent variables, such that in these equations, coefficients of explicit powers of $\e$ on the left- and right-hand sides can be equated.

The order-$\e^0$ orbital-element-equation tells us that $I^A\dcoeff{0}$ depends only on $\tilde\phi$.  Similarly, the order-$\e^{-1}$ time-equation tells us that $t\dcoeff{-1}$ depends only on $\tilde\phi$. Next, the order-$\e$ orbital-element-equation reads
\begin{equation}
\diff{I^A\dcoeff{0}}{\tilde\phi}+\pdiff{I^A\dcoeff{1}}{\phi} = f^A_1(\phi,I^B\dcoeff{0}).\label{A0_A1}
\end{equation}
Since $I^A\dcoeff{1}$ is assumed to be periodic in $\phi$, we have $\left\langle\pdiff{I^A\dcoeff{1}}{\phi}\right\rangle=0$. Thus, taking the average of the above equation yields
\begin{equation}
\diff{I^A\dcoeff{0}}{\tilde\phi} = \left\langle f^A\dcoeff{1}(\phi,I^B\dcoeff{0})\right\rangle.\label{IA0}
\end{equation}
Since an averaged quantity contains no $\phi$-dependence, the term on the right depends only on $I^B\dcoeff{0}$. Hence, this equation can be solved to determine $I^A\dcoeff{0}(\tilde\phi)$. After substituting this back into Eq.~\eqref{A0_A1}, we can solve for $I^A\dcoeff{1}$:
\begin{equation}
I^A\dcoeff{1} = \int\!\!\left[f^A\dcoeff{1}(\phi,I^B\dcoeff{0})-\left\langle f^A\dcoeff{1}(\phi,I^B\dcoeff{0})\right\rangle\right]d\phi+C^A\dcoeff{1}(\tilde\phi).\label{IA1}
\end{equation}
By construction, the integral is a periodic function of $\phi$. The non-oscillatory integration ``constant" $C^A_1(\tilde\phi)$ will be determined only at the next order.

Moving on to the order-$\e^0$ time-equation and following the same procedure of first averaging, then finding the oscillatory terms, we find
\begin{align}
\diff{t\dcoeff{-1}}{\tilde\phi} &= \left\langle P(\phi,I^A\dcoeff{0})\right\rangle\\
&\quad =\frac{p^{3/2}\dcoeff{0}M}{(1-e\dcoeff{0}^2)^{3/2}},\label{tn1}\\
t\dcoeff{0} &= \int \!\!\left[P(\phi,I^A\dcoeff{0})-\left\langle P(\phi,I^A\dcoeff{0})\right\rangle\right]d\phi+C^t\dcoeff{0}(\tilde\phi).\label{t0}
\end{align}
Using the results for $I^A\dcoeff{0}(\tilde\phi)$, the first of these two equations can be solved to completely determine $t\dcoeff{-1}(\tilde\phi)$. As in Eq.~\eqref{IA1}, the integral in Eq.~\eqref{t0} is a periodic function of $\phi$, and the integration ``constant'' $C^t\dcoeff{0}(\tilde\phi)$ will be determined only at the next order.

Proceeding to the order-$\e^2$ orbital-element-equation and again averaging over one period, we find that the non-oscillatory part of $I^A\dcoeff{1}$ is determined by
\begin{align}
\diff{C^A\dcoeff{1}}{\tilde\phi} &= \left\langle f^A\dcoeff{2}(\phi,I^B\dcoeff{0})+\pdiff{f^A\dcoeff{1}}{I^C}(\phi,I^B\dcoeff{0})I^C\dcoeff{1}\right\rangle.\label{CA1}
\end{align}
This equation can be solved for $C^A\dcoeff{1}(\tilde\phi)$, so $I^A\dcoeff{1}$ is now completely determined. We also arrive at an equation for $I^A\dcoeff{2}$, and it will again have an undetermined function $C^A\dcoeff{2}(\tilde\phi)$; but $I^A\dcoeff{2}$ is too high-order a correction to be of interest here.

Lastly, we consider the order-$\e$ time-equation and find, after averaging over a period,
\begin{align}
\diff{C^t\dcoeff{0}}{\tilde\phi} &= \left\langle\pdiff{P}{I^A}(\phi,I^B\dcoeff{0})I^A\dcoeff{1}\right\rangle.\label{Ct0}
\end{align}
This equation can be solved for $C^t\dcoeff{0}(\tilde\phi)$, so $t\dcoeff{0}$ is now completely determined. As I did with the equation for $I^A\dcoeff{2}$, I neglect the equation for $t\dcoeff{1}$.

One can see that this procedure can be carried to arbitrarily high order. But we have already proceeded to sufficiently high order to understand the essential features of the solution. Evaluating the solution at $\tilde\phi=\e\phi$, we have $t(\phi,\e)=\e^{-1}t\dcoeff{-1}(\e\phi)+t\dcoeff{0}(\phi,\e\phi)+o(1)$, and $I^A(\phi,\e)=I^A\dcoeff{0}(\e\phi)+\e I^A\dcoeff{1}(\phi,\e\phi)+o(\e)$, where the errors are presumed to be uniform on the interval $[0,1/\e]$. Let us now consider how each of these terms are determined. From Eqs.~\eqref{IA0} and \eqref{IA1}, we see that both the slowly evolving, leading-order orbital elements $I^A\dcoeff{0}$ and the oscillatory part of the corrections $I^A\dcoeff{1}$ are fully determined by the leading-order force $\bm{F}\dcoeff{1}$. Analogously, from Eq.~\eqref{tn1}, we see that both the leading-order orbital phase, as represented by $t\dcoeff{-1}$, and the oscillatory part of the correction, as represented by $t\dcoeff{0}$, are fully determined by $I^A\dcoeff{0}$, meaning that they too are fully determined by the leading-order force. From Eq.~\eqref{CA1}, we see that the non-oscillatory part of the corrections $I^A\dcoeff{1}$ is determined by (i) the second-order force $\bm{F}\dcoeff{2}$, and (ii) the oscillatory part of $I^A\dcoeff{1}$ (which is required because it can combine with oscillations in $\pdiff{f^A\dcoeff{1}}{I^C}$ to produce stationary terms). Similarly, from Eq.~\eqref{Ct0}, we see that the non-oscillatory part of $t\dcoeff{0}$ is determined by the second-order force and the oscillatory part of $I^A\dcoeff{1}$.

From this analysis, we see that to construct an approximation scheme that keeps errors in the orbital phase uniformly small (i.e., $o(1)$) on a radiation-reaction timescale, one requires both the second-order force and the oscillatory part of the solution---one cannot construct a sufficiently accurate approximation with just the first-order force, or with just the non-oscillatory parts of the solution.

\subsection{Conservative and dissipative effects}
We can make this statement more precise by considering the differing effects of the conservative and dissipative forces. By ``dissipative force," I mean specifically the radiative force $\bm{F}_{\rm rad}=\tfrac{1}{2}(\bm{F}_{\rm ret}-\bm{F}_{\rm adv})$; the conservative force $\bm{F}_{\rm con}=\tfrac{1}{2}(\bm{F}_{\rm ret}+\bm{F}_{\rm adv})$ makes up the remainder of the total force. Following an argument originally made by Mino in the case of orbits in Kerr \cite{Mino}, we can express this radiative force entirely in terms of the retarded force. Consider a particle at position $\mathcal{P}=(t_\mathcal{P},r_\mathcal{P},\theta_\mathcal{P}, \phi_\mathcal{P})$. The coordinate transformation $t\to t'=2t_\mathcal{P}-t$, $\phi\to \phi'=2\phi_\mathcal{P}-\phi$ reverses the direction of time around the particle while leaving its position $\mathcal{P}$ unchanged. The retarded field at $\mathcal{P}$ is generated by the particle's past history in the original coordinates; the advanced field is generated by its future history in the original coordinates, which corresponds to its past history in the new coordinates. (In other words, under this time-reversal, the effect on the Green's functions is that the advanced one becomes retarded, and vice-versa.) Hence, $\bm{F}'_{\rm adv}(t',r',\phi')=\bm{F}_{\rm ret}(t'(t),r'(r),\phi'(\phi))$. But in addition, the components of the force in the time-reversed coordinates are related to the components in the original coordinates via the usual vector transformation law: specifically, $F^r(t,r,\phi)=F'^r(t',r',\phi')$ and $F^\phi(t,r,\phi)=-F'^\phi(t',r',\phi')$.

Combining these equations yields the relationships $F^r_{\rm adv}(t,r,\phi)=F^r_{\rm ret}(2t_\mathcal{P}-t,r,2\phi_\mathcal{P}-\phi)$ and $F^\phi_{\rm adv}(t,r,\phi)=-F^\phi_{\rm ret}(2t_\mathcal{P}-t,r,2\phi_\mathcal{P}-\phi)$. Now, the self-force at $\mathcal{P}$ depends on the position and velocity of the particle, along with the value of the field at its position. The last of these is incorporated into the ``retarded" and ``advanced" labels. The first two can be expressed in terms of our chosen phase space variables $(p,e,w,\phi)$. If $\phi\to2\phi_\mathcal{P}-\phi$, then $w\to 2\phi_\mathcal{P}-w$ while the other orbital elements are unchanged---an elliptical orbit has the same shape in the time-reversed coordinates. Putting all of these results together, we find that the components of the radiative force at the particle's position $\mathcal{P}$ can be written as
\begin{align}
F^r_{\rm rad}(p,e,w,\phi) &= \tfrac{1}{2}\left[F^r_{\rm ret}(p,e,w,\phi)-F^r_{\rm ret}(p,e,2\phi-w,\phi)\right],\\
F^\phi_{\rm rad}(p,e,w,\phi) &= \tfrac{1}{2}\left[F^\phi_{\rm ret}(p,e,w,\phi)+F^\phi_{\rm ret}(p,e,2\phi-w,\phi)\right].
\end{align}
Similarly, the components of the conservative force can be written as
\begin{align}
F^r_{\rm con}(p,e,w,\phi) &= \tfrac{1}{2}\left[F^r_{\rm ret}(p,e,w,\phi)+F^r_{\rm ret}(p,e,2\phi-w,\phi)\right],\\
F^\phi_{\rm con}(p,e,w,\phi) &= \tfrac{1}{2}\left[F^\phi_{\rm ret}(p,e,w,\phi)-F^\phi_{\rm ret}(p,e,2\phi-w,\phi)\right].
\end{align}

Now, in order to determine the effects of these forces, we must determine corresponding expressions for $f^A(p,e,w,\phi)$. Referring to Eqs.~\eqref{fp}--\eqref{fw}, and noting that $\cos[\phi-(2\phi-w)]=\cos(\phi-w)$ and $\sin[\phi-(2\phi-w)]=-\sin(\phi-w)$, we find that 
\begin{align}
f^{p|e}_{\rm rad}(p,e,w,\phi) &= \tfrac{1}{2}\left[f^{p|e}_{\rm ret}(p,e,w,\phi)+f^{p|e}_{\rm ret}(p,e,2\phi-w,\phi)\right],\\
f^{p|e}_{\rm con}(p,e,w,\phi) &= \tfrac{1}{2}\left[f^{p|e}_{\rm ret}(p,e,w,\phi)-f^{p|e}_{\rm ret}(p,e,2\phi-w,\phi)\right],\\
f^w_{\rm rad}(p,e,w,\phi) &= \tfrac{1}{2}\left[f^w_{\rm ret}(p,e,w,\phi)-f^w_{\rm ret}(p,e,2\phi-w,\phi)\right],\\
f^w_{\rm con}(p,e,w,\phi) &= \tfrac{1}{2}\left[f^w_{\rm ret}(p,e,w,\phi)+f^w_{\rm ret}(p,e,2\phi-w,\phi)\right],
\end{align}
where ``$p|e$" is used to indicate that the equation holds for both $p$ and $e$.

We now have sufficient information to determine the effects of the conservative and dissipative pieces of the force. First, note that $\phi$ and $w$ will appear in the force only in the combination $v=\phi-w$, since they appear in no other form in the position and velocity (in the osculating orbits formalism). So in the above expressions we can replace the dependence on $\phi$ and $w$ (or $\phi$ and $2\phi-w$) with a dependence on $v$ (or $-v$). It then follows immediately that\footnote{Note that these equalities do not hold if $I^A\dcoeff{0}$ is replaced with $I^A$, since $I^A$ has oscillatory $\phi$-dependence on the true orbit.}
\begin{align}
\langle f^{p|e}\dcoeff{\emph{n}}{}_{\rm rad}(I^A\dcoeff{0},\phi)\rangle &= \langle f^{p|e}\dcoeff{\emph{n}}{}_{\rm ret}(I^A\dcoeff{0},\phi)\rangle,\\
\langle f^{p|e}\dcoeff{\emph{n}}{}_{\rm con}(I^A\dcoeff{0},\phi)\rangle &= 0,\\
\langle f^w\dcoeff{\emph{n}}{}_{\rm rad}(I^A\dcoeff{0},\phi)\rangle &= 0,\\
\langle f^w\dcoeff{\emph{n}}{}_{\rm con}(I^A\dcoeff{0},\phi)\rangle &= \langle f^w\dcoeff{\emph{n}}{}_{\rm ret}(I^A\dcoeff{0},\phi)\rangle.
\end{align}
In words, these equations state that the secular evolution of the principal elements is governed by the radiative force, while that of $w$ is governed by the conservative force. Combined with the general multiscale analysis, we can now conclude the following: In order to determine the leading-order phase evolution, defined by $\e^{-1}t\dcoeff{-1}(\e\phi)$,\footnote{Hinderer and Flanagan \cite{Hinderer_Flanagan} refer to this leading-order evolution as the adiabatic approximation; they refer to higher-order terms in the multiscale expansion as post-adiabatic corrections.} one requires only the radiative part of the first-order self-force; this is true because $w$ does not appear in $\langle P(I^A\dcoeff{0},\phi)\rangle$. However, the phase evolution defined by such an approximation possesses $O_s(1)$ errors. In order to determine the correction $t\dcoeff{0}(\phi,\e\phi)$, one requires (i) the full first-order self-force, in order to determine $w\dcoeff{0}$ as well as the oscillatory parts of $I^A\dcoeff{1}$, and (ii) the radiative part of the second-order self-force.

The results of this analysis remain valid for the experimentally relevant case of orbits in Kerr \cite{Hinderer_Flanagan}.

\subsection{Choices of initial conditions and averages}
In the foregoing discussion, I have avoided the issue of assigning initial conditions in my multiscale expansion. As one might expect, we have some leeway in how we perform that assignment.  Assume that the initial conditions in the original problem are given by $t(0,\e)=0$ and $I^A(0,\e)=\bar I^A$, with no $\e$-dependence. Then the most natural method of assigning initial conditions in the expansion is $t\dcoeff{-1}(0)=0=t\dcoeff{0}(0)$, $I^A\dcoeff{0}(0)=\bar I^A$, $I^A\dcoeff{1}(0,0)=0$. However, there is no requirement to follow this method. We could instead allow arbitrary initial conditions for the correction terms, and then define ``corrected" initial conditions for $I^A\dcoeff{0}(\e\phi)$ by imposing $I^A\dcoeff{0}(0)=\bar I^A-\e\hat I^A\dcoeff{1}(0,0)$. Suppose that we do not have access to the second-order force (an easy thing to suppose). Then we cannot calculate the entirety of the secular term $C^A\dcoeff{1}$ via Eq.~\eqref{CA1}. However, we can calculate part of it simply by imposing appropriate initial conditions. Writing the oscillatory part of $I^A\dcoeff{1}$ as $\hat I^A\dcoeff{1}$, if we take as our solution $I^A=I^A\dcoeff{0}+\e\hat I^A\dcoeff{1}$, then we can define the corrected initial conditions $I^A\dcoeff{0}(0)=\bar I^A-\e\hat I^A\dcoeff{1}(0,0)$. In practice, with these corrected initial conditions, $I^A\dcoeff{0}$ provides an excellent secular approximation \cite{other_paper}. Of course, one requires the oscillatory pieces of the solution in order to find these corrected initial conditions, even if one disregards the oscillations in the actual evolution.

There is another issue that I have thus far ignored: I have defined all my averages such that they remove oscillations with respect to $\phi$. But in the unperturbed motion, $t(\phi)$ contains oscillations. This means that even at leading order, removing oscillations with respect to $\phi$ is not equivalent to removing oscillations with respect to $t$. In preparation for the following section, let us consider averages over the parameter $\chi$, defined by, for example,
\begin{equation}\label{chiav}
\langle I^A\rangle_{\chi} \equiv 
\frac{1}{2\pi}\int^{\chi + \pi}_{\chi - \pi} I^A(\chi')\, d\chi'.
\end{equation}
This average will differ from that defined by an average over time, such as
\begin{equation}\label{tav}
\langle I^A \rangle_t \equiv 
\frac{\int^{\chi + \pi}_{\chi - \pi} I^A \diff{t}{\chi} d\chi'} 
{\int^{\chi + \pi}_{\chi - \pi} \diff{t}{\chi} d\chi'}. 
\end{equation}
Using the two different averages leads to two different evolutions. More precisely, if we define multiscale expansions based on two different methods of averaging, then the results at any given order may differ significantly. This also affects the choice of initial conditions.

In the next section, I will demonstrate the impact of different choices of initial conditions and  averaging. See Ref.~\cite{other_paper} for further discussion of these points.

%%%%%%%%%%%%%%%%%%%%%%%%%%%%%%%%%%%%%%%%%%%%%%%%%%%%%%%%%%%%%%%%%%%%%%

\section{Post-Newtonian binaries}
\label{PN binaries}
I now move on to the second test case: a post-Newtonian binary system. This system consists of two
gravitationally-bound bodies of mass $m_1$ and $m_2$, with equations
of motion derived to 2.5PN order in a post-Newtonian expansion;
because we are interested in self-force effects, I take the ratio
$m_1/m_2$ to be small, and I neglect the spin of the bodies. In this section I explain how such a system can be analyzed with my
method of osculating orbits.

My analysis is based upon the hybrid equations of motion presented in
Ref.~\cite{Kidder}. These equations begin with the 2.5PN equations of
motion for each one of the two bodies. Within the center-of-mass frame
the relative motion of the bodies is governed by the closed system
of equations \cite{Lincoln}  
\begin{equation}\label{PNeqns}
\ddiff{x^a_h}{t} = -\frac{M}{r_h^2}\left(A\frac{x_h^a}{r_h}
+ B\diff{x_h^a}{t}\right),
\end{equation}
where $x_h^a\equiv x^a_1-x^a_2$ is a Cartesian spatial vector from 
$m_2$ to $m_1$ in harmonic coordinates, $r_h^2=\delta_{ab}x^ax^b$ is
the square of the vector's Euclidean magnitude, $t$ is a harmonic time 
coordinate, and $M = m_1+m_2$ is the total mass of the system. The
functions $A$ and $B$ depend only on the total mass $M$, the reduced
mass $\mu=m_1m_2/M$, and the relative coordinates and
velocities. They can be written as $A=A_M+\epsilon\tilde{A}$ and
$B=B_M+\epsilon\tilde{B}$, where $\epsilon=\mu/M$ and terms with a
subscript $M$ are independent of $\mu$. $\tilde{A}$ and $\tilde{B}$
contain terms independent of and linear in $\epsilon$, and they can be further decomposed into
post-Newtonian orders as $\tilde{A} = \tilde{A}_1 + \tilde{A}_2 
+ \tilde{A}_{2.5}$ and $\tilde{B} = \tilde{B}_1 + \tilde{B}_2 
+ \tilde{B}_{2.5}$.

The hybrid equations are inspired by the fact that when $\epsilon=0$, 
Eq.~\eqref{PNeqns} becomes identical to a 2PN expansion of the
geodesic equation in a Schwarzschild spacetime with mass parameter
$M$. Building on this fact, Kidder, Will, and Wiseman \cite{Kidder}
replaced $A_M$ and $B_M$ with their exact geodesic expressions $A_S$
and $B_S$ in the fictitious Schwarzschild spacetime. In other words,
the hybrid equations of motion are given by Eq.~\eqref{PNeqns} after
substituting $A=A_S+\epsilon\tilde{A}$ and $B=B_S+\epsilon\tilde{B}$,
where  
\begin{eqnarray}
A_S & = & \frac{1-M/r_h}{(1+M/r_h)^3} -\frac{2-M/r_h}{1-M^2/r_h^2}\frac{M}{r_h}
\left(\diff{r_h}{t}\right)^2+v^2, \\
B_S &=&-\frac{4-2M/r_h}{1-M^2/r_h^2}\diff{r_h}{t},
\end{eqnarray}
where $v^2=\delta_{ab}\diff{x_h^a}{t}\diff{x_h^b}{t}$ is the square of the velocity vector in harmonic coordinates. The resulting equations are accurate to 2.5PN order, but in the
test-mass limit $m_1\to 0$ they exactly describe the orbit of the test
mass in the Schwarzschild spacetime of the other body. These equations
form an ideal test case for our method of osculating orbits because,
besides their relative simplicity, they explicitly split into geodesic
terms and perturbation terms. This allows us to construct osculating
orbits as geodesics in the fictitious Schwarzschild spacetime of mass
$M$.

We can then easily derive the perturbing force from the terms
$\tilde{A}$ and $\tilde{B}$; that calculation is presented in Appendix~\ref{hybrid_eqns}. The final expression for the perturbing acceleration is given by
\begin{eqnarray}
a^r & = & -\frac{\mu}{r^2} 
\left[\mathcal{A} + \mathcal{B} \frac{dr}{dt} \right],
\\ 
a^{\phi} & = &-\frac{\mu}{r^2}\mathcal{B} \frac{d\phi}{dt},  
\end{eqnarray}
where the formulas for $\mathcal{A}$ and $\mathcal{B}$ are displayed in Appendix~\ref{hybrid_eqns}. This perturbing force can be substituted into the evolution equations for the orbital elements, which are then straightforwardly integrated numerically.

The force derived in this way is a form of the gravitational self-force, since it is produced by finite-mass effects. However, it differs nontrivially from the post-Newtonian limit of the relativistic self-force: First, the self-force is a gauge-dependent quantity which is typically calculated in the Lorenz gauge, while the hybrid equations of motion are derived within the harmonic gauge. Second, the Lorenz gauge ensures that the coordinates of the small body are defined in relation to the system's center of mass \cite{Eric_Steve}, while here I use coordinates relative to the large mass. And third, my geodesics are in a fictitious Schwarzschild spacetime of mass $M = m_1 + m_2$ and not in the background spacetime of the second body (of mass $m_2$). The last two differences could be easily removed by formulating an alternative set of hybrid equations, but the gauge difference cannot be easily dealt with. Nevertheless, the perturbing force has the same essential features as the gravitational self-force. In particular, the self-force can be expected to have conservative terms at 0PN (the Newtonian level), 1PN, and 2PN orders, etc., and dissipative terms at 2.5PN (corresponding to quadrupole radiation) and 3.5PN orders, etc.; my perturbing force has exactly the same features, except for the Newtonian correction, which is implicitly accounted for by working in terms of total and reduced masses. Thus, I can draw conclusions about the action of the gravitational self-force even from my simplified analysis.

My focus here will be on the magnitude of errors in the radiative approximation. A radiative evolution switches off all conservative terms in the perturbing force ($\tilde{A}_1 = \tilde{A}_2 = \tilde{B}_1 = \tilde{B}_2 = 0$), and retains only the radiative terms at 2.5PN order ($\tilde{A}_{2.5} \neq 0$ and $\tilde{B}_{2.5} \neq 0$). As we shall see, this approximation neglects very considerable secular effects in the orbital evolution. In addition, the radiative approximation is subject to the same ambiguities regarding the choice of initial conditions as the secular approximation. Writing the radiative evolution as the sum of its secular and oscillatory parts, $I_{\rm r}(\chi) = I_{\rm r\ sec}+I_{\rm r\ osc}$, we shall consider three possible candidates for $I_{\rm r}(0)$. The first is $I_{\rm r}(0)=I^A(0)$, the \emph{exact initial data} that is selected for the true evolution of the orbital elements under the action of the full perturbing force. The second is $I_{\rm r\ sec}=\av{I^A}_\chi(0)$, the \emph{$\chi$-averaged initial data}, which identifies the initial secular part of the radiative evolution with the initial $\chi$-averaged part of the true evolution. The third choice is $I_{\rm r\ sec}(0)=\av{I^A}_t(0)$, the \emph{$t$-averaged initial data}, which identifies the initial secular part of the radiative evolution with the initial $t$-averaged part of the true evolution. These three choices of initial data are distinct, and they lead to different evolutions. We shall see that the accuracy of the evolution (relative to the true evolution) depends strongly on the choice of initial data.

%%%%%%%%%%%%%%%%%%%%%%%%%%%%%%%%%%%%%%%%%%%%%%%%%%%%%%%%%%%%%%%%%%%%

\subsection{Results}
\label{results}

\begin{figure}[tb]
\begin{center}
\includegraphics{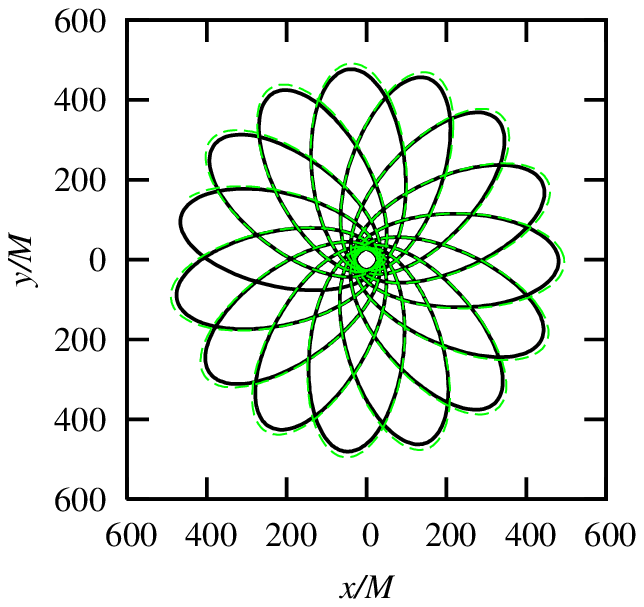}
\includegraphics{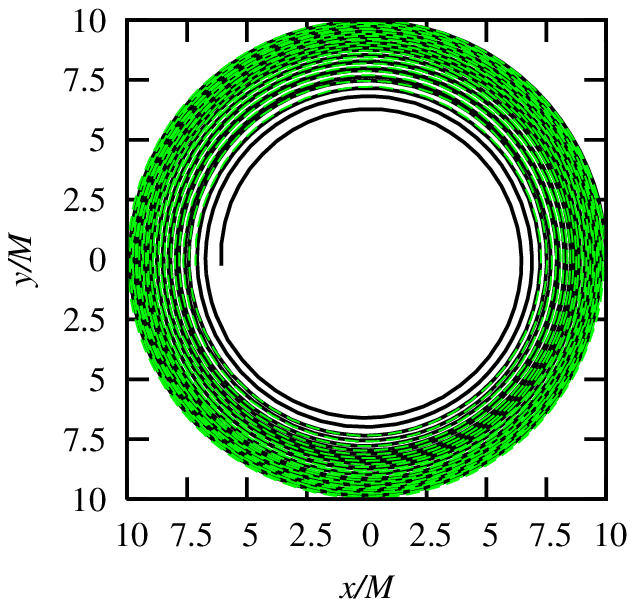}
\end{center}
\caption[Comparison of orbits with identical initial conditions]{Comparisons of true orbits (solid black curves) and radiative
  approximation orbits (dashed green curve) with identical initial
  conditions and with a mass ratio $\mu/M=0.01$. In each case the two
  orbits begin at periapsis and are terminated at the same final
  time. Left plot: highly eccentric orbits with $p_0=50$ and
  $e_0=0.9$. At the end of the simulation the approximate orbit lags
  behind the true orbit by approximately one-half radial cycle out of
  a total of fifteen. RIght plot: quasi-circular orbits with identical
  initial conditions $p_0=10$ and $e_0=0$. Again, the approximate
  orbit lags behind the true orbit.} 
\label{orbits}
\end{figure}

\begin{figure}[tb]
\begin{center}
\includegraphics{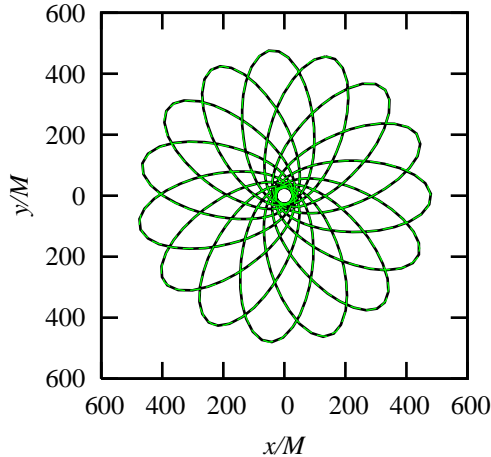}
\end{center}
\caption[Comparison of orbits with time-averaged initial conditions]{The same eccentric orbit as shown in Fig.~\ref{orbits}, but
  now using time-averaged initial conditions for the radiative
  approximation. In this case the approximate orbit is
  indistinguishable from the true orbit on the timescale of the plot
  (fifteen orbital cycles).}  
\label{av_orbit}
\end{figure}

\begin{figure}[tb]
\begin{center}
\includegraphics[trim=0.12in 0in 0.225in 0in,scale=0.925,clip]{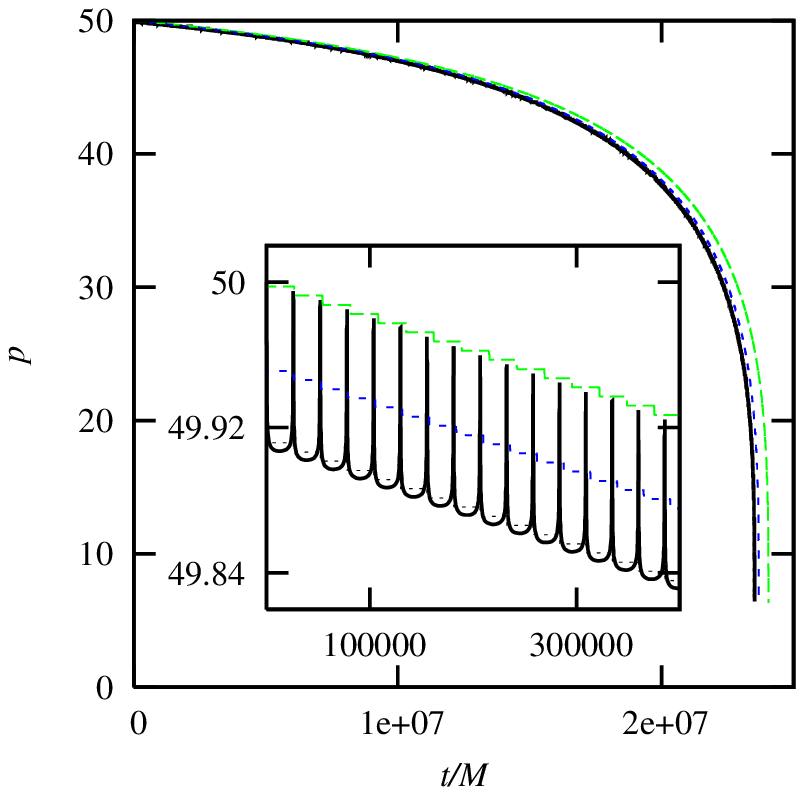}
\includegraphics[trim=0.115in 0in 0.2in 0in,scale=0.925,clip]{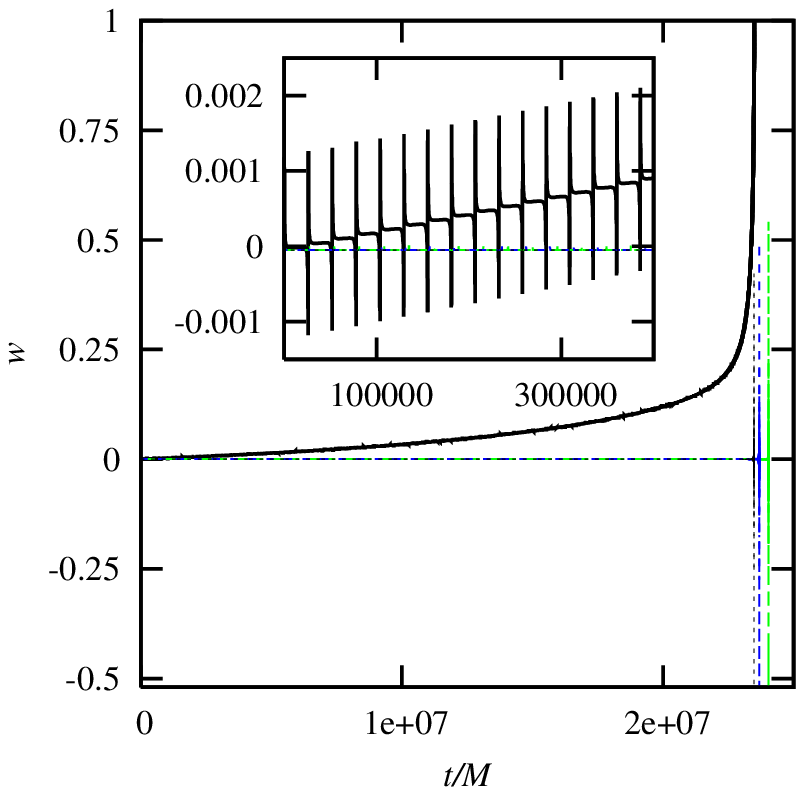}
\end{center}
\caption[Time-evolution of orbital elements]{The principal element $p$ and positional element $w$ as
  functions of time for a complete inspiral, beginning with the
  initial conditions of the eccentric orbit in Fig.~\ref{orbits}. In
  each plot the true curve is in solid black, the radiative curve with
  the same initial conditions is long-dashed in green (the uppermost
  curve in the $p$ plot), the radiative curve with $\chi$-averaged
  initial conditions is short-dashed in blue (middle curve in $p$
  plot), and the radiative curve with time-averaged initial conditions
  is dotted in black (lowest curve in $p$ plot). The insets display
  the early behavior of the curves, covering the same range of time as
  in Fig.~\ref{orbits}.} 
\label{plots_vs_t}
\end{figure}

\begin{figure}[tb]
\begin{center}
\includegraphics{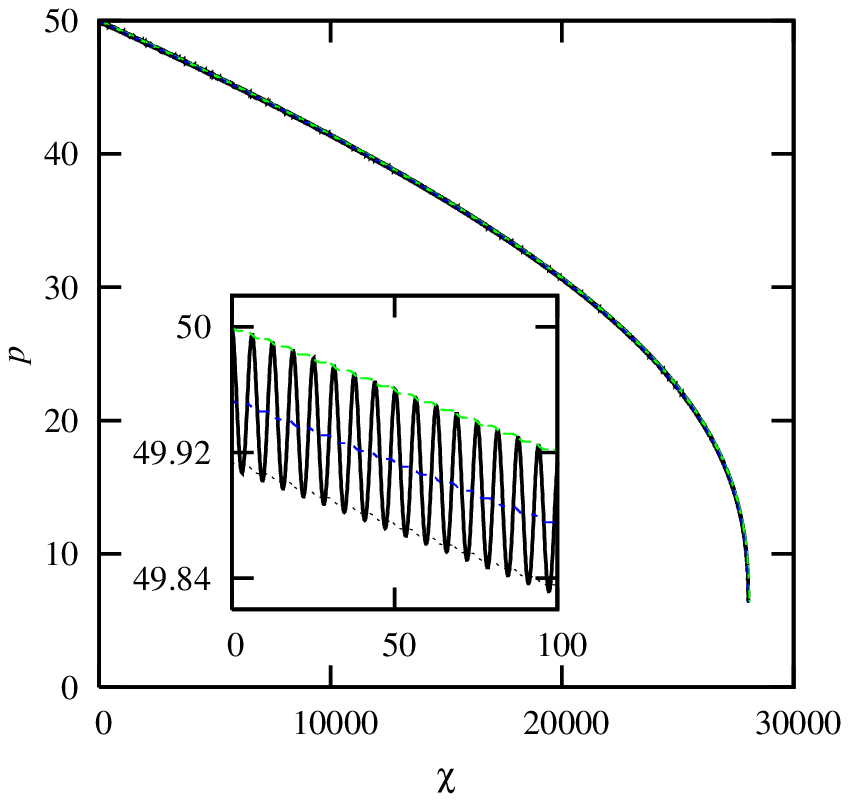}
\end{center}
\caption[Principal element as a function of $\chi$]{The principal element $p$ as a function of the orbital
  parameter $\chi$. The curves are as described in
  Fig.~\ref{plots_vs_t}. The radiative curves do deviate secularly
  from the true curve, but the errors are too small to appear on the
  scale of the graph.} 
\label{plots_vs_chi}
\end{figure}

\subsubsection{Orbital evolution}

A typical inspiral of interest for LISA will form in a highly
eccentric state. Over the course of the inspiral the system will emit
gravitational radiation carrying away energy and angular momentum,
shrinking and circularizing the orbit over time. Thus, the inspiral
will evolve from a highly eccentric orbit to a quasi-circular one,
and it will end in a rapid plunge. We shall now determine the validity 
of the radiative approximation for this class of orbits. Since my
perturbing force is valid only in the post-Newtonian regime, I always  
ensure that $v^2\lesssim 0.1$.  

The general limitations of the radiative approximation are
demonstrated in Fig.~\ref{orbits}, which displays the spatial
trajectories of a highly eccentric orbit and a quasi-circular orbit,
along with corresponding radiative approximations. In each case the
true and approximate orbits are terminated at identical final times,
at which point the radiative approximation lags behind the true
orbit. With a mass ratio of $\mu/M=0.01$, this dephasing of the two
orbits is noticeable after only fifteen radial cycles in the eccentric 
case, while several dozen revolutions are required in the quasi-circular
case. Since the dephasing is apparent before any non-geodesic
precession occurs, it must be caused by conservative effects
in the time-dependence of the orbit. That is, the error in $t(\chi)$
dominates over the errors in $w(\chi)$ and $\phi(\chi)$, such that the
particle lies at the wrong spatial point at a given time, even before
$r(\chi)$ and $\phi(\chi)$ have deviated significantly from the true
orbit. 

For the plots in Fig.~\ref{orbits}, I have chosen exact initial
conditions $I^A(0)$ for the approximate orbit. By choosing averaged
initial conditions we obtain better results in the eccentric case: as
shown in Fig.~\ref{av_orbit}, using time-averaged initial conditions
$\av{I^A}_t(0)$ eliminates the dephasing on the timescale of the
plot. Using $\chi$-averaged initial conditions $\av{I^A}_\chi(0)$ 
results in a smaller improvement, as we will discuss below. However,
in the quasi-circular case all initial conditions fare equally well.\footnote{I note that for both the second and third choices of initial conditions, the
initial value $I_{\mathrm{r}}(0)$ is not fixed by 
$I_{\mathrm{r\ sec}}(0)$ alone, since we also require the initial
value of $I_{\mathrm{r\ osc}}$. Although we do not have 
\textit{a priori} access to this oscillatory part, we can assign it an
approximate initial value based on the results of the radiative
evolution with exact initial conditions. This introduces a negligible
error, since the oscillations in the radiative evolution are extremely
small in practice.}

The evolution of the orbital elements over a complete inspiral,
beginning with the initial conditions of the eccentric orbit in
Fig.~\ref{orbits} and continuing to quasi-circularity, is displayed in
Fig.~\ref{plots_vs_t}. Insets in the plots display the same range of 
time covered by Fig.~\ref{orbits}. The orbit stops before the final
plunge of the small body into the large black hole. There are two
reasons for this truncation. First, my method of osculating orbits
cannot cover the final plunge, because of the underlying restriction
that the orbit must be bounded between a minimum radius $pM/(1+e)$ and
a maximum radius $pM/(1-e)$; this is reflected mathematically by the
condition $p > 6+2e$, which is violated during plunge. Second, we
should in any case leave this portion of the orbit alone, because the
velocities and fields therein are highly relativistic; in this regime
the post-Newtonian expansion of the perturbing force becomes
inaccurate. In Fig.~\ref{plots_vs_t} I display results of the
numerical evolution for the principal element $p$ and positional
element $w$ only; the evolution of $e$ is qualitatively similar to
that of $p$. It is worth noting, however, that the eccentricity never
quite reaches $e \approx 0$; instead, quasi-circularity is manifested
by the condition $\chi - w \approx 0$, which equally well ensures that
$r'\approx 0$. This observation agrees with the results of
Ref.~\cite{Lincoln}.   

The results for all three choices of initial conditions are
plotted in Fig.~\ref{plots_vs_t}. As we see from these plots, the
radiative approximation qualitatively matches the true secular
evolution for the principal element $p$, but neglects all secular
changes in the positional element $w$. This is the expected
result. However, we also see that the radiative approximation deviates
from the true evolution even for the principal element. The extent of
this deviation depends on the choice of initial conditions, with the
time-averaged initial conditions faring the best and exact initial
conditions the worst.

An essential aspect of these results is that the errors in the principal
elements produced by the radiative approximation are mostly due to
errors in $t(\chi)$. As we see in Fig.~\ref{plots_vs_chi}, the errors
almost completely vanish when the principal elements are plotted as
functions of $\chi$; significant errors arise only in the conversion
between $\chi$ and $t$.

\begin{figure}[tb]
\begin{center}
\includegraphics{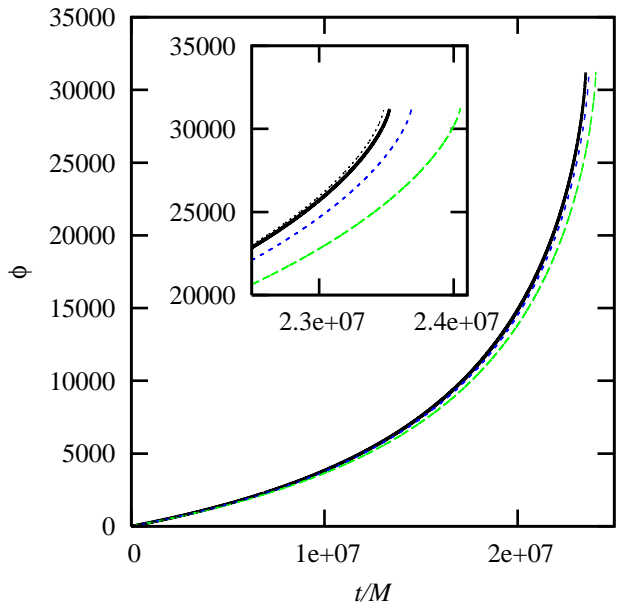}
\end{center}
\caption[Evolution of the orbital phase]{The orbital phase $\phi$, with curves as described in
  Fig.~\ref{plots_vs_t}. The scale of the plot suggests that the
  radiative evolution with time-averaged initial data (the uppermost
  curve in dotted black) gives an accurate approximation of the
  true evolution. The vertical scale, however, is large, and this is a
  false impression. At the late time $t/M = 2.345\times 10^7$, the
  error in phase is $\Delta \phi = 4520$ rad for the exact initial
  data, $\Delta \phi = 1830$ rad for the $\chi$-averaged initial
  data, and $\Delta \phi = 655$ rad for the $t$-averaged initial
  data. This last choice fares best, but its accuracy is poor over a
  complete inspiral.}  
\label{phi_vs_t}
\end{figure}

\begin{figure}[tb]
\begin{center}
\includegraphics[trim=0in 0.24in 0in 0.12in, angle=-90,clip]{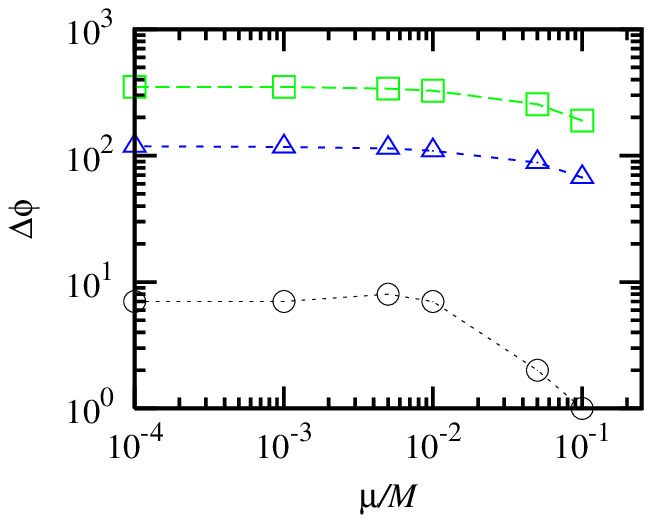}
\includegraphics[trim=0in 0.24in 0in 0.12in, angle=-90,clip]{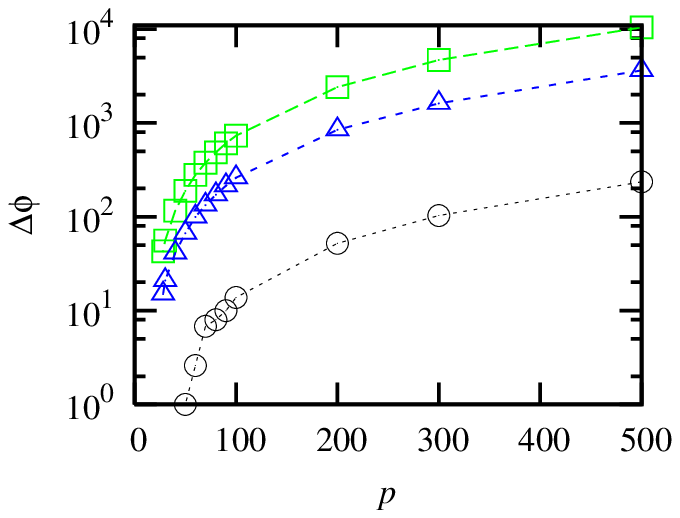}
\includegraphics[angle=-90]{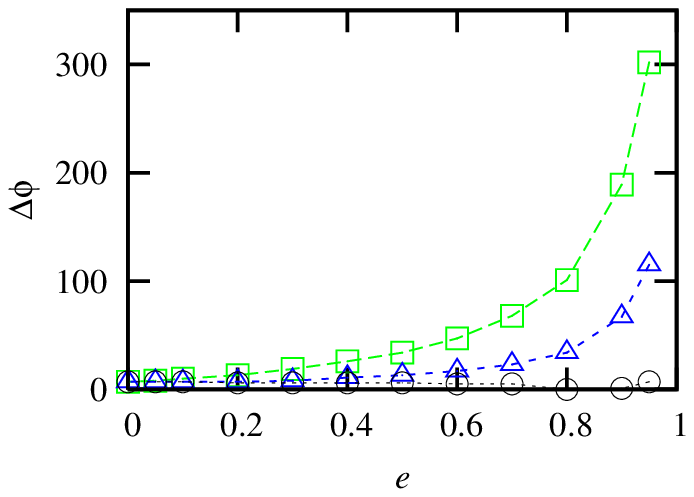}
\end{center}
\caption[Errors in the orbital phase]{The difference in orbital phase $\phi$ between the true orbit
  and approximate orbits after a radiation-reaction time (defined by
  $p\to 0.9p_0$ on the true orbit). Open squares indicate results for
  identical initial conditions, open triangles for matching
  $\chi$-averaged initial conditions, and open circles for matching
  $t$-averaged initial conditions. Top left: dephasing as a function of the
  mass ratio $\mu/M$, with fixed true initial values $p_0=50$ and
  $e_0=0.9$. The dephasing becomes $\mu$-independent for sufficiently
  small $\mu$, when second-order effects become negligible. Top right:
  dephasing as a function of initial value $p_0$, with fixed $e_0=0.9$
  and $\mu/M=0.1$. Bottom: dephasing as a function of initial
  eccentricity $e_0$, with fixed $p_0=50$ and $\mu/M=0.1$. The error
  in the case of time-averaged initial conditions is approximately
  independent of $e$.} 
\label{dephasing}
\end{figure}

\subsubsection{Errors in orbital phase}
The errors in which I am most interested are errors in orbital
phase, since they will lead directly to errors in the phase of the
emitted gravitational radiation. Figure~\ref{phi_vs_t} displays the
phase $\phi$ versus time, again using all three choices of initial
conditions for the radiative approximation. Once again we see that the
time-averaged conditions produce the smallest error, for the same
reasons described in the previous section. 

Figure~\ref{dephasing} shows the dependence of the dephasing
$\Delta\phi=\phi-\phi_{\mathrm{rad}}$ on the parameters of the
problem. I plot the dephasing for a ``radiation-reaction" time
defined by $p\to 0.9p_0$, rather than a complete inspiral, since
gravitational-wave data analysis may require only a portion of a complete inspiral. We see that the
dephasing is independent of $\mu$ for sufficiently small values of
$\mu$. This is an expected result, since the radiation-reaction time
at leading order in $\mu$ varies as $1/\mu$, while the rate of
dephasing varies as $\mu$, leading to a net cancellation in the
total dephasing. However, terms in the perturbing force that are
quadratic in $\mu$ alter this result when $\mu/M$ is sufficiently
large. Somewhat surprisingly, these quadratic terms actually serve to
decrease the dephasing, lowering the impact of conservative terms in
the force.

As expected, the dephasing decreases at lower values of $e$, although 
the eccentricity seems to have negligible impact in the case of
time-averaged initial conditions. Also as expected, the dephasing
varies as $p^{3/2}$, regardless of initial conditions. This scaling
follows from the form of the post-Newtonian force: the leading-order
conservative term enters at 1PN order, which scales as a $p^{-1}$
correction to Newtonian gravitation, while the leading-order
dissipative term enters at 2.5PN order, which scales as a
$p^{-5/2}$ correction. The dephasing is governed by the relative
strength of the conservative terms, leading to a scaling of
$p^{-1}/p^{-5/2}=p^{3/2}$.

In all cases the time-averaged initial conditions yield the best
results. Indeed, the efficacy of these initial conditions is almost
surprising. One way of understanding their impact is to examine the
insets in Fig.~\ref{plots_vs_t}. Peaks in the true curve correspond to
the short periods of time near periapsis, while relatively flat
regions correspond to the long periods of time 
around apoapsis. Thus, choosing exact initial conditions matches the
true and approximate orbits for the minimal amount of time, as well as
in the region of strongest fields, leading to the largest possible
deviation. Choosing time-averaged initial conditions matches the
orbits near apoapsis, for the longest time and with the weakest
fields, leading to the least possible deviation. The $\chi$-averaged
initial conditions are then in some sense the average of all the
incorrect choices. An implication of this is that in some
circumstances the $\chi$-averaged initial conditions could turn out
to be even worse than the exact initial conditions. For example,
choosing exact initial conditions at apoapsis would closely
approximate the time-averaged initial conditions, which would then
fare much better than the $\chi$-averaged initial conditions.

We can understand the long-term impact of the initial conditions by
considering the time-dependence of an orbit. The secular time function 
$\av{t}(\chi)$ can be written in terms of the orbital period 
$P(\chi)=\int_{\chi-\pi}^{\chi+\pi}t'(\tilde\chi)d\tilde\chi$
as $\av{t}(\chi)=\int_0^\chi P(\tilde\chi)d\tilde\chi$. As we see
from the insets in Fig.~\ref{plots_vs_t}, the changes in initial
conditions bring the initial orbital period of the radiative
approximation closer to that of the true orbit; and as we would
intuitively expect, the time-averaged initial conditions best
reproduce the initial temporal period. This correction, $\delta P$,
to the initial period then induces a long-term correction to
$\av{t}(\chi)$ of the form $\delta\av{t}\sim\chi\cdot\delta P$. In essence, the time-averaged initial conditions
carry information about the initial conservative correction to the
true orbital period, and they thus implicitly insert a conservative
correction into the radiative approximation. This serves to remind us 
that we would have difficulty choosing suitable initial conditions
for the radiative approximation if we did not have prior access to
the true evolution. 

Regardless of the choice of initial conditions, one should note that
the errors accumulated over a complete inspiral are much larger than
those shown in Fig.~\ref{dephasing}. (Refer to the caption of 
Fig.~\ref{phi_vs_t} for actual values.) Also, the plots of
$\Delta\phi$ versus $p$ and $e$ are for $\epsilon=0.1$, leading to a
smaller dephasing than would occur if $\epsilon$ were in the region of
linear dominance. Thus, even if ideal initial conditions could be
found without reference to the exact solution, the radiative
approximation would generically fail over a complete inspiral.

			\chapter{Summary and conclusions}\label{conclusion}
In this dissertation I have discussed a variety of approximation schemes in the gravitational self-force problem. In particular, I have emphasized the utility of singular perturbation techniques, which can systematically overcome the limitations of regular expansions. However, these perturbation techniques are not simply powerful tools for finding asymptotic solutions to differential equations: they have a rich underlying geometrical structure, which I have sketched, but which warrants further study. This underlying structure allows us to conceptualize, for example, the meaning of the representative worldline of a black hole. 

Furthermore, in the transition from traditional singular perturbation theory into the diffeomorphism-invariant realm of GR, subtleties arise in the application of singular perturbation techniques. In particular, two types of matching conditions can be formulated in the method of matched asymptotic expansions, and the condition that has been implicitly used in previous derivations of the gravitational self-force is significantly weaker than the condition used in applied mathematics. In order to arrive at unique results with this matching condition, additional assumptions must be made, which weakens the conclusions of a matching calculation. 

Of course, my principal use of singular perturbation techniques has been to formalize an asymptotic expansion that holds the worldline of a small body fixed. This allows one to construct long-term, self-consistent approximations of the motion of the body. My comparison of this approach to a regular expansion, in which the worldline of the body must be expanded as a power series, illuminates the shortcomings of some earlier derivations, and the underlying assumptions in others. In particular, the comparison has made clear that the tail integral should not extend into the infinite past, especially if a regular expansion is used.

But enough about subtleties, formulations and clarifications. I will now summarize what has been accomplished, and what remains to be done.

\section{The self-consistent gravitational self-force}
The core of this dissertation is its new derivation of the gravitational self-force for a small body. The derivation is based on the familiar technique of using two expansions of the metric: an inner expansion that is more accurate near the body, and an outer expansion that is more accurate far from the body. However, unlike in earlier derivations, I have formulated these expansions in terms of a fixed worldline $\gamma$ defined in the external background spacetime. The self-consistent equation of motion of this worldline then follows directly from solving the Einstein equation. When combined with the first-order metric perturbation, the equation of motion defines a solution to the Einstein equation accurate up to order $\e^2$ errors over times $t\lesssim1/\e$. When combined with the second-order perturbation, it defines a solution accurate up to order $\e^3$ errors on the shorter timescale $\sim 1$.

My approach began with a general analysis of the Einstein equation, up to second order in the body's mass, in a buffer region around the small body. Since the buffer region is assumed to be free of matter, my calculation is valid only for bodies that are sufficiently compact to avoid tidal disruption. An equation of motion for the body's worldline was derived from the condition that the body must possess no mass dipole in coordinates centered on the worldline. From this purely local-in-space analysis, we found an expression for the acceleration in terms of irreducible pieces of a homogeneous solution to the wave equation---the Detweiler-Whiting regular field, which is regular on the worldline. This homogeneous, regular field was not determined by the buffer-region expansion, since it can be determined only by boundary conditions.

A formal expression for the metric perturbation was obtained by casting the Einstein equation in a relaxed form, via the imposition of the Lorenz gauge. This relaxed form can be solved iteratively, with the perturbation at each order given by the sum of (1) an integral over a region outside the body and (2) an integral over an initial data surface and a worldtube surrounding the body. Boundary data on the worldtube are provided by the buffer-region expansion. At first order, it can be shown that the integral representation is identically equal to the perturbation produced by a point particle moving on $\gamma$. At higher orders, because of the increasing singularity of the metric perturbation, only parts of it can be simplified in the same way. Because of this limitation, I introduced a method of direct integration. In this method, the Detweiler-Whiting regular field in the neighbourhood of the body is determined, in terms of initial conditions and the body's past history, by expanding the integral representation in the buffer region and demanding its consistency with the boundary data on the tube.

An essential assumption in this derivation is that the acceleration of the fixed, $\e$-dependent worldline possesses an asymptotic expansion beginning in powers of $\e$. This is required to split the Lorenz gauge condition (or the Bianchi identity) into a sequence of exactly solvable equations. It also automatically results in an order-reduced equation of motion, with no spurious runaway solutions. In other words, the requirement that the perturbation equations be exactly solved---rather than approximately solved, as in the traditional gauge-relaxation procedure (or in traditional approaches to the electromagnetic self-force)---also necessarily eliminates the need for an a posteriori order-reduction procedure.

In addition, I made the following assumptions: the exact metric possesses asymptotic expansions of the form given in Sec.~\ref{outline}, there is a smooth coordinate transformation between some internal local coordinates and the external Fermi coordinates in a neighbourhood of the worldtube, the Lorenz gauge condition can be imposed everywhere in the region of interest, and the expansion of the metric perturbation satisfies both the wave equations and (when combined with the expansion of the acceleration) the gauge condition at fixed functional values of the worldline $z^\mu(t)$. While these, especially the last, are strong assumptions, they undoubtedly lead to an eminently useful, systematic approximation scheme. It is worth repeating that while the choice of gauge is not essential in finding an expression for the force in terms of the field in the buffer-region expansion, it \emph{is} essential in my method of determining the field itself. Without making use of the relaxed Einstein equations, no clear method of globally solving the Einstein equation presents itself.

One fruitful avenue of further research might be to explore methods of solving the Einstein equation in alternative gauges but still within the context of a general expansion that holds $\gamma$ fixed. This might require further thought on the behavior of gauge transformations in such an expansion. However, such details of the formalism are most likely to be made sense of not at the level of the field equations, but at the level of the action. Since partial derivatives do not act directly on the worldline, the dependence of the metric perturbations on the worldline does not appear directly in the field equations. But at the level of the action, using functional derivatives, the role of the worldline becomes transparent.

It is also worth noting that the methods used here would work in many other cases. For example, the direct calculation of the boundary integral can be used to completely determine the force even if the source cannot be represented as a distribution. Also, these methods could be used to derive self-consistent equations of motion for a charged body; the expansion of the acceleration in powers of $\e$ would automatically yield an order-reduced equation of motion, with no runaway solutions (c.f. the recent calculation by Gralla, Harte, and Wald \cite{Gralla_Harte_Wald}).

\subsection{Comparison with alternative methods}
One of my goals was to construct an approximation scheme that closely mirrors the extremely successful methods of post-Newtonian theory \cite{DIRE, Futamase_review, Blanchet_review, PN_matching, Futamase_particle1, Futamase_particle2, Racine_Flanagan}. As such, many of the methods used here are similar to those used in post-Newtonian expansions. For example, the expansion with a fixed worldline meshes well with the use of the relaxed form of the Einstein equation \cite{relaxed_EFE1,relaxed_EFE2}, which can be solved without specifying the motion of the source, and which is the starting point for post-Minkowski and post-Newtonian expansions. And the use of an inner limit near the body corresponds to the use of the ``strong-field point particle limit" used by Futamase \cite{Futamase_particle1,Futamase_particle2}. In addition, the calculation of the motion of the body in this dissertation is somewhat similar to the methods used by Futamase and others~\cite{Futamase_review, Futamase_particle1, Futamase_particle2, PN_matching, Racine_Flanagan}, in that it is based on a multipole-expansion of the body's metric in the buffer region. Finally, the direct integration of the relaxed Einstein equation mirrors the approach of Will et al.~\cite{DIRE}.

There are, of course, differences between the two cases. In particular, when the finite size of the body is taken into account in post-Newtonian theory, because the background is flat, finding an equation of motion for the mass dipole of the body is equivalent to finding an equation of motion for its worldline. Although this method, or methods similar to it, has also been used in curved spacetimes \cite{Thorne_Hartle, Mino_matching, Fukumoto}, it is somewhat problematic because the mass dipole corresponds to a displacement from the center of a given coordinate system. But in a curved spacetime, such a displacement is meaningful only when it is infinitesimal. Of course, if at a given instant the coordinate system is mass-centered, then the second time-derivative of the mass dipole is equivalent to the acceleration of the worldline; but since there is no unique global time in a curved spacetime, it is more meaningful to speak of a curve about which the body is centered for its entire history, rather than just at a given time.

A more significant goal in this dissertation was to develop a unified and self-consistent formalism to treat the gravitational self-force problem. Because the problem consists of solving singular perturbation equations, I have emphasized the foundation of the formalism in singular perturbation theory. Because the formalism uses a self-consistent worldline and a finite sized body, it is (potentially) valid on both short and long timescales, and both very near to and far from the small body. As such, it can be used to study (or incorporate studies of) the spacetime near the small body, the long-term motion of the body, and the perturbations produced by it, including the gravitational waves emitted to infinity.

This contrasts with the most recent derivation of the self-force, performed by Gralla and Wald \cite{Gralla_Wald}. In terms of the concrete calculation of the force in the buffer region, my calculation is very similar to theirs, though it differs in many details. (One such difference is that the perturbation I derive satisfies the Lorenz gauge at all orders in $r$ in the local expansion, whereas Gralla and Wald do not impose the Lorenz gauge on the most singular, order-$\e^2/r^2$, term in their calculation.) However, their approach constructs a regular expansion in which both the worldline and the metric perturbation are expanded; they suggest that in order to arrive at a self-consistent set of equations, one must make a ``leap of faith" from the results of their regular expansion. I instead take the stance that the self-consistent equation of motion can, and should, be justified by a more systematic approach; and I have presented one such approach in this dissertation. From the results of this approach, one can easily derive the results of the regular expansion: simply by expanding the $\e$-dependent worldline, one derives a leading-order metric perturbation sourced by a particle on a geodesic (plus secularly growing corrections); and the usual steps involved in deriving the geodesic-deviation equation leads to an equation of motion for the deviation vector ``connecting" the geodesic to the exact worldline. Contrariwise, one cannot derive the results of the general expansion from those of the regular expansion.

Other methods have been developed (or suggested) to accomplish the same goals as my own. One such method is the two-timescale expansion suggested by Hinderer and Flanagan \cite{Hinderer_Flanagan}. As discussed in Secs.~\ref{multiple_scales_GR} and \ref{singular_expansion_point_particle}, their method continuously transitions between regular expansion, resulting in a global, uniform-in-time approximation. One should note that simply patching together a sequence of regular expansions, by shifting to a new geodesic every so often using the deviation vector, would not accomplish this: such a procedure would accumulate a secular error in both the metric perturbation and the force, because the perturbation would be sourced by a worldline secularly deviating from the position of the body, and the force would be calculated from this erroneous perturbation. The error would be proportional to the number of ``shifts" multiplied by a nonlinear factor depending on the time between them. And this error would, formally at least, be of the same magnitude as the solution itself.\footnote{However, such a method would be very similar to one from celestial mechanics known as Encke's method, which uses a deviation vector pointing away from an initial reference orbit, and then switches to new reference orbit (a process called ``rectification") every so often. Although Encke's method is an exact reformulation of the equations of motion, and hence not prone to the same types of errors, it might be a useful point of reference.}

The fundamental difference between the fixed-worldline method and the two-timescale method is the following: In the two-timescale method, the Einstein equation, coupled to the equation of motion of the small body, is reduced to a dynamical system that can be evolved in time. The true worldline of the body then emerges from the evolution of this system. In the method presented here, I have instead sought global, formal solutions to the Einstein equation, written in terms of global integrals; to accomplish this, I have treated the worldline of the body as a fixed structure in the external spacetime. The two timescale method is, perhaps, more practical for concrete calculations, while the global solutions presented here are primarily of formal interest. However, the two methods should agree. Note, though, that Hinderer and Flanagan have identified transient resonances in EMRI systems, which lead to half-integer powers of $\e$ in their asymptotic expansions. It is not clear that such effects are correctly accounted for in the method presented in this dissertation.

%%%%%%%%%%%%%%%%%%%%%%%%%%%%%%%%%%%%%%%%%%%%%%
\subsection{Prospects for a global solution}
%%%%%%%%%%%%%%%%%%%%%%%%%%%%%%%%%%%%%%%%%%%%%%
The principal practical goal of solving the self-force problem is to find the waveform emitted from an EMRI. In order to extract the parameters of an EMRI system from its waveform, we must have a model that tracks the wave's phase to within an error of order $\e$ over a time period $1/\e$. This presents several problems.

First among these problems is the potential for secular errors. For example, secular errors might arise due to ignoring the slow evolution of the background spacetime. Throughout this dissertation, I have assumed that the external background metric is $\e$-independent. However, in practice, it might possess a slow time dependence that would account for the backreaction of the perturbations on the background spacetime; for example, in an EMRI, the large black hole's absorption of gravitational waves slowly alters its mass and spin parameters. Any such effect leaves the expression for the self-force unchanged, and it can be easily incorporated into the perturbations presented here. However, an equation for the slow evolution itself is unknown. Presumably, it can be determined from an averaged version of the Einstein equation, of the form $\av{E_{\mu\nu}[h]} = 2\av{R_{\mu\nu}}+2\av{\delta^2R_{\mu\nu}} + ...$. In an EMRI system, the average of the wave operator will most likely vanish, because the body's orbit is quasi-periodic. The averaged equation will then relate $\delta^2 R$ to the background Ricci tensor $R$, as in the pioneering work of Isaacson \cite{Isaacson}; this corresponds to the effect of quadrupole radiation on the background. In practice, the averaged equation might be solved by using some ansatz for the background metric---e.g., the Kerr metric with slowly varying mass and spin parameters. The feasibility of such a calculation is unclear; the need to perform it will most likely be determined by examining the magnitude of secular growth in a solution that ignores backreaction. It is worth noting that in a two-timescale expansion, the slow-evolution of the background appears naturally as an additional first-order perturbation, on top of the point-particle perturbation \cite{Hinderer_Flanagan}; it is not clear whether this effect is naturally incorporated in the self-consistent scheme. See Refs.~\cite{Galley_backreaction, Hinderer_Flanagan} for more information on the backreaction in the self-force problem.

Putting aside the backreaction problem, other secular errors will also arise due to neglected terms in the acceleration and metric perturbation. Although the approach taken in this dissertation is designed to avoid such errors, a concrete implementation will nevertheless contain them. I have defined the worldline as a fixed curve; proceeding to successively higher orders in perturbation theory yields successively more accurate equations of motion for this curve. However, if we stop at any given order and use any given equation of motion, then the worldline based on that equation of motion will deviate secularly from the true worldline. This in turn implies that the metric perturbation will accumulate secular errors.

Hence, we must have an equation of motion that limits these errors to $\order{\e}$ after a time $1/\e$. If we use the first-order equation of motion, we will be neglecting an acceleration $\sim\e^2$, which will lead to secular errors of order unity after a time $1/\e$. Thus, the second-order self-force is required in order to obtain a sufficiently accurate waveform template.\footnote{Proceeding to second order will also be useful for examining other systems, such as intermediate mass ratio binaries, over shorter timescales.} In order to achieve the correct waveform, we must also obtain the second-order part of the metric perturbation; this can be easily done, at least formally, using the global integral representations outside a worldtube. A practical numerical calculation may prove difficult, however, since one would not wish to excise the small tube from one's numerical domain.

A formal expression for the second-order force has already been derived by Rosenthal~\cite{Eran_field, Eran_force}. However, he expresses the second-order force in a very particular gauge in which the first-order self-force vanishes. This is sensible on short time scales, but not on long timescales, since it forces secular changes into the first-order perturbation, presumably leading to the first-order perturbation becoming large with time. Furthermore, it is not a convenient gauge, since it does not provide what we wish it to: a correction to the nonzero leading-order force in the Lorenz gauge.

Thus, we wish to obtain an alternative to Rosenthal's derivation. Based on the methods developed in this dissertation, there is a clear route to deriving the second-order force. One would construct a buffer-region expansion accurate up to order $\e^3$. Since one would require the order $\e^2 r$ terms in this expansion, in order to determine the acceleration, one would need to increase the order of the expansion in $r$ as well. Specifically, one would need terms up to orders $\e^0 r^3$, $\e r^2$, $\e^2 r$, and $\e^3 r^0$. These could be calculated using the methods presented in this dissertation. In such a calculation, one would expect the following terms to appear: the body's quadrupole moment $Q_{ab}$, corrections $\delta M_i$ and $\delta S_i$ to its mass and spin dipoles, and a second-order correction $\delta^2 m$ to its mass. Although some ambiguity may arise in defining the worldline of the body at this order, a reasonable definition appears to be to guarantee that $\delta M_i$ vanishes. However, at this order one may require some model of the body's internal dynamics, since the equation of motion will involve the body's quadrupole moment, for which the Einstein equation may not yield an evolution equation. But if one seeks only the second-order self-force, one could simply neglect the quadrupole by assuming a Schwarzschild black hole. In any case, the force due to the body's quadrupole moment is already known from various other methods. (See, e.g., the work of Dixon~\cite{Dixon,Dixon1,Dixon2}; more recent methods can be found in Ref.~\cite{multipole_motion} and references therein.)

Unfortunately, beyond these potential difficulties, the calculations involved in such a procedure could be prohibitively lengthy. Hence, we might consider a much simpler alternative: the method of matched asymptotic expansions. Using this method, we would need the buffer region expansion to be accurate to order $\e^2 r$---the order at which the second-order acceleration appears in the background metric---meaning that we would need to extend the buffer region expansion by one order in $r$, but not in $\e$. The equation of motion would then be determined by finding a unique coordinate transformation that makes the external and internal solutions identical in the buffer region. As was discussed in Ch.~\ref{matching}, this method is somewhat problematic. However, it should be possible to overcome its problems, and a calculation of the second order force by this means is entirely feasible. Even if such a calculation must be viewed as assuming, rather than proving, a generalized equivalence principle, it would still serve to determine the explicit form of the second-order force in terms of the first- and second-order metric perturbations.

However, even if we can obtain an approximation with the desired accuracy on the timescale $1/\e$, there remains at least one additional difficulty. The waveform itself is to be calculated at future null infinity, $\mathscr{I}^+$. At first glance, it might seem that we can extend the size of our domain $\Omega$ such that its future null boundary $\mathcal{J}$ is pushed out to $\mathscr{I}^+$ at one end and to the event horizon of the large black hole at the other. However, the size of our domain is intended to be of size $1/\e$. Thus, it cannot be trivially enlarged to infinity. If we wish to enlarge $\Omega$, we must match the solution within it to an outgoing wave solution at its future null boundaries. This essentially amounts to introducing correct initial data on an infinite initial timeslice, a notoriously insoluble problem. The concrete impact of this problem, however, is probably minor.

\section{Relativistic celestial mechanics}
Of course, the goal of the self-force research program is not only to make accurate predictions about the waveforms generated by EMRIs, but to learn something about the orbits of small bodies. Hence, we require some useful means of analyzing accelerated orbits in black hole spacetimes. To this end, in this dissertation I presented a relativistic generalization of the method of osculating orbits, an historically important method of Newtonian celestial mechanics. In this method, the true orbit is parametrized as a smooth transition between tangential geodesics; the orbital parameters of the family of geodesics become functions of time, and their evolution serves to characterize the evolution of the orbit. I implemented this method in the case of bound, accelerated orbits in Schwarzschild spacetime. From the equations in Schwarzschild, we can also recover the results for Keplerian orbits in the Newtonian limit. Because of the simple parametrization of orbits in this method, it provides an attractive conceptual and mathematical foundation for a perturbative approach to weakly accelerated orbits. Furthermore, it is easy to implement in practice in a numerical code.  

I have demonstrated the usefulness of the method in two test cases. First, I used the Newtonian evolution equations to analyze the motion of a charged particle in a weakly curved spacetime. In that case, the method, in conjunction with a two-timescale expansion, allowed me to precisely characterize the effects of the electromagnetic self-force on the orbital evolution. In particular, it allowed me to isolate the short- and long-term dissipative and conservative effects of the force. Next, I used the fully relativistic Schwarzschild evolution equations to analyze the evolution of a post-Newtonian binary. The perturbing force in this case was the hybrid Schwarzschild/post-Newtonian equations of motion of Kidder, Will, and Wiseman \cite{Kidder}; the fact that the method is suitable for PN binaries, rather than solely for orbits in a true Schwarzschild spacetime, shows its flexibility. And again, the method proved itself to be an excellent means of characterizing the effects of the self-force.

In both of these test cases, the secular impact of the conservative part of the self-force was marked. That impact revealed itself both in the form of precession, as evinced by changes in the argument of periapsis, $w$, and in direct changes in the orbital period. Both effects lead to large secular changes in orbital phase. In the test cases, the direct change in the orbital period has a much larger effect than does the orbital precession, though that might not be a generic feature. In addition to these effects, my analysis has shown that the long timescales involved in an inspiral lead to a large, long-term impact of initial conditions.

\section{Adiabatic approximations}
An adiabatic approximation, which uses asymptotic information about wave amplitudes in order to update a trajectory, provides the hope of bypassing lengthy, challenging numerical computations of orbits and waveforms directly from the equations of motion. However, this approximation assumes that those asymptotic wave amplitudes can be calculated as if the particle moved on a geodesic. As I have argued in my analysis of the fixed-worldline expansion, this approximation may be subject to large errors. The approximation is also a radiative approximation, meaning that it discards the conservative effects of the self-force. As I have shown in a multiscale analysis of the Newtonian osculating orbit equations, as well as in numerical integration of post-Newtonian binaries, the conservative part of the self-force generically causes a significant shift in the orbital period, leading to large long-term errors in a radiative approximation. Lastly, the adiabatic approximation neglects the periodic effects of the self-force. This leads to erroneous choices of initial data that cause large secular errors; it also prevents the approximation from systematically incorporating higher-order effects, since periodic effects at first order are required to calculate secular effects at second order.

The test cases that I have presented differ in many respects
from the fully relativistic self-force problem, but they nevertheless
capture many of its essential features. My conclusions, therefore,
might be expected to hold in the fully relativistic case. 

However, the equations of motion that I used in analyzing PN binaries were
calculated within the harmonic gauge of post-Newtonian theory, and the
magnitudes of the conservative effects that I have displayed refer to
this particular gauge choice; different gauges would necessarily lead
to different results. Indeed, Mino has argued in favor of constructing
a ``radiation-reaction gauge'' in which the
conservative effects of the self-force are set to zero over a finite
radiation-reaction time, making the radiative approximation exact over 
that interval \cite{Mino_expansion1, Mino_expansion2}.\footnote{Mino has also argued that his gauge choice induces a change in initial conditions that
partially absorbs conservative effects \cite{Mino_expansion1}, and this
statement agrees with my result that long-term conservative effects
can be mimicked by a small change in initial conditions.} More precisely, Mino has argued that any gauge which preserves the average rate of change of the principal orbital elements should be physically and mathematically sufficient to determine long-term results, because it will preserve the gauge-invariant fluxes at infinity. Hence, he argues that the conservative part of the self-force can be gauged away without losing any long-term accuracy. However, the formula for the average rate of change of the Carter constant, for example, is derived within the Lorenz gauge, using the radiative Green's function for the wave operator $E_{\mu\nu}$. And the force within the Lorenz gauge includes a conservative piece. In order to justify neglecting conservative effects, one would need to explicitly construct the metric perturbation in a ``radiation-reaction"  gauge and show that it is well-behaved; one would need to verify that this gauge is related to the Lorenz gauge by an appropriate transformation (\emph{i.e.}, one generated by a vector $\e\xi\coeff{1}$ that is bounded on the worldline, and which satisfies $\xi\coeff{1}[\gamma]=O_s(1)$ uniformly); and one would need to derive new formulas for the averaged rate of change of the principal orbital elements as functions of asymptotic wave amplitudes and the worldline. Since none of this has been accomplished, we can conclude that the notion of a radiation-reaction gauge, and its usefulness in implementing an adiabatic approximation, is highly tentative. And outside of a finely-tuned gauge choice, one should expect the conservative part of the self-force to produce large secular effects.

I conclude then, that an adiabatic approximation would have to be reformulated in order to provide a model that is sufficiently accurate to extract parameters from EMRI waveforms, though it might provide a model that is sufficiently accurate to detect those waveforms. However, as was shown in the multiscale analyses of Ref.~\cite{Hinderer_Flanagan} and Sec.~\ref{multiscale_osculating}, if the full first-order force is known, then an adiabatic approximation of the second-order force would probably be sufficient for parameter-extraction.

\section{Conclusion}
Throughout this dissertation, I have taken the stance that finding a useful approximate solution to the exact Einstein equation, such as that provided by singular perturbation theory, is more important than finding an exact solution to the approximate Einstein equation, such as that provided by regular perturbation theory. In the gravitational self-force problem, a useful approximate solution is one that remains valid on long timescales, self-consistently incorporates the acceleration of the small body, and accounts for its asymptotically small, but finite, size. The fixed-worldline approximation scheme promises to satisfy these criteria, and it can be systematically extended to any order in perturbation theory.

However, I have also taken the stance that a solution to an approximate equation must be an approximation to an exact solution if it is to render a meaningful test of General Relativity. As such, I have emphasized how the general expansions developed in this dissertation might be related to an exact solution. A far more rigorous, technical, and perhaps altogether unfeasible study would be required to show whether or not the asymptotic solution developed here actually does approximate an exact solution. From this perspective, even well-known results, such as the statement that a test mass moves on a geodesic, are in truth only hypotheses, since the derivations of them rely on the assumed existence of a family of spacetimes with certain properties \cite{Rendall_review,geodesic_hypothesis}. However, existence proofs, even in nonlinear theories such as GR, are certainly possible, and there is no reason not to seek them.

Of course, even if the fixed-worldline solutions are proven to be asymptotic approximations, they remain purely formal. A practical calculation of the motion of a small body will most probably require a numerical implementation, which will require a formulation of the wave equation, coupled to an equation of motion for the source, that is viable for numerical calculations. Unfortunately, it will be difficult to implement such calculations with sufficient accuracy for parameter estimation in an EMRI, which require not only highly precise numerical methods but also higher-order analytical results from perturbation theory. In the meantime, adiabatic approximations offer a means of generating waveform templates for gravitational-wave detection. But if we wish to glean reliable information from those waveforms, there remains much to be done.
			\appendix
			\chapter{Multiscale expansions}\label{multiscale_EFE}
In this appendix, I present an illustrative example of multiscale expansions, along with a resultant discussion of their utility. I then present a concrete calculation of the effects of the electromagnetic self-force in a weak central gravitational field. I conclude by sketching a multiscale expansion of the Einstein equation, beginning with a multiscale expansion of the metric.

\section{Illustrative example}
I consider the following differential equation, adapted from the text by Kevorkian and Cole~\cite{Kevorkian_Cole}
\begin{equation}\label{DE1}
\ddiff{f}{t}+2\e\diff{f}{t}+f=0, \qquad f(0,\e)=0,\ \diff{f}{t}(0,\e)=1.
\end{equation}
Suppose we wish to solve this problem using a regular power series $f(t,\e)=\sum_{n\ge 0}\e^nf\coeff{\emph{n}}(t)$. After substituting this series and equating powers of $\e$, we arrive at the sequence of equations
\begin{equation}
\begin{array}{lll}
\displaystyle\ddiff{f\coeff{0}}{t}+f\coeff{0} & = 0, & \qquad f\coeff{0}(0)=0,\ \displaystyle\diff{f\coeff{0}}{t}(0)=1, \vspace*{5.5 pt}\\ 
\displaystyle\ddiff{f\coeff{1}}{t}+f\coeff{1} & = -2\displaystyle\diff{f\coeff{0}}{t}, & \qquad f\coeff{1}(0)=0,\ \displaystyle\diff{f\coeff{1}}{t}(0)=0.
\end{array}
\end{equation}
The solutions to these equations are easily found to be $f\coeff{0}(t)=\sin t$ and $f\coeff{1}(t)=-t\sin t$, so we have 
\begin{equation}\label{regular}
f(t,\e)=\sin t-\e t\sin t+...
\end{equation}
Based on the unbounded growth of this solution, we surmise that it fails to uniformly approximate the exact solution on any unbounded interval $[0,1/\e^p]$, $p>0$.

To improve on this solution, I adopt the following assumption: there exists a function $F(t,\tilde t,\e)$ satisfying the equality $F(t,\tilde t=\e t, \e)=f(t,\e)$. Substituting this into Eq.~\eqref{DE1} and making use of the chain rule $\frac{d}{dt}=\frac{\partial}{\partial t}+\e\frac{\partial}{\partial \t}$, we arrive at
\begin{equation}\label{DE2}
\pddiff{F}{t}+F+2\e\left(\frac{\partial^2F}{\partial\tilde t\partial t}+\pdiff{F}{t}\right)+\e^2\left(\pddiff{F}{\tilde t}+2\pdiff{F}{\tilde t}\right)=0.
\end{equation}
Now, the fundamental idea in a multiscale expansion is that the function $F$ satisfies this equation not just when $\tilde t=\e t$, but also when $\tilde t$ is treated as an independent coordinate. This means that if one assumes a regular expansion $F(t,\tilde t,\e)=\sum_{n\geq 0}\e^n F\coeff{\emph{n}}(t,\tilde t)$, then the coefficient of each power of $\e$ in Eq.~\eqref{DE2} must vanish; if we insisted on solving the equation only at $\tilde t=\e t$, then the $\e$-dependence embedded in $\tilde t$ would prevent us from concluding that the equation must be satisfied order-by-order in this way. (Of course, we could always solve the equation by setting the coefficient of each power of $\e$ to zero, but we could not deduce that each coefficient \emph{must} vanish.)

So, following this procedure, we arrive at a new sequence of equations 
\begin{align}
\pddiff{F\coeff{0}}{t}+F\coeff{0} &= 0, \\
\pddiff{F\coeff{1}}{t}+F\coeff{1} &= -2\pdiff{F\coeff{0}}{t}-2\frac{\partial^2 F\coeff{0}}{\partial\tilde t\partial t},\label{F1}\\
\pddiff{F\coeff{2}}{t}+F\coeff{2} &= -2\pdiff{F\coeff{1}}{t}-2\frac{\partial^2 F\coeff{1}}{\partial\tilde t\partial t}-\pddiff{F\coeff{0}}{\tilde t}-2\pdiff{F\coeff{0}}{\tilde t},\label{F2}
\end{align}
subject to the initial conditions $F\coeff{\emph{n}}(0,0)=0$ for $n\ge 0$, $\pdiff{F\coeff{0}}{t}(0,0)=1$, and $\pdiff{F\coeff{\emph{n}}}{t}(0,0) =-\pdiff{F\coeff{\emph{n}-1}}{\tilde t}(0,0)$ for $n>0$. The solution to the first equation is
\begin{equation}
F\coeff{0}=A\coeff{0}(\tilde t)\sin t+B\coeff{0}(\tilde t)\cos t,
\end{equation}
where the initial conditions on $F\coeff{0}$ do not fully determine the slow evolution of $A\coeff{0}$ and $B\coeff{0}$, but only impose $A\coeff{0}(0)=1$ and $B\coeff{0}(0)=0$. The general solution to the second equation is
\begin{align}
F\coeff{1} &= A\coeff{1}(\tilde t)\sin t+B\coeff{1}(\tilde t)\cos t-\left(A\coeff{0}(\tilde t)+\pdiff{A\coeff{0}}{\tilde t}(\tilde t)\right)(\cos t+2t\sin t)\nonumber\\
&\quad-\left(B\coeff{0}(\tilde t)+\pdiff{B\coeff{0}}{\tilde t}(\tilde t)\right)t\cos t.
\end{align}
I now make a final assumption, called the \emph{no-secularity condition}: the ratio of successive terms, $F\coeff{\emph{n}+1}/F\coeff{\emph{n}}$, must be bounded. This means that the terms $t\sin t$ and $t\cos t$ are inadmissable, and so their coefficients must vanish. In this case, we have $\diff{A\coeff{0}}{\tilde t}+A\coeff{0}=0$ and $\diff{B\coeff{0}}{\tilde t}+B\coeff{0}=0$, subject to $A\coeff{0}(0)=1$ and $B\coeff{0}(0)=0$. Solving these equations, we find $A\coeff{0}=e^{-\tilde t}$ and $B\coeff{0}=0$. Hence, we have now fully determined $F\coeff{0}$ to be
\begin{equation}
F\coeff{0}=e^{-\tilde t}\sin t.
\end{equation}
And we have
\begin{equation}
F\coeff{1} = A\coeff{1}(\tilde t)\sin t+B\coeff{1}(\tilde t)\cos t,
\end{equation}
where the initial conditions on $F\coeff{1}$ imply that $A\coeff{1}(0)=B\coeff{1}(0)=0$.

If we ceased our work here, there would be no signal that our assumed expansion cannot, in fact, satisfy the non-secularity condition. Following the same procedure for Eq.~\eqref{F2} as we did for Eq.~\eqref{F1}, we find that in order to avoid secular growth in $F\coeff{2}$, the functions $A\coeff{1}$ and $B\coeff{1}$ must satisfy the equations $\diff{A\coeff{1}}{\tilde t}+A\coeff{1}+\tfrac{1}{2}e^{-\tilde t}=0$ and $\diff{B\coeff{1}}{\tilde t}+B\coeff{1}-e^{-\tilde t}=0$, along with the initial conditions $A\coeff{1}(0)=B\coeff{1}(0)=0$. The solutions to these equations are the secularly growing functions $A\coeff{1}=\tilde t e^{-\tilde t}$ and $B\coeff{1}=-\tfrac{1}{2}\tilde t e^{-\tilde t}$. Thus, in order to avoid secular growth in $F\coeff{2}$, we must introduce secular growth into $F\coeff{1}$. In other words, the expansion has failed.

In this case, we can determine the precise reason for the failure. The exact solution to the original ODE is
\begin{equation}
f(t,\e) = \frac{e^{-\e t}}{\sqrt{1-\e^2}}\sin(t\sqrt{1-\e^2}).
\end{equation}
If we expand this in a regular power series, we arrive at $f(t,\e)=\sin t-\e t\sin t+...$, agreeing with the regular expansion given in Eq.~\eqref{regular}. But we find by inspection that $f(t,\e)$ cannot be written as $F(t,\tilde t,\e)$ in such a way that a regular expansion of $F$ satisfies the no-secularity condition. While $e^{-\e t}$ can be written as $e^{-\tilde t}$ to remove secular growth, an expansion of $\sin(t\sqrt{1-\e^2})$ will violate the condition. However, we \emph{can} write $f(t,\e)=\hat F(\phi,\tilde t,\e)$, where $\phi=\Omega(\e)t$, $\Omega(\e)=\sqrt{1-\e^2}$, and $F$ is given by
\begin{equation}
\hat F(\phi,\tilde t,\e) = \frac{e^{-\tilde t}}{\Omega(\e)}\sin\phi.
\end{equation}
This function possesses the regular expansion $\hat F(\phi,\tilde t,\e)=(1+\tfrac{1}{2}\e)e^{-\tilde t}\sin\phi+o(\e)$, which, when expressed in terms of $t$, is a uniform approximation to $f(t,\e)$. One might wonder if we could have discovered this expansion without access to the exact solution. The answer, fortunately, is that we could have: substituting $f=\hat F=\sum\e^n \hat F\coeff{\emph{n}}(\phi,\tilde t)$ and $\Omega(\e)=\sum_{n\ge 0}\e^n\Omega\coeff{\emph{n}}$ into Eq.~\eqref{DE1} and then solving for abritrary $\phi$ and $\tilde t$ yields a sequence of equations that determine the $F\coeff{\emph{n}}$ and $\Omega\coeff{\emph{n}}$ \cite{Kevorkian_Cole}.

There are several points to note from this example. First, while an expansion method might appear to be working, it might still fail at higher order. Second, although we cannot be guaranteed that this failure will reveal itself in the course of our perturbation calculation, that will typically be the case, as it was here. Third, even though my assumptions about $F$ proved to be false, and even though $F\coeff{0}+\e F\coeff{1}$ fails to provide a uniform first-order approximation to $f$, the term $F\coeff{0}$ alone, the only term in $F$ that was fully determined without any obvious contradiction, \emph{does} provide a uniform zeroth-order approximation. (This can easily be checked by calculating the supremum norm of $|f-F\coeff{0}|$.)

\section{A charged particle in a weakly curved spacetime}
As a test case for the adiabatic approximation, I now consider a charged particle orbiting a central object of mass $M$. I take the mass $M$ to be a source of a weak, spherically symmetric gravitational field, and I treat the charged particle as a test mass. Hence, I can use Newtonian spatial vectors and forces, and I can utilize the Newtonian version of osculating orbits. In that context, the equation of motion of the charged particle is 
\begin{equation} 
\bm{a} = \bm{g} + \bm{F}, 
\end{equation}
where $\bm{a} = d^2\bm{r}/dt^2$ is the charge's 3D acceleration vector, $\bm{g}=-\frac{M}{r^2}\bm{\hat r}$ is the Newtonian gravitational acceleration due to the central mass $M$, $\bm{\hat r}=\bm{r}/r$ is a radial unit vector pointing toward the charge, and 
\begin{equation} \label{EM_force}
\bm{F}=\lambda_c \frac{q^2}{m}
\frac{M}{r^3} \bm{\hat{r}} 
+ \lambda_{rr} \frac{2}{3}\frac{q^2}{m}\frac{d \bm{g}}{dt}  
\end{equation}
consists of the leading-order terms in a weak-field expansion of the the electromagnetic self-force (per unit mass) \cite{weakly_curved_spacetime}. Here $\lambda_c$ labels the conservative part of the self-force, and $\lambda_{rr}$ labels the radiation-reaction/dissipative part. In this case, the dissipative term is the (order-reduced) Abraham-Lorentz-Dirac radiation-reaction force \cite{Jackson}, while the conservative piece is due to the backscattering of electromagnetic waves in the weak (rather than strictly vanishing) curvature of the spacetime. Both $\lambda_c$ and $\lambda_{rr}$ are equal to unity, but they allow us to keep track of the source of various contributions in our final results. Making use of the Keplerian parametrization of the orbit, I rewrite the self-force in the form $\bm{F}=\frac{1}{M}F^\phi\bm{\hat\phi}+\frac{1}{M}F^r\bm{\hat r}$, where $F^\phi$ and $F^r$ are dimensionless components of the force, given by
\begin{align}
F^r &= \e(1+e c)^3 p^{-3}(\lambda_c+\tfrac{4}{3}\lambda_{rr}p^{-1/2}e s), \\
F^\phi &= -\tfrac{2}{3}\e\lambda_{rr}(1+e c)^4 p^{-7/2}.
\end{align}
Here I have introduced the shorthand notation $c\equiv\cos v$ and $s\equiv\sin v$, $v=\phi-w$, and the dimensionless small quantity $\e\equiv\frac{q^2}{mM}$. Note that in these expressions, the dissipative pieces of the force are suppressed by a factor of $1/p^{1/2}$ relative to the conservative piece. In the weak field regime (that is, at large distances from the central mass), $p$ is typically large, scaling as $1/(v_i v^i)^2$, where $v_i$ is the orbital velocity. In terms of a post-Newtonian expansion, the conservative force appears at 1PN, while the dissipative piece appears at 1.5PN; in the case of the gravitational self-force, the disparity is larger, with conservative effects appearing at 0PN and dissipative effects appearing at 2.5PN, but the general dominance of conservative effects is common to both types of self-force. 

After substituting the explicit expressions for $F^r$ and $F^\phi$ into the evolution equations \eqref{dphi}, and using simple trigonometric identities, we have
\begin{align}
\diff{p}{\phi} & = -\frac{4\e\lambda_{rr}}{3p^{1/2}}(1+e\cos v), \\
\diff{e}{\phi} & = -\frac{2\e\lambda_{rr}}{3p^{3/2}}\left[\tfrac{3}{2}e+\tfrac{1}{4}(8+5e^2)\cos v+\tfrac{5}{2}e\cos 2v+\tfrac{3}{4}e^2\cos 3v\right]\nonumber\\
&\quad+\frac{\e\lambda_c}{p}(\sin v+\tfrac{1}{2}e\sin 2v),\\
\diff{w}{\phi} & = -\frac{2\e\lambda_{rr}}{3ep^{3/2}}\left[\tfrac{1}{4}(8+3e^2)\sin v +\tfrac{5}{2}\sin 2v+\tfrac{3}{4}e\sin 3v\right]\nonumber\\
&\quad -\frac{\e\lambda_c}{2ep}(1+2\cos v+\cos 2v).\label{w_equation}
\end{align}
In this form, we can easily identify the oscillatory and stationary terms on the right-hand side. For initial conditions, I assume that $p(\phi=0,\e)=\bar p$, $e(0,\e)=\bar e$, $w(0,\e)=0$, and $t(0,\e)=0$.

I now follow the procedure outlined in Sec.~\ref{multiscale_osculating} by assuming multiscale expansions of the form $p(\phi,\e)= p\dcoeff{0}(\tilde\phi)+\e p\dcoeff{1}(\phi,\tilde\phi)+...$ for the orbital elements, and $t(\phi,\e) = \frac{1}{\e}t\dcoeff{-1}(\tilde\phi)+ t\dcoeff{0}(\phi,\tilde\phi)+\e t\dcoeff{1}(\phi,\tilde\phi)+...$ for time.\footnote{In these expansions, I have assumed that the leading-order terms depend only on $\tilde\phi$, since, as shown in Sec.~\ref{multiscale_osculating}, that is always the case for our system of equations.} I assign initial conditions $p\dcoeff{0}(0)=\bar p$,  $e\dcoeff{0}(0)=\bar e$, $w\dcoeff{0}=\bar w$, and $t\dcoeff{-1}(0)=0$ for the leading-order terms in the expansion, and $p\dcoeff{1}(0,0)=e\dcoeff{1}(0,0)=w\dcoeff{1}(0,0)=t\dcoeff{0}(0,0)=0$ for the subleading terms.

From the order-$\e$ orbital-element-equations, we find
\begin{align}
\diff{p\dcoeff{0}}{\tilde\phi} +\pdiff{p\dcoeff{1}}{\phi} & = -\frac{4\e\lambda_{rr}}{3p\dcoeff{0}^{1/2}}(1+e\dcoeff{0}c\dcoeff{0}),\label{p0_equation}\\
\diff{e\dcoeff{0}}{\tilde\phi} +\pdiff{e\dcoeff{1}}{\phi} & =  -\frac{2\e\lambda_{rr}}{3p\dcoeff{0}^{3/2}}\left[\tfrac{3}{2}e\dcoeff{0}+\tfrac{1}{4}(8+5e\dcoeff{0}^2)c\dcoeff{0}+\tfrac{5}{2}e\dcoeff{0}\cos 2v\dcoeff{0}+\tfrac{3}{4}e\dcoeff{0}^2\cos 3v\dcoeff{0}\right]\nonumber\\
&\quad+\frac{\e\lambda_c}{p\dcoeff{0}}(\sin v\dcoeff{0}+\tfrac{1}{2}e\dcoeff{0}\sin 2v\dcoeff{0})\\
\diff{w\dcoeff{0}}{\tilde\phi} +\pdiff{w\dcoeff{1}}{\phi} & =  -\frac{2\e\lambda_{rr}}{3e\dcoeff{0}p\dcoeff{0}^{3/2}}\left[\tfrac{1}{4}(8+3e\dcoeff{0}^2)s\dcoeff{0} +\tfrac{5}{2}\sin 2v\dcoeff{0}+\tfrac{3}{4}e\dcoeff{0}\sin 3v\dcoeff{0}\right]\nonumber\\
&\quad-\frac{\e\lambda_c}{2e\dcoeff{0}p\dcoeff{0}}(1+2c\dcoeff{0}+\cos 2v\dcoeff{0}),\label{w0_equation}
\end{align}
where $c\dcoeff{0}\equiv\cos v\dcoeff{0}$, $s\dcoeff{0}\equiv\sin v\dcoeff{0}$, and $v\dcoeff{0}\equiv\phi-w\dcoeff{0}$. After averaging over one period, we find
\begin{equation}
\diff{p\dcoeff{0}}{\tilde\phi} = -\frac{4}{3}\lambda_{rr}p\dcoeff{0}^{-1/2}, \quad
\diff{e\dcoeff{0}}{\tilde\phi} = -\lambda_{rr}e\dcoeff{0}p\dcoeff{0}^{-3/2}, \quad
\diff{w\dcoeff{0}}{\tilde\phi} = -\tfrac{1}{2}\lambda_cp\dcoeff{0}^{-1}.\label{averaged_elements0}
\end{equation}
These equations can be solved immediately, yielding
\begin{align}
p\dcoeff{0} &= \bar p(1-2\bar p^{-3/2}\lambda_{rr}\tilde\phi)^{2/3},\\
e\dcoeff{0} &= \bar e(1-2\bar p^{-3/2}\lambda_{rr}\tilde\phi)^{1/2},\\
w\dcoeff{0} &= \bar w-\frac{3\lambda_c\bar p^{1/2}}{4\lambda_{rr}}\left[1-(1-2\lambda_{rr}\bar p^{-3/2}\tilde\phi)^{1/3}\right].
\end{align}
Substituting Eq.~\eqref{averaged_elements0} back into Eqs.~\eqref{p0_equation}--\eqref{w0_equation}, we can then easily integrate them to find $p\dcoeff{1}$, $e\dcoeff{1}$, and $w\dcoeff{1}$:
\begin{align}
p\dcoeff{1} &= -\tfrac{4}{3}\lambda_{rr}p\dcoeff{0}^{-1/2}e\dcoeff{0}\sin v\dcoeff{0}+C^p\dcoeff{1}(\tilde\phi),\\
e\dcoeff{1} &= -\tfrac{2}{3}\lambda_{rr}p\dcoeff{0}^{-3/2}\left[(2+\tfrac{5}{4}e\dcoeff{0}^2)\sin v\dcoeff{0}+\tfrac{5}{4}e\dcoeff{0}\sin 2v\dcoeff{0}+\tfrac{1}{4}e\dcoeff{0}^2\sin 3v\dcoeff{0}\right]  \nonumber\\
&\quad  -\lambda_c p\dcoeff{0}^{-1}(\cos v\dcoeff{0}+\tfrac{1}{4}e\dcoeff{0}\cos 2v\dcoeff{0}) +C^e\dcoeff{1}(\tilde\phi),\\
w\dcoeff{1} &= \tfrac{2}{3}\lambda_{rr}e\dcoeff{0}^{-1}p\dcoeff{0}^{-3/2}\left[(2+\tfrac{3}{4}e\dcoeff{0}^2)\cos v\dcoeff{0}+\tfrac{5}{4}e\dcoeff{0}\cos 2v\dcoeff{0}+\tfrac{1}{4}e\dcoeff{0}^2\cos 3v\dcoeff{0}\right]  \nonumber\\
&\quad  -\lambda_c e\dcoeff{0}^{-1}p\dcoeff{0}^{-1}(\sin v\dcoeff{0}+\tfrac{1}{4}e\dcoeff{0}\sin 2v\dcoeff{0}) +C^w\dcoeff{1}(\tilde\phi).
\end{align}

With $p\dcoeff{0}$, $e\dcoeff{0}$, and $w\dcoeff{0}$ determined, we can find $t\dcoeff{0}$ from the order-$\e^0$ time-equation
\begin{equation}
\pdiff{t\dcoeff{-1}}{\tilde\phi}+\pdiff{t\dcoeff{0}}{\phi} = \frac{p\dcoeff{0}^{3/2}M}{(1+e\dcoeff{0}c\dcoeff{0})^2}.\label{t_equation0}
\end{equation}
Averaging over one period yields
\begin{equation}
\pdiff{t\dcoeff{-1}}{\tilde\phi}=\frac{p\dcoeff{0}^{3/2}M}{(1-e\dcoeff{0}^2)^{3/2}},\label{tn1_equation}
\end{equation}
where I have used $\langle(1+e\dcoeff{0}c\dcoeff{0})^{-2}\rangle=(1-e\dcoeff{0}^2)^{-3/2}$. Note that at fixed $\tilde\phi$, the right-hand side is proportional to the Keplerian orbital period $P\coeff{0}=2\pi\frac{p\dcoeff{0}^{3/2}M}{(1-e\dcoeff{0}^2)^{3/2}}$; therefore, this equation describes a slow evolution of the orbital period due to dissipation. After integrating Eq.~\eqref{tn1_equation}, we find
\begin{equation}
t\dcoeff{-1}(\tilde\phi)=\frac{\bar p^3M}{\lambda_{rr}\bar e^4}\frac{\bar e^2-2-2\bar e^2\bar p^{-3/2}\lambda_{rr}\tilde\phi}{(1-\bar e^2+2\bar e^2\bar p^{-3/2}\lambda_{rr}\tilde\phi)^{1/2}}.
\end{equation}
Substituting Eq.~\eqref{tn1_equation} back into Eq.~\eqref{t_equation0} and integrating, we find
\begin{equation}
t\dcoeff{0} = \frac{P\coeff{0}(\tilde\phi)}{2\pi}(E\dcoeff{0}-e\dcoeff{0}\sin E\dcoeff{0}-\phi)+C^t\dcoeff{0}(\tilde\phi).\label{t0_equation}
\end{equation}
In order to evaluate the integral $\int (1+e\dcoeff{0}c\dcoeff{0})^{-2}d\phi$, I have introduced the eccentric anomaly $E\dcoeff{0}$, related to $\phi$ by $\tan\frac{E\dcoeff{0}}{2}=\sqrt{\frac{1-e\dcoeff{0}}{1+e\dcoeff{0}}}\tan\frac{v\dcoeff{0}}{2}$. Note that $E\dcoeff{0}(\phi+2\pi)=E\dcoeff{0}(\phi)+2\pi$, so the sum of terms within parentheses in Eq.~\eqref{t0_equation} form a periodic function of $\phi$.

After averaging, the order-$\e^2$ orbital-element-equations read
\begin{align}
\diff{C^p\dcoeff{1}}{\tilde\phi} & = \frac{2\lambda_{rr}}{3p\dcoeff{0}^{3/2}}(C^p\dcoeff{1}+2\lambda_c), \label{Cp1eqn}\\
\diff{C^e\dcoeff{1}}{\tilde\phi} & = \frac{\lambda_{rr}}{p\dcoeff{0}^{5/2}}(\tfrac{3}{2}e\dcoeff{0}C^p\dcoeff{1}-p\dcoeff{0}C^e\dcoeff{1}+\tfrac{5}{3}\lambda_ce\dcoeff{0}), \\
\diff{C^w\dcoeff{1}}{\tilde\phi} & = \frac{1}{8 p\dcoeff{0}^3}(4\lambda_{rr}^2+\lambda_c^2p\dcoeff{0}+4\lambda_cp\dcoeff{0}C^p\dcoeff{1}). \label{Cw1eqn}
\end{align}
Solving these equations and imposing the initial conditions $p\dcoeff{1}(0,0)=e\dcoeff{1}(0,0)=w\dcoeff{1}(0,0)=0$, we find
\begin{align}
C^p\dcoeff{1} & = 2\lambda_c\left(\frac{\bar p^{1/2}}{p\dcoeff{0}^{1/2}}-1\right), \\
C^e\dcoeff{1} & = \frac{\lambda_ce\dcoeff{0}}{\bar p\bar e}(1-\tfrac{1}{4}\bar e)+\frac{\lambda_c\bar e^2}{\bar pe\dcoeff{0}}\left(\tfrac{3}{2}-\frac{e\dcoeff{0}^{2/3}}{\bar e^{2/3}}\right),\\
C^w\dcoeff{1} & = -\frac{21\lambda_c^2}{16\lambda_{rr}\bar p\dcoeff{0}^{1/2}}-\frac{2\lambda_{rr}}{4p\dcoeff{0}^{3/2}}+\frac{3\lambda_c^2\bar p^{1/2}}{4\lambda_{rr}p\dcoeff{0}}-\frac{\lambda_{rr}(16+13\bar e+8\bar e^2)}{12\bar e\bar p^{3/2}}+\frac{9\lambda_c^2}{16\lambda_{rr}\bar p^{1/2}}.
\end{align}
After averaging, the order-$\e$ time-equation reads
\begin{align}
C^t\dcoeff{0} &= \int_0^{\tilde\phi} \left\langle\frac{\tfrac{3}{2}p\dcoeff{0}^{1/2}p\dcoeff{1}M}{(1+e\dcoeff{0}c\dcoeff{0})^2}-\frac{2p\dcoeff{0}^{3/2}M(e\dcoeff{1}c\dcoeff{0}+w\dcoeff{1}s\dcoeff{0})}{(1+e\dcoeff{0}c\dcoeff{0})^3}\right\rangle d\tilde\phi \nonumber\\
&=\int_0^{\tilde\phi}\left[ \lambda_cp\dcoeff{0}^{1/2}M\frac{8+e\dcoeff{0}^2}{4(1-e\dcoeff{0}^2)^{5/2}}+\frac{3p\dcoeff{0}^{1/2}MC^p\dcoeff{1}}{2(1-e\dcoeff{0}^2)^{3/2}}+\frac{3p\dcoeff{0}^{3/2}e\dcoeff{0}MC^e\dcoeff{1}}{(1-e\dcoeff{0}^2)^{5/2}}\right] d\tilde\phi.\label{Ct_eqn}
\end{align}
This integral could be explicitly evaluated in terms of hypergeometric functions, but it is simpler in its unevaluated form.

We have now arrived at the fully-determined approximation $I^A(\phi,\e)=I^A\dcoeff{0}(\e\phi)+\e I^A\dcoeff{1}(\phi,\e\phi)+o(\e)$, $t(\phi,\e) = \e^{-1}t\dcoeff{-1}(\e\phi)+t\dcoeff{0}(\phi,\e\phi)+o(1)$. The accuracy of these approximations can be confirmed by comparing with a numerical integration of the exact equations. However, this accuracy assumes that the force \eqref{EM_force} is exact. In actuality, the force is a weak-field expansion of the first-order electromagnetic self-force in a spherically-symmetric (i.e., Schwarzschild) spacetime. If we had, for example, the second-order force (either second-order in $\e$, or second-order in the weak-field expansion), then it would appear in  Eqs.~\eqref{Cp1eqn}--\eqref{Cw1eqn} and therefore in \eqref{Ct_eqn}, influencing the secular behavior of the solution. 

As discussed in Sec.~\ref{multiscale_osculating}, and as can be seen by inspecting the location of $\lambda_c$ and $\lambda_{rr}$ in the above results, the dominant secular evolution of the principal elements $p$ and $e$---describing the shrinking and circularizing of the orbit---as well as that of the orbital phase, are determined by the dissipative piece of the force. The dominant secular evolution of $w$---describing the orbital precession---is determined by the conservative piece of the force. At subleading order, both the conservative and dissipative pieces of the force contribute to all the elements and the phase. In particular, the conservative force redefines the orbital period, as represented by the correction $C^t\dcoeff{0}$. We can also see that in this expansion, dissipative effects are suppressed by a factor of $p^{-1/2}$ relative to conservative effects, as we would expect from the form of the force. Hence, in this weak-field regime, the effects of the conservative force on, for example, the orbital period, will be larger than is suggested by the fact that they appear at subleading order in the two-timescale expansion.

\section{Multiscale expansion of the Einstein equation}
I now consider a multiscale expansion of various geometrical quantities, based on a multiscale expansion of the metric. For simplicity, I assume an expansion of the form 
\begin{equation}
\exact{g}(x,\e)=g(x,\zeta)+\sum_{n\ge 1}\e^n \hmn{}{n}(x,\zeta),
\end{equation} 
where $\partial_\mu\zeta=o(1)$. I will occasionally provide details given the simplifying assumption $\partial_\mu\zeta=\e V_\mu(x,\zeta)$ for some $V_\mu=O_s(1)$.

Note that the gauge group in this expansion differs from that of a regular expansion. Gauge transformations are generated by transformations of the form
\begin{equation}
x^\alpha\to x'^\alpha = x^\alpha-\e\xi^\alpha(x,\zeta)+\order{\e^2},
\end{equation}
where $\xi=O_s(1)$. In order to determine the effect of this transformation, we expand the Lie derivative as
\begin{equation}
\Lie{\xi} = \Lie{\xi}\coeff{0}+\Lie{\xi}\coeff{1},
\end{equation}
where, e.g., for a vector $\xi(x,\zeta)$ and a tensor $T^\mu{}_\nu(x,\zeta)$,
\begin{align}
\Lie{\xi}\coeff{0}T^\mu{}_\nu & = \xi^{\rho}\frac{\partial T^\mu{}_\nu}{\partial x^\rho}-T^\rho{}_\nu\frac{\partial\xi^\mu}{\partial x^\rho}+T^\mu{}_\rho\frac{\partial\xi^\rho}{\partial x^\nu},\\
\Lie{\xi}\coeff{1}T^\mu{}_\nu & = \xi^{\rho}\frac{\partial T^\mu{}_\nu}{\partial\zeta}\frac{\partial\zeta}{\partial x^\rho} -T^\rho{}_\nu\frac{\partial\xi^\mu}{\partial\zeta}\frac{\partial\zeta}{\partial x^\rho} +T^\mu{}_\rho\frac{\partial\xi^\rho}{\partial\zeta}\frac{\partial\zeta}{\partial x^\nu}.
\end{align}
These definitions are independent of the behavior of $\zeta$. But in the particular case that $\partial_\mu\zeta=\e V_\mu(x,\zeta)$, the gauge transformation generated by a vector $\e\xi(x,\zeta)$ can be written as
\begin{align}
\Delta \hmn{}{1} & = \Lie{\xi}\coeff{0}g, \\
\Delta \hmn{}{2} & = \tfrac{1}{2}\Lie{\xi}\coeff{0}\Lie{\xi}\coeff{0}g +\Lie{\xi}\coeff{0}\hmn{}{1}+\Lie{\xi}\coeff{1}g.
\end{align}

Similarly, I expand the covariant derivative as
\begin{equation}
\del{\mu}V^{\nu}(x,\zeta) = \left(\del{\mu}\coeff{0}+\del{\mu}\coeff{1}\right)V^{\nu}(x,\zeta),
\end{equation}
where $\del{\mu}$ is compatible with $g(x,\zeta(x,\e))$, $\del{\mu}\coeff{0}$ is compatible with $g$ at fixed $\zeta$, and $\del{\mu}\coeff{1}$ is compatible with $g$ at fixed $x$. Explicitly,
\begin{align}
\del{\mu}\coeff{0}V^{\nu}(x,\zeta) &= \frac{\partial V^\nu}{\partial x^\mu}+\Gamma\coeff{0}{}^{\nu}_{\mu\rho}V^\rho,\\
\del{\mu}\coeff{1}V^{\nu}(x,\zeta) &= \frac{\partial V^\nu}{\partial\zeta}\frac{\partial\zeta}{\partial x^\mu}+\Gamma\coeff{1}{}^{\nu}_{\mu\rho}V^\rho,
\end{align}
where the Christoffel symbols are given by
\begin{align}
\Gamma\coeff{0}{}^{\alpha}_{\beta\gamma} &= \tfrac{1}{2}g^{\alpha\delta}\left(\frac{\partial g_{\delta\beta}}{\partial x^\gamma} +\frac{\partial g_{\delta\gamma}}{\partial x^\beta} -\frac{\partial g_{\beta\gamma}}{\partial x^\delta}\right),\\
\Gamma\coeff{1}{}^{\alpha}_{\beta\gamma} &=\tfrac{1}{2}g^{\alpha\delta}\left(\frac{\partial g_{\delta\beta}}{\partial\zeta}\frac{\partial\zeta}{\partial x^\gamma} +\frac{\partial g_{\delta\gamma}}{\partial\zeta}\frac{\partial\zeta}{\partial x^\beta} -\frac{\partial g_{\beta\gamma}}{\partial\zeta}\frac{\partial\zeta}{\partial x^\delta}\right).
\end{align}
The ``correction" $\del{\mu}\coeff{1}$ ensures that the total covariant derivative $\del{\mu}$ is compatible with the $\zeta$-dependence of $g$. Note that all three derivatives are metric compatible: $\nabla g=0$, $\nabla\coeff{0}g=0$, and $\nabla\coeff{1}g=0$.

By writing the Lie derivative in terms of the covariant derivative, we can express the gauge transformation generated by a vector field $\e\xi(x,\zeta)$ as
\begin{align}
\Delta \hmn{\alpha\beta}{1} & = 2\del{(\alpha}\coeff{0}\xi_{\beta)}, \\
\Delta \hmn{\alpha\beta}{2} & = \xi^\gamma\del{\gamma}\coeff{0}\del{(\alpha}\coeff{0}\xi_{\beta)} +\del{(\gamma}\coeff{0}\xi_{\beta)}\del{\alpha}\coeff{0}\xi^\gamma +\del{(\alpha}\coeff{0}\xi_{\gamma)}\del{\beta}\coeff{0}\xi^\gamma +\xi^\gamma\del{\gamma}\coeff{0}\hmn{\alpha\beta}{1}  \nonumber\\
&\quad+2\hmn{\gamma(\beta}{1}\del{\alpha)}\coeff{0}\xi^\gamma +2\del{(\alpha}\coeff{1}\xi_{\beta)},
\end{align}
assuming that $\partial_\mu\zeta=\e V_\mu(x,\zeta)$.

Note that because the background metric $g$ depends on $\zeta$, the Riemann tensor constructed from it can be expanded in powers of $\e$:
\begin{align}
R_{\mu\rho\nu\sigma}(x,\zeta) &= R\coeff{0}_{\mu\rho\nu\sigma}(x,\zeta) +\e R\coeff{1}_{\mu\rho\nu\sigma}(x,\zeta) +\e^2 R\coeff{2}_{\mu\rho\nu\sigma}(x,\zeta),
\end{align}
where $R\coeff{0}_{\mu\rho\nu\sigma}(x,\zeta)$ is constructed from $g$ and $\nabla\coeff{0}$, $R\coeff{1}_{\mu\rho\nu\sigma}(x,\zeta)$ contains one $\nabla\coeff{1}$ derivative, and $R\coeff{2}_{\mu\rho\nu\sigma}(x,\zeta)$ contains two $\nabla\coeff{1}$ derivatives. The $n$th-order perturbation of the Ricci tensor can be similarly expanded as $\delta^n R_{\mu\nu}[h]=\sum_{m=0}^2\e^m\delta^n R\coeff{n}_{\mu\nu}[h]$. This means that the vacuum Einstein equation $\exact{R}_{\mu\nu}=0$ becomes
\begin{align}
R\coeff{0}_{\mu\nu} &= 0, \\
\delta R\coeff{0}_{\mu\nu}[h\coeff{1}] & = R\coeff{1}_{\mu\nu}, \\
\delta R\coeff{0}_{\mu\nu}[h\coeff{2}] & = R\coeff{2}_{\mu\nu}-\delta R\coeff{1}_{\mu\nu}[h\coeff{1}]-\delta^2 R\coeff{0}_{\mu\nu}[h\coeff{1}], \\
&\ \ \vdots\nonumber
\end{align}

Similarly, the Bianchi identity on the background, $g^{\mu\nu}\del{\mu}G_{\nu\rho}[g]=0$, becomes
\begin{align}
\del{\mu}\coeff{0}\cdot G\coeff{0}_{\nu\rho} &= 0, \\
\del{\mu}\coeff{0}\cdot G\coeff{1}_{\nu\rho} &= -\del{\mu}\coeff{1}\cdot G\coeff{0}_{\nu\rho}, \\
\del{\mu}\coeff{0}\cdot G\coeff{2}_{\nu\rho} &= -\del{\mu}\coeff{1}\cdot G\coeff{1}_{\nu\rho}, \\
\del{\mu}\coeff{1}\cdot G\coeff{2}_{\nu\rho} &= 0,
\end{align}
where a dot indicates contraction over $\mu$ and $\nu$. And the Bianchi identitity on the full spacetime, $\exact{g}^{\mu\nu}{}^{\exact{g}}\del{\mu}\exact{G}_{\nu\rho}=0$, can be expanded schematically as
\begin{align}
\del{}\coeff{0}\cdot\delta G\coeff{0}[\hmn{}{1}] & = 0, \\
\del{}\coeff{0}\cdot\delta G\coeff{0}[\hmn{}{2}] & = -\left(\del{}\coeff{1}\cdot\delta G\coeff{0} +\del{}\coeff{0}\cdot\delta G\coeff{1}\right)[\hmn{}{1}]+\hmn{}{1}\del{}\coeff{0}\left(G\coeff{1} +\delta G\coeff{0}[\hmn{}{1}]\right) \nonumber\\
&\quad -\delta\Gamma\coeff{0}[\hmn{}{1}]\cdot\left(G\coeff{1} +\delta G\coeff{0}[\hmn{}{1}]\right) -\del{}\coeff{0}\cdot\delta^2G\coeff{0}[\hmn{}{1}],
\end{align}
where $\delta\Gamma\coeff{0}{}^{\alpha}_{\beta\gamma}[\hmn{}{1}] =\tfrac{1}{2}g^{\alpha\delta}(\del{\gamma}\coeff{0}\hmn{\delta\beta}{1} +\del{\beta}\coeff{0}\hmn{\delta\gamma}{1} -\del{\delta}\coeff{0}\hmn{\beta\gamma}{1})$, and the leading-order Einstein equation $R\coeff{0}_{\mu\nu} = 0$ and the Bianchi identity $g^{\mu\nu}\del{\mu}G_{\nu\rho}=0$ have already been imposed for compactness.

Note that the $\zeta$-dependence of $g$ allows the background to slowly react to the perturbation. Determining this reaction is the backreaction problem, which has been studied extensively in the past.

			\chapter{Expansions of the Riemann tensor and related quantities}\label{general_expansions}
In the first section of this appendix I review the standard results for an expansion of various geometric quantities in powers of a metric perturbation. In the second section, I present some lengthy results for particular components of the second-order Ricci tensor in the buffer region.

\section{General expansions in powers of the metric perturbation}
I begin by writing the components of the exact metric as $\exact{g}_{\mu\nu}=g_{\mu\nu}+h_{\mu\nu}$, and I use the background metric $g$ to raise and lower indices on $h$ (and on structures constructed from it). The inverse metric then possesses the expansion $\exact{g}^{\mu\nu}=g^{\mu\nu}-h^{\mu\nu}+h^\mu{}_\rho h^{\rho\nu}+O(h^3)$. Next, I define $C^\alpha{}_{\beta\gamma}$ to be the difference between the Christoffel symbols defined by $\exact{g}$ and those defined by $g$: $C^\alpha{}_{\beta\gamma}A^\gamma=({}^{\exact{g}}\del{\beta}-\del{\beta})A^\alpha$. One can easily calculate (by adopting a locally inertial frame, for example), that
\begin{equation}\label{Cabc}
C^\alpha{}_{\beta\gamma}=\tfrac{1}{2}\exact{g}^{\alpha\delta}(h_{\delta\beta;\gamma}+h_{\delta\gamma;\beta}-h_{\beta\gamma;\delta}),
\end{equation}
where $h_{\beta\gamma;\delta}\equiv \del{\delta}h_{\beta\gamma}$. This quantity will be used to determine the expansion of the Riemman tensor and related quantities. Using the definition $\exact{R}^\alpha{}_{\beta\gamma\delta}A^\beta =({}^{\exact{g}}\del{\gamma}{}^{\exact{g}}\del{\delta}-{}^{\exact{g}}\del{\delta}{}^{\exact{g}}\del{\gamma})A^\alpha$ (or the explicit expression for the Riemann tensor in terms of Christoffel symbols), we find
\begin{equation}\label{Riemann_expansion}
\exact{R}^\alpha{}_{\beta\gamma\delta} = R^\alpha{}_{\beta\gamma\delta}+2C^\alpha{}_{\beta[\delta;\gamma]}+2C^\alpha{}_{\rho[\gamma}C^\rho{}_{\delta]\beta}
\end{equation}
Now, I define the expansion $\exact{R}^\alpha{}_{\beta\gamma\delta}=R^\alpha{}_{\beta\gamma\delta}+\delta  R^\alpha{}_{\beta\gamma\delta}+\delta^2 R^\alpha{}_{\beta\gamma\delta}+...$ such that $\delta R^\alpha{}_{\beta\gamma\delta}$ is linear in $h$, $\delta^2 R^\alpha{}_{\beta\gamma\delta}$ is quadratic in $h$, etc. Analogously, I define $\exact{R}_{\alpha\beta}=R_{\alpha\beta}+\delta R_{\alpha\beta}+\delta^2 R_{\alpha\beta}+...$ and $\exact{G}_{\alpha\beta}=G_{\alpha\beta}+\delta G_{\alpha\beta}+\delta^2 G_{\alpha\beta}+...$, where $\exact{R}_{\alpha\beta}\equiv \exact{R}^\mu{}_{\alpha\mu\beta}$ and $\exact{G}_{\alpha\beta}\equiv \exact{R}_{\alpha\beta} -\tfrac{1}{2}\exact{g}_{\alpha\beta}\exact{g}^{\mu\nu}\exact{R}_{\mu\nu}$.

Using Eq.~\eqref{Riemann_expansion}, one can straightforwardly calculate the following results:
\begin{align}
\delta R_{\alpha\beta} &= -\tfrac{1}{2}(\Box h_{\alpha\beta}+g^{\mu\nu}h_{\mu\nu;\alpha\beta})+h_{\mu(\alpha;\beta)}{}^\mu,\\
\delta^2R_{\alpha\beta} &=-\tfrac{1}{2}\bar h^{\mu\nu}{}_{;\nu}\left(2h_{\mu(\alpha;\beta)}-h_{\alpha\beta;\mu}\right) +\tfrac{1}{4}h^{\mu\nu}{}_{;\alpha}h_{\mu\nu;\beta}+\tfrac{1}{2}h^{\mu}{}_{\beta}{}^{;\nu}\left(h_{\mu\alpha;\nu} -h_{\nu\alpha;\mu}\right)\nonumber\\
&\quad-\tfrac{1}{2}h^{\mu\nu}\left(2h_{\mu(\alpha;\beta)\nu}-h_{\alpha\beta;\mu\nu}-h_{\mu\nu;\alpha\beta}\right).
\end{align}
where $\Box\equiv g^{\mu\nu}\del{\mu}\del{\nu}$ and $\bar h_{\mu\nu}\equiv h_{\mu\nu}-\tfrac{1}{2}g_{\mu\nu}g^{\rho\sigma}h_{\rho\sigma}$. From these quantities, we can calculate $\delta G_{\alpha\beta}$, which is given by $\delta R_{\alpha\beta}-\tfrac{1}{2}(h_{\alpha\beta}R-g_{\alpha\beta}h^{\mu\nu}R_{\mu\nu}$), and $\delta^2 G_{\alpha\beta}$, which is given below. In a Ricci-flat background ($R_{\mu\nu}=0$), we find
\begin{align}
\delta G_{\alpha\beta} &= -\tfrac{1}{2}(\Box h_{\alpha\beta}+g^{\mu\nu}h_{\mu\nu;\alpha\beta}) + h_{\mu(\alpha}{}^{;\mu}{}_{\beta)}+\tfrac{1}{2}g_{\alpha\beta}(g^{\mu\nu}\Box h_{\mu\nu}-h^{\mu\nu}{}_{;\mu\nu})-R_{\alpha\mu\beta\nu}h^{\mu\nu},\\
\delta^2 G_{\alpha\beta} &= \delta^2 R_{\alpha\beta}-\tfrac{1}{2}g_{\alpha\beta}g^{\mu\nu}\delta^2 R_{\mu\nu}-\tfrac{1}{2}h_{\alpha\beta}g^{\mu\nu}\delta R_{\mu\nu}+\tfrac{1}{2}g_{\alpha\beta}h^{\mu\nu}\delta R_{\mu\nu}.
\end{align}

If the background is Ricci-flat and the metric perturbation is in the Lorenz gauge, then the linearized Ricci and Einstein tensors can be written as
\begin{equation}
\delta R_{\alpha\beta}= -\tfrac{1}{2}E_{\alpha\beta}[h],\quad \delta G_{\alpha\beta}= -\tfrac{1}{2}E_{\alpha\beta}[\bar h],
\end{equation}
where the wave-operator $E_{\mu\nu}$ is defined by $E_{\mu\nu}[f]=\left(g^\rho_\mu g^\sigma_\nu\nabla^\gamma\del{\gamma} +2R\indices{_\mu^\rho_\nu^\sigma}\right)\!f_{\rho\sigma}$. (In more usual form, this reads $E_{\mu\nu}[f]=\Box f_{\mu\nu}+2R\indices{_\mu^\rho_\nu^\sigma}f_{\rho\sigma}$.)

\section{Buffer-region expansions required for the second-order wave equation}\label{second-order expansions}
I present here various expansions used in solving the second-order Einstein equation in Sec.~\ref{buffer_expansion2}.

I require an expansion of $\delta^2 R\coeff{0}_{\alpha\beta}[\hmn{}{1}]$ in powers of the Fermi radial coordinate $r$, where for a function $f$, $\delta^2 R\coeff{0}_{\alpha\beta}[f]$ consists of $\delta^2 R_{\alpha\beta}[f]$ with the acceleration $a^\mu$ set to zero. Explicitly, I require the coefficients in the expansion
\begin{align}
\delta^2R\coeff{0}_{\alpha\beta}[\hmn{}{1}] &= \frac{1}{r^4}\ddR{\alpha\beta}{0,-4}{\hmn{}{1}}\!\! +\frac{1}{r^3}\ddR{\alpha\beta}{0,-3}{\hmn{}{1}}\ \ \nonumber\\
&\quad+\frac{1}{r^2}\ddR{\alpha\beta}{0,-2}{\hmn{}{1}} +\order{1/r},
\end{align}
where the second superscript index in parentheses denotes the power of $r$. Making use of the expansion of $\hmn{}{1}$, obtained by setting the acceleration to zero in the results for $\hmn{E}{1}$ found in Sec.~\ref{buffer_expansion1}, one finds
\begin{align}
\ddR{\alpha\beta}{0,-4}{\hmn{}{1}} &= 2m^2\left(7\nhat_{ab}+\tfrac{4}{3}\delta_{ab}\right)x^a_\alpha x^b_\beta -2m^2t_\alpha t_\beta,\label{ddR0n4}
\end{align}
and
\begin{align}
\ddR{tt}{0,-3}{\hmn{}{1}} &= 3m\H{ij}{1,0}\nhat^{ij},\label{ddR0n3_tt}\\
\ddR{ta}{0,-3}{\hmn{}{1}} &= 3m\C{i}{1,0}\nhat_{a}^i,\\
\ddR{ab}{0,-3}{\hmn{}{1}} &= 3m\big(\A{}{1,0}+\K{}{1,0}\big)\nhat_{ab}-6m\H{i\langle a}{1,0}\nhat_{b\rangle}^i+m\delta_{ab}\H{ij}{1,0}\nhat^{ij},\label{ddR0n3_ab}
\end{align}
and
{\allowdisplaybreaks\begin{align}
\ddR{tt}{0,-2}{\hmn{}{1}} &= -\tfrac{20}{3}m^2\etide_{ij}\nhat^{ij}+3m\H{ijk}{1,1}\nhat^{ijk}
+\tfrac{7}{5}m\A{i}{1,1}n^i+\tfrac{3}{5}m\K{i}{1,1}n^i\nonumber\\
&\quad-\tfrac{4}{5}m\partial_t\C{i}{1,0}n^i,\label{ddR0n2_tt}\\
\ddR{ta}{0,-2}{\hmn{}{1}}&= -m\partial_t\K{}{1,0}n_a+3m\C{ij}{1,1}\nhat_{a}{}^{ij}
+m\Big(\tfrac{6}{5}\C{ai}{1,1}-\partial_t\H{ai}{1,0}\Big)n^i\nonumber\\
 &\quad+2m\epsilon_a{}^{ij}\D{i}{1,1}n_j+\tfrac{4}{3}m^2\epsilon_{aik}\btide^k_j\nhat^{ij},\\
\ddR{ab}{0,-2}{\hmn{}{1}} &= \delta_{ab}m\left(\tfrac{16}{15}\partial_t\C{i}{1,0} -\tfrac{13}{15}\A{i}{1,1} -\tfrac{9}{5}\K{i}{1,1}\right)n^i\nonumber\\
&\quad+\delta_{ab}\left(-\tfrac{50}{9}m^2\etide_{ij}\nhat^{ij}+m\H{ijk}{1,1}\nhat^{ijk}\right) -\tfrac{14}{3}m^2\etide_{ij}\nhat_{ab}{}^{ij}\nonumber\\
&\quad+m\left(\tfrac{33}{10}\A{i}{1,1}+\tfrac{27}{10}\K{i}{1,1} -\tfrac{3}{5}\partial_t\C{i}{1,0}\right)\nhat_{ab}{}^i\nonumber\\
&\quad+m\left(\tfrac{28}{25}\A{\langle a}{1,1}-\tfrac{18}{25}\K{\langle a}{1,1}-\tfrac{46}{25}\partial_t\C{\langle a}{1,0}\right)\nhat^{}_{b\rangle}\nonumber\\
&\quad-\tfrac{8}{3}m^2\etide_{i\langle a}\nhat_{b\rangle}{}^i-6m\H{ij\langle a}{1,1}\nhat^{}_{b\rangle}{}^{ij}+3m\epsilon_{ij(a}\nhat_{b)}{}^{jk}\I{ik}{1,1} \nonumber\\
&\quad+\tfrac{2}{45}m^2\etide_{ab}-\tfrac{2}{5}m\H{abi}{1,1}n^i +\tfrac{8}{5}m\epsilon^i{}_{j(a}^{}\I{b)i}{1,1}n^j.\label{ddR0n2_ab}
\end{align}}

I require an analogous expansion of $E\coeff{0}_{\alpha\beta}\left[\frac{1}{r^2}\hmn{}{2,-2} +\frac{1}{r}\hmn{}{2,-1}\right]$, where $E\coeff{0}_{\alpha\beta}[f]$ is defined for any $f$ by setting the acceleration to zero in $E_{\alpha\beta}[f]$. The coefficients of the $1/r^4$ and $1/r^3$ terms in this expansion can be found in Sec.~\ref{buffer_expansion2}; the coefficient of $1/r^2$ will be given here. For compactness, I define  this coefficient to be $\tilde E_{\alpha\beta}$. The $tt$-component of this quantity is given by  
\begin{align}
\tilde E_{tt} &= 2\partial^2_tM_in^i+\tfrac{8}{5}S^j\btide_{ij}n^i-\tfrac{2}{3}M^j\etide_{ij}n^i +\tfrac{82}{3}m^2\etide_{ij}\nhat^{ij} + 24S_{\langle i}\btide_{jk\rangle}\nhat^{ijk}\nonumber\\
&\quad-20M_{\langle i}\etide_{jk\rangle}\nhat^{ijk}.\label{E_tt}
\end{align}
The $ta$-component is given by
\begin{align}
\tilde E_{ta} &= \tfrac{44}{15}\epsilon_{aij}M^k\btide^j_kn^i -\tfrac{2}{15}\left(11S^i\etide^j_k+18M^i\btide^j\right)\epsilon_{ija}n^k +\tfrac{2}{15}\left(41S^j\etide^k_a -10M^j\btide^k_a\right)\epsilon_{ijk}n^i\nonumber\\
&\quad +4\epsilon_{aij}\left(S^j\etide_{kl} +2M_k\btide^j_l\right)\nhat^{ikl}+4\epsilon^{}_{ij\langle k}\etide_{l\rangle}^jS^i\nhat_a{}^{kl} +\tfrac{68}{3}m^2\epsilon_{aij}\btide^j_k\nhat^{ik}.
\end{align}
This can be decomposed into irreducible STF pieces via the identities
\begin{align}
\epsilon_{aij}S^i\etide^j_k & = S^i\etide^j_{(k}\epsilon_{a)ij}+\tfrac{1}{2}\epsilon_{akj}S^i\etide_i^j,\\
\epsilon_{aj\langle i}\etide_{kl\rangle}S^j&=\mathop{\STF}_{ikl}\!\left[\epsilon^j{}_{al}S_{\langle i}\etide_{jk\rangle} -\tfrac{2}{3}\delta_{al}S^p\etide^j_{(i}\epsilon^{}_{k)jp}\right],\\
\epsilon_{aj\langle i}M_l\btide_{k\rangle}{}^j&=\mathop{\STF}_{ikl}\!\left[\epsilon^j{}_{al}M_{\langle i}\btide_{jk\rangle}\! +\!\tfrac{1}{3}\delta_{al}M^p\btide^j_{(i}\epsilon^{}_{k)jp}\right],
\end{align}
which follow from Eqs.~\eqref{decomposition_1} and \eqref{decomposition_2}, and which lead to
\begin{align}
\tilde E_{ta} &=\tfrac{2}{5}\epsilon_{aij}\Big(6M^k\btide^j_k-7S^k\etide^j_k\Big)n^i+\tfrac{4}{3}\left(2M^l\btide^k_{(i} -5S^l\etide^k_{(i}\right)\epsilon^{}_{j)kl}\nhat_a{}^{ij} \nonumber\\
&\quad+\left(4S^j\etide^k_{(a} -\tfrac{56}{15}M^j\btide^k_{(a}\right)\epsilon_{i)jk}n^i + 4\epsilon_{ai}{}^l\left(S_{\langle j}\etide_{kl\rangle}+2M_{\langle j}\btide_{kl\rangle}\right)\nhat^{ijk}\nonumber\\
&\quad+\tfrac{68}{3}m^2\epsilon_{aij}\btide^j_k\nhat^{ik}.\label{E_ta}
\end{align}
The $ab$-component is given by
\begin{align}
\tilde E_{ab} &= \tfrac{56}{3}m^2\etide_{ij}\nhat_{ab}{}^{ij} +\tfrac{52}{45}m^2\etide_{ab}-\delta_{ab}\left[\left(2\partial^2_tM_i+\tfrac{8}{5}S^j\btide_{ij} +\tfrac{10}{9}M^j\etide_{ij}\right)n^i+\tfrac{100}{9}m^2\etide_{ij}\nhat^{ij}\right] \nonumber\\
&\quad -\delta_{ab}\left(\tfrac{20}{3}M_{\langle i}\etide_{jk\rangle} -\tfrac{8}{3}S_{\langle i}\btide_{jk\rangle}\right)\nhat^{ijk}+\tfrac{8}{15}M_{\langle a}\etide_{b\rangle i}n^i+\tfrac{8}{15}M^i\etide_{i\langle a}n_{b\rangle}+\tfrac{56}{3}m^2\etide_{i\langle a}\nhat^{}_{b\rangle}{}^i\nonumber\\
&\quad+16M_i\etide_{j\langle a}\nhat_{b\rangle}{}^{ij}-\tfrac{32}{5}S_{\langle a}\btide_{b\rangle i}n^i+\tfrac{4}{15}\left(10S_i\btide_{ab}+27M_i\etide_{ab}\right)n^i\nonumber\\
&\quad+\tfrac{16}{3}S^i\btide_{i\langle a}n_{b\rangle}-8\epsilon_{ij\langle a}\epsilon_{b\rangle kl}S^j\btide^l_m\nhat^{ikm}+\tfrac{16}{15}\epsilon_{ij\langle a}\epsilon_{b\rangle kl}S^j\btide^{il}n^k.
\end{align}
Again, this can be decomposed, using the identities
\begin{align}
S_{\langle a}\btide_{b\rangle i} &=S_{\langle a}\btide_{bi\rangle} +\mathop{\STF}_{ab}\tfrac{1}{3}\epsilon_{ai}{}^j\epsilon_{kl(b} \btide_{j)}{}^lS^k +\tfrac{1}{10}\delta_{i\langle a}\btide_{b\rangle j}S^j,\\
S_i\btide_{ab} &=S_{\langle a}\btide_{bi\rangle} -\mathop{\STF}_{ab}\tfrac{2}{3}\epsilon_{ai}{}^j\epsilon_{kl(b} \btide_{j)}{}^lS^k +\tfrac{3}{5}\delta_{i\langle a}\btide_{b\rangle j}S^j,\\
\epsilon_{ij\langle a}\epsilon_{b\rangle kl}S^j\btide^{il} & = \mathop{\STF}_{ab}\epsilon_{akj}S^l\btide^i_{(j}\epsilon^{}_{b)il} -\tfrac{1}{2}\delta_{k\langle a}\btide_{b\rangle i}S^i,\\
\mathop{\STF}_{ikm}\epsilon_{ij\langle a}\epsilon_{b\rangle kl}S^j\btide^l_m & = \mathop{\STF}_{ikm}\mathop{\STF}_{ab}\Big(2\delta_{ai}S_{\langle b}\btide_{km\rangle}+\tfrac{1}{3}\delta_{ai}\epsilon^l{}_{bk}S^j \btide^p_{(l}\epsilon^{}_{m)jp}\nonumber\\
&\quad-\tfrac{3}{10}\delta_{ai}\delta_{bk}\btide_{mj}S^j\Big),
\end{align}
which lead to
\begin{align}
\tilde E_{ab} & = -2\delta_{ab}\left[\left(\partial^2_tM_i+\tfrac{4}{5}S^j\btide_{ij} +\tfrac{5}{9}M^j\etide_{ij}\right)n^i +\left(\tfrac{10}{3}M_{\langle i}\etide_{jk\rangle} -\tfrac{4}{3}S_{\langle i}\btide_{jk\rangle} \right)\nhat^{ijk}\right]\nonumber\\
 &\quad-\tfrac{100}{9}\delta_{ab}m^2\etide_{ij}\nhat^{ij}+\tfrac{1}{5}\left(8M^j\etide_{ij} +12S^j\btide_{ij}\right)\nhat_{ab}{}^i +\tfrac{56}{3}m^2\etide_{ij}\nhat_{ab}{}^{ij}\nonumber\\
&\quad +\tfrac{4}{75}\left(92M^j\etide_{j\langle a} +108S^j\btide_{j\langle a}\right)n_{b\rangle}^{} +\tfrac{56}{3}m^2\etide_{i\langle a}\nhat_{b\rangle}{}^i \nonumber\\
&\quad+16\mathop{\STF}_{aij}\left(M_i\etide_{j\langle a}-S_i\btide_{j\langle a}\right)\nhat^{}_{b\rangle}{}^{ij} -\tfrac{8}{3}\epsilon^{pq}{}_{\langle j}\left(2\etide_{k\rangle p}M_q+\btide_{k\rangle p}S_q\right)\epsilon^k{}_{i(a}\nhat_{b)}{}^{ij}  \nonumber\\
&\quad +\tfrac{16}{15}m^2\etide_{ab}+\tfrac{4}{15}\left(29M_{\langle a}\etide_{bi\rangle}-14S_{\langle a}\btide_{bi\rangle}\right)n^i \nonumber\\
&\quad-\tfrac{16}{45}\mathop{\STF}_{ab}\epsilon_{ai}{}^jn^i\epsilon^{pq}{}_{(b} \left(13\etide_{j)q}M_p+14\btide_{j)q}S_p\right).\label{E_ab}
\end{align}

			\chapter{Local covariant expansion methods}\label{local_expansions}
In the first section in this appendix, I review the method of expanding bitensors at $x$ and $x'$ in the coincidence limit $x\to x'$. A bitensor lives in the tangent space of the Cartesian product of the spacetime manifold with itself, $\man\times\man$, such that, for example, $T^\alpha{}_{\beta'\gamma'}(x,x')$ is a vector at $x$ and a rank-2 tensor at $x'$. The method also applies to the expansion of a tensor at a point $x$ about a nearby point $x'$. See Ref.~\cite{Eric_review} for a detailed, pedagogical review of the subject.

In the second section of this appendix, I present a method of expanding bitensors with one point on a worldtube and one point off the worldtube. The end result is written in terms of tensors on the worldline at the center of the tube's interior.

In both sections, I present explicit expressions for the expansion of various important bitensors.

\section{Important bitensors and near-coincidence expansions}
I begin by defining the convex normal neighbourhood of a point $x'$ to be the set of points that are linked to $x'$ by a unique geodesic. Now consider a bitensor $A_{PQ'}(x,x')$, where $P=i_1...i_p$ and $Q'=i'_i..i'_q$, and where $x$ is within the convex normal neighbourhood of $x'$. The two points are connected by a unique geodesic with coordinates $z^\alpha(\lambda)$, where the parameter $\lambda$ is arbitrary, and the endpoints are given by $x'=z(\lambda_0)$ and $x=z(\lambda_1)$. The near-coincidence expansion of $A$ will consist of an expansion ``along" that geodesic. Such an expansion relies on two fundamental bitensors. The first is Synge's world function, defined as 
\begin{equation}
\sigma(x,x') \equiv \tfrac{1}{2}(\lambda_1-\lambda_0)\int^{\lambda_1}_{\lambda_0}g_{\mu\nu}(z(\lambda))\diff{z^\mu}{\lambda}\diff{z^\nu}{\lambda}d\lambda,
\end{equation}
which is one-half the square of the geodesic-distance between $x'$ and $x$. The derivative $\sigma_\alpha\equiv \partial_\alpha\sigma(x,x')$ is a dual vector at $x$, tangential to the geodesic, pointing away from $x'$, and with a magnitude equal to the geodesic-distance between the two points. The derivative $\sigma_{\alpha'}\equiv \partial_{\alpha'}\sigma(x,x')$ is a dual vector at $x'$, again tangential to the geodesic, now pointing away from $x$, with the same magnitude as $\sigma_\alpha$. Because of their magnitudes, these quantities satisfy the relationships
\begin{equation}
g^{\mu\nu}\sigma_\mu\sigma_\nu =2\sigma= g^{\mu'\nu'}\sigma_{\mu'}\sigma_{\nu'}.
\end{equation}
Taking derivatives of these equalities yields
\begin{equation}
\sigma_\mu = \sigma^\nu\sigma_{\mu\nu},\qquad \sigma_{\mu'} = \sigma^{\nu'}\sigma_{\mu'\nu'}.\label{magnitude}
\end{equation}
For higher derivatives of $\sigma$, I introduce analogous notation (e.g., $\sigma_{\alpha\beta'}=\sigma_{;\alpha\beta'}$, $\sigma_{\alpha\beta\gamma}=\sigma_{;\alpha\beta\gamma}$). Note that covariant derivatives at $x$ and $x'$ commute, such that, for example, $A_{;P'Q}=A_{;QP'}$.

The second bitensor of fundamental importance is the parallel propagator $g^\alpha_{\alpha'}(x,x')$, which parallel-transports a dual vector at $x$ to a dual vector at $x'$, or a vector at $x'$ to a vector at $x$; similarly, $g_\alpha^{\alpha'}(x,x')$ parallel-transports a vector at $x$ to a vector at $x'$, or a dual vector at $x'$ to a dual vector at $x$. Suppose we construct a tetrad $e^\alpha_I$ that is parallel transported on the geodesic connecting $x$ to $x'$. Then the parallel propagator is given by
\begin{equation}
g^\alpha_{\alpha'}=e^\alpha_I e^I_{\alpha'},\qquad g^{\alpha'}_{\alpha}=e^{\alpha'}_I e^I_{\alpha}.
\end{equation}
Since the metric $g$ is parallel-transported between any two points, we have $g_{\alpha\beta}=g^{\alpha'}_{\alpha}g^{\beta'}_{\beta}g_{\alpha'\beta'}$. Because the tetrad is parallel-transported along the geodesic, its covariant derivative in the direction $\sigma_\alpha$ or $\sigma_{\alpha'}$ vanishes, from which it follows that
\begin{equation}
g^\alpha_{\alpha';\beta'}\sigma^{\beta'}= g^\alpha_{\alpha';\beta}\sigma^{\beta}=0.
\end{equation}
We also have 
\begin{equation}
\sigma_\alpha= -g^{\alpha'}_\alpha\sigma_{\alpha'},\qquad \sigma_{\alpha'}= -g^\alpha_{\alpha'}\sigma_{\alpha}.\label{g_sigma relations}
\end{equation}

Now consider taking the limit $x\to x'$. For the coincidence limit of a bi-tensorial quantity, I introduce the notation $[A_{PQ'}]=\lim_{x\to x'}A_{PQ'}(x,x')$, which is a tensor at $x'$. First, Synge's world function satisfies the equalities
\begin{align}
[\sigma]=[\sigma_\alpha]=[\sigma_{\alpha'}]=0,
\end{align}
since the magnitudes of these quantities vanish when $x\to x'$. From Eq.~\eqref{magnitude}, it also satisfies
\begin{equation}
[\sigma_{\mu\nu}]=[\sigma_{\mu'\nu'}]=g_{\mu'\nu'},\qquad [\sigma_{\mu\nu'}]=[\sigma_{\mu'\nu}]=-g_{\mu'\nu'}.
\end{equation}
In order to determine analogous equations for higher derivatives, we introduce Synge's rule, which tells us how to perform derivatives of coincidence limits:
\begin{equation}
[A_{PQ'}]_{;\alpha'}=[A_{PQ';\alpha'}]+[A_{PQ';\alpha}].
\end{equation}
By taking a derivative of Eq.~\eqref{magnitude} and using Synge's rule, we can establish the following identities:
\begin{equation}
[\sigma_{\alpha\beta\gamma}]=[\sigma_{\alpha\beta\gamma'}]=[\sigma_{\alpha\beta'\gamma'}]=[\sigma_{\alpha'\beta'\gamma'}]=0.
\end{equation}
The  parallel propagator satisfies
\begin{equation}
[g^\alpha_{\beta'}]=\delta^{\alpha'}_{\beta'},\qquad [g^\alpha_{\beta';\gamma}]=[g^\alpha_{\beta';\gamma'}]=0,
\end{equation}
the latter of of which follows from taking the derivative of Eq.~\eqref{g_sigma relations}. Coincidence limits of higher derivatives of $\sigma$ and $g^\alpha_{\beta'}$ can be obtained from repeated differentiation of Eqs.~\eqref{magnitude} and \eqref{g_sigma relations}, respectively.

Finally, we write the near-coincidence expansion of bitensors in powers of $-\sigma^{\alpha'}$; this is the covariant generalization of an expansion in powers of $x^{\alpha}-x^{\alpha'}$. For a bitensor  $T_{Q'}(x,x')$, the expansion consists of
\begin{equation}\label{x_expansion}
T_{Q'}(x,x') = \sum_{m\ge0}\frac{(-1)^m}{m!}{}^mt_{Q'\mu'_1...\mu'_m}(x')\sigma^{\mu'_1}\cdots\sigma^{\mu'_m},
\end{equation}
where the coefficients ${}^m t_{Q'\mu'_1...\mu'_m}(x')$ are defined by the recurrence relation \cite{quasilocal}
\begin{align}
{}^0t_{Q'}(x') &= [T_{P'}],\\
{}^mt_{Q'\mu'_1...\mu'_m}(x') & =  [T_{Q';\mu'_1...\mu'_m}]- \sum_{\ell=0}^{m-1}{m\choose\ell}{}^\ell t_{\mu'_1...\mu'_\ell;\mu'_{\ell+1}...\mu'_m}(x').
\end{align}
These equations can be proved by induction. For a bitensor $T_{PQ'}$, we introduce the auxiliary quantity $\widetilde T_{P'Q'}=g^{i_1}_{i'_1}...g^{i_p}_{i'_p}T_{PQ'}$. This auxiliary quantity can be expanded using the above equation, and then the expansion of $T_{PQ'}$ can be retrieved using $T_{PQ'}=g^{i'_1}_{i_1}...g^{i'_p}_{i_p}\widetilde T_{P'Q'}$.

For an ordinary tensor field, the above method reduces to the simple expansion
\begin{equation}
T_{\alpha\beta}(x)=g^{\alpha'}_\alpha g^{\beta'}_\beta\left(T_{\alpha'\beta'}-T_{\alpha'\beta';\gamma'}\sigma^{\gamma'}+\tfrac{1}{2}T_{\alpha'\beta';\gamma'\delta'}\sigma^{\gamma'}\sigma^{\delta'}\right)+O(\zeta^3),
\end{equation}
where $\zeta$ is the geodesic distance between $x$ and $x'$. The generalization to tensors of other ranks is obvious.

\subsection{Near-coincidence expansions of important bitensors}
The expansions of the second derivatives of Synge's world function are given by
\begin{align}
\sigma_{\alpha'\beta'} &= g_{\alpha'\beta'}-\tfrac{1}{3}R_{\alpha'\gamma'\beta'\delta'}\sigma^{\gamma'}\sigma^{\delta'}+O(\zeta^3),\\
\sigma_{\alpha'\beta} &= -g^{\beta'}_\beta\left(g_{\alpha'\beta'}+\tfrac{1}{6}R_{\alpha'\gamma'\beta'\delta'}\sigma^{\gamma'}\sigma^{\delta'}\right)+O(\zeta^3),\\
\sigma_{\alpha\beta} &= g^{\alpha'}_\alpha g^{\beta'}_\beta\left(g_{\alpha'\beta'}-\tfrac{1}{3}R_{\alpha'\gamma'\beta'\delta'}\sigma^{\gamma'}\sigma^{\delta'}\right)+O(\zeta^3),
\end{align}
The expansions of the derivatives of the parallel propagators are given by
\begin{equation}
g^\alpha_{\beta';\gamma'} = \tfrac{1}{2}g^\alpha_{\alpha'}R^{\alpha'}{}_{\beta'\gamma'\delta'}\sigma^{\delta'}+O(\zeta^2),\qquad 
g^\alpha_{\beta';\gamma} = \tfrac{1}{2}g^\alpha_{\alpha'}g^{\gamma'}_{\gamma}R^{\alpha'}{}_{\beta'\gamma'\delta'}\sigma^{\delta'}+O(\zeta^2).
\end{equation}
In a vacuum spacetime, the expansions of the bitensors $U_{\alpha\beta\alpha'\beta'}$ and $V_{\alpha\beta\alpha'\beta'}$, which appear in the Green's functions for the gravitational wave operator, are given by
\begin{align}
U^{\alpha\beta}{}_{\gamma'\delta'} &= g^{(\alpha}_{\gamma'}g^{\beta)}_{\delta'}+O(\zeta^3),\\
V^{\alpha\beta}{}_{\gamma'\delta'} &= g^{(\alpha}_{\alpha'}g^{\beta)}_{\beta'}R^{\alpha'}{}_{\gamma'}{}^{\beta'}{}_{\delta'}+O(\zeta),\\
U^{\alpha\beta}{}_{\gamma'\delta';\mu'} &= g^{(\alpha}_{\gamma'}g^{\beta)}_{\delta'}\delta^{\beta'}_{(\gamma'}R^{\alpha'}{}_{\delta')\mu'\nu'}\sigma^{\nu'}+O(\zeta^2),\\
U^{\alpha\beta}{}_{\gamma'\delta';\mu} &= g^{(\alpha}_{\gamma'}g^{\beta)}_{\delta'}g^{\mu'}_\mu\delta^{\beta'}_{(\gamma'}R^{\alpha'}{}_{\delta')\mu'\nu'}\sigma^{\nu'}+O(\zeta^2).
\end{align}

\section{Expansions of bitensors near the worldtube}\label{worldtube expansions}
I present here the expansions of various important bitensors of the form $T(x,x')$, where $x$ is a point in the exterior of the worldtube $\Gamma$, and $x'$ is a nearby point on the worldtube. The expansions are based on the methods just presented. To keep track of the orders of the expansions, I use the quantity $\zeta\sim r\sim \rad\sim\Delta t$, where $r$ is the radial coordinate at $x$, $\rad$ is the radius of $\Gamma$, and $\Delta t$ is the proper-time difference between $x'$ and $x$. For brevity, I introduce the shorthand notation $x^{ab}\equiv x^a x^b$, $\spb\equiv\sigma(x',\bar x)$, $\sb\equiv\sigma(x,\bar x)$, and $\spp\equiv\sigma(x,x'')$, where $\bar x$ and $x''$ are points on the worldline $\gamma$. The relationship between the various points will eventually be identified with that depicted in Fig.~\ref{tube}.

For completeness, the expansions in this section allow for an arbitrary, non-vacuum background and an arbitrarily accelerating worldline. In the calculations in Sec.~\ref{integral_expansion1}, both the acceleration and the Ricci tensor can be set to zero.

%%%%%%%%%%%%
\subsection{General expansions}
%%%%%%%%%%%%

Consider a bitensor $T_{\alpha'}(x,x')$. We can expand this in a sequence of steps: First, we expand along a geodesic that connects the point $x'$ on the worldtube to a point $\bar x=\gamma(t')$ on the worldline; this is an expansion in powers of $\rad$. Next, we expand up along the worldline, about a point $x''=\gamma(t)$; this is an expansion in powers of the proper-time difference $\Delta t\equiv t-t'$. These two expansions leave us with bitensors that depend on $x$ and $x''$. The final step in our procedure is a near-coincidence expansion of these bitensors; this last step is an expansion in powers of $r$. This procedure does not rely on any particular relationship between $x'$ and $\bar x$ or between $x$ and $x''$. It becomes a coordinate expansion by fixing these relationships: for example, by connecting $x$ to $x''$ with a geodesic perpendicular to the worldline (and doing likewise for $x'$ and $\bar x$), the covariant expansion becomes an expansion in Fermi coordinates.

We first hold $x$ fixed and expand the $x'$-dependence about $\bar x$:
\begin{equation}\label{xprime_expansion}
T_{\alpha'}(x,x') = g^{\bar\alpha}_{\alpha'}\sum_{k\ge0}\frac{(-1)^k}{k!} T_{\bar\alpha;\bar\gamma_1...\bar\gamma_k}(x,\bar x)\spb^{\bar\gamma_1}\cdots\spb^{\bar\gamma_k}.
\end{equation}
where the reader is reminded that the parallel propagator is given by $g^{\bar\alpha}_{\alpha'}=e^{\bar\alpha}_Ie^I_{\alpha'}$. Next, still holding $x$ fixed, we expand each of the bitensors $T_{\bar\alpha;\bar\gamma_1...\bar\gamma_k}(x,\bar x)$ about $x''$. Since we do not possess a convenient expression for the parallel propagator between $\bar x$ and $x''$, we perform this expansion along the worldline. We do this by expressing $T_{\bar\alpha}$ in terms of its tetrad components, converting to tetrad components via the relationship $T_{\bar\alpha}=T_I(t')e^I_{\bar\alpha}$, and then expanding the tetrad components in powers of the proper time interval $\Delta t$. The time-derivatives along the worldline are evaluated covariantly by re-expressing the tetrad components in terms of the coordinate basis, leading to 
\begin{align}\label{t_expansion}
T_{\bar\alpha;\bar\gamma_1...\bar\gamma_k} & = e^I_{\bar\alpha}e^{J_1}_{\bar\gamma_1}\cdots e^{J_k}_{\bar\gamma_k}\sum_{n\ge0}\frac{(-1)^n}{n!}(\Delta t)^n\bigg(\frac{D}{dt''}\bigg)^{\!\!n}\!\!\Big(T_{\alpha'';\gamma''_1...\gamma''_k} e^{\alpha''}_Ie_{J_1}^{\gamma''_1}\cdots e_{J_k}^{\gamma''_k}\Big).
\end{align}
This can be expressed in terms of covariant derivatives of $T$ and combinations of tetrad and acceleration vectors by using the identity
\begin{align}\label{t_expansion2}
\left(\frac{D}{dt''}\right)^{\!\!n}\!\! \left(T_{\alpha'';\gamma''_1...\gamma''_k}e^{\alpha''}_Ie_{J_1}^{\gamma''_1} \cdots e_{J_k}^{\gamma''_k}\right)
&=\sum_{i=0}^n{n\choose i}\left(\frac{D}{dt''}\right)^{\!\!n-i}\!\!\! \Big(e^{\alpha''}_Ie_{J_1}^{\gamma''_1}\cdots e_{J_k}^{\gamma''_k}\Big)\nonumber\\
&\times\sum_{j=0}^iT_{\alpha'';\gamma''_1...\gamma''_k\delta''_1...\delta''_j}A^{\delta''_1...\delta''_j}(i,j),
\end{align}
where the derivative of the tetrad is given by $\frac{D}{dt}e^{\alpha}_I=(u^\alpha a_\beta-a^\alpha u_\beta)e^\beta_I$. The indexed tensor $A^{\delta''_1...\delta''_j}(i,j)$ is constructed from the four-velocity $u^{\alpha''}$ and its derivatives. Explicitly, 
\begin{align}
A(0,0) & = 1, \\
A^{\delta''_1...\delta''_j}(i,j) & = \frac{D}{dt''}A^{\delta''_1...\delta''_j}(i-1,j)+A^{\delta''_1...\delta''_{j-1}}(i-1,j-1)u^{\delta''_j},
\end{align}
where $0\leq j\leq i$, and $A(i,j)\equiv 0$ for $j<0$ and $j>i$.

Finally, the bitensors $T_{\alpha'';\gamma''_1...\gamma''_k\delta''_1...\delta''_j}(x,x'')$ can be expanded about $x''$ using the near-coincidence expansion method presented in the previous section (with $x'$ replaced by $x''$ in Eq.~\eqref{x_expansion}). Substituting Eq. \eqref{x_expansion} into Eq.~\eqref{t_expansion2}, Eq.~\eqref{t_expansion2} into Eq.~\eqref{t_expansion}, and Eq.~\eqref{t_expansion} into Eq.~\eqref{xprime_expansion}, we arrive at an expression for $T_{\alpha'}(x,x')$ in terms of tensors at $x''$ and the small expansion quantities $\spb^{\bar\alpha}$, $\Delta t$, and $\spp^{\alpha''}$. This procedure is valid in any coordinate system. It can be made into a coordinate expansion in terms of Fermi coordinates $(t,x^a)$ by using the identities $\spb^{\bar\beta}=-e^{\bar\beta}_bx'^b$ and $\spp^{\beta''}=-e^{\beta''}_bx^b$, and identifying $t$ with Fermi time. Analogous identities would generate an expansion in terms of retarded coordinates $(u,x_{\text{ret}}^a)$ or advanced coordinates $(v,x_{\text{adv}}^a)$

The generalization of this procedure to tensors of other ranks is obvious.

%%%%%%%%%%%
\subsection{Expansions of $\Delta t$, $\sigma_{\mu'}(x,x')$, and related quantities}
%%%%%%%%%%%

From this point on, I restrict myself to the case in which $x$ and $x'$ are connected by a unique null geodesic. In other words, $x'$ lies on the surface $\mathcal{S}$ in Fig.~\ref{tube}. I begin by expanding $\sigma(x,x')$; since $x'\in\mathcal{S}$, we have $\sigma(x,x')=0$. We can make use of this fact to find $\Delta t$, which is required for all other expansions.

So, following the procedure outlined above, we first expand about $\bar x$:
\begin{align}\label{sigma_expansion}
\sigma(x,x') & = \sb-\sb_{\bar\alpha}\spb^{\bar\alpha} +\tfrac{1}{2}\sb_{\bar\alpha\bar\beta}\spb^{\bar\alpha}\spb^{\bar\beta}
-\tfrac{1}{6}\sb_{\bar\alpha\bar\beta\bar\gamma}\spb^{\bar\alpha} \spb^{\bar\beta}\spb^{\bar\gamma}\nonumber\\
&\quad+\tfrac{1}{24}\sb_{\bar\alpha\bar\beta\bar\gamma\bar\delta} \spb^{\bar\alpha}\spb^{\bar\beta}\spb^{\bar\gamma}\spb^{\bar\delta} +\order{\zeta^5}.
\end{align}
We next expand $\sb_{...}$ about $x''$: for example,
\begin{align}
\sb & = \spp-\spp_{\mu''}u^{\mu''}\Delta t +\tfrac{1}{2}\big(\spp_{\mu''}u^{\mu''}\big)_{\!;\nu''}u^{\nu''}(\Delta t)^2-\tfrac{1}{6} \big(\big(\spp_{\mu''}u^{\mu''}\big)_{\!;\nu''}u^{\nu''}\big)_{\!;\rho''} u^{\rho''}(\Delta t)^3\nonumber\\
&\quad+\tfrac{1}{24} \big(\big(\big(\spp_{\mu''}u^{\mu''}\big)_{\!;\nu''}u^{\nu''}\big)_{\!;\rho''} u^{\rho''}\big)_{\!;\upsilon''}u^{\upsilon''}(\Delta t)^4+\order{\zeta^5}.
\end{align}
Using $\spp=\frac{1}{2}r^2$, $\spp_{\mu''}=-e^a_{\mu''}x_a$, and the standard near-coincidence expansion $\spp_{\mu''\nu''}=g_{\mu''\nu''}-\frac{1}{3}R_{\mu''\gamma''\nu''\delta''} \spp^{\gamma''}\spp^{\delta''}+\mathcal{O}(\zeta^3)$, and dropping terms of order $a^2$, we arrive at the expansion
\begin{align}
\sb & =\tfrac{1}{2}r^2-\tfrac{1}{2}\left(1+a_c(t)x^c +\tfrac{1}{3}R_{0c0d}(t)x^{cd}\right)(\Delta t)^2+\tfrac{1}{6}\left(\dot a_c(t)x^c\right)(\Delta t)^3+\order{\zeta^5}.
\end{align}
The same procedure yields
\begin{align}
\sb_{\bar\alpha} &= -x_ae^a_{\bar\alpha}+\big(e^0_{\bar\alpha}+a_a(t)x^ae^0_{\bar\alpha} +\tfrac{1}{3}R_{0a0b}(t)x^{ab}e^0_{\bar\alpha}+\tfrac{1}{3}R_{ca0b}(t)x^{ab}e^c_{\bar\alpha}\big)\Delta t\nonumber\\
&\quad-\tfrac{1}{2}\left(\tfrac{1}{3}R_{0a0b}(t)x^ae^b_{\bar\alpha} +a_a(t)e^a_{\bar\alpha}\right)(\Delta t)^2+\tfrac{1}{3}\dot a_a(t)e^a_{\bar\alpha}(\Delta t)^3+\order{\zeta^4},\\
\sb_{\bar\alpha\bar\beta} & = g_{\bar\alpha\bar\beta} -\tfrac{1}{3}R_{IaJb}(t)x^{ab}e^I_{\bar\alpha}e^J_{\bar\beta} +\tfrac{2}{3}R_{0IJb}(t)x^be^I_{(\bar\alpha}e^J_{\bar\beta)}\Delta t\nonumber\\
&\quad -\tfrac{1}{3}R_{a0b0}(t)e^a_{\bar\alpha}e^b_{\bar\beta}(\Delta t)^2+\order{\zeta^3},\\
\sb_{\bar\alpha\bar\beta\bar\gamma} & = -\tfrac{2}{3}R_{KIJa}(t)x^ae^{I}_{(\bar\alpha}e^{J}_{\bar\beta)} e^{K}_{\bar\gamma} -\tfrac{2}{3}R_{KIJ0}(t)e^{I}_{(\bar\alpha} e^{J}_{\bar\beta)}e^{K}_{\bar\gamma}\Delta t+\order{\zeta^2},\\
\sb_{\bar\alpha\bar\beta\bar\gamma\bar\delta} & = \tfrac{2}{3}R_{KIJL}(t)e^I_{(\bar\alpha}e^J_{\bar\beta)} e^K_{\bar\gamma}e^L_{\bar\delta} +\order{\zeta}.
\end{align}

Substituting these expansions into \eqref{sigma_expansion} and setting the result equal to zero, we get
\begin{align}\label{sigma_equation}
0 & = \sigma(x,x') \nonumber\\
 & = \tfrac{1}{2}r^2+\tfrac{1}{2}\rad^2-x_ax'^a-\tfrac{1}{6}R_{acbd}(t)x^{cd}x'^{ab} +\tfrac{1}{3}R_{0abc}(t)(x'^{ab}x^c+x'^bx^{ac})\Delta t \nonumber\\
&\quad -\tfrac{1}{2}\Big[1+a_c(t)(x^c+x'^c) +\tfrac{1}{3}R_{0c0d}(t)(x^{cd}+x^cx'^d+x'^{cd})\Big](\Delta t)^2 \nonumber\\
&\quad +\tfrac{1}{6}\dot a_c(t)(x^c+2x'^c)(\Delta t)^3+\order{\zeta^5}.
\end{align}

We next expand $\Delta t$ as 
\begin{equation}
\Delta t = \zeta\left(\Delta t_0+\zeta\Delta t_1+\zeta^2\Delta t_2\right)+\order{\zeta^4}.
\end{equation}
Substituting this into \eqref{sigma_equation} and solving order by order, we find
\begin{equation}\label{Delta t}
\begin{split}
\Delta t_0 & = \r_0, \\
\Delta t_1 & = -\tfrac{1}{2}\r_0a_c(t)(x^c+x'^c),\\
\Delta t_2 & = -\tfrac{1}{6}\r_0^{-1}R_{acbd}(t)x^{cd}x'^{ab} -\tfrac{1}{6}\r_0R_{0a0b}(t)(x^{ab}+x^ax'^b+x'^{ab})\\
&\quad+\tfrac{1}{3}R_{0abc}(t)(x^{ac}x'^b-x'^{ac}x^b).
\end{split}
\end{equation}
where
\begin{equation}
\r_0=\sqrt{r^2+\rad^2-2r\rad n^an'_a}
\end{equation}
is the flat-spacetime luminosity distance between $x$ and $x'$.

Using this result for $\Delta t$, we can now find an explicit expansion for any bitensor at $x$ and $x'$. In particular, $\r\equiv\sigma_{\mu'}\frac{\partial x^{\mu'}}{\partial t'}$ can be expanded as
\begin{equation}
\r = \zeta(\r_0+\zeta\r_1+\zeta^2\r_2)+\order{\zeta^4},
\end{equation}
where $\r_0$ is given above, and
\begin{equation}
\begin{split}
\r_1  &= \tfrac{1}{2}\r_0a_c(t)(x^c+x'^c),\\
\r_2  &= -\tfrac{1}{6}\r_0^{-1}R_{acbd}(t)x^{cd}x'^{ab}-\r_0^2\dot a_a(t)x'^a +\tfrac{1}{6}\r_0R_{0a0b}(t)(x^{ab}+x^ax'^b+x'^{ab}).
\end{split}
\end{equation}
Note that $\r_1=-\Delta t_1$, which is what we would expect in flat spacetime.

The time-derivative of $\r$ can similarly be expanded to find
\begin{equation}
\partial_{t'}\r = \dot{\r_0}+\zeta\dot{\r_1}+\zeta^2\dot{\r_2}+\order{\zeta^3},
\end{equation}
where
\begin{align}
\dot{\r_0} &\equiv -1 \nonumber\\
\dot{\r_1} &\equiv -a_a(t)(x'^a+x^a)\\
\dot{\r_2} &\equiv 2\r_0\dot a_a(t)x'^a-\tfrac{1}{3}R_{0a0b}(t)(x^{ab}+x^ax'^b+x'^{ab})\nonumber
\end{align}

Other useful expansions are
\begin{align}
\sigma_{\mu'}(x,x')n^{\mu'} &= \rad-x_an'^a-\tfrac{1}{2}\r_0a_a(t)n'^a(\r_0-2\r_1)+\tfrac{1}{3}\dot a_a(t)n'^a\r_0^3\nonumber\\
&\quad-\tfrac{1}{6}\r_0^2R_{0a0b}(t)(x^a+2x'^a)n'^b+\tfrac{1}{3}\r_0R_{0acb}(t)(x^{ab}+2x'^ax^b)n'^c\nonumber\\
&\quad-\tfrac{1}{3}\rad R_{acbd}(t)x^{ab}n'^{cd}+\order{\zeta^4}
\end{align}
and
\begin{align}
\sigma_{\mu'\nu'}(x,x')n^{\mu'}u^{\nu'} &= \tfrac{1}{3}\r_0R_{0a0b}(t)(x^b-x'^b)n'^a+\tfrac{1}{3}R_{0abc}(t)(x^a-x'^a)x^bn'^c\nonumber\\
&\quad+\order{\zeta^3}.
\end{align}

%%%%%%%%%%%
\subsection{Expansions of Green's function}
%%%%%%%%%%%
I present here the expansion of part of the Green's function for the case in which $x$, $x'$, $\bar x$, and $x''$ lie within one another's convex normal neighbourhood. By following the same procedure and making use of the near-coincidence expansions given above, one finds the following expansion for the direct part of the Green's function:
\begin{align}
U\indices{_{\alpha\beta}^{\alpha'\beta'}}  & = e^{(\alpha'}_{I}e^{\beta')}_{J}\Big[\U^1_{\alpha\beta}{}^{IJ} +\r_0\U^2_{\alpha\beta}{}^{IJ} +\r_0^2\U^3_{\alpha\beta}{}^{IJ} +\U^4_{\alpha\beta}{}^{IJ}{}_cx'^c+\r_0\U^5_{\alpha\beta}{}^{IJ}{}_cx'^c \nonumber\\
&\quad +\U^6_{\alpha\beta}{}^{IJ}{}_{cd}x'^{cd}+\order{\zeta^3}\Big],
\end{align}
where
\begin{align}
\U^1_{\alpha\beta}{}^{IJ} &= e_{(\alpha}^{I}e_{\beta)}^{J}\left(1+\tfrac{1}{12}R_{kl}(t)x^{kl}\right),\\
\U^2_{\alpha\beta}{}^{IJ} &= e_{(\alpha}^{I}e_{\beta)}^{K}R_K{}^J{}_{0l}(t)x^l +\tfrac{1}{6}e_{(\alpha}^{I}e_{\beta)}^{J}R_{0k}(t)x^k +2e^I_{(\alpha}\left(e^b_{\beta)}\delta^J_0 +e^0_{\beta)}\delta^{Jb}\right)a_b(t),\\
\U^3_{\alpha\beta}{}^{IJ} &= \tfrac{1}{12}e_{(\alpha}^{I}e_{\beta)}^{J}R_{00}(t) -e^I_{(\alpha}\left(e^b_{\beta)}\delta^J_0+e^0_{\beta)}\delta^{Jb}\right)\dot a_b(t),\\
\U^4_{\alpha\beta}{}^{IJ}{}_c &= -e_{(\alpha}^{I}e_{\beta)}^{M}R_M{}^J{}_{cd}(t)x^d -\tfrac{1}{6}e_{(\alpha}^{I}e_{\beta)}^{J}R_{cd}(t)x^d,\\
\U^5_{\alpha\beta}{}^{IJ}{}_c &= -e_{(\alpha}^{I}e_{\beta)}^{M}R_M{}^J{}_{c0}(t) -\tfrac{1}{6}e_{(\alpha}^{I}e_{\beta)}^{J}R_{c0}(t),\\
\U^6_{\alpha\beta}{}^{IJ}{}_{cd} &= \tfrac{1}{12}e_{(\alpha}^{I}e_{\beta)}^{J}R_{cd}(t).
\end{align}
And the expansion of its covariant derivative is given by
\begin{align}
U\indices{_{\alpha\beta}^{\alpha'\beta'}_{;\delta'}} & = e^{(\alpha'}_{I}e^{\beta')}_{J}e^K_{\delta'}\Big[\U^7_{\alpha\beta}{}^{IJ}{}_K +\r_0\U^8_{\alpha\beta}{}^{IJ}{}_K +\U^9_{\alpha\beta}{}^{IJ}{}_{Kc}x'^c+\order{\zeta^2}\Big],
\end{align}
where
\begin{align}
\U^7_{\alpha\beta}{}^{IJ}{}_K  &= -e_{(\alpha}^{I}e_{\beta)}^{L}R_L{}^J{}_{Kc}(t)x^c -\tfrac{1}{6}e_{(\alpha}^{I}e_{\beta)}^{J}R_{Kc}(t)x^c,\\
\U^8_{\alpha\beta}{}^{IJ}{}_K  &= -e_{(\alpha}^{I}e_{\beta)}^{L}R_L{}^J{}_{K0}(t) -\tfrac{1}{6}e_{(\alpha}^{I}e_{\beta)}^{J}R_{K0}(t),\\
\U^9_{\alpha\beta}{}^{IJ}{}_{Kc}  &= e_{(\alpha}^{I}e_{\beta)}^{L}R_L{}^J{}_{Kc}(t) +\tfrac{1}{6}e_{(\alpha}^{I}e_{\beta)}^{J}R_{Kc}(t).
\end{align}

			\chapter{Local coordinate systems}\label{coordinates}
In this chapter I review the construction of coordinates centered on an arbitrary worldline $\gamma$. The construction relies on near-coincidence expansions and bitensors introduced in Appendix~\ref{local_expansions}. The coordinates exist only in the convex normal neighbourhood of $\gamma$. 

I present the metric in both Fermi and retarded coordinates, along with the transformation between them. For more detail, refer to Ref.~\cite{Eric_review}.

\begin{figure}
\begin{center}
\includegraphics{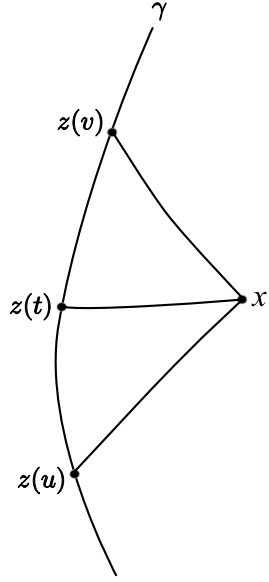}
\end{center}
\caption[Points involved in local coordinates]{The point $x$ off the worldline is connected by unique geodesics to three relevant points on the worldline $\gamma$, which has coordinates $z(\tau)$. The points $z(u)$ and $z(v)$, where $u$ is the retarded time and $v$ is the advanced time, are connected to $x$ by null geodesics. The point $z(t)$ is connected to $x$ by a spatial geodesic that is perpendicular to $\gamma$.}
\label{coord_points}
\end{figure}

\section{Fermi coordinates}
Let $z^\mu(\tau)$ be the coordinate representation of $\gamma$, where $\tau$ is proper time, and let $u^\mu=dz^\mu/d\tau$ and $a^\mu=Du^\mu/d\tau$ be the four-velocity and acceleration on the worldline. The construction of Fermi coordinates is based on Fermi-Walker transport, which differs from parallel transport but which still preserves inner products. A vector field $v^\mu$ is said to be Fermi-Walker transported along the worldline if
\begin{equation}
\frac{Dv^\mu}{d\tau} = v_\nu a^\nu u^\mu-v_\nu u^\nu a^\mu.
\end{equation}
Now we erect an orthonormal tetrad along $\gamma$ by choosing a tetrad $e^\mu_I=(u^\mu,e_i^\mu)$ at an arbitrary point, then Fermi-Walker transporting it along $\gamma$. (Note that $u^\mu$ is automatically Fermi-Walker transported.) At each point on $\gamma$, we then have
\begin{equation}
\frac{De^\mu_i}{d\tau}=a_i u^\mu,\qquad g_{\mu\nu}u^\mu u^\nu = -1,\qquad g_{\mu\nu}e^\mu_iu^\nu=0,\qquad g_{\mu\nu}e^\mu_i e^\nu_j = \delta_{ij},
\end{equation}
where $a_i\equiv a_\nu e^\nu_i$. The dual tetrad, defined by $e^I_\mu=\eta^{IJ}g_{\mu\nu}e^\nu_J$, is given by $e^0_\mu=-u_\mu$ and $e^i_\mu=\delta^{ij}g_{\mu\nu}e^\nu_j$. (Here $\eta_{IJ}={\rm diag}(-1,1,1,1)$.) The metric can be written in terms of these quantities as 
\begin{equation}
g_{\mu\nu}=\eta_{IJ}e^I_\mu e^J_\nu = -e^0_\mu e^0_\nu+\delta_{ij}e^i_\mu e^j_\nu,\qquad
g^{\mu\nu}=\eta^{IJ}e_I^\mu e_J^\nu = -u^\mu u^\nu+\delta^{ij}e_i^\mu e_j^\nu.
\end{equation}
Note that spatial triad indices $i,j,k$ are raised and lowered with $\delta_{ij}$, tetrad indices $I,J,K$ are raised and lowered with $\eta_{IJ}$, and coordinate indices $\alpha,\beta,\gamma$ are raised and lowered with $g_{\mu\nu}$.

Fermi coordinates are now defined as follows. For a point $x$, we select the unique geodesic that passes through $x$ and intersects $\gamma$ orthogonally. The intersection point is denoted $\bar x=z(t)$, where $t$ is the value of proper time at that point. The Fermi coordinates $x^\mu$ at $x$ are defined as
\begin{equation}
x^0=t,\qquad x^a=-e^a_{\bar\alpha}\sigma^{\bar\alpha}(x,\bar x).
\end{equation} 
Along with these definitions, we have the constraint $\sigma_{\bar\alpha}u^{\bar\alpha}=0$, which enforces the condition that the geodesic connecting $x$ to $\bar x$ is perpendicular to $\gamma$. Along with the Cartesian-type coordinates $x^i$, I will frequently make use of the radial coordinate
\begin{equation}
r\equiv\sqrt{\delta_{ij}x^i x^j}=\sqrt{2\sigma(x,\bar x)},
\end{equation}
which is the geodesic distance between $x$ and $\bar x$. From this, we define the unit vector $n^i\equiv x^i/r$, which can be written covariantly as
\begin{equation}
n_\mu=\partial_\mu r.
\end{equation}
This one-form has the convenient property that its indices can be raised and lowered with either $\delta_{ab}$ or $g_{\alpha\beta}$: $n_\alpha$ has components $(0,n_a)$ in Fermi coordinates, where $n_a=\delta_{ab}n^b$, and with a raised index $n^\alpha\equiv g^{\alpha\beta}n_\beta$ has components $(0,n^a)$.

In order to construct the metric in these coordinates, consider taking exterior derivatives of the above identities. Note that $\bar x$ is implicitly a function of $x$, so $d\bar x=\partial_\mu \bar x dx^\mu$; since $\bar x$ is constrained to lie on $\gamma$, we have $\partial_\mu \bar x^\alpha dx^\mu = u^{\bar\alpha}dt$. Now, from the definition of $x^a$, we have $dx^a=-e^a_{\bar\alpha}\sigma^{\bar\alpha}{}_{\beta}dx^\beta-e^a_{\bar\alpha}\sigma^{\bar\alpha}{}_{\bar\beta}u^{\bar\beta}dt$. Taking the exterior derivative of the orthogonality constraint yields another such relationship. Putting the two together, we find
\begin{equation}
dt=\mu\sigma_{\bar\alpha\beta}u^{\bar\alpha}dx^\beta,\qquad dx^a=-e^a_{\bar\alpha}(\sigma^{\bar\alpha}{}_\beta+\mu\sigma^{\bar\alpha}{}_{\bar\beta}u^{\bar\beta}\sigma_{\beta\bar\gamma}u^{\bar\gamma})dx^\beta,
\end{equation}
where $\mu\equiv-(\sigma_{\bar\alpha\bar\beta}u^{\bar\alpha}u^{\bar\beta}+\sigma_{\bar\alpha}a^{\bar\alpha})^{-1}$.

We expand the above differential equalities using the near-coincidence expansions presented in Appendix~\ref{local_expansions}. In the expansions, we substitute $\sigma^{\bar\alpha}=-e^{\bar\alpha}_ax^a$ and $g^{\bar\alpha}_{\alpha}=e^{\bar\alpha}_I e^I_\alpha$, where the dual tetrad at $x$ is defined by parallel transporting the tetrad at $\bar x$ along the geodesic connecting them. The result of this expansion is
\begin{align}
dt &= \left[1-a_a x^a+(a_ax^a)^2-\tfrac{1}{2}R_{0c0d}x^c x^d\right]e^0_\beta dx^\beta-\tfrac{1}{6}R_{0cbd}x^c x^d e^b_\beta dx^\beta+O(r^3),\\
dx^a &= \tfrac{1}{2}R^a_{c0d}x^cx^de^0_\beta dx^\beta+(\delta^a_b+\tfrac{1}{6}R_{0cbd}x^c x^d)e^b_\beta dx^\beta,
\end{align}
where $a_a$ is evaluated at time $t$, and $R_{IJKL}(t)\equiv R_{\bar\alpha\bar\beta\bar\gamma}e^{\bar\alpha}_Ie^{\bar\alpha}_Je^{\bar\alpha}_Ke^{\bar\alpha}_L$ are frame components of the Riemann tensor evaluated on the worldline. Note that from the above equations we can infer that frame components on $\gamma$ are equal to Fermi coordinate components there.

Inverting the expressions for the differentials, we find
\begin{align}
e^0_\alpha &= \left(1+ra_an^a+\tfrac{1}{2}r^2R_{0c0d}n^{cd}\right)t_\alpha+\tfrac{1}{6}r^2R_{0cbd}n^{cd}x^b_\alpha+\order{\zeta^3}, \label{Fermi_tetrad1}\\
e^a_\alpha &= \left(\delta^a_b-\tfrac{1}{6}r^2R^a{}_{cbd}n^{cd}\right)x^b_\alpha-\tfrac{1}{2}r^2R^a{}_{c0d}n^{cd}t_\alpha+\order{\zeta^3},\label{Fermi_tetrad2}
\end{align}
where I have introduced $t_\alpha\equiv\partial_\alpha t$ and $x^a_\alpha\equiv\partial_\alpha x^a$, which are the components of $dt$ and $dx^a$ in an arbitrary coordinate system. Using the identity $g_{\alpha\beta}=\eta_{IJ}e^I_\alpha e^J_\beta$, we arrive at the components of the metric:
\begin{align}
g_{tt} &= -(1+ra_in^i)^2-r^2R_{0i0j}n^i n^j+O(r^3),\label{Fermi_tt}\\
g_{ta} &= -\frac{2}{3}r^2R_{0iaj}n^in^j+O(r^3),\label{Fermi_ta}\\
g_{ab} &= \delta_{ab}-\frac{1}{3}r^2R_{aibj}n^i n^j+O(r^3).\label{Fermi_ab}
\end{align}

I also define the electric-type tidal field 
\begin{equation}
\etide_{ab}\equiv R_{a0b0},
\end{equation}
and the magnetic-type tidal field
\begin{equation}
\btide_{ab}\equiv \tfrac{1}{2}\epsilon_a{}^{cd}R_{0bcd}.
\end{equation}
(If the spacetime is not vacuum, then these tidal fields are defined in terms of the Weyl tensor rather than the complete Riemann tensor.) In vacuum, each of these fields is symmetric trace-free (STF) with respect to the Euclidean spatial metric $\delta_{ab}$, and they satisfy the identities
\begin{equation}
\delta^{cd}R_{acbd}=\etide_{ab},\qquad R_{0abc} = \epsilon_{bcd}\btide^d_a,
\end{equation}
and
\begin{equation}
R_{acbd} = \delta_{ab}\etide_{cd}+\delta_{cd}\etide_{ab}-\delta_{ad}\etide_{bc}-\delta_{bc}\etide_{ab}.
\end{equation}
In Eqs.~\eqref{Fermi background tt}--\eqref{Fermi background ab}, the background metric is written in terms of these quantities. Throughout most of the thesis, the metric perturbation is written in terms of them.

\section{Retarded coordinates}\label{retarded coordinates}
Retarded coordinates are constructed in a very similar manner. We define a tetrad $b^\alpha_I$ on $\gamma$; this tetrad will be identified with the Fermi tetrad on the worldline, but it will be transported off the worldline differently. Given a point $x$ off $\gamma$, we connect it to a point $x'=z(u)$ on $\gamma$ via the unique future-directed null-geodesic that goes from the worldline to $x$. The coordinates at $x$ are defined to be
\begin{equation}
x^0=u,\qquad x^a_{\rm ret}= -e^a_{\alpha'}\sigma^{\alpha'}(x,x'),
\end{equation}
subject to the constraint $\sigma(x,x')=0$. The retarded distance between $x$ and $x'$ is given by
\begin{equation}
\retr\equiv (\delta_{ij} x^i_{\rm ret} x^j_{\rm ret})^{1/2}=u_{\alpha'}\sigma^{\alpha'},
\end{equation}
from which we define the unit vector $\Omega^i\equiv x^i_{\rm ret}/\retr$.

I now define the tetrad at $x$ by parallel-propagating the tetrad at $x'$ along the null geodesic connecting the two points. Following the same steps as in the previous section, one can then derive an expression for the components of the metric. I convert it to polar coordinates $(u,\retr,\theta^A)$ by defining 
\begin{equation}
\Omega^i_A\equiv \pdiff{\Omega^i}{\theta^A}.
\end{equation}
In terms of this quantity, the transformation is accomplished via $\pdiff{x^a_{\rm ret}}{\retr}=\Omega^a$ and $\pdiff{x^a_{\rm ret}}{\theta^A}=\retr\Omega^a_A$. The final result is
\begin{align}
g_{uu} &= -(1+\retr a_i(u)\Omega^i)^2+\retr^2 a_i(u)a^i(u)-\retr^2\etide^*(u)+\order{\retr^3}, \label{ret_coords1}\\
g_{ur} & = -1, \\
g_{uA} & = \retr^2 a_a(u)\Omega^a_A+\tfrac{2}{3}\retr^3(\etide^*_A+\btide^*_A)+\order{\retr^4}, \\
g_{rr} & = g_{rA} = 0, \\
g_{AB} & = \retr^2\Omega_{AB}-\tfrac{1}{3}\retr^4(\etide^*_{AB}+\btide^*_{AB}) +\order{\retr^5},\label{ret_coords5}
\end{align}
where I have defined the quantities
\begin{align}
\etide^* &\equiv \etide_{ab}\Omega^{ab},\label{Estar}\\
\etide^*_A &\equiv \etide_{ab}\Omega^a_A\Omega^b,\\
\etide^*_{AB} &\equiv 2\etide_{ab}\Omega^a_A\Omega^b_B+\etide^*\Omega_{AB},\\
\btide^*_A &\equiv \epsilon_{abc}\Omega^a_A\Omega^b\btide^c_d\Omega^d,\\
\btide^*_{AB} &\equiv 2\epsilon_{acd}\Omega^c\btide^d_b\Omega^a_{(A}\Omega^b_{B)}.\label{Bstar_AB}
\end{align}
In these expressions, $\Omega_{AB}=\delta_{ab}\Omega^a_A\Omega^b_B$ is the metric of a unit two-sphere. It serves to raise and lower indices on the subspace spanned by $\theta^A$. From it, we can define $\Omega^A_a=\delta_{ab}\Omega^{AB}\Omega^b_B$, which satisfies the useful identities
\begin{equation}
\Omega^A_a\Omega^a_B=\delta^A_B,\qquad \Omega^a_A\Omega^A_b=\delta^a_b-\Omega^a_b.
\end{equation}

The transformation from angular coordinates back to Cartesian coordinates is accomplished via $\pdiff{\retr}{x^a_{\rm ret}}=\Omega_a$ and $\pdiff{\theta^A}{x^a_{\rm ret}}=\frac{1}{\retr}\Omega^A_a$.

\section{Transformations}
In various places in this dissertation, I arrive at an expression in terms of retarded quantities. In this section, I present the transformations from those quantities into Fermi-coordinate quantities. The basic method is to expand all quantities at $x'=z(u)$ about $\bar x=z(t)$. For example, $\sigma(x,z(u))=\sigma(x,z(t))-\Delta t\diff{}{t} \sigma(x,z(t))+(\Delta t)^2\ddiff{}{t}\sigma(x,z(t))-...$, where $\Delta t\equiv t-u$. We evaluate these expansions by using the near-coincidence expansions for $\sigma_{\bar\alpha\bar\beta}$ (presented in Appendix \ref{local_expansions}) and its derivatives (obtainable by the same method). In those expansions, we substitute $\sigma^{\bar\alpha}=-re^{\bar\alpha}_{a}n^a$. Solving for $\Delta t$ yields
\begin{align}
u &= t-r\left[1-\tfrac{1}{2}ra_a(t)n^a+\tfrac{3}{8}r^2a_a(t)a_b(t)n^{ab}+\tfrac{1}{24}r^2\dot a_0+\tfrac{1}{6}r^2\dot a_a(t)n^a-\tfrac{1}{6}\etide_{ab}(t)n^{ab}\right]\nonumber\\
&\quad+\order{r^3}.
\end{align}

With the time-difference determined, one can expand the retarded distance $\retr(u)=\sigma_{\alpha'}(x,z(u))u^{\alpha'}(u)$ and the Cartesian-type coordinates $x^a_{\rm ret}$ in powers of $\Delta t$ to find
\begin{align}
\retr &= r\left[1+\tfrac{1}{2}ra_i(t)n^i-\tfrac{1}{8}r^2a_i(t)a_j(t)n^{ij}-\tfrac{1}{8}r^2\dot a_0(t)-\tfrac{1}{3}r^2\dot a_i(t) n^i+\tfrac{1}{6}r^2\etide_{ij}(t)n^{ij}\right]\nonumber\\
&\quad+\order{r^3},\\
r_{\text{ret}}\Omega^a &= r\left[n^a+\tfrac{1}{2}ra^a(t)-\tfrac{1}{3}r^2\dot a^a(t)-\tfrac{1}{3}r^2R^a{}_{b0c}(t)n^{bc}+\tfrac{1}{6}r^2R^a{}_{0b0}(t)n^b\right]+\order{r^4}.
\end{align}

We also have two tetrads defined at any point $x$ off the worldline: the Fermi tetrad $e^\alpha_I$ that is parallel-propagated along a spacelike geodesic, and the retarded tetrad $b^\alpha_I$ that is parallel-propagated along a null geodesic. These two quantities can be related by expanding $g^\alpha_{\beta'}(x,z(u))b^{\beta'}_I(u)$ about $z(t)$ and making use of the near-coincidence expansions for the derivatives of the parallel propagator. The result is
\begin{align}
b^\alpha_0 &= \left[1-\tfrac{1}{2}r^2\dot a_0(t)\right]e^\alpha_0+\left[-r(1-\tfrac{1}{2}ra_b(t)n^b)a^a(t)+\tfrac{1}{2}r^2 \dot a^a-\tfrac{1}{2}r^2\etide^a_b(t)n^b\right]e^\alpha_a \nonumber\\
&\quad+\order{r^3},\label{tetrad_transformation1}\\
b^\alpha_a &= \left[-r(1-\tfrac{1}{2}ra_b(t)n^b)a_a(t)+\tfrac{1}{2}r^2\dot a_a(t) -\tfrac{1}{2}r^2\etide_{ab}(t)n^b\right]e_0^\alpha \nonumber\\
&\quad +\left[\delta^b_a+\tfrac{1}{2}r^2a^b(t)a_a(t)+\tfrac{1}{2}r^2 R^b{}_{a0c}(t)n^c \right]e_b^\alpha  +\order{r^3}. \label{tetrad_transformation2}
\end{align}
			\chapter{STF tensors}\label{STF tensors}
This appendix briefly reviews the use of STF decompositions and collects several useful formulas. Refer to Ref.~\cite{STF_1,STF_2,STF_3} for thorough reviews.

\section{STF multipole decompositions}
All formulas in this section are either taken directly from Refs.~\cite{STF_1} and \cite{STF_2} or are easily derivable from formulas therein.

Any Cartesian tensor field depending on two angles $\theta^A$ spanning a sphere can be expanded in a unique decomposition in terms of symmetric trace-free tensors. Such a decomposition is equivalent to a decomposition in terms of tensorial harmonics, but it is sometimes more convenient. It begins with the fact that the angular dependence of a Cartesian tensor $T_{S}(\theta^A)$ can be expanded in a series of the form
\begin{equation}\label{nhat_expansion}
T_S(\theta^A)=\sum_{\ell\geq 0}T_{S\langle L\rangle}\nhat^L,
\end{equation}
where $S$ and $L$ denote multi-indices $S=i_1...i_s$ and $L=j_1...j_\ell$, angular brackets denote an STF combination of indices, $n^a$ is a Cartesian unit vector, $n^L\equiv n^{j_1}\ldots n^{j_\ell}$, and $\nhat^L\equiv n^{\langle L\rangle}$. This is entirely equivalent to an expansion in spherical harmonics. Each coefficient $T_{S\langle L\rangle}$ can be found from the formula
\begin{equation}
T_{S\langle L\rangle} = \frac{(2\ell+1)!!}{4\pi\ell!}\int T_S(\theta^A)\nhat_L d\Omega,
\end{equation}
where $!!$ is a double factorial, defined by $x!!=x(x-2)...1$. These coefficients can then be decomposed into irreducible STF tensors. For example, for $s=1$, we have
\begin{equation}\label{decomposition_1}
T_{a\langle L\rangle} = \hat T^{(+1)}_{aL}+\epsilon^j{}_{a\langle i_\ell}\hat T^{(0)}_{L-1\rangle j}+\delta_{a\langle i_\ell}\hat T^{(-1)}_{L-1\rangle},
\end{equation}
where the $\hat T^{(n)}$'s are STF tensors given by
\begin{align}
\hat T^{(+1)}_{L+1} & \equiv T_{\langle L+1\rangle}, \\
\hat T^{(0)}_{L} & \equiv \frac{\ell}{\ell+1}T_{pq\langle L-1}\epsilon_{i_\ell\rangle}{}^{pq}, \\
\hat T^{(-1)}_{L-1} & \equiv \frac{2\ell-1}{2\ell+1}T^j{}_{jL-1}.
\end{align} 
Similarly, for a symmetric tensor $T_S$ with $s=2$, we have
\begin{align}\label{decomposition_2}
T_{ab\langle L\rangle} & = \mathop{\STF}_L\mathop{\STF}_{ab}\Big( \epsilon^p{}_{ai_\ell}\hat T^{(+1)}_{bpL-1} + \delta_{ai_\ell}\hat T^{(0)}_{b L-1} +\delta_{a i_\ell}\epsilon^p{}_{bi_{\ell-1}}\hat T^{(-1)}_{pL-2} +\delta_{ai_\ell}\delta_{bi_{\ell-1}}\hat T^{(-2)}_{L-2}\Big) \nonumber\\
&\quad +\hat T^{(+2)}_{abL}+\delta_{ab}\hat K_L,
\end{align}
where
\begin{align}
\hat T^{(+2)}_{L+2} & \equiv T_{\langle L+2\rangle}, \\
\hat T^{(+1)}_{L+1} & \equiv \frac{2\ell}{\ell+2}\mathop{\STF}_{L+1}(T_{\langle pi_\ell\rangle qL-1}\epsilon_{i_{\ell+1}}{}^{pq}), \\
\hat T^{(0)}_L & \equiv \frac{6\ell(2\ell-1)}{(\ell+1)(2\ell+3)}\mathop{\STF}_L(T_{\langle ji_\ell\rangle}{}^j{}_{L-1}), \\
\hat T^{(-1)}_{L-1} & \equiv \frac{2(\ell-1)(2\ell-1)}{(\ell+1)(2\ell+1)}\mathop{\STF}_{L-1}(T_{\langle jp\rangle q}{}^j{}_{L-2}\epsilon_{i_{\ell-1}}{}^{pq}), \\
\hat T^{(-2)}_{L-2} & \equiv \frac{2\ell-3}{2\ell+1}T_{\langle jk\rangle}{}^{jk}{}_{L-2} \\
\hat K_L & \equiv \tfrac{1}{3}T^j{}_{jL}.
\end{align}
These decompositions are equivalent to the formulas for addition of angular momenta, $J=S+L$, which results in terms with angular momentum $\ell-s\leq j\leq \ell+s$; the superscript labels $(\pm n)$ in these formulas indicate by how much each term's angular momentum differs from $\ell$.

By substituting Eqs.~\eqref{decomposition_1} and \eqref{decomposition_2} into Eq.~\eqref{nhat_expansion}, we find that a scalar, a Cartesian 3-vector, and the symmetric part of a rank-2 Cartesian 3-tensor can be decomposed as, respectively,
\begin{align}
T(\theta^A) &= \sum_{\ell\ge0}\hat A_L\nhat^L, \label{generic_STF tt}\\
T_a(\theta^A) &= \sum_{\ell\ge0}\hat B_L\nhat_{aL}+\sum_{\ell\ge1}\left[\hat C_{aL-1}\nhat^{L-1} + \epsilon^i{}_{aj}\hat D_{iL-1}\nhat^{jL-1}\right],\label{generic_STF ta}\\
T_{(ab)}(\theta^A) &= \delta_{ab}\sum_{\ell\ge0}\hat K_L\nhat^L+\sum_{\ell\ge0}\hat E_L\nhat_{ab}{}^L +\sum_{\ell\ge1}\left[\hat F_{L-1\langle a}\nhat^{}_{b\rangle}{}^{L-1} +\epsilon^{ij}{}_{(a}\nhat_{b)i}{}^{L-1}\hat G_{jL-1}\right] \nonumber\\
&\quad+\sum_{\ell\ge2}\left[\hat H_{abL-2}\nhat^{L-2}+\epsilon^{ij}{}_{(a}\hat I_{b)jL-2}\nhat_{i}{}^{L-2}\right].\label{generic_STF ab}
\end{align}
Each term in these decompositions is algebraically independent of all the other terms.

We can also reverse a decomposition to ``peel" a fixed index from an STF expression:
\begin{align}
(\ell+1)\mathop{\STF}_{iL}T_{i\langle L\rangle} & = T_{i\langle L\rangle} + \ell\mathop{\STF}_LT_{i_\ell \langle iL-1\rangle} -\frac{2\ell}{2\ell+1}\mathop{\STF}_L T^j{}_{\langle jL-1\rangle}\delta_{i_\ell i}.
\end{align}

In evaluating the action of the wave operator on a decomposed tensor, the following formulas are useful:
\begin{align}
n^c\nhat^L &= \nhat^{cL}+\frac{\ell}{2\ell+1}\delta^{c\langle i_1}\nhat^{i_2...i_\ell\rangle},\\
n_c\nhat^{cL} & = \frac{\ell+1}{2\ell+1}\nhat^L, \\
r\partial_c\nhat_L &= -\ell\nhat_{cL}+\frac{\ell(\ell+1)}{2\ell+1}\delta_{c\langle i_1}\nhat_{i_2...i_\ell\rangle}, \\
\partial^c\partial_c\nhat^L & = -\frac{\ell(\ell+1)}{r^2}\nhat^L, \\
n^c\partial_c\nhat^L &= 0, \\
r\partial_c\nhat^{cL} &= \frac{(\ell+1)(\ell+2)}{(2\ell+1)}\nhat^L.
\end{align}

In evaluating the $t$-component of the Lorenz gauge condition, the following formula is useful for finding the most divergent term (in an expansion in $r$):
\begin{align}\label{gauge_help1}
r\partial^c\hmn{tc}{\emph{n,m}} &= \sum_{\ell\geq 0}\frac{(\ell+1)(\ell+2)}{2\ell+1}\B{L}{\emph{n,m}}\nhat^L  -\sum_{\ell\geq 2}(\ell-1)\C{L}{\emph{n,m}}\nhat^L.
\end{align}
And in evaluating the $a$-component, the following formula is useful for the same purpose:
\begin{align}\label{gauge_help2}
& r\partial^b\hmn{ab}{\emph{n,m}}-\tfrac{1}{2}r\eta^{\beta\gamma}\partial_a\hmn{\beta\gamma}{\emph{n,m}}
\nonumber\\
&=
\sum_{\ell\geq0}\left[\tfrac{1}{2}\ell(\K{L}{\emph{n,m}}-\A{L}{\emph{n,m}}) +\frac{(\ell+2)(\ell+3)}{2\ell+3}\E{L}{\emph{n,m}} -\tfrac{1}{6}\ell\F{L}{\emph{n,m}}\right]\nhat_a{}^L \nonumber\\
&\quad +\sum_{\ell\geq 1}\bigg[\frac{\ell(\ell+1)}{2(2\ell+1)}(\A{aL-1}{\emph{n,m}} -\K{aL-1}{\emph{n,m}}) +\frac{(\ell+1)^2(2\ell+3)}{6(2\ell+1)(2\ell-1)}\F{aL-1}{\emph{n,m}} \nonumber\\
&\quad-(\ell-2)\H{aL-1}{\emph{n,m}}\bigg]\nhat^{L-1}  +\sum_{\ell\geq 1}\left[\frac{(\ell+2)^2}{2(2\ell+1)}\G{dL-1}{\emph{n,m}} -\tfrac{1}{2}(\ell-1)\I{dL-1}{\emph{n,m}}\right]\epsilon_{ac}{}^d\nhat^{cL-1}
\end{align}
where I have defined $\H{a}{\emph{n,m}}\equiv 0$ and $\I{a}{\emph{n,m}}\equiv 0$.

%%%%%%%%%%%
\section{Angular integrals}\label{angular integrals}
%%%%%%%%%%%
The calculation of the boundary integral in Sec.~\ref{integral_expansion1} requires the evaluation of numerous integrals over a cross-section of the worldtube $\Gamma$. This subsection compiles these integrals. Let $x^\alpha=(t,r,\theta,\phi)$ be a point lying outside the worldtube, and let $x^{\alpha'}=(t',\rad, \theta',\phi')$ be a point on the worldtube; here $r$ and $\rad$ denote the Fermi coordinate distances to $x^\alpha$ and $x^{\alpha'}$. Let $n^a$ and $n^{a'}$ be Cartesian unit vectors defined by the angles $(\theta,\phi)$ and $(\theta',\phi')$, respectively. The quantity $\r_0\equiv\sqrt{r^2+\rad^2-2r\rad n_an^{a'}}$ is the leading-order flat-spacetime luminosity distance between $x^\alpha$ and $x^{\alpha'}$. For brevity's sake, I introduce the notation $\av{f}\equiv\frac{1}{4\pi}\int f(\theta',\phi')d\Omega'$ to denote an average over the primed angles, where $d\Omega'\equiv\sin\theta'd\theta'd\phi'$.

First, I present standard results from Ref.~\cite{STF_1}:
\begin{align}
\langle \nhat_L\rangle &=0 {\rm\ if\ } \ell>0, \label{n_integral}\\
\langle n_L\rangle & = 0 {\rm\ if\ } \ell {\rm\ is\ odd}, \\
\langle n_L\rangle & = \frac{\delta_{\lbrace i_1 i_2}...\delta_{i_{\ell-1}i_{\ell\rbrace}}}{(\ell+1)!!} {\rm\ if\ } \ell {\rm\ is\ even}, \label{nhat_integral}
\end{align}
where the curly braces indicate the smallest set  of permutations of indices that make the result symmetric. For example, $\delta_{\lbrace ab}n_{c\rbrace}=\delta_{ab}n_c+\delta_{bc}n_a+\delta_{ca}n_b$.

I now present the integrals involving $\r_0$, grouping them according to the power of $\r_0$ that appears in their integrand:
{\allowdisplaybreaks\begin{align}
\av{\r_0} &= r+\frac{\rad^2}{3r},\\
\av{\frac{1}{\r_0}} & = \frac{1}{r},\\
\av{\frac{n'^{a}}{\r_0}} & = \frac{\rad n^a}{3r^2}, \\
\av{\frac{\nhat'^{ab}}{\r_0}} & = \frac{\rad^2\nhat^{ab}}{5r^3},\\
\av{\frac{\nhat'^{abc}}{\r_0}} & = \frac{\rad^3\nhat^{abc}}{7r^4},\\
\av{\frac{1}{\r_0^2}} & = \frac{1}{2r\rad}\ln\left(\frac{r+\rad}{r-\rad}\right), \\
\av{\frac{n'^{a}}{\r_0^2}} & = \frac{1}{2}\left[\frac{r^2+\rad^2}{2r^2\rad^2} \ln\left(\frac{r+\rad}{r-\rad}\right) -\frac{1}{r\rad}\right]n^a, \\
\av{\frac{\nhat'^{ab}}{\r_0^2}} & = \frac{3}{8}\frac{r^4+\rad^4+\frac{2}{3}r^2\rad^2}{2r^3\rad^3} \ln\left(\frac{r+\rad}{r-\rad}\right)\nhat^{ab} -\frac{3}{8}\frac{r^2+\rad^2}{r^2\rad^2}\nhat^{ab},\\
\av{\frac{1}{\r_0^3}} & = \frac{1}{r(r^2-\rad^2)}, \\
\av{\frac{n'^{a}}{\r_0^3}} & = \frac{\rad n^a}{r^2(r^2-\rad^2)}, \\
\av{\frac{\nhat'^{ab}}{\r_0^3}} & = \frac{\rad^2\nhat^{ab}}{r^3(r^2-\rad^2)}, \\
\av{\frac{\nhat'^{abc}}{\r_0^3}} & = \frac{\rad^3\nhat^{abc}}{r^4(r^2-\rad^2)},\\
\av{\frac{1}{\r_0^4}} & = \frac{1}{(r^2-\rad^2)^2},\\
\av{\frac{n'^{a}}{\r_0^4}} & = \frac{r^2+\rad^2}{2r\rad(r^2-\rad^2)^2}n^a-\frac{1}{4r^2\rad^2} \ln\left(\frac{r+\rad}{r-\rad}\right)n^a, \\
\av{\frac{\nhat'^{ab}}{\r_0^4}} & = \frac{3}{4}\frac{r^4+\rad^4-\frac{2}{3}r^2\rad^2}{r^2\rad^2(r^2-\rad^2)^2}\nhat^{ab} -\frac{3}{8}\frac{r^2+\rad^2}{r^3\rad^3} \ln\left(\frac{r+\rad}{r-\rad}\right)\nhat^{ab},\\
\av{\frac{\nhat'^{abc}}{\r_0^4}} & = \frac{15}{16}\frac{(r^2+\rad^2)(r^2+\rad^2 -\frac{22}{15}r^2\rad^2)}{r^3\rad^3(r^2-\rad^2)^2}\nhat^{abc}\nonumber\\
&\quad -\frac{15}{32}\frac{r^4+\rad^4+\frac{6}{5}r^2\rad^2}{r^4\rad^4} \ln\left(\frac{r+\rad}{r-\rad}\right)\nhat^{abc},\\
\av{\frac{1}{\r_0^5}} & = \frac{3r^2+\rad^2}{3r(r^2-\rad^2)^3}, \\
\av{\frac{n'^{a}}{\r_0^5}} & = \frac{\rad(5r^2-\rad^2)n^a}{3r^2(r^2-\rad^2)^3}, \\
\av{\frac{\nhat'^{ab}}{\r_0^5}} & = \frac{\rad^2(7r^2-3\rad^2)\nhat^{ab}}{3r^3(r^2-\rad^2)^3}, \\
\av{\frac{\nhat'^{abc}}{\r_0^5}} & = \frac{\rad^3(9r^2-5\rad^2)\nhat^{abc}}{3r^4(r^2-\rad^2)^3}.
\end{align}}

			\chapter{Green's Functions}\label{Greens_functions}
I follow the notation and conventions of Ref.~\cite{Eric_review}. The Green's function for the tensor wave operator is defined by the equation
\begin{equation}\label{tensor Green}
\left(g^\rho_\mu g^\sigma_\nu\Box + 2R_\mu{}^\rho{}_\nu{}^\sigma\right) G_{\rho\sigma}{}^{\mu'\nu'}(x,x') = -4\pi g_{(\mu}^{\mu'}g_{\nu)}^{\nu'}\delta(x,x'),
\end{equation}
where $\delta(x,x')=\delta^4(x^\alpha-x^{\alpha'})/\sqrt{|g|}$, and $g_{\mu}^{\mu'}$ is the parallel propagator from $x$ to $x'$; in terms of the wave-operator $E_{\mu\nu}$, the left-hand side is $E_{\mu\nu}[G]$. The Green's function for the vector wave operator is defined by
\begin{equation}\label{vector Green}
\left(g^\nu_\mu\Box -R_\mu{}^{\nu}\right)G_\nu{}^{\mu'}(x,x') = -4\pi g_{\mu}^{\mu'}\delta(x,x'),
\end{equation}
and the Green's function for the scalar wave operator is defined by 
\begin{equation}\label{scalar Green}
\left(\Box - \lambda R\right)G(x,x') = -4\pi\delta(x,x'),
\end{equation}
where $\lambda$ is an arbitrary coupling constant. All quantities are defined with respect to the background metric $g$.

We can define retarded and advanced solutions to these equations, yielding retarded and advanced Green's functions. The retarded Green's function $G^{\rm ret}_{\alpha\beta\alpha'\beta'}(x,x')$ is nonzero only if $x$ is in the causal future of $x'$; the advanced Green's function $G^{\rm adv}_{\alpha\beta\alpha'\beta'}(x,x')$ is nonzero only if $x$ is in the causal past of $x'$. If these two Green's functions fall off sufficiently fast when one of the points moves toward asymptotic infinity, then they satisfy the reciprocity relation
\begin{equation}
G^{\rm adv}_{\alpha'\beta'\alpha\beta}(x',x) = G^{\rm ret}_{\alpha\beta\alpha'\beta'}(x,x').
\end{equation}

We can also define singular and regular Green's functions, as first formulated by Detweiler and Whiting \cite{Detweiler_Whiting}. The singular Green's function is defined by 
\begin{equation}
G^S_{\alpha\beta\alpha'\beta'}(x,x') \equiv \tfrac{1}{2}\left[G^{\rm ret}_{\alpha\beta\alpha'\beta'}(x,x')+G^{\rm adv}_{\alpha\beta\alpha'\beta'}(x,x')-H_{\alpha\beta\alpha'\beta'}(x,x')\right],
\end{equation}
where $H$ satisfies the following: the homogeneous wave equation $E_{\mu\nu}[H]=0$; the symmetry relation $H_{\alpha\beta\alpha'\beta'}(x,x')=H_{\alpha'\beta'\alpha\beta}(x',x)$; identity with $G^{\rm ret}_{\alpha\beta\alpha'\beta'}(x,x')$ if $x$ lies in the chronological future of $x'$; and identity with $G^{\rm adv}_{\alpha\beta\alpha'\beta'}(x,x')$ if $x$ lies in the chronological past of $x'$. It follows from these properties that the singular Green's function $G^S$ satisfies the same inhomogeneous wave equation as $G^{\rm ret}$ and $G^{\rm adv}$, as well as the symmetry relation $G^S_{\alpha\beta\alpha'\beta'}(x,x')=G^S_{\alpha'\beta'\alpha\beta}(x',x)$. It also follows that $G^S$ vanishes if $x$ is in the chronological past or future of $x'$. Therefore, it corresponds to acausal propagation of waves. The regular Green's function is defined by 
\begin{equation}
G^R_{\alpha\beta\alpha'\beta'}(x,x') \equiv \tfrac{1}{2}\left[G^{\rm ret}_{\alpha\beta\alpha'\beta'}(x,x')-G^{\rm adv}_{\alpha\beta\alpha'\beta'}(x,x')+H_{\alpha\beta\alpha'\beta'}(x,x')\right],
\end{equation}
such that $G^{\rm ret}=G^R+G^S$. The regular Green's function satisfies the homogeneous wave equation, it is identical to the retarded Green's function if $x$ is in the chronological future of $x'$, and it vanishes if $x$ is in the chronological past of $x'$. Because it satisfies the homogeneous wave equation, it is not a Green's function in the sense of being an inverse of the wave operator.

If the point $x'$ lies in the convex normal neighbourhood of $x$, then the retarded gravitational Green's functions can be written in the Hadamard decomposition
\begin{equation}
G^{\rm ret}_{\alpha\beta\alpha'\beta'}=U_{\alpha\beta\alpha'\beta'}\delta_+(\sigma) +V_{\alpha\beta\alpha'\beta'}\theta_+(-\sigma),
\end{equation}
where $\delta_+(\sigma)$ is a delta function with support on the past light cone of $x$, $\theta_+(-\sigma)$ is a Heaviside function with support in the interior of the past light cone of $x$, $\sigma$ is Synge's world function, equal to one-half the squared geodesic distance between $x$ and $x'$, and $U_{\alpha\beta\alpha'\beta'}$ and $V_{\alpha\beta\alpha'\beta'}$ are smooth bitensors. We can see that in the convex normal neighbourhood, the retarded Green's function has support on and within the past lightcone of $x$. Similarly, the advanced Green's function has an analogous decomposition, with $\delta_+(\sigma)$ and $\theta_+(-\sigma)$ replaced by $\delta_-(\sigma)$ and $\theta_-(-\sigma)$, which, respectively, have support on and within the future lightcone of $x$. Along with the retarded and advanced Green's functions, we can define the singular and regular Green's functions $G^S$ and $G^R$. In the convex normal neighbourhood, the bitensor $H$ agrees with the smooth part of the retarded Green's function: that is, $H_{\alpha\beta\alpha'\beta'}(x,x')=V_{\alpha\beta\alpha'\beta'}(x,x')$. It follows that in this neighbourhood, the singular Green's function is given by
\begin{equation}
G^S_{\alpha\beta\alpha'\beta'}=\tfrac{1}{2}U_{\alpha\beta\alpha'\beta'}\delta(\sigma) -\tfrac{1}{2}V_{\alpha\beta\alpha'\beta'}\theta(\sigma),\label{GS_Hadamard}
\end{equation}
which has support on and outside the past and future light cones of $x$, and which does not distinguish between past and future; and the regular Green's function is given by
\begin{equation}
G^R_{\alpha\beta\alpha'\beta'}=\tfrac{1}{2}U_{\alpha\beta\alpha'\beta'}\left[\delta_+(\sigma)+\delta_-(\sigma)\right] +V_{\alpha\beta\alpha'\beta'}\left[\theta_+(-\sigma)+\tfrac{1}{2}\theta(\sigma)\right],
\end{equation}
which has support on the past and future light cones of $x$, outside the light cones, and within the past light cone.

The equations that determine the bitensors $U_{\mu\nu\mu'\nu'}$ and $V_{\mu\nu\mu'\nu'}$ can be derived by substituting the Hadamard decomposition into the original equation \eqref{tensor Green}. In this dissertation, I only ever explicitly evaluate these bitensors when expanding them near coincidence. The relevant near-coincidence expansions are provided in Appendix~\ref{local_expansions}.

If $g$ is a vacuum metric, such that $R=0=R_{\mu\nu}$, then one can easily derive the following identities:
\begin{align}
G^{\mu\nu}{}_{\mu'\nu';\nu} & = - G^\mu{}_{(\mu';\nu')}, \label{Green1}\\
G^\mu{}_{\mu';\mu} & = -G_{;\mu'}, \label{Green2}\\
G_{\mu\nu}{}^{\mu'\nu'}g_{\mu'\nu'} & = g_{\mu\nu}G. \label{Green3}
\end{align}
Equation~\eqref{Green1} follows from taking the divergence of Eq.~\eqref{tensor Green} and the covariant derivative of Eq.~\eqref{vector Green}. Equation~\eqref{Green2} follows from taking the divergence of Eq.~\eqref{vector Green} and the covariant derivative of Eq.~\eqref{scalar Green}. Equation~\eqref{Green3} follows from contracting the primed indices in Eq.~\eqref{tensor Green}. In each case, these operations show that the two relevant bitensors satisfy the same differential equation; the equations \eqref{Green1}--\eqref{Green3} hold when the bitensors on the left and right satisfy identical boundary conditions. Equation \eqref{Green2} appears in Ref.~\cite{DeWitt_Brehme}. To the best of my knowledge, Eqs.~\eqref{Green1} and \eqref{Green3} were first presented in Ref.~\cite{my_paper}.

Throughout most of this dissertation, I assume that the Green's function being used is the retarded one, which I write simply as $G_{\mu\nu\mu'\nu'}$, with no ``${\rm ret}$" label.

			\chapter{The linear perturbation due to a point particle}\label{point_particle_soln}
In this appendix, I present the solution to the wave equation with a point particle source; I also present the Detweiler-Whiting decomposition of the solution into its singular and regular pieces.

\section{The metric perturbation}
The solution to the wave equation is
\begin{equation}
\hmn{E\alpha\beta}{1} = 2m\int_\gamma G_{\alpha\beta\alpha'\beta'}(2u^{\alpha'}u^{\beta'}+g^{\alpha'\beta'})dt'.
\end{equation}
(One can use Eq.~\eqref{Green3} show that this formula is equivalent to the solution displayed in Eq.~\eqref{point_particle_solution}.)  I seek an expansion of this equation in Fermi normal coordinates, in the case that $x$ is near to a point on $\gamma$. The domain of integration can be split into two: the points in the convex normal neighbourhood $\mathcal{N}(x)$---that is, the points that are connected to $x$ by a unique geodesic---and the points in the complement of $\mathcal{N}(x)$. In the convex normal neighbourhood, the Green's function can be decomposed into a divergent part $U_{\alpha\beta\alpha'\beta'}\delta(\sigma(x,x'))$ and a smooth part $V_{\alpha\beta\alpha'\beta'}\theta(-\sigma(x,x'))$ (see Appendix~\ref{Greens_functions}). After performing a change of variables using $dt'=\frac{d\sigma}{\sigma_{\alpha'}u^{\alpha'}}$ to evaluate the delta function, the metric perturbation becomes
\begin{equation}
\hmn{E\alpha\beta}{1} =\frac{2m}{r_{\text{ret}}}U_{\alpha\beta\alpha'\beta'}(2u^{\alpha'}u^{\beta'}+g^{\alpha'\beta'}) +\tail_{\alpha\beta}(u).
\end{equation}
where primed indices now refer to the point on $\gamma$ connected to $x$ by a null geodesic; this point is given by $z^\alpha(u)$, where $u$ is the retarded time; $r_{\text{ret}}$ is the retarded distance between $x$ and $z^\alpha(u)$, given by $\sigma_{\alpha'}u^{\alpha'}$; and the tail integral is given by
\begin{align}
\tail_{\alpha\beta}(u) & = 2m\int_{t^<}^{u} V_{\alpha\beta\alpha'\beta'}(2u^{\alpha'}u^{\beta'}+g^{\alpha'\beta'})dt'+ 2m\int^{t^<}_{-\infty} G_{\alpha\beta\alpha'\beta'}(2u^{\alpha'}u^{\beta'}+g^{\alpha'\beta'})dt'\nonumber\\
&= 2m\int_{-\infty}^{u^-} G_{\alpha\beta\alpha'\beta'}(2u^{\alpha'}u^{\beta'}+g^{\alpha'\beta'})dt',
\end{align}
where $t^<$ is the first time at which the worldline enters $\mathcal{N}(x)$. The two formulas for the tail are equivalent because the upper limit of integration $u^-=u-0^+$ falls short of the past light cone of $x$, avoiding the divergent behavior of the Green's function there. 
 
The first term in $\hmn{E}{1}$, sometimes called the ``direct term", can be expanded in powers of $r$ using the following: the near-coincidence expansion $U_{\alpha\beta\alpha'\beta'}=g^{\alpha'}_\alpha g^{\beta'}_\beta(1+\order{r^3})$; the transformation between $r_{\text{ret}}$ and the Fermi radial distance $r$, given by
\begin{equation}
\retr=r(1+\tfrac{1}{2}ra_in^i-\tfrac{1}{8}r^2a_ia_jn^{ij}-\tfrac{1}{8}r^2\dot a_{\bar\alpha}u^{\bar\alpha}-\tfrac{1}{3}r^2\dot a_i n^i+\tfrac{1}{6}r^2\etide_{ij}n^{ij}+\order{r^3});
\end{equation}
and the coordinate expansion of the parallel-propagators, obtained from the formula $g^{\alpha'}_\alpha=b^{\alpha'}_I b^I_\alpha$, where the retarded tetrad $b^\alpha_I$ is given in terms of the Fermi tetrad in Eqs.~\eqref{tetrad_transformation1}--\eqref{tetrad_transformation2}, and the coordinate expansion of the Fermi tetrad is given in Eqs.~\eqref{Fermi_tetrad1}--\eqref{Fermi_tetrad2}. The tail integral can be similarly expanded as follows: noting that $u=t-r+\order{r^2}$, we can expand $\tail(u)$ about $t$ as $\tail(t)-r\partial_t \tail(t)+...$; each term can then be expanded using the near-coincidence expansions $V_{\alpha\beta}^{\alpha''\beta''}=g^{\gamma''}_{(\alpha} g^{\delta''}_{\beta)}R^{\alpha''}{}_{\gamma''}{}^{\beta''}{}_{\delta''}+\order{r}$ and $\tail_{\alpha\beta}(t)=g^{\bar\alpha}_\alpha g^{\bar\beta}_\beta(\tail_{\bar\alpha\bar\beta}+r\tail_{\bar\alpha\bar\beta i}n^i)+\order{r^2}$, where barred indices correspond to the point $\bar x=z(t)$, connected to $x$ by a spatial geodesic perpendicular to $\gamma$, and $\tail_{\bar\alpha\bar\beta\bar\gamma}$ is given by
\begin{equation}
\tail_{\bar\alpha\bar\beta\bar\gamma}=2m\int_{-\infty}^{t^-} \del{\bar\gamma}G_{\bar\alpha\bar\beta\alpha'\beta'}(2u^{\alpha'}u^{\beta'}+g^{\alpha'\beta'})dt'.
\end{equation}
This yields the expansion
\begin{equation}
\tail_{\alpha\beta}(u)=g^{\bar\alpha}_\alpha g^{\bar\beta}_\beta(\tail_{\bar\alpha\bar\beta}+r\tail_{\bar\alpha\bar\beta i}n^i-4mr\etide_{\bar\alpha\bar\beta}) +\order{r^2}.
\end{equation}
As with the direct part, the final coordinate expansion is found by substituting $g^{\bar\alpha}_\alpha=e^{\bar\alpha}_I e^I_\alpha$, where the tetrads are given in Eqs.~\eqref{Fermi_tetrad1}--\eqref{Fermi_tetrad2}.

Combining the expansions of the direct and tail parts of the perturbation, we arrive at the expansion in Fermi coordinates:
\begin{align}
\hmn{Ett}{1} & = \frac{2m}{r}(1+\tfrac{3}{2}ra_in^i +\tfrac{3}{8}r^2a_ia_jn^{ij} -\tfrac{15}{8}r^2\dot a_{\bar\alpha}u^{\bar\alpha} +\tfrac{1}{3}r^2\dot a_i n^i +\tfrac{5}{6}r^2\etide_{ij}n^{ij})\nonumber\\
&\quad +(1+2ra_in^i)\tail_{00}+r\tail_{00i}n^i+\order{r^2},\\
\hmn{Eta}{1} & = 4ma_a-\tfrac{2}{3}mrR_{0iaj}n^{ij}+2mr\etide_{ai}n^i-2mr\dot a_a \nonumber\\
&\quad +(1+ra_in^i)\tail_{0a}+r\tail_{0ai}n^i+\order{r^2}, \\
\hmn{Eab}{1} & = \frac{2m}{r}(1-\tfrac{1}{2}ra_in^i +\tfrac{3}{8}r^2a_ia_jn^{ij} +\tfrac{1}{8}r^2\dot a_{\bar\alpha}u^{\bar\alpha} +\tfrac{1}{3}r^2\dot a_i n^i -\tfrac{1}{6}r^2\etide_{ij}n^{ij})\delta_{ab}\nonumber\\
&\quad +4mra_aa_b-\tfrac{2}{3}mrR_{aibj}n^{ij}-4mr\etide_{ab}+\tail_{ab} +r\tail_{abi}n^i+\order{r^2}.
\end{align}
As the final step, each of these terms is decomposed into irreducible STF pieces using the formulas \eqref{nhat_expansion}, \eqref{decomposition_1}, and \eqref{decomposition_2}, to yield
\begin{align}
\hmn{Ett}{1} &= \frac{2m}{r}+\A{}{1,0}+3ma_in^i+r\left[4ma_ia^i+\A{i}{1,1}n^i+m\left(\tfrac{3}{4}a_{\langle i}a_{j\rangle} +\tfrac{5}{3}\etide_{ij}\right)\nhat^{ij}\right]\nonumber\\
&\quad+O(r^2),\\
\hmn{Eta}{1} &= \C{a}{1,0}+r\big(\B{}{1,1}n_a-2m\dot a_a+\C{ai}{1,1}n^i+\epsilon_{ai}{}^j\D{j}{1,1}n^i +\tfrac{2}{3}m\epsilon_{aij}\btide^j_k\nhat^{ik}\big)\nonumber\\
&\quad+O(r^2), \\
\hmn{Eab}{1} &= \frac{2m}{r}\delta_{ab}+(\K{}{1,0}-ma_in^i)\delta_{ab}+\H{ab}{1,0} +r\Big\lbrace\delta_{ab}\big[\tfrac{4}{3}ma_ia^i+\K{i}{1,1}n^i \nonumber\\
&\quad+\left(\tfrac{3}{4}ma_{\langle i}a_{j\rangle}-\tfrac{5}{9}m\etide_{ij}\right)\nhat^{ij}\big] +\tfrac{4}{3}m\etide^i_{\langle a}\nhat_{b\rangle i} +4ma_{\langle a}a_{b\rangle}-\tfrac{38}{9}m\etide_{ab} \nonumber\\
&\quad+\H{abi}{1,1}n^i +\epsilon_i{}^j{}_{(a}\I{b)j}{1,1}n^i+\F{\langle a}{1,1}n^{}_{b\rangle}\Big\rbrace+O(r^2),
\end{align}
where the uppercase script tensors are specified in Table~\ref{STF wrt tail}. The naming convention for those tensors follows that in Eqs.~\eqref{generic_STF tt}--\eqref{generic_STF ab}. 
\begin{table}[tb]
\caption[The regular field in terms of $\etide_{ab}$ and $\tail_{\alpha\beta}$]{Symmetric trace-free tensors in the first-order metric perturbation in the buffer region, written in terms of the electric-type tidal field $\etide_{ab}$, the acceleration $a_i$, and the tail of the perturbation.}  
\begin{tabular*}{\textwidth}{@{\hspace{0.05\textwidth}}c@{\hspace{0.05\textwidth}}|@{\hspace{0.05\textwidth}}r}
\hline\hline
$\begin{array}{ll}
\A{}{1,0} &= \tail_{00}\\
\C{a}{1,0} &= \tail_{0a}+ma_a\\
\K{}{1,0} &= \tfrac{1}{3}\delta^{ab}\tail_{ab}\\
\H{ab}{1,0} &= \tail_{\langle ab\rangle}\\
\A{a}{1,1} &= \tail_{00a}+2\tail_{00}a_a+\tfrac{2}{3}m\dot a_a\\
\B{}{1,1} &= \tfrac{1}{3}\tail_{0ij}\delta^{ij}+\tfrac{1}{3}\tail_{0i}a^i
\end{array}$
& 
$\begin{array}{ll}
\C{ab}{1,1} &= \tail_{0\langle ab\rangle}+2m\etide_{ab}+\tail_{0\langle a}a_{b\rangle} \\
\D{a}{1,1} &= \tfrac{1}{2}\epsilon_a{}^{bc}(\tail_{0bc}+\tail_{0b}a_c)\\
\K{a}{1,1} &= \frac{1}{3}\delta^{bc}\tail_{bca}+\tfrac{2}{3}m\dot a_a\\
\H{abc}{1,1} &= \tail_{\langle abc\rangle}\\
\F{a}{1,1} &= \tfrac{3}{5}\delta^{ij}\tail_{\langle ia\rangle j}\\
\I{ab}{1,1} &= \tfrac{2}{3}\displaystyle{\mathop{\STF}_{ab}} \left(\epsilon_b{}^{ij}\tail_{\langle ai\rangle j}\right)
\end{array}$\\
\hline\hline
\end{tabular*}
\label{STF wrt tail}
\end{table}  

\section{Singular and regular pieces}
The Detweiler-Whiting singular field is given by
\begin{equation}
h^S_{\alpha\beta} = 2m\int G^S_{\alpha\beta\alpha'\beta'}(2u^{\alpha'}u^{\beta'}+g^{\alpha'\beta'})dt',
\end{equation}
where $G^S$ is the singular Green's function defined in Appendix~\ref{Greens_functions}. Using the Hadamard decomposition \eqref{GS_Hadamard}, we can write this as
\begin{align}
h^S_{\alpha\beta} &= \frac{m}{\retr}U_{\alpha\beta\alpha'\beta'}(2u^{\alpha'}u^{\beta'}+g^{\alpha'\beta'})+ \frac{m}{\advr}U_{\alpha\beta\alpha''\beta''}(2u^{\alpha''}u^{\beta''}+g^{\alpha''\beta''}) \nonumber\\
&\quad-2m\int^v_u V_{\alpha\beta\bar\alpha\bar\beta}(u^{\bar\alpha}u^{\bar\beta}+\tfrac{1}{2}g^{\bar\alpha\bar\beta})d\bar t, \label{hS}
\end{align}
where primed indices now refer to the retarded point $x'=z(u)$, where $u$ is the retarded time; double-primed indices refer to the advanced point $x''=z(v)$, where $v$ is advanced time; $\advr$ is the advanced distance between $x$ and $z^\alpha(v)$, given by $-\sigma_{\alpha''}u^{\alpha''}$; barred indices refer to points in the segment of the worldline between $z(u)$ and $z(v)$. The first term in Eq.~\eqref{hS} can be read off from the calculation of the retarded field. The other terms are expanded using the identities $v=u+2r+O(r^2)$ and $\advr=\retr(1+\tfrac{2}{3}r^2\dot a_in^i)$; see Ref.~\cite{Eric_review} for details (though the expansion therein is for $h^S_{\alpha\beta;\gamma}$, rather than $h^S_{\alpha\beta}$). The final result is
\begin{align}
h^S_{tt} &= \frac{2m}{r}+3ma_in^i+r\left[4ma_ia^i+m\left(\tfrac{3}{4}a_{\langle i}a_{j\rangle} +\tfrac{5}{3}\etide_{ij}\right)\nhat^{ij}\right]+O(r^2),\\
h^S_{ta} &=r\big(-2m\dot a_a +\tfrac{2}{3}m\epsilon_{aij}\btide^j_k\nhat^{ik}\big)+O(r^2), \\
h^S_{ab} &= \frac{2m}{r}\delta_{ab}-ma_in^i\delta_{ab} +r\Big\lbrace\delta_{ab}\big[\tfrac{4}{3}ma_ia^i+\left(\tfrac{3}{4}ma_{\langle i}a_{j\rangle}-\tfrac{5}{9}m\etide_{ij}\right)\nhat^{ij}\big] \nonumber\\
&\quad +\tfrac{4}{3}m\etide^i_{\langle a}\nhat_{b\rangle i} +4ma_{\langle a}a_{b\rangle}-\tfrac{38}{9}m\etide_{ab} \Big\rbrace+O(r^2).
\end{align}

The regular field could be calculated from the regular Green's function. But it is more straightforwardly calculated using $h^R_{\alpha\beta}=\hmn{\alpha\beta}{1}-h^S_{\alpha\beta}$. The result is
\begin{align}
h^R_{tt} &= \A{}{1,0}+r\A{i}{1,1}n^i+\order{r^2}, \\
h^R_{ta} &= \C{a}{1,0} +r\Big(\B{}{1,1}n_a+\C{ai}{1,1}n^i +\epsilon_{ai}{}^j\D{j}{1,1}n^i\Big)+\order{r^2},\\
h^R_{ab} &= \delta_{ab}\K{}{1,0}+\H{ab}{1,0}+r\Big(\delta_{ab}\K{i}{1,1}n^i+\H{abi}{1,1}n^i +\epsilon\indices{_i^j_{(a}}\I{b)j}{1,1}n^i+\F{\langle a}{1,1}n^{}_{b\rangle}\Big) \nonumber\\
&\quad+\order{r^2}.
\end{align}

			\chapter{The metric of a tidally perturbed black hole}\label{perturbed_BH}
In this appendix, I present some general results for perturbations of Schwarzschild in a light cone gauge. Over the course of the calculation, I highlight the restrictions that must be imposed on the metric in order to arrive at the usual result for a tidally perturbed black hole. In the final section of the appendix, I present that metric, along with its expansion in the buffer region. The notation and definitions in the first section mostly follows that of Ref.~\cite{Eric_perturbations}.

\section{The metric, perturbation equations, and gauge condition}
The exact metric $\exact{g}$ is expanded as $g_I(\tilde X,\e)=g_B(\tilde X)+H(\tilde X,\e)$, where $H(\tilde X,\e)=\e H\coeff{1}(\tilde X)+\e^2 H\coeff{2}(\tilde X)+...$, and $\tilde X^\alpha=(U,\tilde R,\Theta^A)$ are (scaled) retarded Eddington-Finkelstein coordinates adapted to the background metric $g_B$, where $\tilde R\equiv R/\e$. As described in Sec.~\ref{outline}, the terms in the inner expansion of $\exact{g}$ must satisfy the sequence of equations \eqref{inner_eqn0}--\eqref{inner_eqn2}, which I rewrite here:
\begin{align}
G\coeff{0}_I{}^{\mu\nu}[g_B] &= 0,\label{inner_eqn0alt}\\
\delta G\coeff{0}_I{}^{\mu\nu}[H\coeff{1}] &= -G\coeff{1}_I{}^{\mu\nu}[g_B],\label{inner_eqn1alt}\\
\delta G\coeff{0}_I{}^{\mu\nu}[H\coeff{2}] &= -\delta^2 G\coeff{0}_I{}^{\mu\nu}[H\coeff{1}] - \delta G\coeff{1}_I{}^{\mu\nu}[H\coeff{1}]-G\coeff{2}_I{}^{\mu\nu}[g_B],\label{inner_eqn2alt}\\
&\vdots\nonumber
\end{align}
In these equations, $G_I\coeff{\emph{n}}$ and $\delta^kG_I\coeff{\emph{n}}$ consist of the terms in $G_I$ and $\delta^k G_I$ that contain $n$ derivatives with respect to $U$. The first equation, \eqref{inner_eqn0alt}, is the ordinary Einstein equation for $g_B$, except that all derivatives with respect to $U$ are removed. As the solution to this equation, I take the Schwarzschild metric
\begin{equation}
g_B = -f(U,\tilde R) dU^2 -2dUdR+R^2d\Omega^2, 
\end{equation}
where $f=1-2M(U)/\tilde R$, and $M(U)$ is the Bondi mass of $g_B$ at time $U$ divided by the initial mass. The dependence on $U$ can not be determined at this stage, because time-derivatives appear only in the higher-order equations.

Rather than fully solving the perturbation equations \eqref{inner_eqn1alt} and \eqref{inner_eqn2alt}, I will solve only certain parts of them, in order to pinpoint several key points about the general solution. First, I adopt the light cone gauge. This gauge choice consists of setting $H^{(n)}_{UR}=H^{(n)}_{RR}=H^{(n)}_{RA}=0$, which preserves the geometrical meaning of the retarded coordinates in the perturbed spacetime: $U$ remains constant on each outgoing light cone, and $R$ remains an affine parameter on outgoing light rays. Second, as a boundary condition, I insist that the perturbations must be regular on the event horizon.

Because of the spherical symmetry of the background, it is convenient to expand the perturbations in tensorial harmonics:
\begin{align}
H^{(n)}_{\mathsf{ab}} &= \sum_{\ell m}P^{(n)\ell m}_{\mathsf{ab}}Y^{\ell m},\\
H^{(n)}_{\mathsf{a}A} &= R\sum_{\ell m}\left(J_{\mathsf{a}}^{(n)\ell m}Y^{\ell m}_A+H_{\mathsf{a}}^{\ell m}X^{\ell m}_A\right),\\
H^{(n)}_{AB} &= R^2\sum_{\ell m}\left(K^{(n)\ell m}\Omega_{AB}Y^{\ell m}+G^{(n)\ell m}Y^{\ell m}_{AB}+H_2^{(n)\ell m}X^{\ell m}_{AB}\right), 
\end{align} 
where I have split the coordinates into the two sets $X^{\mathsf{a}}=(U,R)$ and $\Theta^A$, the various harmonics will be defined below, and the coefficients of the harmonics are functions of $U$ and $\tilde R$. In the context of this expansion, the light cone gauge is imposed by setting $P^{(n)\ell m}_{UR}=P^{(n)\ell m}_{RR}=P^{(n)\ell m}_{RA}=0$.

I define the various harmonics as follows: The scalar functions $Y^{\ell m}(\Theta^A)$ are the usual orthonormal spherical harmonics, which satisfy $[\Omega^{AB}D_AD_B+\ell(\ell+1)]Y^{\ell m}=0$, where $\Omega_{AB}$ is the metric of a unit 2-sphere, and $D_A$ is the covariant derivative compatible with $\Omega_{AB}$. The even-parity vector harmonics $Y^{\ell m}_A$ and odd-parity vector harmonics $X^{\ell m}_A$ are defined as
\begin{equation}
Y^{\ell m}_A\equiv D_AY^{\ell m},\quad X^{\ell m}_A\equiv-\epsilon_A{}^BD_BY^{\ell m},
\end{equation}
where $\epsilon_{AB}$ is the Levi-Civita tensor on the unit two-sphere. The even-parity and odd-parity tensor harmonics $Y^{\ell m}_{AB}$ and $X^{\ell m}_{AB}$ are defined as
\begin{align}
Y^{\ell m}_{AB} &= \left[D_A D_B+\tfrac{1}{2}\ell(\ell+1)\Omega_{AB}\right]Y^{\ell m},\\
X^{\ell m}_{AB} &= -\tfrac{1}{2}\left(\epsilon_A{}^CD_B+\epsilon_B{}^CD_A\right)D_CY^{\ell m}.
\end{align}
The tensor harmonics are trace-free with respect to the metric $\Omega_{AB}$. In addition, the various harmonics satisfy the following orthogonality relations:
\begin{equation}
\int\bar Y^A_{\ell m}Y_A^{\ell' m'}d\Omega =\ell(\ell+1)\delta_{\ell\ell'}\delta_{mm'} = \int\bar X^A_{\ell m}X_A^{\ell' m'}d\Omega,
\end{equation}
and
\begin{equation}
\int\bar Y^{AB}_{\ell m}Y_{AB}^{\ell' m'}d\Omega =\tfrac{1}{2}(\ell-1)\ell(\ell+1)(\ell+2)\delta_{\ell\ell'}\delta_{mm'} = \int\bar X^{AB}_{\ell m}X_{AB}^{\ell' m'}d\Omega,
\end{equation}
where in these equations an overbar indicates complex conjugation. We also have
\begin{equation}
\int\bar Y^A_{\ell m}X_A^{\ell' m'}d\Omega = 0 = \int\bar Y^{AB}_{\ell m}X_{AB}^{\ell' m'}d\Omega.
\end{equation}
In this appendix, I will forgo any discussion of the odd-parity terms, since the even-parity terms are sufficient for my purpose of delineating the types of restrictions required to arrive at the usual form of a tidally perturbed metric.

In order to determine the effect of a gauge transformation, I write an even-parity gauge vector $\Xi\coeff{1}_\alpha=(\Xi\coeff{1}_{\mathsf{a}},\Xi\coeff{1}_A)$ as
\begin{equation}
\Xi\coeff{1}_{\mathsf{a}} = \sum_{\ell m}\xi\coeff{1}_{\mathsf{a}}{}^{\ell m}Y^{\ell m},\quad \Xi\coeff{1}_A = R\sum_{\ell m}\xi\coeff{1}{}^{\ell m}Y_A^{\ell m}.
\end{equation}
This vector has the following first-order effects:
\begin{align}
\Delta P\coeff{1}_{UU} &= -\frac{2M}{\R^2}\xi\coeff{1}_U+\frac{2Mf}{\R^2}\xi\coeff{1}_R,\label{gauge1}\\
\Delta P\coeff{1}_{UR} &= -\pdiff{}{\R}\xi\coeff{1}_U+\frac{2M}{\R^2}\xi\coeff{1}_R,\\
\Delta P\coeff{1}_{RR} &= -2\pdiff{}{\R}\xi\coeff{1}_R,\\
\Delta J_U\coeff{1} &= -\frac{1}{\R}\xi\coeff{1}_U,\\
\Delta J_R\coeff{1} &= -\pdiff{}{\R}\xi\coeff{1}-\frac{1}{\R}\xi\coeff{1}_R+\frac{1}{\R}\xi\coeff{1},\\
\Delta K\coeff{1} &= -\frac{2f}{\R}\xi\coeff{1}_R+\frac{2}{\R}\xi\coeff{1}_U+\frac{\ell(\ell+1)}{\R}\xi\coeff{1},\\
\Delta G\coeff{1} &= -\frac{2}{\R}\xi\coeff{1},\label{gauge2}
\end{align}
where for simplicity I have omitted the harmonic labels $\ell m$. I neglect $U$-derivatives in the transformation, since they are second-order effects in the present scheme.

Now, due to the spherical symmetry of the background metric, each mode in the harmonic expansion decouples from all the others in $\delta G_I$. Likewise, the even- and odd-parity sectors decouple. I write the even-parity terms in $\delta G\coeff{0}_I$ as
\begin{align}
Q^{\mathsf{ab}}_{\ell m} &= \int \delta G\coeff{0}_I{}^{\mathsf{ab}}Y^{\ell m} d\Omega,\\
Q^{\mathsf{a}}_{(\ell m} &= \frac{2R^2}{\ell(\ell+1)}\int\delta G\coeff{0}_I{}{\mathsf{a}A} Y_A^{\ell m}d\Omega,\\
Q^\flat_{\ell m} &= R^2\int\delta G\coeff{0}_I{}^{AB}\Omega_{AB}Y^{\ell m}d\Omega,\\
Q^\sharp_{\ell m} &= \frac{4R^4}{(\ell-1)\ell(\ell+1)(\ell+2)}\int \delta G\coeff{0}_I{}^{AB} Y_{AB}^{\ell m}d\Omega.
\end{align}
Explicitly, these gauge-invariant quantities are given by~\cite{Eric_perturbations}
{\allowdisplaybreaks \begin{align}
Q^{UU} & = -\pddiff{}{\tilde R}\tilde K-\frac{2}{\tilde R}\pdiff{}{\tilde R}\tilde K+\frac{f}{\tilde R}\pdiff{}{\tilde R}\tilde P_{RR}-\frac{2}{\tilde R}\pdiff{}{\tilde R}\tilde P_{UR}+\frac{\ell(\ell+1)\tilde R+4M}{2\tilde R^3}\tilde P_{RR},\\
Q^{UR} & = -\frac{\R-M}{\R^2}\pdiff{}{\R}\tilde K+\frac{1}{\R}\pdiff{}{\R}\tilde P_{UU}+\frac{1}{\R^2}\tilde P_{UU}-\frac{\ell(\ell+1)+4}{2\R^2}\tilde P_{UR}+\frac{f}{\R^2}\tilde P_{RR}+\frac{\mu}{2\R^2}\tilde K,\\
Q^{RR} & = \frac{(\R-M)f}{\R^2}\pdiff{}{\R}\tilde K -\frac{f}{\R}\pdiff{}{\R}\tilde P_{UU}+\frac{\mu\R+4M}{2\R^3}\tilde P_{UU}+\frac{2f}{\R^2}\tilde P_{UR}-\frac{f^2}{\R^2}\tilde P_{RR}-\frac{\mu f}{2\R^2}\tilde K,\\
Q^U & = -\pdiff{}{\tilde R}\tilde P_{UR}+\pdiff{}{\tilde R}\tilde K+\frac{2}{\tilde R}\tilde P_{UR}-\frac{\tilde R-M}{\tilde R^2}\tilde P_{RR},\\
Q^R & = \pdiff{}{\tilde R}\tilde P_{UU}-f\pdiff{}{\tilde R}\tilde K-\frac{2(\tilde R-M)}{\tilde R^2}\tilde P_{UR}+\frac{(\tilde R-M)f}{\tilde R^2}\tilde P_{RR},\\
Q^\flat & = -\pddiff{}{\R}\tilde P_{UU}+f\pddiff{}{\R}\tilde K -\frac{2}{\R}\pdiff{}{\R}\tilde P_{UU}+\frac{2(\R-M)}{\R^2}\pdiff{}{\R}\tilde P_{UR}-\frac{(\R-M)f}{\R^2}\pdiff{}{\R}\tilde P_{RR}\nonumber\\
&\quad+\frac{2(\R-M)}{\R^2}\pdiff{}{\R}\tilde K+\frac{\ell(\ell+1)}{\R^2}\tilde P_{UR}-\frac{\ell(\ell+1)\R^2-2\mu M\tilde R-4M^2}{2\R^4}\tilde P_{RR},\\
Q^\sharp & = 2\tilde P_{UR}-f\tilde P_{RR},
\end{align}}
where $\mu\equiv \ell(\ell+1)-2$, and I have introduced the gauge-invariant combinations
\begin{align}
\tilde P_{UU} &= P_{UU}-\frac{2M}{\R}J_U+\frac{2Mf}{\R}J_R-Mf\pdiff{}{\R}G,\\
\tilde P_{UR} &= P_{UR}-\pdiff{}{\R}(\R J_U)+\frac{2M}{\R}J_R-M\pdiff{}{\R}G,\\
\tilde P_{RR} &= P_{RR}-2\pdiff{}{\R}(\R J_R)+\R^2\pddiff{}{\R}G+2\R\pdiff{}{\R}G,\\
\tilde K &= K+2J_U-2fJ_R+\R f\pdiff{}{\R}G+\tfrac{1}{2}\ell(\ell+1)G.
\end{align}
For simplicity, I have omitted the indices $(n)\ell m$ in all of the above expressions.

\section{First-order solution}
The first-order equation reads $\delta G\coeff{0}_I{}^{\alpha\beta}[H\coeff{1}]=-G\coeff{1}_I[g_B]$. The source term in this equation has a single non-vanishing component,
\begin{equation}
G\coeff{1}_I{}^{RR}[g_B] = \frac{2}{\R^2}\diff{M}{U}.
\end{equation}
Thus, in the notation introduced above, the $RR$, $\ell=0$ equation reads $Q^{RR}_{(0)00} = 4\sqrt{\pi}\R^{-2}\diff{M}{U}$, while all the other equations are source-free. For $\ell\geq 2$, the equations can be solved for arbitrary $\ell$. However, because various quantities are defined only for $\ell\geq 2$, the equations for the low multipoles $\ell=0$ and $\ell=1$ must be dealt with individually. I will write undetermined functions of $U$ as $A^{(n)\ell m}_k(U)$.

I begin by solving the $\ell=0$ equations. For $\ell=0$, the quantities $J_{\mathsf{a}}$, $G$, $Q^{\mathsf{a}}$, and $Q^\sharp$ are undefined. So the only equations are $Q^{UU}_{(0)00}[H\coeff{1}]=Q^{UR}_{(0)00}[H\coeff{1}]=Q^\flat_{(0)00}[H\coeff{1}]=0$ and $Q^{RR}_{(0)00}[H\coeff{1}] = 4\sqrt{\pi}\R^{-2}\diff{M}{U}$, in which $J_{\mathsf{a}}$ and $G$ are set to zero. Given the gauge condition, the only functions appearing in these equations are $K^{(1)00}$ and $P^{(1)00}_{UU}$. I first solve $Q^{UU}_{(1)00}=0$, which reads explicitly $-\pddiff{K^{(1)00}}{\tilde R}-\frac{2}{\tilde R}\pdiff{K^{(1)00}}{\tilde R}=0$. The solution to this equation is
\begin{equation}\label{K100}
K^{(1)00} = A^{(1)00}_1+\frac{1}{\R}A^{(1)00}_2.
\end{equation}
Substituting this into $Q^{UR}_{(0)00}[H\coeff{1}]=0$ yields an equation for $P^{(1)00}_{UU}$ that can be readily solved to find
\begin{equation}\label{P100}
P^{(1)00}_{UU} = A^{(1)00}_1-\frac{M}{\R^2}A^{(1)00}_2+\frac{1}{\tilde R}A^{(1)00}_3.
\end{equation}
Substituting these results into $Q^\flat_{(1)00}[H\coeff{1}]$ and $Q^{RR}_{(1)00}[H\coeff{1}]$, we find that both quantities are identically zero. Hence, from the equation $Q^{RR}_{(0)00}[H\coeff{1}] = 4\sqrt{\pi}\R^{-2}\diff{M}{U}$ we can conclude
\begin{equation}
\diff{M}{U} = 0;
\end{equation}
that is, the mass of the internal background is constant. The functions $A^{(1)00}_k$, $k=1,2,3$, can be be determined only by solving the second-order EFE.

Next, I proceed to the $\ell=1$ equations. For $\ell=1$, the quantities $G$ and $Q^\sharp$ are undefined, and the field equations read $Q^{\mathsf{ab}}_{(0)1m}[H\coeff{1}]=Q^{\mathsf{a}}_{(0)1m}[H\coeff{1}]=Q^\flat_{(0)1m}[H\coeff{1}]=0$, in which $G$ is set to zero. Solving $Q^{UU}_{(0)1m}=0$ yields
\begin{equation}
K^{(1)1m} = A^{(1)1m}_1+\frac{1}{\R}A^{(1)1m}_2.
\end{equation}
Substituting this into $Q^U_{(0)1m}=0$ and solving then yields
\begin{equation}
J_U^{(1)1m} = -\frac{1}{2\R}A^{(1)1m}_2+\frac{1}{\R^2}A^{(1)1m}_3+\R A^{(1)1m}_4.
\end{equation}
Substituting these results into $Q^R_{(0)1m}=0$ and solving then yields  
\begin{equation}
P_{UU}^{(1)1m} = -\frac{M}{\R^2}A^{(1)1m}_2+\frac{1}{\R^2}A^{(1)1m}_3-2\R A^{(1)1m}_4+A^{(1)1m}_5.
\end{equation}
Two of the remaining equations, $Q^{UR}_{(0)1m}=0=Q^{RR}_{(0)1m}$ fixes several of the free functions in these solutions: $A^{(1)1m}_4=0$ and $A^{(1)1m}_5=0$. The final equation, $Q^\flat_{(0)1m}=0$, yields no further information. Putting these results together, we find
\begin{align}
P\coeff{1}_{UU}{}^{1m} & = -\frac{M}{\R^2}A^{(1)1m}_2+\frac{1}{\R^2}A^{(1)1m}_3,\label{P11m}\\
J\coeff{1}_U{}^{1m} & = -\frac{1}{2\R}A^{(1)1m}_2+\frac{1}{\R^2}A^{(1)1m}_3,\\
K\coeff{1}{}^{1m} & = A^{(1)1m}_1+\frac{1}{\R}A^{(1)1m}_2.\label{K11m}
\end{align}

Finally, I proceed to the $\ell\geq 2$ equations. As with $\ell=0,1$, the equation $Q^{UU}_{(0)\ell m}=0$ can be immediately solved to find
\begin{equation}
K^{(1)\ell m} = A^{(1)\ell m}_1+\frac{1}{\R}A^{(1)\ell m}_2.
\end{equation}
Using this result and $Q^{R}_{(0)\ell m}=0$, we can express $P^{(1)\ell m}_{UU}$ in terms of $G^{(1)\ell m}$ and $J^{(1)\ell m}_U$; next, we can use $Q^U_{(0)\ell m}=0$ to express $G^{(1)\ell m}$ in terms of $J^{(1)\ell m}_U$; finally, substituting these results into $Q^\sharp_{(0)\ell m}=0$, we can solve for $J^{(1)\ell m}_U$ to find
\begin{align}
J^{(1)\ell m}_U &= -\frac{1}{2\R}A^{(1)\ell m}_2 +\frac{4M+\mu\R}{\R^2}A^{(1)\ell m}_4+A^{(1)\ell m}_5\R^\ell (-f)^{\ell-1}\ {}_2F_1\!\!\left(2-\ell, 1-\ell;-2\ell;-\frac{2M}{\R f}\right) \nonumber\\
&\quad +\frac{A^{(1)\ell m}_6}{\R^{\ell+1}(-f)^{2+\ell}}\ {}_2F_1\!\!\left(2+\ell, 3+\ell;2+2\ell;-\frac{2M}{\R f}\right),\label{Jlm}
\end{align}
where ${}_2F_1$ is a hypergeometric function. The term next to $A^{(1)\ell m}_6$ diverges at the unperturbed event horizon, $\R=2M$, violating the boundary condition. Therefore, we have $A^{(1)\ell m}_6=0$.\footnote{The term next to $A^{(1)\ell m}_6$ corresponds to an induced multipole moment. The fact that  it must vanish agrees with the no-hair theorem.} Next, I make my first restriction of the solution: if $J^{(1)\ell m}_U$ is expressed in terms of $R=\e\R$, then the term next to $A^{(1)\ell m}_5$ behaves as $\e^{-\ell}$. Since $H\coeff{1}$ is accompanied by a factor of $\e$ in the metric, this term behaves as $\e^{1-\ell}$. If I assume that $R\sim r$, where $r$ is, for example, the Fermi radial coordinate centered on the body's worldline in the external spacetime, then these negative powers of $\e$ would also appear in the outer expansion. By assumption, no negative powers do appear in the outer expansion; therefore, I set $A^{(1)\ell m}_5=0$. So $J^{(1)\ell m}_U$ simplifies to 
\begin{equation}
J^{(1)\ell m}_U = -\frac{1}{2\R}A^{(1)\ell m}_2 +\frac{4M+\mu\R}{\R^2}A^{(1)\ell m}_4.
\end{equation}

Recall that we had expressed $P^{(1)\ell m}_{UU}$ and $G^{(1)\ell m}$ in terms of $J^{(1)\ell m}_U$. With $J^{(1)\ell m}_U$ determined, we now have
\begin{align}
P^{(1)\ell m}_{UU} &= -\frac{M}{\R^2}\left(A^{(1)\ell m}_2-2\ell(\ell+1)A^{(1)\ell m}_4\right)+A^{(1)\ell m}_3,\\
G^{(1)\ell m} &= -\frac{4}{\R}A^{(1)\ell m}_4 + A^{(1)\ell m}_7.
\end{align}
Substituting these results into $Q^{UR}_{(0)\ell m}=0$ and then $Q^{RR}_{(0)\ell m}=0$, we determine
\begin{equation}
A^{(1)\ell m}_7 = -\frac{2A^{(1)\ell m}_1}{\ell(\ell+1)},\quad A^{(1)\ell m}_3=0.
\end{equation}
The sole remaining equation, $Q^\flat_{(1)\ell m}=0$, yields no new information. Hence, the first-order calculation is now complete. We have found that $\diff{M}{U}=0$; the $\ell=0$ modes in $H\coeff{1}$ are given by Eqs.~\eqref{K100} and \eqref{P100}; the $\ell=1$ modes are given by Eqs.~\eqref{P11m}--\eqref{K11m}; and the $\ell\geq2$ modes are given by
\begin{align}
P^{(1)\ell m}_{UU} &= -\frac{M}{\R^2}\left(A^{(1)\ell m}_2-2\ell(\ell+1)A^{(1)\ell m}_4\right),\\
J^{(1)\ell m}_U &= -\frac{1}{2\R}A^{(1)\ell m}_2 +\frac{4M+\mu\R}{\R^2}A^{(1)\ell m}_4,\\
K^{(1)\ell m} &= A^{(1)\ell m}_1+\frac{1}{\R}A^{(1)\ell m}_2,\\
G^{(1)\ell m} &= -\frac{4}{\R}A^{(1)\ell m}_4-\frac{2A^{(1)\ell m}_1}{\ell(\ell+1)}.
\end{align}

These results can be simplified by a refinement of the lightcone gauge. For $\ell=0$, the function $A^{(1)00}_2$ can be removed via
\begin{align}
\xi_R^{(1)00} & = \xi^{(1)00}_R(U),\\
\xi_U^{(1)00} & = f\xi^{(1)00}_R - \tfrac{1}{2}A^{(1)00}_2,
\end{align} 
leaving
\begin{align}
P^{(1)00}_{UU} &= A^{(1)00}_1+\frac{1}{\tilde R}A^{(1)00}_3,\\
K^{(1)00} &= A^{(1)00}_1.
\end{align}
Although this does not exhaust the residual freedom within the lightcone gauge, since $\xi_R(U)$ is arbitrary, the remaining freedom cannot be used to remove either $A^{(1)00}_1$ or $A^{(1)00}_3$. However, if we were to transform out of the lightcone gauge, $A^{(1)00}_1$ could be removed, leaving only a $1/\R$ mass monopole term, corresponding to a time-dependent correction to the mass.

For $\ell\geq1$, we can removed \emph{all} $A^{(1)\ell m}_k$ via
\begin{align} 
\xi^{(1)1 m}_U &= -\tfrac{1}{2}A^{(1)1m}_2+\frac{1}{\R}A^{(1)1m}_3,\\
\xi^{(1)1 m}_R &= -\frac{1}{2M}A^{(1)1m}_3,\\
\xi^{(1)1 m} &= -\tfrac{1}{2}\R A^{(1)1m}_2-\frac{1}{2M}-\tfrac{1}{2}A^{(1)1m}_3,
\end{align}
and
\begin{align}
\xi^{(1)\ell m}_U & = -\tfrac{1}{2}A^{(1)\ell m}_2+\left(\mu+\frac{4M}{\R}\right)A^{(1)\ell m}_4,\\
\xi^{(1)\ell m}_R & = -2A^{(1)\ell m}_4,\\
\xi^{(1)\ell m} & = -2A^{(1)\ell m}_4-\frac{\R A^{(1)\ell m}_1}{\ell(\ell+1)}.
\end{align}
This exhausts the residual freedom in the gauge condition.

In order to arrive at the usual form of a tidally perturbed metric, we must go beyond these gauge refinements by setting the entirety of $H\coeff{1}$ to zero. That is, we must choose $A^{(1)00}_1=A^{(1)00}_3=0$. If the odd-parity calculation had been performed, we would find that $\ell=1$ terms, corresponding to time-dependent spin terms, must be set to zero. (Recall also that I have restricted the possible perturbations by disallowing terms that would contain negative powers of $\e$ in the unscaled coordinates.) Hence, we can conclude that to arrive at the usual metric of a tidally perturbed black hole, we must restrict the perturbation by setting numerous functions to zero, without any evident justification.

\section{Second-order solution}
With $H\coeff{1}$ set to zero, and with $M$ determined to be a constant, the second-order EFE becomes the homogeneous, linear equation
\begin{equation}
\delta G\coeff{0}_I{}^{\alpha\beta}[H\coeff{2}]=0.
\end{equation}
This equation is solved in the same manner as the first-order equation. The calculation of the low multipoles $\ell=0$ and $\ell=1$ proceeds just as at first order, yielding
\begin{align}
P^{(2)00}_{UU} &= A^{(2)00}_1-\frac{M}{\R}A^{(2)00}_2+\frac{1}{\tilde R}A^{(2)00}_3,\\
K^{(2)00} &= A^{(2)00}_1+\frac{1}{\R}A^{(2)00}_2,
\end{align}
and
\begin{align}
P_{UU}^{(2)1m} & = -\frac{M}{\R^2}A^{(2)1m}_2+\frac{1}{\R^2}A^{(2)1m}_3,\\
J_U^{(2)1m} & = -\frac{1}{2\R}A^{(2)1m}_2+\frac{1}{\R^2}A^{(2)1m}_3,\\
K^{(2)1m} & = A^{(2)1m}_1+\frac{1}{\R}A^{(2)1m}_2.
\end{align}

For $\ell\geq 2$, the calculation proceeds just as at first order, up until the point marked by Eq.~\eqref{Jlm}. When the term that diverges on the event horizon is removed, the analogue of that equation reads
\begin{equation}
J^{(2)\ell m}_U = -\frac{1}{2\R}A^{(2)\ell m}_2 +\frac{4M+\mu\R}{\R^2}A^{(2)\ell m}_4+A^{(2)\ell m}_5\R^\ell (-f)^{\ell-1}\ {}_2F_1\!\!\left(2-\ell, 1-\ell;-2\ell;-\frac{2M}{\R f}\right).
\end{equation}
At first order, this solution was simplified by setting the coefficient of $\R^\ell$ to zero, because it led to negative powers of $\e$ when written in terms of $R$. However, at second order, this term will be multiplied by $\e^2$ in the metric, so it will scale as $\e^{2-\ell}$. Hence, the term is acceptable for $\ell=2$, but must be set to zero for $\ell>2$. I will hence deal with these two cases separately.

For $\ell=2$, we have
\begin{equation}
{}_2F_1\!\!\left(2-\ell, 1-\ell;-2\ell;-\frac{2M}{\R f}\right) = 1,
\end{equation}
and so 
\begin{equation}
J^{(2)2m}_U = -\frac{1}{2\R}A^{(2)2m}_2 +\frac{4M+4\R}{\R^2}A^{(2)2m}_4-A^{(2)2m}_5\R^2f.
\end{equation}
As at first order, this determines $P^{(2)2m}_{UU}$ and $G^{(2)2m}$:
\begin{align}
P^{(2)\ell m}_{UU} &= -\frac{M}{\R^2}\left(A^{(2)2 m}_2-12A^{(2)2m}_4\right)+A^{(2)2m}_3+3\R(\R-4M)A^{(2)2m}_5,\\
G^{(2)\ell m} &= -\frac{4}{\R}A^{(2)2 m}_4 + \R^2 A^{(2)2m}_5+A^{(2)2 m}_7.
\end{align}
And as at first order, we determine $A^{(2)2 m}_3$ and $A^{(2)2 m}_7$ from $Q^{UR}_{(0)\ell m}=0$ and then $Q^{RR}_{(0)\ell m}=0$, which yield
\begin{equation}
A^{(2)2 m}_7 = -\tfrac{1}{3}A^{(2)2m}_1-2M^2A^{(2)2m}_5,\quad A^{(2)2m}_3=12M^2A^{(2)2m}_5.
\end{equation}
Putting these results together, we have the solution
\begin{align}
P_{UU}^{(2)2 m} & = 3\R^2f^2 A_5^{(2)2m}-\frac{M}{\R^2}A_2^{(2)2 m}+\frac{12M}{\R^2}A_4^{(2)2 m},\\
J_U^{(2)2 m} & = -\frac{1}{2\R}A_2^{(2)2m}+\frac{4}{\R}\left(1+\frac{M}{\R}\right)A_4^{(2)2 m}-\R^2fA_5^{(2)2 m},\\
K^{(2)2 m} & = A^{(2)2 m}_1+\frac{1}{\R}A^{(2)2m}_2,\\
G^{(2)2 m} & = -\tfrac{1}{3}A_1^{(2)2m}+\R^2\left(1-\frac{2M^2}{\R^2}\right)A_5^{(2)2 m}-\frac{4}{\R}A_4^{(2)2 m}.
\end{align}

For $\ell>2$, the calculation proceeds just as at first order, leading to the solution
\begin{align}
P_{UU}^{(2)\ell m} & = -\frac{M}{\R^2}A_2^{(2)\ell m}+\frac{2M}{\R^2}\ell(\ell+1)A_4^{(2)\ell m},\\
J_U^{(2)\ell m} & = -\frac{1}{2\R}A_2^{(2)\ell m}+\frac{\mu\R+4M}{\R^2}A_4^{(2)\ell m},\\
K^{(2)\ell m} & = A^{(2)\ell m}_1+\frac{1}{\R}A^{(2)\ell m}_2,\\
G^{(2)\ell m} & = -\frac{2}{\ell(\ell+1)}A_1^{(2)\ell m}-\frac{4}{\R}A_4^{(2)\ell m}.
\end{align}

Just as at first order, with an appropriate gauge refinement, we can remove the functions $A^{(2)00}_2$, $A^{(2)1 m}_k$, $A_1^{(2)\ell m}$, $A_2^{(2)\ell m}$, and $A_4^{(2)\ell m}$, where $\ell\geq2$, thereby exhausting the freedom within the lightcone gauge. In order to arrive at the usual form of the tidally perturbed black hole metric, we must then set $A_1^{(2)00}=A_3^{(2)00}=0$. (And I again remind the reader that I have disallowed terms that would possess a negative power of $\e$ when expressed in unscaled coordinates.) This leaves only one undetermined function: $A_5^{(2)\ell m}$. Although I do not show the odd-parity calculation, it yields an analogous result: all but one of the possible undetermined functions must be set to zero to yield the usual form of the metric. After imposing these restrictions, the only non-vanishing components of the metric perturbation are
\begin{align}
H\coeff{2}_{UU} &= \sum_m 3\R^2f^2 A_5^{(2)2m}Y^{2m},\\
H\coeff{2}_{UA} &= -R\sum_m\left(\R^2f A_5^{(2)2 m}Y_A^{2m}+\R^2 f B^{(2)2m}X^{2m}_A\right),\\
H\coeff{2}_{AB} &= R^2\sum_m\left[\R^2\left(1-\frac{2M^2}{\R^2}\right)A_5^{(2)2 m}Y^{2m}_{AB}+\R^2 B^{(2)2m}X^{2m}_{AB}\right].
\end{align}

\section{Tidally perturbed black hole metric and its expansion in the buffer region}
The metric perturbations have been restricted by the following assumptions: the local coordinates are related to the external Fermi coordinates by a small transformation; and there are no monopole or dipole perturbations. Given these restrictions, the inner expansion has the following form:
\begin{align}
g_{IUU} &= -f+\e^2\sum_m 3\R^2f^2 A_5^{(2)2m}Y^{2m}+O(\e^3),\\
g_{IUA} &= -R\left[\e^2\R^2f\sum_mA_5^{(2)2 m}Y_A^{2m}+\e^2\R^2f\sum_mB^{(2)2m}X^{2m}_A+O(\e^3)\right],\\
g_{IAB} &= R^2\left[\Omega_{AB}+\e^2\R^2\left(1-\frac{2M^2}{\R^2}\right)\sum_mA_5^{(2)2 m}Y^{2m}_{AB}+\e^2\R^2 \sum_mB^{(2)2m}X^{2m}_{AB}+O(\e^3)\right],
\end{align}
along with the exact results $g_{IUR}=-1$ and $g_{IRR}=g_{IRA}=0$. Although it is written in a slightly different form, this is the usual metric of a tidally perturbed black hole. It is characterized by (i) having only quadrupole perturbations, and (ii) those perturbations scaling as $\R^2$ for large $\R$.

To write the metric in terms of a pair of tidal fields $\tilde\etide_{ab}$ and $\tilde\btide_{ab}$, I follow Appendix~A of Ref.~\cite{Eric_Igor}; I also make use of notation and identities from the end of Sec.~\ref{retarded coordinates}. First, I define the quantities
\begin{equation}
\tilde\etide^*\equiv -3\sum_m A_5^{(2)2m}Y^{2m},\quad \tilde\btide^*\equiv -3\sum_m B^{(2)2m}Y^{2m}.
\end{equation}
Next, I define the derived quantities
\begin{equation}\label{tide_A def}
\tilde\etide^*_A\equiv\tfrac{1}{2}D_A\tilde\etide^*,\quad \tilde\btide^*_A\equiv-\tfrac{1}{2}\epsilon_A{}^B D_B\tilde\btide^*
\end{equation}
and
\begin{equation}\label{tide_AB def}
\tilde\etide^*_{AB}\equiv\left(D_AD_B+3\Omega_{AB}\right)\tilde\etide^*,\quad \tilde\btide^*_{AB}\equiv-\tfrac{1}{2}\left(\epsilon_A{}^C D_B+\epsilon_B{}^C D_A\right)D_C\tilde\btide^*.
\end{equation}
Using the definitions of $\tilde\etide^*$, $\tilde\btide^*$, $Y^{\ell m}_A$, $Y^{\ell m}_{AB}$, $X^{\ell m}_A$, and $X^{\ell m}_{AB}$, we can express the derived quantities as
\begin{equation}
\tilde\etide^*_A\equiv-\tfrac{3}{2}\sum_m A_5^{(2)2m}Y^{2m}_A,\quad \tilde\btide^*_A\equiv -\tfrac{3}{2}\sum_m B^{(2)2m}X^{2m}_A,
\end{equation}
and
\begin{equation}
\tilde\etide^*_{AB}\equiv-3\sum_m A_5^{(2)2m}Y^{2m}_{AB},\quad \tilde\btide^*_{AB}\equiv-3\sum_mB^{(2)2m}X^{2m}_{AB}.
\end{equation}
In terms of these quantities, we can write the metric as
\begin{align}
g_{IUU} &= -f-\e^2\R^2f^2\tilde\etide^* +O(\e^3),\\
g_{IUA} &= R\left[\tfrac{2}{3}\e^2\R^2f\left(\tilde\etide^*+\tilde\btide^*\right)+O(\e^3)\right],\\
g_{IAB} &= R^2\left[\Omega_{AB}-\tfrac{1}{3}\e^2\R^2\left(1-\frac{2M^2}{\R^2}\right)\tilde\etide^*_{AB}-\tfrac{1}{3}\e^2\R^2 \tilde\btide^*_{AB}+O(\e^3)\right].
\end{align}
This is the usual form of the metric of a tidally perturbed black hole.

Now, because $\tilde\etide^*$ and $\tilde\btide^*_a$ are, respectively, even- and odd-parity quadrupole terms, they can be written in terms of an STF decomposition:
\begin{equation}
\tilde\etide^* = \tilde\etide_{ab}\Omega^{\langle ab\rangle},\quad \tilde\btide^*_a = \epsilon_{acd}\tilde\btide^d_b\Omega^{\langle bc\rangle}.
\end{equation}
By applying the definitions \eqref{tide_A def} and \eqref{tide_AB def}, we find
\begin{equation}
\tilde\etide^*_A = \Omega^a_A\tilde\etide_{ab}\Omega^b,\quad \tilde\btide^*_A = \Omega^a_A\epsilon_{acd}\tilde\btide^d_b\Omega^{\langle bc\rangle},
\end{equation}
and
\begin{align}
\tilde\etide^*_{AB} &= 2\Omega^a_A\Omega^b_B\tilde\etide_{ab}+\Omega_{AB}\tilde\etide_{ab}\Omega^{\langle ab\rangle},\\
\tilde\btide^*_{AB} &= \Omega^a_A\Omega^b_B\epsilon_{acd}\tilde\btide^d_b\Omega^c+\Omega^a_A\Omega^b_B\epsilon_{bcd}\tilde\btide^d_a\Omega^c.
\end{align}
When written in more explicit form, these expressions agree with those of Eqs.~\eqref{Estar}--\eqref{Bstar_AB}. In order to arrive at these results, one requires the identities
\begin{equation}
\epsilon_{AB} = \epsilon_{abc}\Omega^a_A\Omega^b_B\Omega^c,\quad
\epsilon_A{}^B\Omega^b_B = -\Omega^a_A\epsilon_{ac}{}^b\Omega^c,\quad
D_AD_B\Omega^a = -\Omega^a\Omega_{AB}.
\end{equation}

Finally, I convert to Cartesian coordinates $(U,Y^a)$, adapting the identities of Sec.~\ref{retarded coordinates}. The result is
\begin{align}
g_{IUU} &= -f-\e^2 f^2\R^2\ein^*+O(\e^3),\\
g_{IUa} &= -\Omega_a+\tfrac{2}{3}\e^2\R^2 f(\ein^*_a+\bin^*_a)+O(\e^3),\\
g_{Iab} &= \delta_{ab}-\Omega_{ab}-\tfrac{1}{3}\e^2\R^2\left(1-\frac{2 M^2}{\R^2}\right)\ein^*_{ab}-\tfrac{1}{3}\e^2\R^2\bin^*_{ab}+O(\e^3).
\end{align}
where $\ein_a^*=\ein_A^*\Omega^A_a$, $\bin_a^*=\bin_A^*\Omega^A_a$, $\ein_{ab}^*=\ein_{AB}^*\Omega^A_a\Omega^B_b$, and $\bin_{ab}^*=\bin_{AB}^*\Omega^A_a\Omega^B_b$. When this is expanded in the buffer region, by rewriting it in terms of the unscaled radial function $R$ and then re-expanding for small $\e$, it becomes
\begin{align}
g_{IUU} & = -1+\e\frac{2M}{R}-R^2\etide^*+4\e MR\etide^*+\order{\e^2,\e R^2,R^3}, \\
g_{IUa} & = -N_a+\tfrac{2}{3}R^2(\etide_a^*+\btide_a^*)-\tfrac{4}{3}MR(\etide_a^*+\btide_a^*)+\order{\e^2,\e R^2,R^3}\\
g_{Iab} & = \delta_{ab}-N_{ab}-\tfrac{1}{3}R^2(\etide^*_{ab}+\btide^*_{ab})+\order{\e^2,\e R^2,R^3}.
\end{align}

In order to agree with the external Lorenz gauge, I switch to harmonic coordinates via the transformation $Y^a = X^{a'}+\e MN^{a'}$, where $N^{a'}=X^{a'}/R'$,\footnote{Note that, in the buffer region at first order in $\e$, this transformation to harmonic coordinates cannot be distinguished from a transformation to isotropic coordinates.} and then switch from retarded coordinates to Fermi-type coordinates $(T,X^a)$ via the transformation
\begin{align}
U & = T-R-2\e M(\ln R + \tfrac{1}{6}R^2\etide_{ij}N^{ij}) +\tfrac{1}{6}R^3\etide_{ij}N^{ij}, \\
X^{a'} & = X^a - \tfrac{1}{3}R^3R^a{}_{b0c}N^{bc}+\tfrac{1}{6}R^3\etide^a_bN^b,
\end{align}
where $N^a=X^a/R$. After performing these transformations and decomposing the result into irreducible STF pieces, we arrive at  
\begin{align}
g_{ITT} & = -1+\e\frac{2M}{R}+\tfrac{5}{3}\e MR\ein_{ij}\hat N^{ij}-R^2\ein_{ij}\hat N^{ij}+\order{\e^2,\e R^2, R^3}, \\
g_{ITa} & = 2\e MR\ein_{ai}N^i+\tfrac{2}{3}\e MR\epsilon_{aij}\bin^j_k\hat N^{ik}+\tfrac{2}{3}R^2\epsilon_{aik}\bin^k_j\hat N^{ij}+\order{\e^2,\e R^2, R^3}, \\
g_{Iab} & = \delta_{ab}\left(1+\e\frac{2M}{R}-\tfrac{5}{9}\e MR\ein_{ij}\hat N^{ij}-\tfrac{1}{9}R^2\ein_{ij}\hat N^{ij}\right) +\tfrac{64}{21}\e MR\ein_{i\langle a}\hat N_{b\rangle}{}^i \nonumber\\
&\quad-\tfrac{46}{45}\e MR\ein_{ab}-\tfrac{1}{9}R^2\ein_{ab}+\tfrac{2}{3}\e MR\ein_{ij}\hat N_{ab}{}^{ij}+\tfrac{2}{3}R^2\ein_{i\langle a}\hat N^i_{b\rangle}\nonumber\\
&\quad -\tfrac{4}{3}\e MR\epsilon_{jk(a}\bin_{b)}^kN^j+\order{\e^2,\e R^2, R^3}.
\end{align}
			\chapter{Evaluation of the boundary integral}\label{boundary_integral}
In this appendix, I present the explicit evaluation of the integral over the worldtube $\Gamma$, the results of which are required in Ch.~\ref{perturbation calculation}.

%%%%%%%%%%%%%%
\section{Integral over the past light cone}
%%%%%%%%%%%%%%
Each of the quantities in the bitensor $\h^{\text{dir}}_{\alpha\beta}$ can be expanded in powers of $\zeta$ by first expanding the $x'$-dependence about the point $\bar x=\gamma(t')$ and then expanding the $\bar x$-dependence about the point $x''=\gamma(t)$. See Fig.~\ref{tube} for a depiction of the relationship between these points. The expansions are provided in Appendix~\ref{worldtube expansions}. Most significantly, the distance $\r$ and its time-derivative $\partial_{t'}\r\equiv\dot{\r{}_{\ }}$ are expanded as $\r=\zeta(\r_0+\zeta\r_1+\zeta^2\r_2+...)$ and $\dot{\r{}_{\ }}=\dot{\r_0}+\zeta\dot{\r_1}+\zeta^2\dot{\r_2}+...$, where the leading-order terms are the flat-spacetime values
\begin{align}
\r_0 &= \sqrt{r^2+\rad^2-2r\rad n^an'_a},\\
\dot{\r_0} &= -1.
\end{align}

After making use of these two expansions, and expressing $\hmn{}{1,0}(t')$ and $\hmn{}{1,1}(t')$ in terms of their values at $t$, we can express $\h^{\text{dir}}$ explicitly as
\begin{equation}\label{hdir}
\begin{split}
\h^{\text{dir}}_{\alpha\beta} & =  \frac{1}{\r_0}\left(1-\frac{\r_2}{\r_0}\right)U_{\alpha\beta}{}^{\gamma'\delta'} \hmn{\gamma'\delta'}{1,-1} -\frac{\rad}{\r_0^3}\bigg(1-\dot{\r_2}-\frac{3\r_2}{\r_0}\bigg) U_{\alpha\beta}{}^{\gamma'\delta'}\hmn{\gamma'\delta'}{1,-1} \sigma_{\mu'}n^{\mu'} \\&\quad
-\frac{\rad}{\r_0}\Big(U_{\alpha\beta}{}^{\gamma'\delta'}\del{n'} \hmn{\gamma'\delta'}{1,-1} -\del{n'}U_{\alpha\beta}{}^{\gamma'\delta'}\hmn{\gamma'\delta'}{1,-1} +V_{\alpha\beta}{}^{\gamma'\delta'}\hmn{\gamma'\delta'}{1,-1} \sigma_{\mu'}n^{\mu'}\Big)\\
&\quad -\frac{\rad}{\r_0^2}\del{u'}\Big(U_{\alpha\beta}{}^{\gamma'\delta'} \sigma_{\mu'}n^{\mu'} \hmn{\gamma'\delta'}{1,-1}\Big) -\frac{\rad}{\r_0^3}U_{\alpha\beta}{}^{\gamma'\delta'}e^I_{\gamma'}e^J_{\delta'} \hmn{IJ}{1,0}\!(t)\sigma_{\mu'}n^{\mu'}\\
&\quad -\frac{\rad^2}{\r_0}U_{\alpha\beta} {}^{\gamma'\delta'}\hmn{\gamma'\delta'}{1,1}-\frac{\rad^3}{\r_0^3}U_{\alpha\beta} {}^{\gamma'\delta'}e^I_{\gamma'}e^J_{\delta'}\hmn{IJ}{1,1}\!(t) \sigma_{\mu'}n^{\mu'} +\order{\zeta^2,\e}.
\end{split}
\end{equation}
Each term in this expression is further expanded using the results of Appendix~\ref{worldtube expansions}, which details the expansion of the Green's function and $\sigma_{\mu'}(x,x')$. In order to integrate the final, fully expanded expression for $\h^{\text{dir}}$ over the surface $\mathcal{S}$, we make use of the angular integrals displayed in Appendix~\ref{angular integrals}.

The end results of these calculations are as follows: The $\hmn{\alpha\beta}{1,-1}$ terms in $\h^{\text{dir}}$ contribute
\begin{equation}
\begin{split}
&\frac{1}{4\pi}\oint\limits_\mathcal{S}\Big(\hmn{\alpha\beta}{1,-1}\text{ terms in \eqref{hdir}}\Big)Nd\Omega' \\
&= 4m\etide_{ab}x^be^a_{(\alpha}e^0_{\beta)}+\frac{2m}{r}\big(1 -\tfrac{1}{6}\etide_{ab}x^{ab}\big)\big(e^0_\alpha e^0_\beta+e_{i\alpha}e^i_\beta\big)-\frac{26m\rad^2}{9r}\etide_{ab}e^a_\alpha e^b_\beta \\
&\quad +\frac{m\rad^4}{15r^3}\Big(-5\etide_{cd}n^{cd}e^0_\alpha e^0_\beta+\tfrac{5}{3}\etide_{cd}n^{cd}e_{i\alpha}e^i_\beta-4\etide_{c\langle a}\nhat_{b\rangle}^c e^a_\alpha e^b_\beta \\
&\quad -4\epsilon_{bdc}\btide^c_a\nhat^{ad}e^0_{(\alpha}e^b_{\beta)}  -\tfrac{4}{3}\etide_{cd}n^{cd}e_{i\alpha}e^i_\beta\Big)+\order{\zeta^2,\e}.
\end{split}
\end{equation}
The $\hmn{\alpha\beta}{1,0}$ terms integrate to zero at the orders of interest. And the $\hmn{\alpha\beta}{1,1}$ terms contribute
\begin{equation}
\begin{split}
&\frac{1}{4\pi}\oint\limits_\mathcal{S} \Big(\hmn{\alpha\beta}{1,1} \text{ terms in \eqref{hdir}}\Big)Nd\Omega' \\
& =  \frac{m\rad^4}{3r^3}\etide_{ab}n^{ab}e^0_\alpha e^0_\beta +\frac{4m\rad^4}{15r^3}\epsilon_{abc} \btide^c_d\nhat^{bd}e^0_{(\alpha}e^a_{\beta)} +\frac{38m\rad^2}{9r}\etide_{ab}e^a_\alpha e^b_\beta \\
&\quad +\frac{m\rad^4}{r^3}\Big(-\tfrac{1}{9}\delta_{ab}\etide_{cd}n^{cd} +\tfrac{4}{15}\etide_{c\langle a}\nhat_{b\rangle}^c\Big)e^a_\alpha e^b_\beta +\order{\zeta^2,\e}
\end{split}
\end{equation}
Note that the $\hmn{\alpha\beta}{1,1}$ terms are all $\rad$-dependent, and they are necessary to cancel $\rad$-dependent terms arising from $\hmn{\alpha\beta}{1,-1}$. All the actual terms that appear in the buffer region expansion arise from the most singular part of the perturbation, but the regular terms, such as $\hmn{\alpha\beta}{1,1}$, are required on the boundary to ensure the consistency of the solution.

Putting these results together, we arrive at
\begin{align}
\frac{1}{4\pi}\oint\limits_\mathcal{S}\h^{\text{dir}}_{\alpha\beta}Nd\Omega' &= \frac{2m}{r}\big(1-\tfrac{1}{6}\etide_{ab}x^{ab}\big)\big(e^0_\alpha e^0_\beta+e_{i\alpha}e^i_\beta\big)+4m\etide_{ab}x^be^a_{(\alpha}e^0_{\beta)}\nonumber\\
&\quad +\frac{4m\rad^2}{3r}\etide_{ab}e^a_\alpha e^b_\beta+\order{\zeta^2,\e},
\end{align}
where all tensors are evaluated at time $t$. By expressing the tetrad $(e^0_\alpha,e^a_\alpha)$ in terms of the coordinate one-forms $(t_\alpha,x^a_\alpha)$, we can write this result in Fermi coordinates as 
\begin{align}
\frac{1}{4\pi}\oint\limits_\mathcal{S}\h^{\text{dir}}_{\alpha\beta}Nd\Omega' &= \frac{2m}{r}\left(t_\alpha t_\beta + \delta_{ab}x^a_\alpha x^b_\beta\right) +\tfrac{5}{3}mr\etide_{ij}\nhat^{ij}t_\alpha t_\beta\nonumber\\
&\quad +4mr\left(\etide_{bi}n^i+\tfrac{1}{3}\epsilon_{bij}\btide^j_k\nhat^{ik}\right) t_{(\alpha}x^b_{\beta)}\nonumber\\
&\quad +\tfrac{1}{9}mr\Big[12\etide_{i\langle a}\nhat_{b\rangle}^i-5\delta_{ab}\etide_{ij}\nhat^{ij} +\left(12\rad^2/r^2-2\right)\etide_{ab}\Big]x^a_\alpha x^b_\beta\nonumber\\
&\quad+\order{\zeta^2,\e}.
\end{align}

We now  proceed to the calculation of the tail terms.

%%%%%%%%%%%%%%
\section{Integral over the interior of the past light cone}
%%%%%%%%%%%%%%
The interior of the light cone covers the worldtube from the lower limit $t'=0$, where the worldtube intersects the surface $\Sigma$, to the upper limit defined by the surface $\mathcal{S}$. Because the Fermi time varies over $\mathcal{S}$, I split the integration into two regions: one region from $t'=0$ to $t'=t_{\text{min}}\equiv\displaystyle\min_{\mathcal{S}}\{t'\}$, and another from $t'=t_{\text{min}}$ to $t'=t_{\text{max}}\equiv\displaystyle\max_{\mathcal{S}}\{t'\}$. The integral over the worldtube is then expressed as
\begin{align}
\int\limits_{\Gamma\cap\past}\!\!\!d\Omega'dt' = \int_0^{t_{\text{min}}}\!\!\!\!\!\!\!\!dt'\!\int_0^{4\pi}\!\!\!d\Omega'\!
&+\int_{t_{\text{min}}}^{t_{\text{max}}}\!\!\!\!\!\!\!\!dt' \int_0^{\theta_{\text{max}}(t')}\!\!\!\!\!\!\!\!\!\!\!\!\!\!\!d\theta'\ \int_0^{\phi_{\text{max}}(t',\theta')}\!\!\!\!\!\!\!\!\!\!\!\!\!\!\!\!\!\!\!\!\! d\phi'.\ \ 
\end{align}
In the integral from $t_{\text{min}}$ to $t_{\text{max}}$, the angles $\theta'^A=(\theta',\phi')$ are cut off at some maximum values defined by $\mathcal{S}$.

Because $\h^{\text{tail}}$ is of order $\zeta^0$, and we only seek terms up to order $\zeta$, we can further simplify the integral. For $x'\in\mathcal{S}$, we can write the time difference $t-t'$ as $t-t'=\r_0+\order{\zeta^2}$, where the $\order{\zeta^2}$ error term consists of acceleration and curvature terms (see Eq.~\eqref{Delta t}). I choose $\theta'$ to be the angle between $x'^a$ and $x^a$, such that $t'=t-\sqrt{r^2+\rad^2-2r\rad\cos\theta'}+\order{\zeta^2}$. From this we infer that the maximum and minimum times on $\mathcal{S}$ are given by $t_{\text{max}}=t-(r-\rad)+\order{\zeta^2}$ and $t_{\text{min}}=t-(r+\rad)+\order{\zeta^2}$. The value for $t_{\text{max}}$ corresponds to the time at the point on $\mathcal{S}$ closest to $x$; the value at $t_{\text{min}}$ is the time at the point furthest from $x$. Since the maximum value of $\theta'$ at a given value of $t'$ is determined by the intersection with $\mathcal{S}$, it is given by
\begin{equation}
\cos\theta_{\text{max}}=\frac{r^2+\rad^2-(t-t')^2}{2r\rad} +\order{\zeta}.
\end{equation}
Since $\phi'$ runs from 0 to $2\pi$ everywhere on $\mathcal{S}$, its maximum value is $2\pi$, independent of $t'$ and $\theta'$.

Making use of these approximations, we can expand the first integral, running from $t'=0$ to $t'=t_{\text{min}}$, about the upper limit $t'=t$. This enables us to write the integral over the worldtube as
\begin{align}
\int\limits_{\Gamma\cap\past}\!\!\!d\Omega'dt' & = \int_0^t\!\!\!dt'\!\int_0^{4\pi}\!\!\!d\Omega' -(r+\rad)\!\!\int_0^{4\pi}\!\!\!d\Omega'\Big|_{t'=t}\nonumber\\
&\quad+\int_0^{2\pi}\!\!\!d\phi'\!\int_{t_{\text{min}}}^{t_{\text{max}}}\!\!\!dt' \!\int_{\cos\theta_{\text{max}}}^1\!\!\!\!\!\!\!\!\!d\cos\theta' +\order{\zeta^2}.
\end{align}

With the simplification $\del{n'}\hmn{\gamma'\delta'}{1,-1} = \order{\zeta^2,\zeta\e} = \del{n'}\hmn{\gamma'\delta'}{1,0}$, along with the expansion
\begin{equation}
G\indices{_{\alpha\beta}^{\alpha'\beta'}} = e^{(\alpha'}_{I}e^{\beta')}_{J}\left[G\indices{_{\alpha\beta}^{IJ}}(t') +G\indices{_{\alpha\beta}^{IJ}_{|c}}(t')x'^c+\order{\zeta^2}\right],
\end{equation}
$\h^{\text{tail}}$ can be written as
\begin{align}
\h^{\text{tail}}_{\alpha\beta} &= \hmn{\gamma'\delta'}{1,-1}e_I^{\gamma'}e_J^{\delta'}\left( G\indices{_{\alpha\beta}^{IJ}}(t') +2\rad G\indices{_{\alpha\beta}^{IJ}_{|c}}n'^c\right)+\order{\zeta^2}\nonumber\\
&=2m(t')\left(\delta^0_I\delta^0_J+\delta_{ij}\delta^i_I\delta^j_J\right)\Big( G\indices{_{\alpha\beta}^{IJ}}(t') +2\rad G\indices{_{\alpha\beta}^{IJ}_{|c}}n'^c\Big)+\order{\zeta^2}.
\end{align}
In addition, in the second and third integral, which lie within the normal neighbourhood of $x$, the Green's function can be replaced with $V\indices{_{\alpha\beta}^{\alpha'\beta'}}$ and we can use the near-coincidence expansion
\begin{equation}
V\indices{_{\alpha\beta}^{\alpha'\beta'}} = e_{(\alpha}^{K}e_{\beta)}^{L}e^{(\alpha'}_{I}e^{\beta')}_{J}R^I{}_K{}^J{}_L(t) +\order{\zeta}
\end{equation}
for $x'$ near $x''=\gamma(t)$.

Substituting these expressions into the integral, and noting that $\int n'^ad\Omega'=0$, yields the result 
\begin{align}
\frac{1}{4\pi}\!\!\!\int\limits_{\Gamma\cap\past}\!\!\! \h^{\text{tail}}_{\alpha\beta}Nd\Omega'dt' & =  \int_0^{t^-}\!\!\!2m\Big(G_{\alpha\beta}{}^{00} +G_{\alpha\beta}{}^{ij}\delta_{ij}\Big)dt' -4m\bigg(r+\frac{\rad^2}{3r}\bigg)e^a_\alpha e^b_\beta \etide_{ab}\nonumber\\
&\quad+\order{\zeta^2}.
\end{align}
Note that the integrals are cut off at $t^-\equiv t-0^+$ to avoid the singular behavior of the Green's function at coincidence. 

Now, since $x$ is near the point $x''$ on the worldline, we can expand the integrand as
\begin{align}
G_{\alpha\beta}{}^{00}+G_{\alpha\beta}{}^{ij}\delta_{ij} &= G_{\alpha\beta\bar\alpha\bar\beta}(2u^{\bar\alpha}u^{\bar\beta} +g^{\bar\alpha\bar\beta})\nonumber\\
&=\left(G_{\alpha''\beta''\bar\alpha\bar\beta} +\del{\gamma''}G_{\alpha''\beta''\bar\alpha\bar\beta} e^{\gamma''}_cx^c\right) g^{\alpha''}_{\alpha}g^{\beta''}_{\beta}\left(2u^{\bar\alpha}u^{\bar\beta} +g^{\bar\alpha\bar\beta}\right).
\end{align}
Substituting this into the integral results in the expansion
\begin{align}
\frac{1}{4\pi}\!\int\limits_{\Gamma\cap\past}\!\!\! \h^{\text{tail}}_{\alpha\beta}Nd\Omega'dt' & =  g^{\alpha''}_{\alpha}g^{\beta''}_{\beta}\!\left(\tail_{\Gamma \alpha''\beta''} +\tail_{\Gamma \alpha''\beta''\gamma''}e^{\gamma''}_cx^c\right) \nonumber\\
&\quad-4m\left(r+\frac{\rad^2}{3r}\right)e^a_\alpha e^b_\beta \etide_{ab}+\order{\zeta^2,\e},
\end{align}
where I have defined
\begin{align}
\tail_{\Gamma\alpha''\beta''}&=\int_0^{t^{-}}\!\!\!\! 2mG_{\alpha''\beta''\bar\alpha\bar\beta} \Big(2u^{\bar\alpha}u^{\bar\beta} +g^{\bar\alpha\bar\beta}\Big)d\bar t,\\
\tail_{\Gamma\alpha''\beta''\gamma''}&= \int_0^{t^{-}}\!\!\!\!2m\del{\gamma''} G_{\alpha''\beta''\bar\alpha\bar\beta}\Big(2u^{\bar\alpha}u^{\bar\beta} +g^{\bar\alpha\bar\beta}\Big)d\bar t.
\end{align}
By making use of the identity \eqref{Green3}, we can express these tail terms in their more usual form:
\begin{align}
\tail_{\Gamma\alpha''\beta''}&= \int_0^{t^{-}}\!\!\!\!4m\big(G_{\alpha''\beta''\bar\alpha\bar\beta} -\tfrac{1}{2}g_{\alpha''\beta''} G^{\delta''}{}_{\!\!\delta''\bar\alpha\bar\beta}\big) u^{\bar\alpha}u^{\bar\beta}d\bar t,\\
\tail_{\Gamma\alpha''\beta''\gamma''}&= \int_0^{t^{-}}\!\!\!\!4m\del{\gamma''} \big(G_{\alpha''\beta''\bar\alpha\bar\beta} -\tfrac{1}{2}g_{\alpha''\beta''} G^{\delta''}{}_{\!\!\delta''\bar\alpha\bar\beta}\big) u^{\bar\alpha}u^{\bar\beta}d\bar t.
\end{align}
The complete tail term will consist of the sum of the $\tail_\Gamma$ terms and the $\hmn{\Sigma}{1}$ terms.

In Fermi coordinates, the final result of this section is
\begin{align}
\frac{1}{4\pi}\!\!\!\int\limits_{\Gamma\cap\past}\!\!\! \h^{\text{tail}}_{\alpha\beta}Nd\Omega'dt' & =  \Big(\tail_{\Gamma 00}+\tail_{\Gamma 00c}x^c\Big)t_\alpha t_\beta+2\Big(\tail_{\Gamma 0b}+\tail_{\Gamma 0bc}x^c\Big)t_{(\alpha}x^b_{\beta)}\nonumber\\
&\quad+\Big(\tail_{\Gamma ab}+\tail_{\Gamma abc}x^c\Big)x^a_\alpha x^b_\beta  -4m\bigg(r+\frac{\rad^2}{3r}\bigg)\etide_{ab}x^a_\alpha x^b_\beta\nonumber\\
&\quad +\order{\zeta^2,\e}.
\end{align}

			\chapter{The hybrid equations of motion}\label{hybrid_eqns}
In this appendix, I present the hybrid equations of motion devised by Kidder, Will, and Wiseman~\cite{Kidder}. I also present the reformulation of these equations suitable for the method of osculating orbits in Sec.~\ref{PN binaries}.

As presented in that section, the hybrid equations for a post-Newtonian binary begin with the equations
\begin{equation}\label{PNeqns2}
\ddiff{x^a_h}{t} = -\frac{M}{r_h^2}\left(A\frac{x_h^a}{r_h}
+ B\diff{x_h^a}{t}\right),
\end{equation}
in the center of mass frame. The
functions $A$ and $B$ can be written as $A=A_M+\epsilon\tilde{A}$ and
$B=B_M+\epsilon\tilde{B}$, where $\epsilon=\mu/M$ and terms with a
subscript $M$ are independent of $\mu$. The $\mu$-dependent terms can be further decomposed into
post-Newtonian orders as $\tilde{A} = \tilde{A}_1 + \tilde{A}_2 
+ \tilde{A}_{2.5}$ and $\tilde{B} = \tilde{B}_1 + \tilde{B}_2 
+ \tilde{B}_{2.5}$. Explicitly, these have the form 
{\allowdisplaybreaks\begin{eqnarray}
A_M & = &
1-4\frac{M}{r_h}+v^2+9\left(\frac{M}{r_h}\right)^2-2\frac{M}{r_h}
\left(\diff{r_h}{t}\right)^2,\\ 
\epsilon\tilde{A_1} & = & -\epsilon\left[2\frac{M}{r_h}-3v^2
+\frac{3}{2}\left(\diff{r_h}{t}\right)^2\right], \\
\epsilon\tilde{A_2} & = &
\epsilon\bigg[\frac{87}{4}\left(\frac{M}{r_h}\right)^2 
+ (3-4\epsilon)v^4-\frac{1}{2}(13-4\epsilon) 
\frac{M}{r_h}v^2-\frac{3}{2}(3-4\epsilon)v^2
\left(\diff{r_h}{t}\right)^2\nonumber\\
 && \phantom{\epsilon\bigg[}+\frac{15}{8}(1-3\epsilon)
\left(\diff{r_h}{t}\right)^4 -(25+2\epsilon)\frac{M}{r_h}
\left(\diff{r_h}{t}\right)^2\bigg],\\
\epsilon\tilde A_{2.5} & = & -\frac{8}{5}\epsilon\frac{M}{r_h}
\diff{r_h}{t}\left[3v^2+\frac{17}{3}\frac{M}{r_h}\right],\\
\nonumber\\
B_M & = &  -\diff{r_h}{t}\left(4-2\frac{M}{r_h}\right),\\ 
\epsilon\tilde{B}_1 & = & 2\epsilon\diff{r_h}{t},\\
\epsilon\tilde{B}_2 & = & -\frac{1}{2}\epsilon\diff{r_h}{t}
\bigg[(15+4\epsilon)v^2-(41+8\epsilon)\frac{M}{r_h} 
    -3(3+2\epsilon)\left(\diff{r_h}{t}\right)^2\bigg],\\ 
\epsilon\tilde{B}_{2.5} & = &
\frac{8}{5}\epsilon\frac{M}{r_h}\left[v^2+3\frac{M}{r_h}\right], 
\end{eqnarray}}
where $v^2 \equiv \delta_{ab} \diff{x^a_h}{t} \diff{x^b_h}{t}$ is the
square of the velocity vector in harmonic coordinates.   
 
The hybrid equations of motion are given by Eq.~\eqref{PNeqns2} after
substituting $A=A_S+\epsilon\tilde{A}$ and $B=B_S+\epsilon\tilde{B}$,
where $A_S$ and $B_S$ are the exact $\mu$-independent terms for a geodesic in a Schwarzschild spacetime of mass $M$.
We derive the perturbing force from the terms
$\tilde{A}$ and $\tilde{B}$. The first step in this process is to write the equations of motion in
plane polar coordinates $(r_h,\phi)$, which are defined by
$x^1_h=r_h\cos(\phi)$ and $x^2_h=r_h\sin(\phi)$. In terms of these
coordinates, Eq.~\eqref{PNeqns2} becomes 
\begin{eqnarray}
\ddiff{r_h}{t} & = & -\frac{M}{r_h^2}\left(A+B\diff{r_h}{t}\right)
+ r_h\left(\diff{\phi}{t}\right)^2,\label{r acceleration}\\ 
\ddiff{\phi}{t} & = & -\frac{M}{r_h^2}B\diff{\phi}{t}
- \frac{2}{r_h}\diff{r_h}{t}\diff{\phi}{t}.\label{phi acceleration}
\end{eqnarray}
The harmonic coordinates used here are related to Schwarzschild
coordinates by the simple transformation $r_h = r-M$. Since $M$ is
constant, the subscript $h$ can be safely dropped within 
derivatives. Expressing $r_h$ in terms of $r$, the above equations are 
transformed into Schwarzschild coordinates. 

We derive $a^{\alpha}$ from these equations as follows. From
Eq.~\eqref{eq mot} we have 
\begin{equation}
a^{\alpha} = \dot t^2\left(\ddiff{\orbit}{t}
+\Chr{\alpha}{\beta}{\gamma}\diff{z^{\beta}}{t}\diff{z^{\gamma}}{t}
-\kappa(t) \diff{\orbit}{t}\right).\label{force 1}
\end{equation}
Although we could calculate $\kappa(t)$ directly from its definition,
the result would be unwieldy. We instead use the equation of motion
for $t$, 
\begin{equation}
\ddiff{t}{t}+\Chr{t}{\beta}{\gamma}\diff{z^{\beta}}{t}\diff{z^{\gamma}}{t}  
= a^t\dot t^{-2} + \kappa\diff{t}{t}, 
\end{equation}
to replace $\kappa$ with
\begin{equation}
\kappa = \Chr{t}{\beta}{\gamma}\diff{z^{\beta}}{t}\diff{z^{\gamma}}{t}
- a^t\dot t^{-2}.
\end{equation}
Substituting this expression for $\kappa$ into Eq.~\eqref{force 1}, we
find  
\begin{equation}
a^{\alpha} = \dot t^2 a^\alpha_p + \diff{\orbit}{t}a^t, 
\label{force from d^2xdt^2} 
\end{equation}
where
\begin{equation}
a^\alpha_p \equiv
\ddiff{\orbit}{t}+\Big(\Chr{\alpha}{\beta}{\gamma} -
\diff{\orbit}{t}\Chr{t}{\beta}{\gamma}\Big)
\diff{z^{\beta}}{t}\diff{z^{\gamma}}{t}. 
\end{equation}
The subscript $p$ refers to the fact that $a^\alpha_p$ involves only
the perturbative terms in $d^2\orbit/dt^2$. Indeed, a
simple calculation based on the preceding equations for $d^2r/dt^2$
and $d^2\phi/dt^2$, as well as the Christoffel symbols obtained from
the Schwarzschild metric, reveals that 
\begin{eqnarray}
a^r_p &=& -\frac{M}{r_h^2}\left(\epsilon\tilde A
+\epsilon\tilde B\diff{r}{t}\right), \\
a^\phi_p &=& -\frac{M}{r_h^2}\epsilon\tilde B\diff{\phi}{t}. 
\end{eqnarray}

Equation~\eqref{force from d^2xdt^2} determines $a^r$ and $a^{\phi}$ 
in terms of $a^t$. The orthogonality condition
\eqref{orthogonality} then allows us to find all three components of
the acceleration. The result is 
\begin{eqnarray}
a^t\!&=&\!\displaystyle\frac{\dot t^2\!
\left[a^r_p \diff{r}{t} + a^\phi_p r^2f \diff{\phi}{t}\right]}
{f^2 - \big(\diff{r}{t}\big)^{2} 
- fr^2\big(\diff{\phi}{t}\big)^{2}},\\
a^r\! &=& \!\frac{\dot t^2\!
\left[a^r_p \Big(f - r^2\big(\diff{\phi}{t}\big)^{2}\Big) 
+ a^\phi_p r^2\diff{r}{t}\diff{\phi}{t}\right]}
  {f^{-1}\Big(F^2 - \big(\diff{r}{t}\big)^{\!2} 
- fr^2\big(\diff{\phi}{t}\big)^{2}\Big)},\\
a^{\phi}\! &=& \!\frac{\dot t^2\!
\left[a^r_p \diff{r}{t}\diff{\phi}{t} 
+ a^\phi_p \Big(f^2 - \big(\diff{r}{t}\big)^2\Big)\right]}
  {f^2 - \big(\diff{r}{t}\big)^{2} 
- fr^2\big(\diff{\phi}{t}\big)^{2}}. 
\end{eqnarray}
Substituting $a^\alpha_p$ into the above results, and using the
normalization condition $-1=\dot z^{\alpha} \dot z_{\alpha} 
=-f\dot t^2 + f^{-1}\dot r^2 +r^2 \dot\phi^2$, leads to 
\begin{eqnarray}
a^r & = & -\frac{\epsilon M \dot{t}^4}{r_h^2} \Bigl\{ 
\bigl[ f - r^2 (d\phi/dt)^2 \bigr] \tilde{A} 
+ f (dr/dt) \tilde{B} \Bigr\}, 
\nonumber \\ 
a^\phi & = & -\frac{\epsilon M \dot{t}^4}{r_h^2} \frac{d\phi}{dt} 
\Bigl\{ f^{-1} (dr/dt) \tilde{A} + f \tilde{B} \Bigr\}.
\end{eqnarray} 
Since $a^t$ is not required in my formalism, I will not provide an 
explicit expression for it.  

We can recast these equations in a form analogous to that of
Eqs.~\eqref{r acceleration} and \eqref{phi acceleration},  
\begin{eqnarray}
a^r & = & -\frac{\mu}{r^2} 
\left[\mathcal{A} + \mathcal{B} \frac{dr}{dt} \right],
\\ 
a^{\phi} & = &-\frac{\mu}{r^2}\mathcal{B} \frac{d\phi}{dt},  
\end{eqnarray}
by defining $\mathcal{A}$ and $\mathcal{B}$ as
\begin{eqnarray}
\mathcal{A} & = & \frac{\dot{t}^2}{(1-M/r)^2} \tilde{A}, 
\label{script A} \\ 
\mathcal{B} & = & \frac{\dot{t}^4}{(1-M/r)^2}
\left( \frac{1}{f} \frac{dr}{dt} \tilde{A} + f \tilde{B} \right). 
\label{script B} 
\end{eqnarray}
The factors of $\dot t$ convert the ``time" variable in the
acceleration from coordinate time to proper time; this is given by 
\begin{equation} 
\dot{t}^2 = \frac{1}{f - f^{-1} (dr/dt)^2 - r^2 (d\phi/dt)^2}, 
\end{equation}
where, we recall, $f = 1-2M/r$. The factors of $1/(1-M/r)^2$, on the
other hand, convert from harmonic coordinates to Schwarzschild
coordinates. One could incorporate these factors into each 
$\tilde A_i$ and $\tilde B_i$ and then re-expand these in powers
of $M/r$ to find new expressions for $\mathcal{A}_i$ and
$\mathcal{B}_i$, neglecting terms of 3PN order and higher; but since
the hybrid equations already introduce errors above 2.5PN order, doing
so is unnecessary. Thus, for simplicity we shall use the force in its
above form.  

The final expression for the perturbing force is obtained by
substituting the post-Newtonian expansions for $\tilde{A}$ and
$\tilde{B}$ into Eqs.~(\ref{script A}) and (\ref{script B}); the
relevant equations are listed near the beginning of
Sec.~\ref{PN binaries}. In these equations we must make the substitution
$r_h = r-M$, and convert $t$-derivatives into $\chi$-derivatives by
employing Eq.~(\ref{tprime}). In these final forms, the expressions
for $f^r$ and $f^\phi$ are ready to be inserted within the evolution
equations for the orbital elements.

			\chapter{Evolution equations from Killing vectors}
\label{Killing_osculating}

It is possible to derive Eqs.~\eqref{pprime}--\eqref{wprime} for the 
derivatives of the osculating elements from Eq.~\eqref{diffI 1b} and
the Killing vectors of the Schwarzschild spacetime, without reference
to Eq.~\eqref{diffI 2b}. Although this derivation is equivalent to
that given in Sec.~\ref{evolution}, its physical significance is more
intuitive. We begin by defining energy and angular momentum (per unit
mass) as $E = -\xi_{(t)}^{\alpha}u_{\alpha}$ and
$L = \xi_{(\phi)}^{\alpha}u_{\alpha}$, where
$\xi_{(t)}=\pdiff{}{t}$ and $\xi_{(\phi)}=\pdiff{}{\phi}$ are Killing
vectors corresponding to the spacetime's invariance under time
translations and spatial rotations. From these definitions we find  
\begin{eqnarray}
-\dot{E} & = &
u^{\beta}(\xi_{(t)}^{\alpha}u_{\alpha})_{;\beta} 
\nonumber\\
& = & \xi^{\alpha}_{(t);\beta}u_{\alpha}u^{\beta}
+ \xi^{\alpha}_{(t)}u^{\beta}u_{\alpha;\beta} \nonumber\\
& = & \xi^{\alpha}_{(t)}a_{\alpha}.
\end{eqnarray}
The first term on the second line vanishes due to the antisymmetry of
$\xi_{\alpha;\beta}$ for any Killing vector $\xi$, and the final line
then follows from the equation of motion
$u^{\alpha}u^{\beta}{}_{;\alpha}=a^{\beta}$. An analogous
result holds for $\dot L$. From the definitions of $\xi_{(t)}$ and
$\xi_{(\phi)}$ we then find 
\begin{equation}
\dot E = fa^t, \quad \dot L = r^2a^{\phi}.
\end{equation}

These results can be used to find $\dot e$ and $\dot p$ using 
Eqs.~\eqref{E} and \eqref{L}, which define $E(p,e)$ and
$L(p,e)$. Using these relationships, we write $\dot E =
\pdiff{E}{p}\dot p + \pdiff{E}{e}\dot e$ and $\dot L =
\pdiff{L}{p}\dot p + \pdiff{L}{e}\dot e$, which can be rearranged to 
find 
\begin{equation} 
\dot p = \frac{\pdiff{E}{e}\dot L 
- \pdiff{L}{e}\dot E}{\pdiff{L}{p}\pdiff{E}{e}
- \pdiff{L}{e}\pdiff{E}{p}}, \quad
\dot e = \frac{\pdiff{L}{p}\dot E 
- \pdiff{E}{p}\dot L}{\pdiff{L}{p}\pdiff{E}{e}
- \pdiff{L}{e}\pdiff{E}{p}}.
\end{equation}
The equation for $\dot w$ can then be found from Eq.~\eqref{diffI 1b},
which leads to Eq.~(\ref{wprime_raw}), or 
\begin{equation}
\dot w = \frac{1}{r'}\left(\pdiff{r}{e}\dot e + \pdiff{r}{p}\dot p\right).
\end{equation}

The explicit results of these calculations are
{\allowdisplaybreaks\begin{eqnarray}
\dot p & = & -\frac{2p^{1/2}(p-2-2e)^{1/2}(p-2+2e)^{1/2}
(p-3-e^2)^{1/2}}{(p-6+2e)(p-6-2e)}\nonumber\\
&&\times(p-2-2e\cos v)a^t\nonumber\\
&&+\frac{2p^2M(p-4)^4(p-3-e^2)^{1/2}}{(p-6+2e)(p-6-2e)
(1+e\cos v)^2}a^{\phi}, \\
\nonumber\\
\dot e & = & \frac{(p-6-2e^2)(p-2-2e)^{1/2}
(p-2+2e)^{1/2}(p-3-e^2)^{1/2}}{p^{1/2}e(p-6+2e)(p-6-2e)}\nonumber\\
&&\times(p-2-2e\cos v)a^t
\nonumber\\
&&-\frac{pM(1-e^2)(p^2-8p+12+4e^2)(p-3-e^2)^{1/2}}{e(p-6+2e)
(p-6-2e)(1+e\cos v)^2}a^{\phi},  \\
\nonumber\\
\dot w & = & -\frac{(p-2-2e)^{1/2}(p-2+2e)^{1/2}(p-3-e^2)^{1/2}}{p^{1/2}e^2(p-6+2e)(p-6-2e)}\nonumber\\
&&\times\frac{(2e+(p-6)\cos v)(p-2-2e\cos v)}{\sin v}a^t
\nonumber\\
&&+\frac{pM(p-3-e^2)^{1/2}}{e^2(p-6+2e)(p-6-2e)}\Bigg\lbrace\frac{2e(p^2-8p+32)}{\sin v(1+e\cos v)^2}\nonumber\\
&&+\frac{[(p^2-8p)(1+e^2)+4e^2(6-e^4)]\cos v} {\sin v(1+e\cos v)^2}\Bigg\rbrace a^{\phi}.
\end{eqnarray}}
When accompanied by the auxiliary equation \eqref{chidot} for
$\diff{\chi}{\tau}$, these equations form a closed, autonomous system
for the orbital elements. 

The results in this section are equivalent to those in
Sec.~\ref{evolution}, which can be easily shown by using
Eq.~\eqref{orthogonality} to replace $a^t$ with $a^r$. But they are
numerically ill-behaved. Specifically, $\dot e$ appears to diverge in
the limit $e\to 0$, and $\dot w$ appears to diverge when $\sin v = 0$
(i.e., at every turning point in the orbit). Although these divergences
are canceled analytically by the numerators in each case, they are
serious obstacles in a numerical integration. Thus, the equations
given in Sec.~\ref{evolution} are more practical, though slightly
lengthier.  

			\bibliography{thesis}
			\bibliographystyle{unsrt}
			\end{document}